%% We use the memoir class because it offers a many easy to use features.
\documentclass[11pt,a4paper,titlepage,openany,twoside]{memoir}

%% `usepackage` commands.
%% Packages
%% ========

%% LaTeX Font encoding -- DO NOT CHANGE
\usepackage[OT1]{fontenc}

%% Babel provides support for languages.  'english' uses British
%% English hyphenation and text snippets like "Figure" and
%% "Theorem". Use the option 'ngerman' if your document is in German.
%% Use 'american' for American English.  Note that if you change this,
%% the next LaTeX run may show spurious errors.  Simply run it again.
%% If they persist, remove the .aux file and try again.
\usepackage[english]{babel}

%% Input encoding 'utf8'. In some cases you might need 'utf8x' for
%% extra symbols. Not all editors, especially on Windows, are UTF-8
%% capable, so you may want to use 'latin1' instead.
\usepackage[utf8]{inputenc}

%% This changes default fonts for both text and math mode to use Herman Zapfs
%% excellent Palatino font. Do not change this.
% \let\temp\rmdefault
\usepackage[sc]{mathpazo}
% \let\rmdefault\temp

%% The AMS-LaTeX extensions for mathematical typesetting.  Do not
%% remove.
\usepackage{amsmath,amssymb,amsfonts,mathrsfs}

%% NTheorem is a reimplementation of the AMS Theorem package. This
%% will allow us to typeset theorems like examples, proofs and
%% similar.  Do not remove.
%% NOTE: Must be loaded AFTER amsmath, or the \qed placement will
%% break
\usepackage[amsmath,thmmarks]{ntheorem}

%% LaTeX' own graphics handling
\usepackage{graphicx}

%% We unfortunately need this for the Rules chapter.  Remove it
%% afterwards; or at least NEVER use its underlining features.
\usepackage{soul}

%% This allows you to add .pdf files. It is used to add the
%% declaration of originality.
\usepackage{pdfpages}

%% Extra Packages
%% ==============

%% [OPT] Multi-rowed cells in tabulars
%\usepackage{multirow}

%% [REC] Intelligent cross reference package. This allows for nice
%% combined references that include the reference and a hint to where
%% to look for it.
\usepackage{varioref}

%% [OPT] Easily changeable quotes with \enquote{Text}
%\usepackage[german=swiss]{csquotes}

%% [REC] Format dates and time depending on locale
\usepackage{datetime}

%% [OPT] Provides a \cancel{} command to stroke through mathematics.
%\usepackage{cancel}

%% [NEED] This allows for additional typesetting tools in mathmode.
%% See its excellent documentation.
\usepackage{mathtools}

%% [ADV] Conditional commands
%\usepackage{ifthen}

%% [OPT] Manual large braces or other delimiters.
%\usepackage{bigdelim, bigstrut}

%% [REC] Alternate vector arrows. Use the command \vv{} to get scaled
%% vector arrows.
\usepackage[h]{esvect}

%% [NEED] Some extensions to tabulars and array environments.
\usepackage{array}

%% [OPT] Postscript support via pstricks graphics package. Very
%% diverse applications.
%\usepackage{pstricks,pst-all}

%% [?] This seems to allow us to define some additional counters.
%\usepackage{etex}

%% [ADV] XY-Pic to typeset some matrix-style graphics
%\usepackage[all]{xy}

%% [OPT] This is needed to generate an index at the end of the
%% document.
%\usepackage{makeidx}

%% [OPT] Fancy package for source code listings.  The template text
%% needs it for some LaTeX snippets; remove/adapt the \lstset when you
%% remove the template content.
\usepackage{listings}
\lstset{language=TeX,basicstyle={\normalfont\ttfamily}}

%% [REC] Fancy character protrusion.  Must be loaded after all fonts.
\usepackage[activate]{pdfcprot}

%% [REC] Nicer tables.  Read the excellent documentation.
\usepackage{booktabs}

%% Bibliography package.
\usepackage[
style=authoryear,
hyperref=true,
url=true,
sorting=ynt,
natbib=true,
mincitenames=1,
maxcitenames=2,
backend=bibtex]{biblatex}
\renewbibmacro{in:}{}
\addbibresource{reference.bib}

\renewbibmacro*{name:andothers}{% Based on name:andothers from biblatex.def
  \ifboolexpr{
    test {\ifnumequal{\value{listcount}}{\value{liststop}}}
    and
    test \ifmorenames
  }
  {\ifnumgreater{\value{liststop}}{1}
    {\finalandcomma}
    {}%
    \andothersdelim\bibstring[\emph]{andothers}
  }
  {}
}

%% Side-by-Side Figures
\usepackage{caption}
\usepackage{subcaption}

%% Bold math symbols.
 % remove error "too many math alphabets used"

\usepackage{bm}
\usepackage{bbm}

%% Color for table cells.
\usepackage{xcolor,colortbl}
\definecolor{matlab-blue}{rgb}{0,0.4470,0.7410}
\definecolor{matlab-orange}{rgb}{0.8500,0.3250,0.0980}
\definecolor{matlab-yellow}{rgb}{0.9290,0.6940,0.1250}
\def\cyellow{{\cellcolor{matlab-yellow!25}}}
\definecolor{matlab-green}{rgb}{0.4660,0.6740,0.1880}
\def\cgreen{{\cellcolor{matlab-green!25}}}
\definecolor{matlab-red}{rgb}{0.6350,0.0780,0.1840}
\def\cred{{\cellcolor{matlab-red!25}}}

%% Table multiple rows.
\usepackage{multirow}

%% Algorithms environment
\usepackage[ruled,vlined,linesnumbered]{algorithm2e}

%% Mathematical notation package.
\usepackage[intoc]{nomencl}

% \renewcommand{\nompreamble}{This section provides a reference describing the notation used throughout this report.}
% center the labels

\makenomenclature
 
%% Groups
% -----------------------------------------
\usepackage{etoolbox}
\renewcommand\nomgroup[1]{%
  \item[\bfseries
  \ifstrequal{#1}{A}{Numbers and Arrays}{%
  \ifstrequal{#1}{B}{Sets and Graphs}{%
  \ifstrequal{#1}{C}{Indexing and Slicing}{%
  \ifstrequal{#1}{D}{Linear Algebra Operations}{%
  \ifstrequal{#1}{E}{Probability and Information Theory}{%
  \ifstrequal{#1}{F}{Functions and Calculus}{%
  \ifstrequal{#1}{G}{Datasets and Distributions}{%
  \ifstrequal{#1}{H}{Financial Signal Processing}{%
  \ifstrequal{#1}{I}{Reinforcement Learning}{}}}}}}}}}
]}
% -----------------------------------------

%% Numbers and Arrays
\nomenclature[A, 01]{$a$}{A scalar}
\nomenclature[A, 02]{$\va$}{A vector}
\nomenclature[A, 03]{$\mA$}{A matrix}
% \nomenclature[A, 04]{$\tA$}{A tensor}
\nomenclature[A, 05]{$\mathbf{I}_{n}$}{Identity matrix with n rows and n columns}
\nomenclature[A, 06]{$\mathbf{I}$}{Identity matrix with, one-hot-encoding, with a $1$ at position $i$}
\nomenclature[A, 07]{$\text{diag}(\va)$}{A square diagonal matrix with diagonal entries given by $\va$}
\nomenclature[A, 08]{$\text{diagonal}(\mA)$}{The vector defined by the diagonal entries of the square matrix $\mA$}
% \nomenclature[A, 09]{$\ra$}{A scalar random variable}
% \nomenclature[A, 10]{$\rva$}{A vector-valued random variable}
% \nomenclature[A, 11]{$\rmA$}{A matrix-valued random variable}

%% Sets and Operators
\nomenclature[B, 01]{$\sA$}{A set}
\nomenclature[B, 02]{$\sR$}{The set of real numbers}
\nomenclature[B, 03]{$\sR_{+}$}{The set of non-negative real numbers}
\nomenclature[B, 04]{$\sR^{N}$}{The $N$-dimensional space of real numbers}
\nomenclature[B, 05]{$\sC$}{The set of complex numbers}
\nomenclature[B, 06]{$\{0,1\}$}{The set containing $0$ and $1$}
\nomenclature[B, 07]{$\{0,1,\ldots,n\}$}{The set of all integers between $0$ and $n$}
\nomenclature[B, 08]{$[a,b]$}{The real interval including $a$ and $b$}
\nomenclature[B, 09]{$(a,b]$}{The real interval excluding $a$ but including $b$}
\nomenclature[B, 10]{$\mathfrak{F}$}{Fourier Operator}
\nomenclature[B, 11]{$\mathfrak{B}$}{Bellman Operator}

%% Indexing and Slicing
\nomenclature[C, 01]{$\eva_{i}$}{Element $i$ of vector $\va$, with indexing starting at $1$}
\nomenclature[C, 02]{$\eva_{-i}$}{All elements of vector $\va$ except for element $i$}
\nomenclature[C, 03]{$\emA_{i j}$}{Element $(i,j)$ of matrix $\mA$, with indexing starting at $(1,1)$}
\nomenclature[C, 04]{$\mA_{i :}$}{Row $i$ of matrix $\mA$}
\nomenclature[C, 05]{$\mA_{: j}$}{Column $j$ of matrix $\mA$}
% \nomenclature[C, 06]{$\etA_{i j k}$}{Element $(i,j,k)$ of 3-D tensor $\tA$}
% \nomenclature[C, 07]{$\tA_{: : k}$}{2-D slice of 3-D tensor $\tA$}
% \nomenclature[C, 08]{$\erva_{i}$}{Element $i$ of the random vector $\rva$}

%% Linear Algebra Operations
\nomenclature[D, 01]{$\mA^{T}$}{Transpose of matrix $\mA$}
\nomenclature[D, 02]{$\mA^{\dagger}$}{Moore-Penrose pseudoinverse of matrix $\mA$}
\nomenclature[D, 03]{$\mA \circ \mB$}{Element-wise (Hadamard) product of matrices $\mA$ and $\mB$}
\nomenclature[D, 04]{$\text{det}(\mA)$}{Determinant of matrix $\mA$}

%% Probability and Information Theory
\nomenclature[E, 01]{$\ra \perp \rb$}{The random variables $\ra$ and $\rb$ are independent}
\nomenclature[E, 02]{$\ra \perp \rb\ |\ c$}{The random variables $\ra$ and $\rb$ are conditionally independent given $c$}
\nomenclature[E, 03]{$P(\ra)$}{A probability distribution over a discrete random variable $\ra$}
\nomenclature[E, 04]{$p(\ra)$}{A probability distribution over a continuous random variable $\ra$}
\nomenclature[E, 05]{$\ra \sim P$}{The random variable $\ra$ has distribution $P$}
\nomenclature[E, 06]{$\E_{\rx \sim P}[f(x)]$}{Expectation of $f(x)$ with respect to $P(\rx)$}
\nomenclature[E, 07]{$\Var[f(x)]$}{Variance of $f(x)$ under $P(\rx)$}
\nomenclature[E, 08]{$\Cov[f(x), g(x)]$}{Covariance of $f(x)$ and $g(x)$ under $P(\rx)$}
\nomenclature[E, 09]{$\KL(p\;\delimsize\|\;q)$}{Kullback-Leibler divergence of $p$ and $q$ distributions}
\nomenclature[E, 10]{$\mathcal{N}(\vx;\vmu,\mSigma)$}{Gaussian distribution over $\vx$ with mean $\vmu$ and covariance $\mSigma$}

%% Functions and Calculus
\nomenclature[F, 01]{$f : \sA \rightarrow \sB$}{The function $f$ with domain $\sA$ and range $\sB$}
\nomenclature[F, 02]{$f \circ g$}{Composition of the function $f$ and $g$}
\nomenclature[F, 03]{$f(\vx; \vtheta)$}{A function of $\vx$ parametrized by $\vtheta$}
\nomenclature[F, 04]{$\norm{\vx}_{p}$}{$\normlp$ norm of $\vx$}
\nomenclature[F, 05]{$\text{sign}(x)$}{is 1 if $x \geq 0$, 0 otherwise}
\nomenclature[F, 06]{$\mathbf{1}_{\text{condition}}$}{is 1 if the condition is true, 0 otherwise}
\nomenclature[F, 07]{$\frac{df}{dx}$}{Derivative of $f$ with respect to $x$}
\nomenclature[F, 08]{$\frac{\partial f}{\partial x}$}{Partial derivative of $f$ with respect to $x$}
\nomenclature[F, 09]{$\nabla_{x}f$}{Gradient of $f$ with respect to $x$}
\nomenclature[F, 10]{$\int f(\vx) d\vx$}{Definite integral with respect to $\vx$ over the entire domain of $f$}
\nomenclature[F, 11]{$\int_{\sS} f(\vx) d\vx$}{Definite integral with respect to $\vx$ over the set $\sS$}

%% Datasets and Distributions
\nomenclature[G, 01]{$\pdata$}{The data generating distribution}
\nomenclature[G, 02]{$\ptrain$}{The empirical distribution defined by the training set}
\nomenclature[G, 03]{$\sX$}{A set of training examples}
\nomenclature[G, 04]{$\vx^{(i)}$}{The $i$-th example (input) from a dataset}
\nomenclature[G, 05]{$y^{(i)} \text{ or } \vy^{(i)}$}{The target associated with $\vx^{(i)}$}
\nomenclature[G, 06]{$\mX$}{The $m \times n$ matrix with input example $\vx^{(i)}$ in row $\mX_{i,:}$}

%% Financial Engineering
\nomenclature[H, 01]{$t$}{Discrete time index}
\nomenclature[H, 02]{$p_{i, t}$}{Price of asset $i$ at time $t$}
\nomenclature[H, 03]{$\vvp_{i, 1:T}$ or $\vvp_{i, T}$}{Price time-series of asset $i$ between time steps $1$ and $T$}
\nomenclature[H, 04]{$\vp_{t}$}{Price vector of portfolio assets at time $t$}
\nomenclature[H, 05]{$\mvP_{1:T}$ or $\mvP_{T}$}{Price matrix of portfolio assets between time steps $1$ and $T$}
\nomenclature[H, 06]{$r_{i, t}$}{Simple return of asset $i$ at time $t$}
\nomenclature[H, 07]{$\vr_{t}$}{Simple returns vector of portfolio assets at time $t$}
\nomenclature[H, 08]{$\mvR_{1:T}$ or $\mvR_{T}$}{Simple returns matrix of portfolio assets between time steps $1$ and $T$}
\nomenclature[H, 09]{$\rho_{i, t}$}{Log return of asset $i$ at time $t$}
\nomenclature[H, 10]{$\vrho_{t}$}{Log returns vector of portfolio assets at time $t$}
\nomenclature[H, 11]{$w_{i, t}$}{Portfolio weight of asset $i$ at time $t$}
\nomenclature[H, 12]{$\vw_{t}$}{Portfolio vector at time $t$}
\nomenclature[H, 13]{$\text{D}_{t}$}{Differential Sharpe Ratio at time $t$}

%% Reinforcement Learning
\nomenclature[I, 01]{$\sA$}{Action space}
\nomenclature[I, 02]{$\va_{t}$}{Action at time $t$}
\nomenclature[I, 03]{$\sS$}{State space}
\nomenclature[I, 04]{$\vs_{t}$}{State at time $t$}
\nomenclature[I, 05]{$\sO$}{Observation space}
\nomenclature[I, 06]{$\vo_{t}$}{Observation (observed state) at time $t$}
\nomenclature[I, 07]{$\mathcal{R}_{\vs}^{\va}$}{Reward generating function}
\nomenclature[I, 08]{$\mathcal{P}_{\vs \vs'}^{\va}$}{Probability transition function}
\nomenclature[I, 09]{$\pi$}{A policy}
\nomenclature[I, 10]{$u_{\pi}$}{State-value function under policy $\pi$}
\nomenclature[I, 11]{$q_{\pi}$}{Action-value function under policy $\pi$}
\nomenclature[I, 12]{$\gamma$}{A discount factor}

%% API
%% ===

%% Make document internal hyperlinks wherever possible. (TOC, references)
%% This MUST be loaded after varioref. We just load it last to be safe.
\definecolor{darkblue}{rgb}{0.0,0.0,0.5}
\usepackage[
linkcolor=black,
colorlinks=true,
citecolor=black,
urlcolor=blue,
filecolor=black]{hyperref}

% \let\oldcitet=\citet
% \let\oldcitep=\citep 
% \renewcommand{\citet}[1]{\textcolor[rgb]{0,0,1}{\oldcitet{#1}}}
% \renewcommand{\citep}[1]{\textcolor[rgb]{0,0,1}{\oldcitep{#1}}}

%% Dummy command to non-existing assets folder to allow \renewcommand to work
\newcommand{\assets}{assets}

%% Remove indentation from `enumerate` and `itemize`.
\usepackage{enumitem}

%% Our layout configuration.
%% Memoir layout setup

%% NOTE: You are strongly advised not to change any of them unless you
%% know what you are doing.  These settings strongly interact in the
%% final look of the document.

% Turn extra space before chapter headings off.
\setlength{\beforechapskip}{0pt}

\nonzeroparskip
\parindent=0pt
\defaultlists

% Chapter style redefinition
\makeatletter

\if@twoside
  \pagestyle{Ruled}
  \copypagestyle{chapter}{Ruled}
\else
  \pagestyle{ruled}
  \copypagestyle{chapter}{ruled}
\fi
\makeoddhead{chapter}{}{}{}
\makeevenhead{chapter}{}{}{}
\makeheadrule{chapter}{\textwidth}{0pt}
\copypagestyle{abstract}{empty}

\makechapterstyle{bianchimod}{%
  \chapterstyle{default}
  \renewcommand*{\chapnamefont}{\normalfont\Large\sffamily}
  
  \renewcommand*{\printchaptername}{%
    \chapnamefont\centering\@chapapp}

  }

% Use the newly defined style
\chapterstyle{bianchimod}

\setsecheadstyle{\Large\bfseries\sffamily}
\setsubsecheadstyle{\large\bfseries\sffamily}
\setsubsubsecheadstyle{\bfseries\sffamily}
\setparaheadstyle{\normalsize\bfseries\sffamily}
\setsubparaheadstyle{\normalsize\itshape\sffamily}
\setsubparaindent{0pt}

% Set captions to a more separated style for clearness
\captionnamefont{\sffamily\bfseries\footnotesize}
\captiontitlefont{\sffamily\footnotesize}
\setlength{\intextsep}{16pt}
\setlength{\belowcaptionskip}{1pt}

% Set section and TOC numbering depth to subsection
\setsecnumdepth{subsection}
\settocdepth{subsection}

%% Titlepage adjustments
\pretitle{\vspace{0pt plus 0.7fill}\begin{center}\HUGE\sffamily\bfseries}
\posttitle{\end{center}\par}
\preauthor{\par\begin{center}\let\and\\\Large\sffamily}
\postauthor{\end{center}}
\predate{\par\begin{center}\Large\sffamily}
\postdate{\end{center}}

\def\@supervisor{}
\newcommand{\supervisor}[1]{\def\@supervisor{#1}}
\def\@secondmarker{}
\newcommand{\secondmarker}[1]{\def\@secondmarker{#1}}
\def\@advisors{}
\newcommand{\advisors}[1]{\def\@advisors{#1}}
\def\@department{}
\newcommand{\department}[1]{\def\@department{#1}}
\def\@thesistype{}
\newcommand{\thesistype}[1]{\def\@thesistype{#1}}

\renewcommand{\maketitlehookb}{\vspace{1in}%
  \par\begin{center}\Large\sffamily\@thesistype\vspace{0.25in}\end{center}}

\renewcommand{\maketitlehookd}{%
  \vfill\par
  \begin{flushright}
    \sffamily
    Supervisor: \@supervisor\par
    Second Marker: \@secondmarker\par
    Advisors: \@advisors\par
    \@department, Imperial College London
  \end{flushright}
}

\checkandfixthelayout

\setlength{\droptitle}{-48pt}

\makeatother

% This defines how theorems should look. Best leave as is.
\theoremstyle{plain}
\setlength\theorempostskipamount{0pt}

%%% Local Variables:
%%% mode: latex
%%% TeX-master: "thesis"
%%% End:

%% Theorem environments.
%% Theorem-like environments

\numberwithin{equation}{chapter}

%% English variants
\newtheorem{theorem}{Theorem}[chapter]

\newtheorem{remark}[theorem]{Remark}

\newtheorem{hypothesis}[theorem]{Hypothesis}
\newtheorem{assumption}[theorem]{Assumption}

%% Expected Properties
\newtheorem{property}{Expected Properties}[chapter]

%% Investment Advices
\newtheorem{advice}{Investor Advice}[chapter]

%% Proof environment with a small square as a "qed" symbol
\theoremstyle{nonumberplain}
\theorembodyfont{\normalfont}
\theoremsymbol{\ensuremath{\square}}

%\newtheorem{beweis}{Beweis}

%% Helpful macros.
%% Custom commands
%% ===============

%% logo command

%% Ian Goodfellow LaTeX commands for math.
%%%%% NEW MATH DEFINITIONS %%%%%

\def\1{\bm{1}}

% Random variables

\def\ra{{\textnormal{a}}}
\def\rb{{\textnormal{b}}}

\def\re{{\textnormal{e}}}

\def\rr{{\textnormal{r}}}

\def\rx{{\textnormal{x}}}

% Random vectors

\def\rve{{\mathbf{e}}}

\def\rvr{{\mathbf{r}}}

\def\rvx{{\mathbf{x}}}

% Random vectors with arrow

\def\rvvx{{\vec{\mathbf{x}}}}

% Elements of random vectors

% Random matrices

\def\rmX{{\mathbf{X}}}

% Elements of random matrices

% Vectors

\def\va{{\bm{a}}}
\def\vb{{\bm{b}}}
\def\vc{{\bm{c}}}

\def\ve{{\bm{e}}}

\def\vh{{\bm{h}}}

\def\vo{{\bm{o}}}
\def\vp{{\bm{p}}}
\def\vq{{\bm{q}}}
\def\vr{{\bm{r}}}
\def\vs{{\bm{s}}}

\def\vv{{\bm{v}}}
\def\vw{{\bm{w}}}
\def\vx{{\bm{x}}}
\def\vy{{\bm{y}}}

\def\vmu{{\bm{\mu}}}

\def\vtheta{{\bm{\theta}}}

\def\vphi{{\bm{\phi}}}

\def\vrho{{\bm{\rho}}}

\def\vphi{{\bm{\phi}}}

% Vectors with arrow

\def\vvh{{\vec{\bm{h}}}}

\def\vvp{{\vec{\bm{p}}}}

\def\vvr{{\vec{\bm{r}}}}

\def\vvx{{\vec{\bm{x}}}}

\def\vvrho{{\vec{\bm{\rho}}}}

% Elements of vectors

\def\eva{{a}}

\def\evw{{w}}

% Matrix
\def\mA{{\bm{A}}}
\def\mB{{\bm{B}}}

\def\mL{{\bm{L}}}

\def\mS{{\bm{S}}}

\def\mV{{\bm{V}}}
\def\mW{{\bm{W}}}
\def\mX{{\bm{X}}}

\def\mSigma{{\bm{\Sigma}}}

% Matrix with arrow

\def\mvP{{\vec{\bm{P}}}}

\def\mvR{{\vec{\bm{R}}}}

\def\mvW{{\vec{\bm{W}}}}
\def\mvX{{\vec{\bm{X}}}}

% Tensor
% Got this from here: http://tex.stackexchange.com/questions/77640/bold-italic-and-sans-serif-math-symbols
\DeclareMathAlphabet{\mathsfit}{\encodingdefault}{\sfdefault}{m}{sl}
\SetMathAlphabet{\mathsfit}{bold}{\encodingdefault}{\sfdefault}{bx}{n}

% Graph

\def\gP{{\mathcal{P}}}

\def\gR{{\mathcal{R}}}

\def\gZ{{\mathcal{Z}}}

% Sets
\def\sA{{\mathbb{A}}}
\def\sB{{\mathbb{B}}}
\def\sC{{\mathbb{C}}}

% Don't use a set called E, because this would be the same as our symbol
% for expectation.

\def\sN{{\mathbb{N}}}
\def\sO{{\mathbb{O}}}
\def\sP{{\mathbb{P}}}

\def\sR{{\mathbb{R}}}
\def\sS{{\mathbb{S}}}

\def\sX{{\mathbb{X}}}

% Matrix elements
% NOTE: we changed these from lower case to upper case to avoid
% clashes between vectors and matrices with same letter name.
% For example, we don't want elements of the hidden layer vector
% h to have the same name as elements of the Hessian matrix H

\def\emA{{A}}

% Tensor elements
% Same font as tensor, without \bm wrapper

% The true underlying data-generating distribution
\newcommand{\pdata}{p_{\text{data}}}
% The empirical distribution defined by the training set
\newcommand{\ptrain}{\hat{p}_{\text{data}}}
 % Accidental duplicate

% The model distribution

% Stochastic autoencoder distributions

 % Laplace distribution, see first use in prob.tex

\newcommand{\E}{\mathbb{E}}

 % Note: main text often hardcodes \sigma still, also
% many equations would likely frame bust if expanded to say \mathrm{sigmoid}
 % Duplicate command, this one is deprecated
 % Note: some figures may not use this macro, but main text consistently does
\newcommand{\KL}{D_{\mathrm{KL}}}
\newcommand{\Var}{\mathrm{Var}}

\newcommand{\Cov}{\mathrm{Cov}}
\newcommand{\corr}{\mathrm{corr}}
\newcommand{\skewness}{\mathrm{skew}}
\newcommand{\kurtosis}{\mathrm{kurt}}
% Put these here so we can keep same syntax throughout the book but also change
% it easily.
% Wolfram Mathworld says $L^2$ is for function spaces and $\ell^2$ is for vectors
% But then they seem to use $L^2$ for vectors throughout the site, and so does
% wikipedia.
% No one really seems to use subscripts (l_2, L_2) on general math reference
% sites, though subscripts seem popular in machine learning.

\newcommand{\normlone}{L^1}
\newcommand{\normltwo}{L^2}
\newcommand{\normlp}{L^p}

% NOTE: This one was not done with a macro originally, many uses need to converted
% to \parents rather than Pa if we want to change it.
 % See usage in notation.tex. Chosen to match Daphne's book.

\DeclareMathOperator*{\argmax}{argmax}

%% Fixed/scaling delimiter examples (see mathtools documentation)

\DeclarePairedDelimiter\norm{\lVert}{\rVert}

%% Use the alternative epsilon per default and define the old one as \oldepsilon

\renewcommand{\epsilon}{\ensuremath\varepsilon}

%% Also set the alternate phi as default.

\renewcommand{\phi}{\ensuremath{\varphi}}

%% Text over sign.
\newcommand{\textoversign}[2]{\overset{\mathclap{\strut\text{#2}}}#1}

%% Statistics symbols
\def\E{{\mathbb{E}}}

%% Document information
%% ====================
\title{Reinforcement Learning\protect\\for Portfolio Management}
\author{Angelos Filos\protect\\CID: 00943119}
\thesistype{MEng Dissertation}
\supervisor{Professor Danilo Mandic}
\secondmarker{Professor Pier Luigi Dragotti}
\advisors{Bruno Scalzo Dees, Gregory Sidier}
\department{Department of Electrical and Electronic Engineering}
\date{June 20, 2018}
%% ====================

% % Page margins
% \setlrmarginsandblock{3.5cm}{3.5cm}{*}
% \setulmarginsandblock{3.5cm}{3.5cm}{1}
%% Line spacing rule

\checkandfixthelayout

\begin{document}

\frontmatter

%% Title page is autogenerated from document information above.
\begin{titlingpage}
  \calccentering{\unitlength}
  \begin{adjustwidth*}{\unitlength-24pt}{-\unitlength-24pt}
    \maketitle
  \end{adjustwidth*}
\end{titlingpage}

% Acknowledgement
\chapter*{Acknowledgement} \label{ch:acknowledgement}

I would like to thank Professor Danilo Mandic for agreeing to
supervise this self-proposed project, despite the uncertainty about the viability
of the topic. His support and guidance contributed to
the delivery of a challenging project.

I would also like to take this opportunity and thank
Bruno Scalzo Dees for his helpful comments, suggestions and enlightening discussions,
which have been instrumental in the progress of the project.

Lastly, I would like to thank Gregory Sidier for spending
time with me, out of his working hours. His experience,
as a practitioner, in Quantitative Finance helped me demystify
and engage with topics, essential to the project.

% Notation
\printnomenclature[1.5in]

% Abstract of thesis.
\chapter*{Abstract} \label{ch:abstract}

The challenges of modelling the behaviour of financial markets, such
as nonstationarity, poor predictive behaviour, and  weak historical coupling, have
attracted attention of the scientific community over the last 50 years,
and has sparked a permanent strive to employ engineering methods
to address and overcome these challenges. Traditionally, mathematical formulations of
dynamical systems in the context of Signal Processing and Control Theory
have been a lynchpin of today's Financial Engineering. More recently,
advances in sequential decision making, mainly  through the concept of Reinforcement Learning,
have been instrumental in the development of multistage stochastic optimization,
a key component in sequential portfolio optimization (asset allocation) strategies.
In this thesis, we develop a comprehensive account of the
expressive power, modelling efficiency, and performance advantages of so called
trading agents (i.e., Deep Soft Recurrent Q-Network (DSRQN) and Mixture of Score
Machines (MSM)), based on both traditional system identification (model-based approach)
as well as on context-independent agents (model-free approach).
The analysis provides a conclusive support for the ability of
model-free reinforcement learning methods to act as universal trading agents,
which are not only capable of reducing the computational and
memory complexity (owing to their linear scaling with size of
the universe), but also serve as generalizing strategies across assets
and markets, regardless of the trading universe on which they have
been trained. The relatively low volume of daily returns in
financial market data is addressed via data augmentation (a generative approach)
and a choice of pre-training strategies, both of which are
validated against current state-of-the-art models. For rigour, a risk-sensitive framework
which includes transaction costs is considered, and its performance advantages
are demonstrated in a variety of scenarios, from synthetic time-series
(sinusoidal, sawtooth and chirp waves), simulated market series (surrogate data
based), through to real market data (S\&P 500 and EURO
STOXX 50). The analysis and simulations confirm the superiority of
universal model-free reinforcement learning agents over current portfolio management model
in asset allocation strategies, with the achieved performance advantage of
as much as 9.2\% in annualized cumulative returns and 13.4\% in annualized Sharpe Ratio.

%% TOC with the proper setup.
\cleartorecto
\tableofcontents*
\mainmatter

\chapter{Introduction} \label{ch:Introduction}

Engineering methods and systems are routinely used in financial market applications,
including signal processing, control theory and advanced statistical methods.
The computerization of the markets \citep{Schinckus2018} encourages automation and algorithmic solutions,
which are now well-understood and addressed by the engineering communities.
Moreover, the recent success of Machine Learning has attracted interest
of the financial community, which permanently seeks for the successful
techniques from other areas, such as computer vision and natural language
processing to enhance modelling of financial markets. In this thesis,
we explore how the asset allocation problem can be addressed by Reinforcement Learning, a branch of Machine Learning that
optimally solves sequential decision making problems via direct interaction with the environment in an episodic manner.

In this introductory chapter, we define the objective of the thesis and
highlight the research and application domains from which we draw inspiration.

\section{Problem Definition} \label{sec:problem-definition}

The aim of this report is to investigate the effectiveness
of Reinforcement Learning agents on asset allocation\footnote{The terms \textit{Asset
Allocation} and \textit{Portfolio Management} are used interchangeably throughout the report.}.
A finite universe of financial instruments,
assets, such as stocks, is selected and the role of an agent is to construct an internal
representation (model) of the market, allowing it to determine how to
optimally allocate funds of a finite budget to those assets.
The agent is trained on both synthetic and real market data.
Then, its performance is compared with standard portfolio management algorithms on
an out-of-sample dataset; data that the agent has not been trained on (i.e., test set).

\section{Motivations} \label{sec:motivations}

From the IBM TD-Gammon \citep{tesauro1995temporal} and the IBM Deep Blue \citep{campbell2002deep} to the Google
DeepMind Atari \citep{deepmind:atari} and the Google DeepMind AlphaGo \citep{silver2016alphago}, reinforcement learning
is well-known for its effectiveness in board and video games.
Nonetheless, reinforcement learning applies to many more domains, including Robotics, Medicine and Finance,
applications of which align with the mathematical formulation of portfolio
management. Motivated by the success of some of these applications,
an attempt is made to improve and adjust the underlying
methods, such that they are applicable to the asset allocation
problem settings. In particular special attention is given to:

\begin{itemize}
    \item \textbf{Adaptive Signal Processing}, where Beamforming has been successfully addressed via
    reinforcement learning by \citet{sp:beamforming};
    \item \textbf{Medicine}, where a data-driven medication dosing system \citep{paper:medication-dosing} has been made
    possible thanks to model-free reinforcement agents;
    \item \textbf{Algorithmic Trading}, where the automated execution \citep{loxm} and market making \citep{spooner2018market} have
    been recently revolutionized by reinforcement agents.
\end{itemize}

Without claiming equivalence of portfolio management with any of
the above applications, their relatively similar optimization problem formulation encourages
the endeavour to develop reinforcement learning agents for asset allocation.

\section{Report Structure} \label{sec:report-structure}

The report is organized in three Parts: the \textbf{Background} (Part \ref{part:background}),
the \textbf{Innovation} (Part \ref{part:innovation}) and the \textbf{Experiments} (Part \ref{part:experiments}).
The readers are advised to follow the sequence
of the parts as presented, however, if comfortable with Modern
Portfolio Theory and Reinforcement Learning, they can focus on
the last two parts, following the provided references to background material
when necessary. A brief outline of the project structure and chapters is provided below:

\begin{itemize}[label={}]
    \item \textbf{Chapter 2: Financial Signal Processing}\space\space
        The objective of this chapter is to introduce essential financial terms
        and concepts for understanding the methods developed later in
        the report.
    \item \textbf{Chapter 3: Portfolio Optimization}\space\space
        Providing the basics of Financial Signal Processing, this chapter proceeds with
        the mathematical formulation of static Portfolio Management, motivating the use
        of Reinforcement Learning to address sequential Asset Allocation via multi-stage
        decision making.
    \item \textbf{Chapter 4: Reinforcement Learning}\space\space
        This chapter serves as an important step toward demystifying
        Reinforcement Learning concepts, by highlighting their analogies to Optimal
        Control and Systems Theory. The concepts developed in this chapter
        are essential to the understanding of the trading algorithms and
        agents developed later in the report.
    \item \textbf{Chapter 5: Financial Market as Discrete-Time Stochastic Dynamical System}\space\space
        This chapter parallels Chapters \ref{ch:portfolio-optimization} and \ref{ch:reinforcement-learning},
        introducing a unified, versatile framework for training agents and investment strategies.
    \item \textbf{Chapter 6: Trading Agents}\space\space
        This objectives of this chapter are to: (1) introduce traditional
        model-based (i.e., system identification) reinforcement learning trading agents;
        (2) develop model-free reinforcement learning trading agents;
        (3) suggest a flexible universal trading agent architecture that enables pragmatic
        applications of Reinforcement Learning for Portfolio Management;
        (4) assess performance of developed trading agents on a small scale
        experiment (i.e., $12$-asset S\&P 500 market) 
    \item \textbf{Chapter 7: Pre-Training}\space\space
        In this chapter, a pre-training strategy is suggested, which addresses
        the local optimality of the Policy Gradient agents, when only
        a limited number of financial market data samples is available.
    \item \textbf{Chapter 8: Synthetic Data}\space\space
        In this chapter, the effectiveness of the trading agents of Chapter \ref{ch:trading-agents}
        is assessed on synthetic data - from deterministic time-series (sinusoidal, sawtooth
        and chirp waves) to simulated market series (surrogate data based).
        The superiority of model-based or model-free agents is highlighted in each scenario.
    \item \textbf{Chapter 9: Market Data}\space\space
        This chapter parallels Chapter \ref{ch:market-data}, evaluating the performance of the
        trading agents of Chapter \ref{ch:trading-agents} on real market data, from two
        distinct universes: (1) the underlying U.S. stocks of S\P 500 and (2) the underlying
        European stocks of EURO STOXX 50.
\end{itemize}

\part{Background} \label{part:background}

%% assets folder to reference
\renewcommand{\assets}{report/background/financial-signal-processing/assets}

\chapter{Financial Signal Processing} \label{ch:financial-signal-processing}

Financial applications usually involve the manipulation and analysis of sequences of
observations, indexed by time order, also known as time-series.
Signal Processing, on the other hand, provides a rich toolbox for
systematic time-series analysis, modelling and forecasting \citep{mandic:rnn}.
Consequently, signal processing methods can be employed to mathematically formulate
and address fundamental economics and business problems. In addition, Control Theory
studies discrete dynamical systems, which form the basis of Reinforcement Learning,
the set of algorithms used in this report to solve the asset allocation problem.
The links between signal processing algorithms, systems and control theory motivate
their integration with finance, to which we refer as \textbf{Financial Signal Processing} or \textbf{Financial Engineering}.

\begin{figure}[h]
    \centering
    \includegraphics[width=0.5\textwidth]{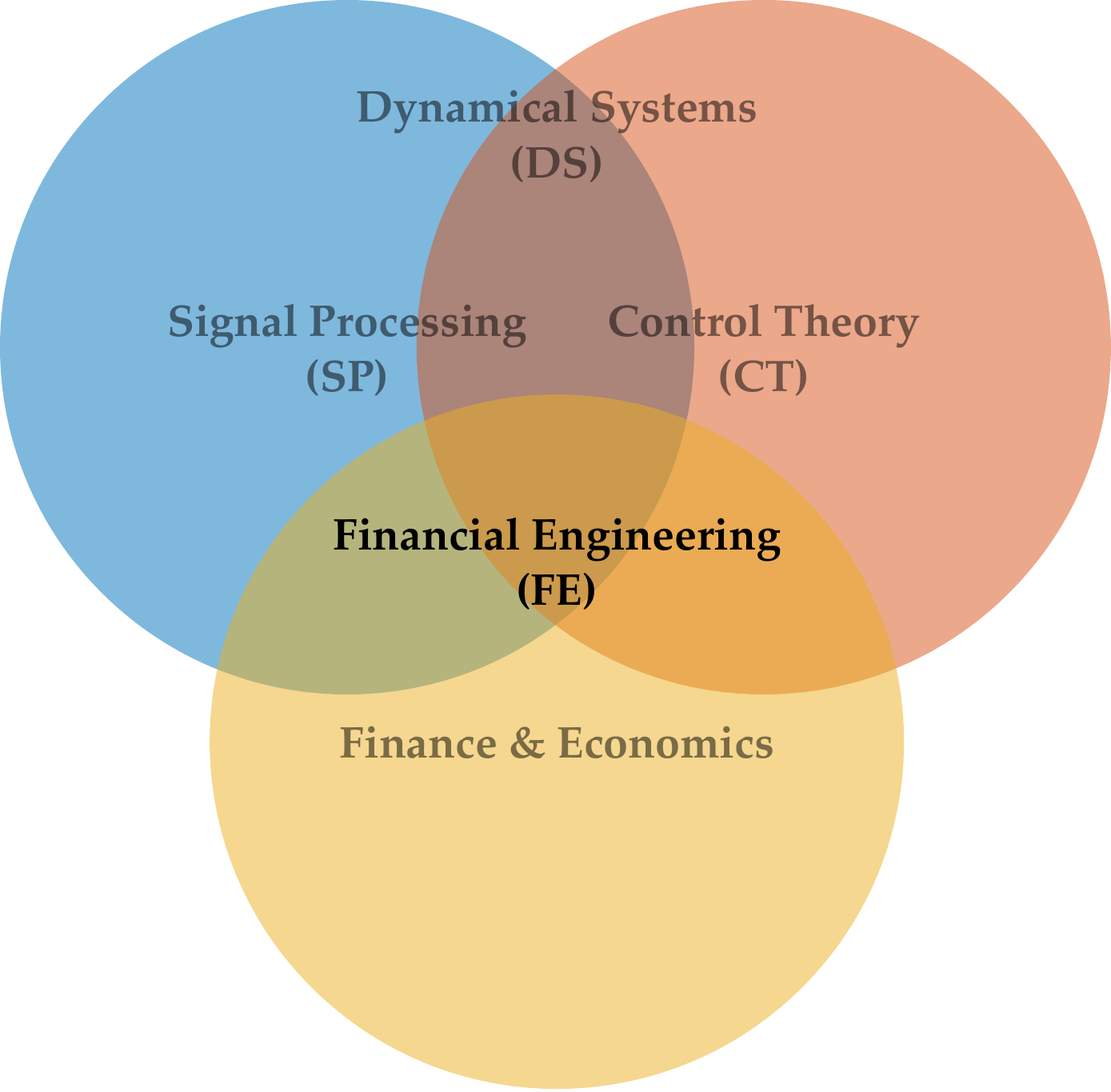}
    \caption{Financial Engineering relative to Signal Processing and Control Theory.}
    \label{fig:fields-map}
\end{figure}

In this chapter, the overlap between signal processing and control theory
with finance is explored, attempting to bridge their gaps and highlight their similarities.
Firstly, in Section \ref{sec:terms-concepts}, essential
financial terms and concepts are introduced, while In Section \ref{sec:financial-time-series},
the time-series in the context of finance are formalized.
In Section \ref{sec:evaluation-criteria} the evaluation criteria used throughout the report
to assess the performance of the different algorithms and strategies are explained,
while in Section \ref{sec:time-series-analysis} signal processing methods for modelling sequential data are studied.

\section{Financial Terms \& Concepts} \label{sec:terms-concepts}

In order to better communicate ideas and gain insight into the economic problems,
basic terms are defined and explained in this section. However,
useful definitions are also provided by \citet{sp:risk-management}.

\subsection{Asset} \label{sub:asset}

An \textbf{asset} is an item of economic value.
Examples of assets are cash (in hand or in a bank), stocks, loans and advances,
accrued incomes etc. Our main focus on this report is on cash and stocks,
but general principles apply to all kinds of assets.

\begin{assumption}
The assets under consideration are liquid, hence they can be converted
into cash quickly, with little or no loss in value.
Moreover, the selected assets have available historical data in order
to enable analysis.
\end{assumption}

\subsection{Portfolio} \label{sub:portfolio}

A \textbf{portfolio} is a collection of multiple financial assets,
and is characterized by its:
\begin{itemize}
\item \textbf{Constituents}: $M$ assets of which it consists;
\item \textbf{Portfolio vector}, $\vw_{t}$:
its $i$-th component represents the ratio of the total budget
invested to the $i$-th asset, such that:
\begin{equation}
\vw_{t} =
    \renewcommand\arraystretch{1.5}
    \begin{bmatrix}
        \evw_{1, t}, & \evw_{2, t}, & \ldots, & \evw_{M, t}
    \end{bmatrix}
    ^{T} \in \sR^{M}
\quad
\text{and}
\quad
\sum_{i=1}^{M} \evw_{i, t} = 1
\label{def:portfolio-vector}
\end{equation}
\end{itemize}
For fixed constituents and portfolio vector $\vw_{t}$, a portfolio can be treated as
a single master asset. Therefore, the analysis of single simple assets can be applied to
portfolios upon determination of the constituents and the corresponding portfolio vector.

\begin{figure}[h]
    \centering
    \begin{subfigure}[t]{0.48\textwidth}
        \includegraphics[width=\textwidth]{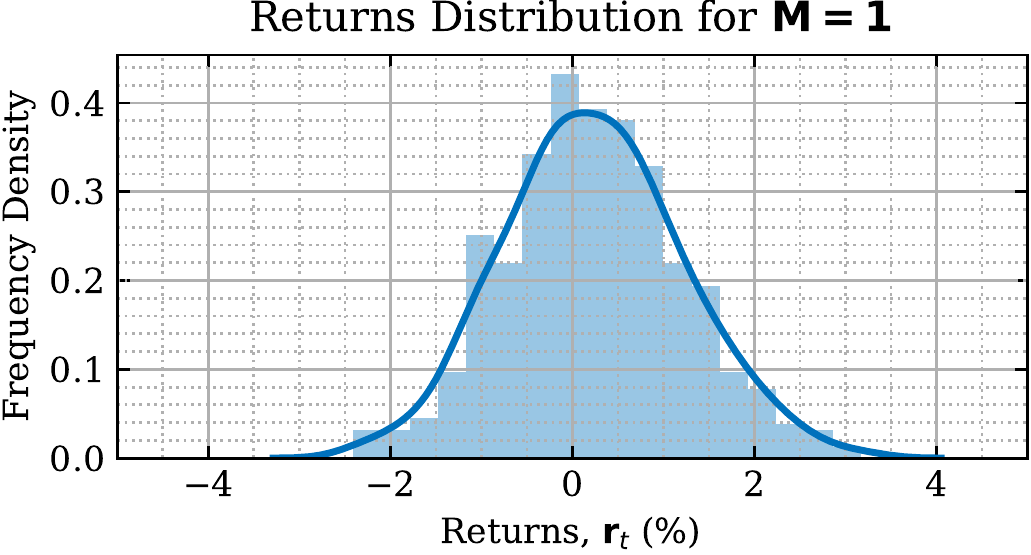}
    \end{subfigure}
    ~ 
    \begin{subfigure}[t]{0.48\textwidth}
        \includegraphics[width=\textwidth]{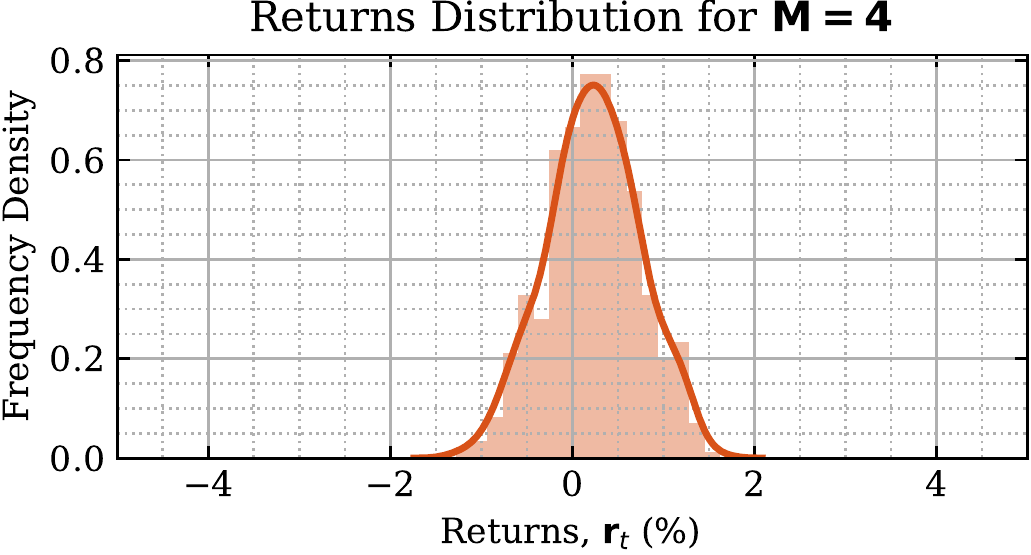}
    \end{subfigure}
    
    \vspace{0.5cm}
    
    \begin{subfigure}[t]{0.48\textwidth}
        \includegraphics[width=\textwidth]{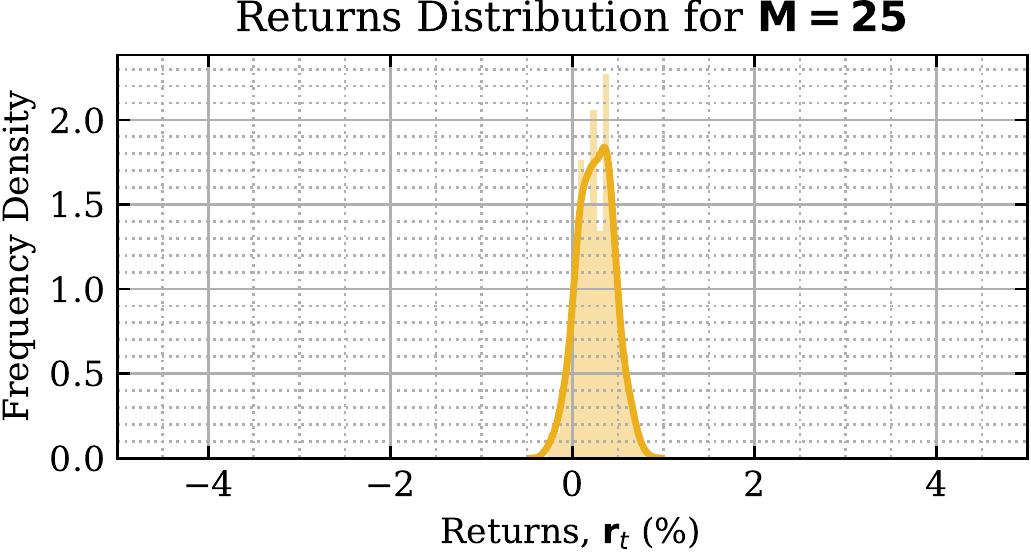}
    \end{subfigure}
    ~    
    \begin{subfigure}[t]{0.48\textwidth}
        \includegraphics[width=\textwidth]{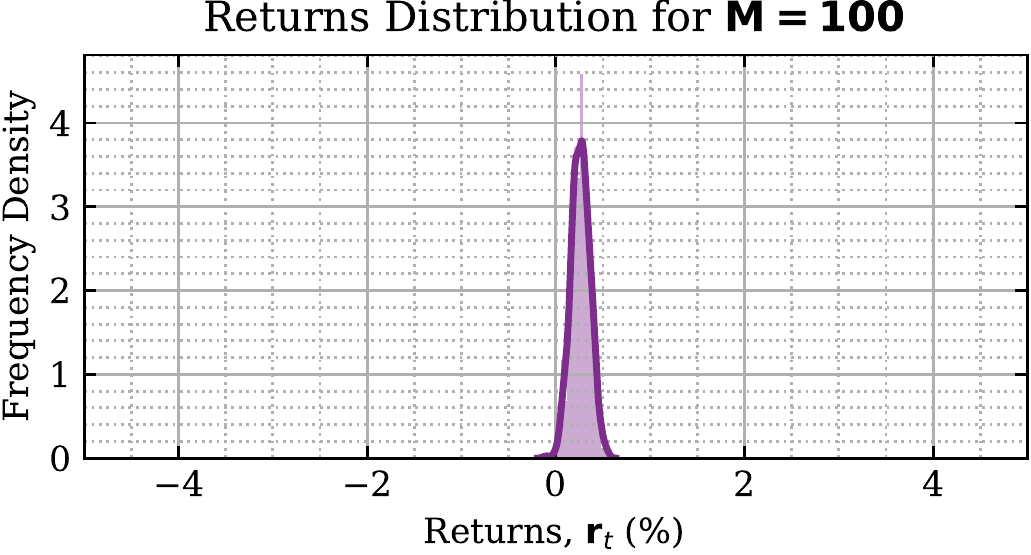}
    \end{subfigure}
    \caption{Risk for a single asset and a number of uncorrelated portfolios.
    Risk is represented by the standard deviation or the width of the distribution curves,
    illustrating that a large portfolio ($M=100$) can be significantly less risky than a
    single asset ($M=1$).}
    \label{fig:asset-portfolio-risk}
\end{figure}

Portfolios are more powerful, general representation of financial assets since
the single asset case can be represented by a portfolio;
the $j$-th asset is equivalent to the portfolio with vector
$\ve^{(j)}$, where the $j$-th term is equal to unity and the
rest are zero. Portfolios are also preferred over single assets
in order to minimize risk, as illustrated in Figure \ref{fig:asset-portfolio-risk}.

\subsection{Short Sales} \label{sub:short-sales}

Sometimes is it possible to sell an asset that we do not own.
This process is called \textbf{short selling} or \textbf{shorting}
\citep{finance:investment-science}. The exact shorting mechanism varies between markets,
but it can be generally summarized as:
\begin{enumerate}
    \item \textbf{Borrowing} an asset $i$ from someone who owns it at time $t$;
    \item \textbf{Selling} it immediately to someone else at price $p_{i, t}$;
    \item \textbf{Buying back} the asset at time $(t+k)$, where $k>0$, at price $p_{i, t+k}$;
    \item \textbf{Returning} the asset to the lender
\end{enumerate}
Therefore, if one unit of the asset is shorted, the \textbf{overall} absolute {return}
is $p_{i, t} - p_{i, t+k}$  and as a result short selling is profitable only if the asset
price declines between time $t$ and $t+k$ or $p_{i, t+k} < p_{i, t}$.
Nonetheless, note that the potential loss of short selling is unbounded,
since asset prices are not bounded from above ($0 \leq p_{i, t+k} < \infty$).

\begin{remark}
If short selling is allowed, then the portfolio vector satisfies
(\ref{def:portfolio-vector}), but $\evw_{i}$ can be negative, if the $i$-th
asset is shorted. As a consequence, $\evw_{j}$ can be greater than $1$,
such that $\sum_{i=1}^{M} \evw_{i} = 1$.
\end{remark} 

For instance, in case of a two-assets
portfolio, the portfolio vector
$
\vw_{t} = 
    \begin{bmatrix}
    -0.5, & 1.5
    \end{bmatrix}
    ^{T}
$
is valid and can be interpreted as:
$50\%$ of the budget is short sold on the first asset ($\evw_{1, t} = -0.5$) and
$150\%$ of the budget is invested on the second asset ($\evw_{2, t} = 1.5$).
Note that the money received from shorting asset 1 are used in the investment on asset 2,
enabling $\evw_{2, t} > 1$.

Usually the terms \textbf{long} and \textbf{short} position to an asset
are used to refer to investments where we buy or short sell the asset, respectively.

\section{Financial Time-Series} \label{sec:financial-time-series}

The dynamic nature of the economy, as a result of the non-static supply and demand balance,
causes prices to evolve over time. This encourages to treat market dynamics as time-series and
employ technical methods and tools for analysis and modelling.

In this section, asset prices are introduced, whose definition immediately reflect our intuition,
as well as other time-series, derived to ease analysis and evaluation.

\subsection{Prices} \label{sub:prices}

Let $p_{t} \in \sR$ be the \textbf{price} of an asset at discrete time index $t$
\citep{sp:financial-engineering},
then the sequence $p_{1}, p_{2}, \ldots, p_{T}$ is a univariate time-series.
The equivalent notations $p_{i, t}$ and $p_{\text{asset}_{i}, t}$
are also used to distinguish between the prices of the different assets.
Hence, the $T$-samples price time-series of an asset $i$,
is the column vector $\vvp_{i, 1:T}$, such that:
\begin{equation}
    \vvp_{i, 1:T} =
        \renewcommand\arraystretch{1.5}
        \begin{bmatrix}
            p_{i, 1} \\ p_{i, 2} \\ \vdots \\ p_{i, T}
        \end{bmatrix}
        \in \sR_{+}^{T}
\end{equation}
where the arrow highlights the fact that it is a time-series.
For convenience of portfolio analysis, we define the \textbf{price vector}
$\vp_{t}$, such that:
\begin{equation}
    \vp_{t} =
        \renewcommand\arraystretch{1.5}
        \begin{bmatrix}
            p_{1, t}, & p_{2, t}, & \ldots, & p_{M, t}
        \end{bmatrix}
        \in \sR_{+}^{M}
\label{def:price-vector}
\end{equation}
where the $i$-th element is the asset price of the $i$-th asset in the portfolio
at time $t$. Extending the single-asset time-series notation to the multivariate case,
we form the asset \textbf{price matrix} $\mvP_{1:T}$ by stacking column-wise the
$T$-samples price time-series of the $M$ assets of the portfolio, then:
\begin{equation}
    \mvP_{1:T} =
        \renewcommand\arraystretch{1.5}
        \begin{bmatrix}
            \vvp_{1, 1:T}, & \vvp_{2, 1:T}, & \ldots, & \vvp_{M, 1:T}
        \end{bmatrix}
        =
        \begin{bmatrix}
            p_{1, 1}    & p_{2, 1}  & \cdots    & p_{M,1}   \\
            p_{1, 2}    & p_{2, 2}  & \cdots    & p_{M, 2}  \\
            \vdots      & \vdots    & \ddots    & \vdots    \\
            p_{1, T}    & p_{2, T}  & \cdots    & p_{M, T}
        \end{bmatrix}
        \in \sR_{+}^{T \times M}
\label{def:price-matrix}
\end{equation}
This formulation enables cross-asset analysis and consideration of the inter-dependencies
between the different assets. We usually relax notation by omitting subscripts 
when they can be easily inferred from context.

\begin{figure}[h]
    \centering
    \begin{subfigure}[t]{0.48\textwidth}
        \includegraphics[width=\textwidth]{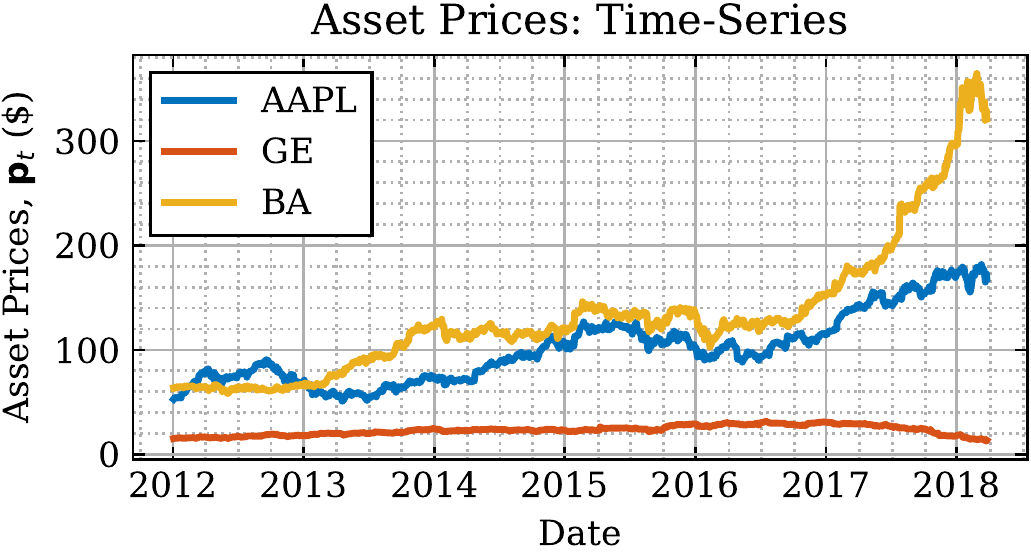}
    \end{subfigure}
    ~ 
    \begin{subfigure}[t]{0.48\textwidth}
        \includegraphics[width=\textwidth]{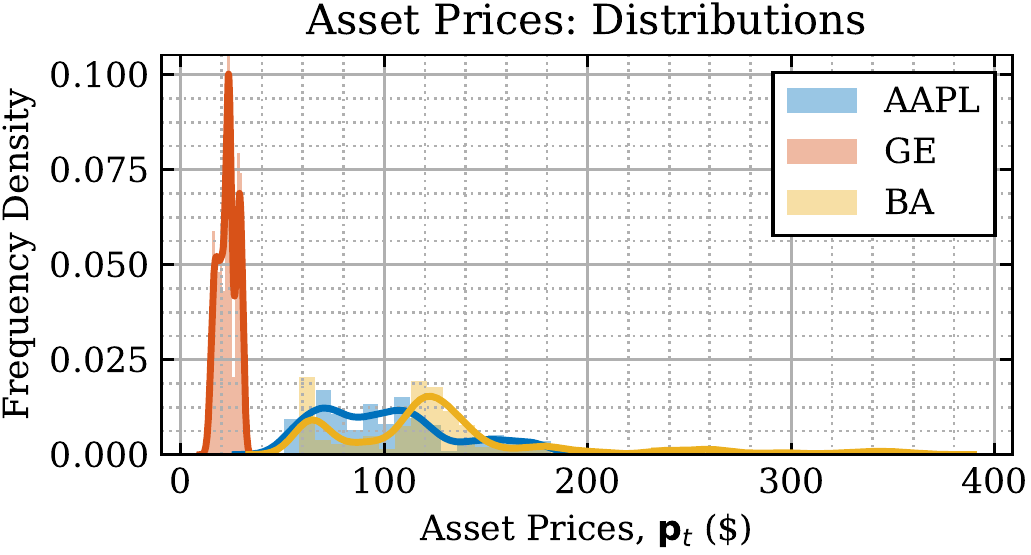}
    \end{subfigure}
    \caption{Asset prices time-series (left) and distributions (right)
    for \texttt{AAPL} (Apple), \texttt{GE} (General Electric) and \texttt{BA} (Boeing Company).}
    \label{fig:prices}
\end{figure}

Figure \ref{fig:prices} illustrates examples of asset prices time-series
and the corresponding distribution plots. At a first glance, note the highly non-stationary
nature of asset prices and hence the difficulty to interpret distribution plots.
Moreover, we highlight the unequal scaling between prices, where for example,
\texttt{GE} (General Electric) average price at $23.14\$$ and
\texttt{BA} (Boeing Company) average price at $132.23\$$ are of different order and difficult to compare.

\subsection{Returns} \label{sub:returns}

Absolute asset prices are not directly useful for an investor.
On the other hand, prices changes over time are of great importance,
since they reflect the investment profit and loss, or more compactly, its \textbf{return}.

%%%
\subsubsection{Gross Return} \label{subsub:gross-return}
The \textbf{gross return} $R_{t}$ of an asset
% represents the change in wealth level because of the investment in the asset.
represents the scaling factor of an investment
in the asset at time $(t-1)$ \citep{sp:financial-engineering}.
For example, a $B$ dollars investment in an asset at time $(t-1)$
will worth $B R_{t}$ dollars at time $t$.
It is given by the ratio of its prices at times $t$ and $(t-1)$, such that:
\begin{equation}
    R_{t} \triangleq \frac{p_{t}}{p_{t-1}} \in \sR
\label{def:gross-returns}
\end{equation}

\begin{figure}[h]
    \centering
    \begin{subfigure}[t]{0.48\textwidth}
        \includegraphics[width=\textwidth]{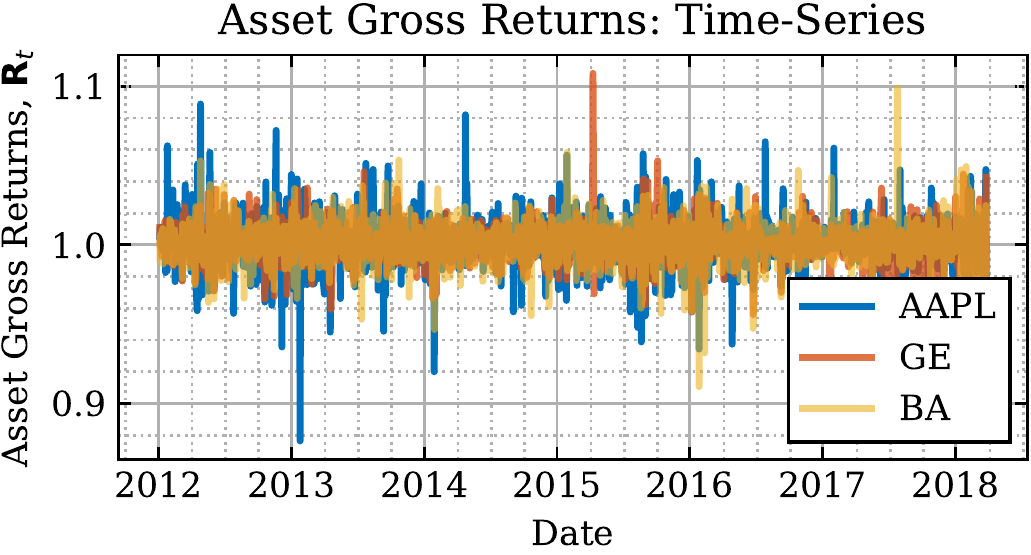}
    \end{subfigure}
    ~ 
    \begin{subfigure}[t]{0.48\textwidth}
        \includegraphics[width=\textwidth]{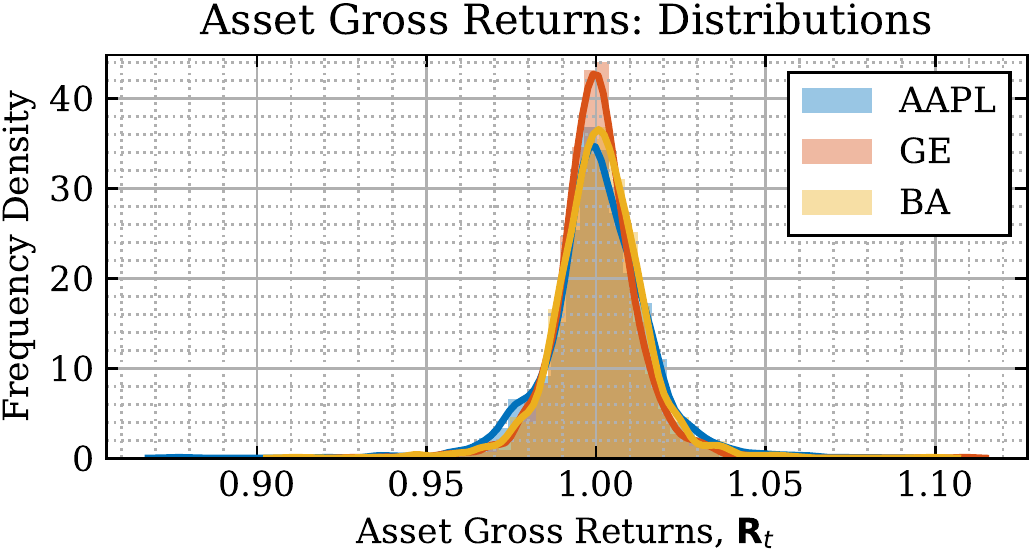}
    \end{subfigure}
    \caption{Asset gross returns time-series (left) and distributions (right).}
    \label{fig:gross-returns}
\end{figure}

Figure \ref{fig:gross-returns} illustrates the benefit of using gross returns over asset prices.

\begin{remark}
The asset gross returns are concentrated around unity and their behaviour
does not vary over time for all stocks, making them attractive candidates for stationary
autoregressive (AR) processes \citep{mandic:asp-arma}.
\end{remark}

%%%
\subsubsection{Simple Return} \label{subsub:simple-return}

A more commonly used term is the \textbf{simple return}, $r_{t}$,
which represents the percentage change in asset price from time $(t-1)$ to time $t$, such that:
\begin{equation}
    r_{t} \triangleq \frac{p_{t} - p_{t-1}}{p_{t-1}} = \frac{p_{t}}{p_{t-1}} - 1 
        \textoversign{=}{(\ref{def:gross-returns})} R_{t} - 1 \in \sR
\label{def:simple-returns}
\end{equation}
The gross and simple returns are straightforwardly connected, but the latter is more interpretable,
and thus more frequently used.

\begin{figure}[h]
    \centering
    \begin{subfigure}[t]{0.48\textwidth}
        \includegraphics[width=\textwidth]{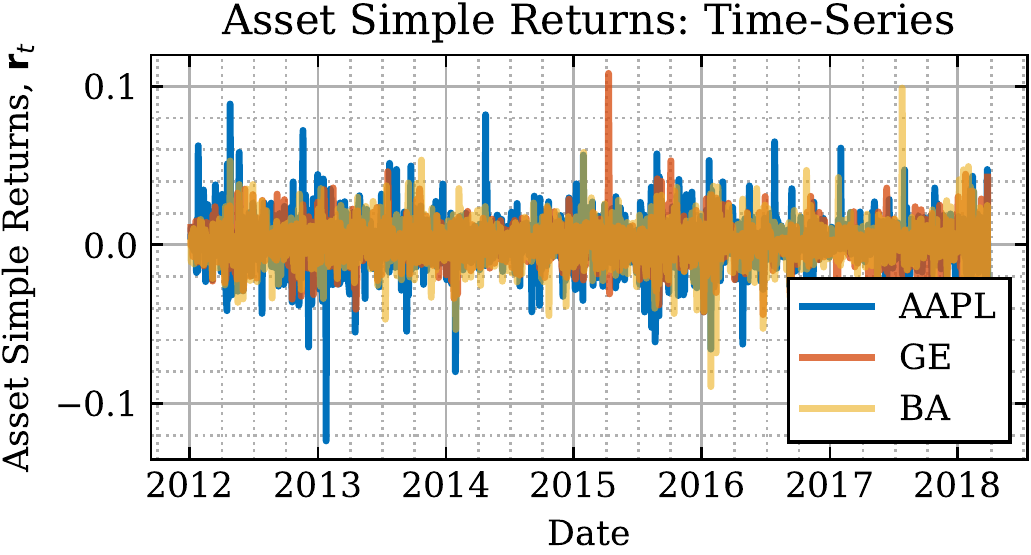}
    \end{subfigure}
    ~ 
    \begin{subfigure}[t]{0.48\textwidth}
        \includegraphics[width=\textwidth]{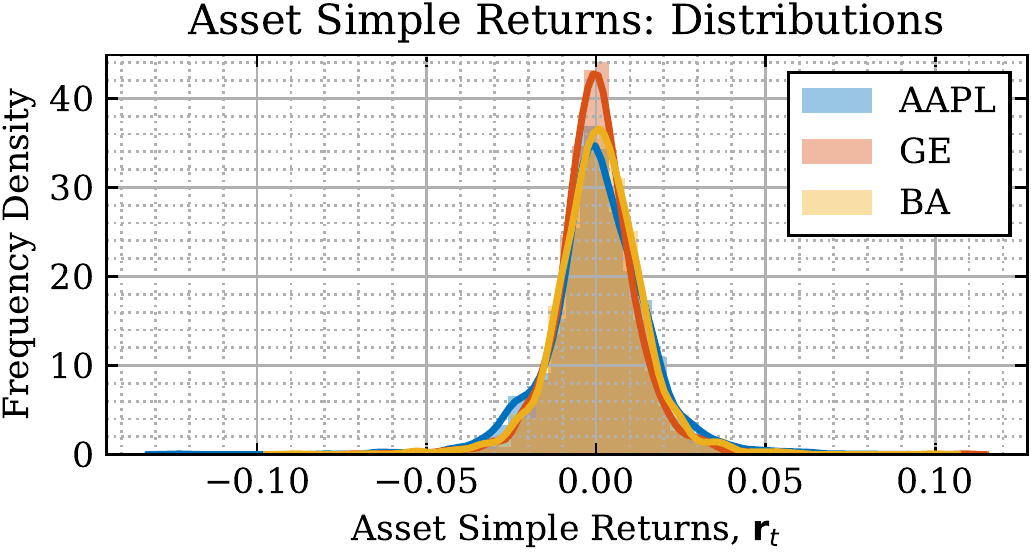}
    \end{subfigure}
    \caption{Single Assets Simple Returns}
    \label{fig:simple-returns}
\end{figure}

Figure \ref{fig:simple-returns} depicts the example asset simple returns time-series and
their corresponding distributions. Unsurprisingly, simple returns possess the
representation benefits of gross returns, such as stationarity and normalization.
Therefore, we can use simple returns as a comparable metric for all assets, thus enabling
the evaluation of analytic relationships among them, despite originating from asset prices
of different scale.

% %%% START: CUMULATIVE SIMPLE RETURNS %%%

% Holding the asset for $k$ periods between time indexes $t-k$ and $t$ gives a
% \textbf{$k$-period simple return}
% \begin{align}
% R_{t}[k]    \triangleq \frac{p_{t} - p_{t-k}}{p_{t-k}}
%              &= \frac{p_{t}}{p_{t-k}} - 1\nonumber\\
%                 &= -1 + \frac{p_{t}}{p_{t-1}} \frac{p_{t-1}}{p_{t-2}} \cdots \frac{p_{t-k+1}}{p_{t-k}}\nonumber\\
%              &\textoversign{=}{(\ref{def:simple-returns})}
%                 -1 + (1 + R_{t}) (1 + R_{t-1}) \cdots (1 + R_{t-k+1})\nonumber\\
%              &= -1 + \prod_{i=0}^{k-1} (1 + R_{t-i})
% \label{def:multi-period-simple-returns}
% \end{align}
% Therefore the k-period simple return involves the product of the one-period
% gross returns.

% %%% END: CUMULATIVE SIMPLE RETURNS %%%

The $T$-samples simple returns time-series of the $i$-th asset is given by the column vector
$\vvr_{i, 1:T}$, such that:
\begin{equation}
    \vvr_{i, 1:T} =
        \renewcommand\arraystretch{1.5}
        \begin{bmatrix}
            r_{i, 1} \\ r_{i, 2} \\ \vdots \\ r_{i, T}
        \end{bmatrix}
        \in \sR^{T}
\end{equation}
while the \textbf{simple returns vector} $\vr_{t}$:
\begin{equation}
    \vr_{t} =
        \renewcommand\arraystretch{1.5}
        \begin{bmatrix}
            r_{1, t} \\ r_{2, t} \\ \vdots \\ r_{M, t}
        \end{bmatrix}
        \in \sR^{M}
\end{equation}
where $r_{1, t}$ the simple return of the $i$-th asset at time index $t$.

\begin{remark}
Exploiting the representation advantage of the portfolio over single assets,
we define the \textbf{portfolio simple return} as the linear combination of the simple
returns of each constituents, weighted by the portfolio vector.
\end{remark}

Hence, at time index $t$, we obtain:
\begin{align}
    r_{t} \triangleq \sum_{i=1}^{M} \evw_{i, t} r_{i, t} = \vw_{t}^{T} \vr_{t} \in \sR
\label{def:portfolio-simple-returns}
\end{align}

Combining the price matrix in (\ref{def:price-matrix}) and the definition of
simple return (\ref{def:simple-returns}), we construct the \textbf{simple return matrix}
$\mvR_{1:T}$ by stacking column-wise the $T$-samples simple returns
time-series of the $M$ assets of the portfolio, to give:

\begin{equation}
    \mvR_{1:T} =
        \renewcommand\arraystretch{1.5}
        \begin{bmatrix}
            \vvr_{1, 1:T} & \vvr_{2, 1:T} & \cdots & \vvr_{M, 1:T}
        \end{bmatrix}
        =
        \begin{bmatrix}
            r_{1, 1}    & r_{2, 1}  & \cdots    & r_{M,1}   \\
            r_{1, 2}    & r_{2, 2}  & \cdots    & r_{M, 2}  \\
            \vdots      & \vdots    & \ddots    & \vdots    \\
            r_{1, T}    & r_{2, T}  & \cdots    & r_{M, T}  \\
        \end{bmatrix}
        \in \sR^{T \times M}
\label{def:simple-return-matrix}
\end{equation}

Collecting the portfolio (column) vectors for the time interval $t \in [1, T]$ into a
\textbf{portfolio weights matrix} $\mvW_{1:T}$, we obtain the portfolio returns time-series
by multiplication of $\mvR_{1:T}$ with $\mvW_{1:T}$ and extraction of the $T$ diagonal elements
of the product, such that:

\begin{equation}
    \vr_{1:T} = \text{diag}(\mvR_{1:T} \mvW_{1:T}) \in \sR^{T}
\end{equation}

%%%
\subsubsection{Log Return} \label{subsub:log-return}

Despite the interpretability of the simple return as the percentage change in asset price
over one period, it is asymmetric and therefore practitioners tend to use log returns instead
\citep{finance:stochastic-financial-models}, in order to preserve interpretation and to yield a symmetric measure.
Using the example in Table \ref{tab:log-returns}, a $15\%$ increase in price followed
by a $15\%$ decline does not result in the initial price of the asset. On the contrary,
a $15\%$ log-increase in price followed by a $15\%$ log-decline returns to the initial asset price,
reflecting the symmetric behaviour of log returns.

\begin{table}[h]
\centering
\begin{tabular}{|c||c|c||c|} 
\hline
time $t$ & simple return & price $(\$)$ & log return \\
\hline\hline
0 & - & \cellcolor{matlab-yellow!25}100 & - \\
\hline
1 & \cellcolor{matlab-red!25}+0.15 & 110 & +0.13 \\
\hline
2 & \cellcolor{matlab-red!25}-0.15 & \cellcolor{matlab-yellow!25}99 & -0.16 \\
\hline\hline
3 & +0.01 & \cellcolor{gray!25}100 & +0.01 \\
\hline
4 & -0.14 & 86 & \cellcolor{matlab-green!25}-0.15 \\
\hline
5 & +0.16 & \cellcolor{gray!25}100 & \cellcolor{matlab-green!25}+0.15 \\
\hline
\end{tabular}
\caption{Simple Return Asymmetry \& Log Return Symmetry}
\label{tab:log-returns}
\end{table}

\begin{figure}[h]
    \centering
    \begin{subfigure}[t]{0.48\textwidth}
        \includegraphics[width=\textwidth]{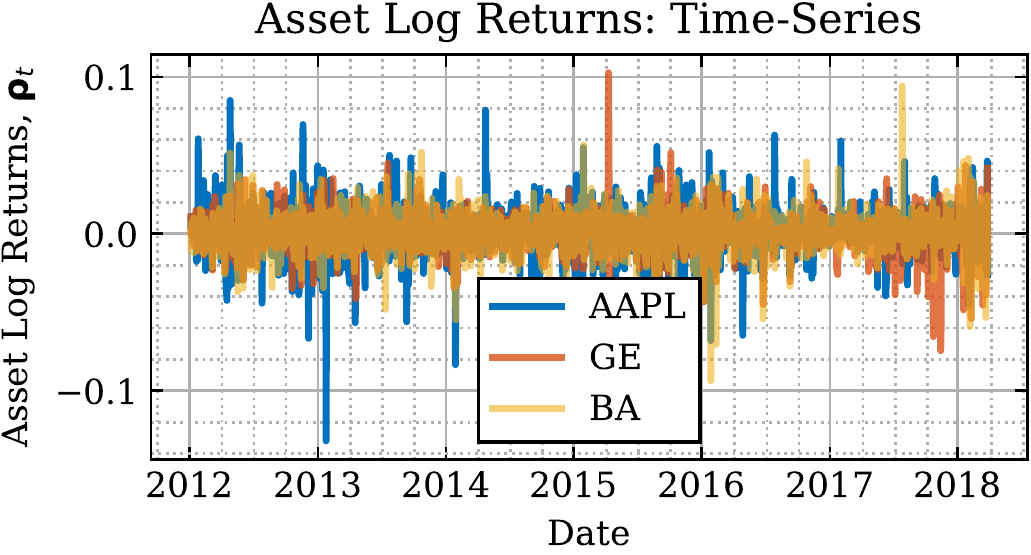}
    \end{subfigure}
    ~ 
    \begin{subfigure}[t]{0.48\textwidth}
        \includegraphics[width=\textwidth]{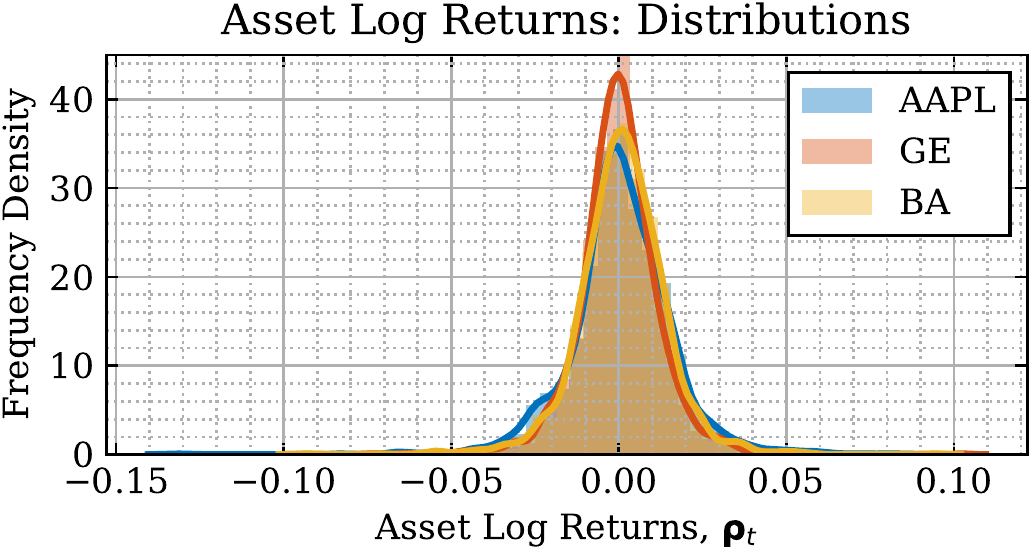}
    \end{subfigure}
    \caption{Single Assets Log Returns}
    \label{fig:log-returns}
\end{figure}

Let the \textbf{log return} $\rho_{t}$ at time $t$ be:
\begin{equation}
    \rho_{t} \triangleq ln(\frac{p_{t}}{p_{t-1}}) \textoversign{=}{(\ref{def:gross-returns})} ln(R_{t}) \in \sR
\label{def:log-returns}
\end{equation}
Note the very close connection of gross return to log return. Moreover, since gross return is centered around unity,
the logarithmic operator makes log returns concentrated around zero, clearly observed in Figure \ref{fig:log-returns}.   

% %%% START: CUMULATIVE LOG RETURNS %%%

% Considering the \textbf{multi-period log returns}
% by taking the natural logarithm of simple gross return $1 + R_{t}[k]$
% \begin{align}
% r_{t}[k] \triangleq ln(1 + R_{t}[k])
%             &\textoversign{=}{(\ref{def:multi-period-simple-returns})}
%                   ln \prod_{i=0}^{k-1} (1 + R_{t-i})\nonumber\\
%             &= \sum_{i=0}^{k-1} ln(1 + R_{t-i})
%             \textoversign{=}{(\ref{def:log-returns})}
%                 \sum_{i=0}^{k-1} r_{t-i}
% \label{def:multi-period-log-returns}
% \end{align}
% Thus, the multi-period log (gross) return of an asset
% is simply the sum of one period log returns.

% %%% END: CUMULATIVE SIMPLE RETURNS %%%

Comparing the definitions of simple and log returns in (\ref{def:simple-returns}) and (\ref{def:log-returns}),
respectively, we obtain the relationship:
\begin{equation}
    \rho_{t} = ln(1 + r_{t})
\label{eq:log-to-simple-returns}
\end{equation}
hence we can define all time-series and convenient portfolio representations of log returns
by substituting simple-returns in (\ref{eq:log-to-simple-returns}).
For example, the \textbf{portfolio log return} is given by substitution of (\ref{eq:log-to-simple-returns})
into (\ref{def:portfolio-simple-returns}), such that:

\begin{align}
    \rho_{t} \triangleq ln(1 + \vw_{t}^{T} \vr_{t}) \in \sR
\label{def:portfolio-log-returns}
\end{align}

\section{Evaluation Criteria} \label{sec:evaluation-criteria}

The end goal is the construction of portfolios, linear combinations of individual assets,
whose properties (e.g., returns, risk) are optimal under provided conditions and constraints.
As a consequence, a set of evaluation criteria and metrics is necessary in order to evaluate
the performance of the generated portfolios. Due to the uncertainty of the future dynamics of
the financial markets, we study the statistical properties of the assets returns, as well as
other risk metrics, motivated by signal processing.

\subsection{Statistical Moments} \label{sub:statistical-moments}

Future prices and hence returns are inherently unknown and uncertain \citep{finance:stochastic-financial-models}.
To mathematically capture and manipulate this stochasticity, we treat future market dynamics
(i.e., prices, cross-asset dependencies) as random variables and study their properties.
Qualitative visual analysis of probability density functions is a labour-intensive process
and thus impractical, especially when high-dimensional distributions (i.e., 4D and higher)
are under consideration. On the other hand, quantitative measures, such as moments,
provide a systematic way to analyze (joint) distributions \citep{meucci2009risk}.

%%%
\subsubsection{Mean, Median \& Mode}

Suppose that we need to summarize all the information regarding a random variable $X$ in only one number,
the one value that best represents the whole range of possible outcomes. We are looking for a location
parameter that provides a fair indication of where on the real axis the random variable
$X$ will end up taking its value.

An immediate choice for the location parameter is the center of mass of the distribution,
i.e., the weighted average of each possible outcome, where the weight of each outcome is
provided by its respective probability. This corresponds to computing the \textbf{expected value}
or \textbf{mean} of the random variable:

\begin{equation}
    \E[X] = \mu_{X} \triangleq \int_{- \infty}^{+ \infty} \rx f_{X}(\rx) d\rx \in \sR
\label{def:univariate-mean}
\end{equation}

Note that the mean is also the first order statistical moment of the distribution $f_{X}$.
When a finite number of observations $T$ is available, and there is no closed form expression
for the probability density function $f_{X}$, the \textbf{sample mean} or \textbf{empirical mean}
is used as an unbiased estimate of the expected value, according to:

\begin{equation}
    \E[X] \approx \frac{1}{T} \sum_{t=1}^{T} \rx_{t}
\label{def:univariate-empirical-mean}
\end{equation}

\begin{advice}[Greedy Criterion]
    For the same level of risk, choose the portfolio that maximizes the expected returns \citep{finance:quant-finance}.
\label{adv:greedy-criterion}
\end{advice}

\begin{figure}[h]
    \centering
    \begin{subfigure}[t]{0.48\textwidth}
        \includegraphics[width=\textwidth]{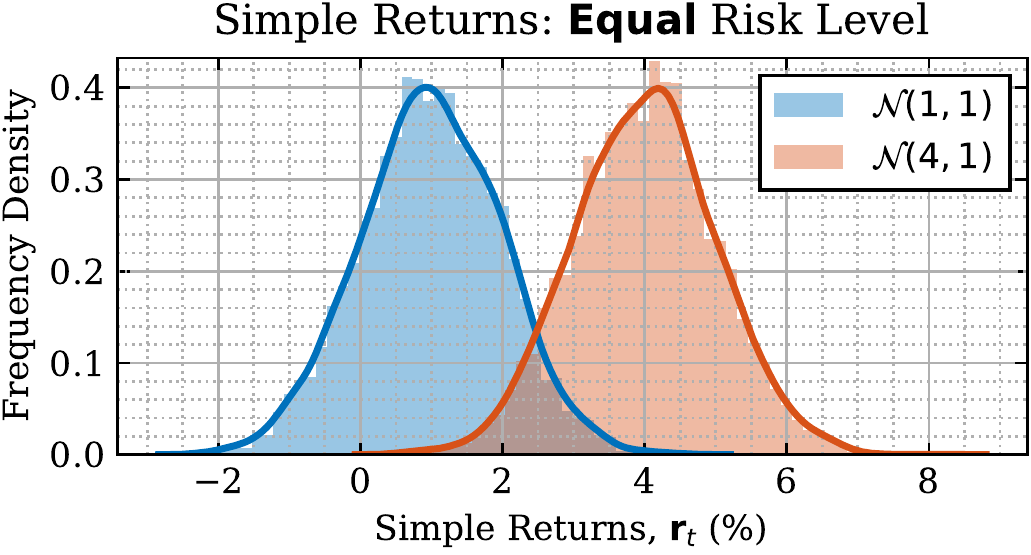}
    \end{subfigure}
    ~ 
    \begin{subfigure}[t]{0.48\textwidth}
        \includegraphics[width=\textwidth]{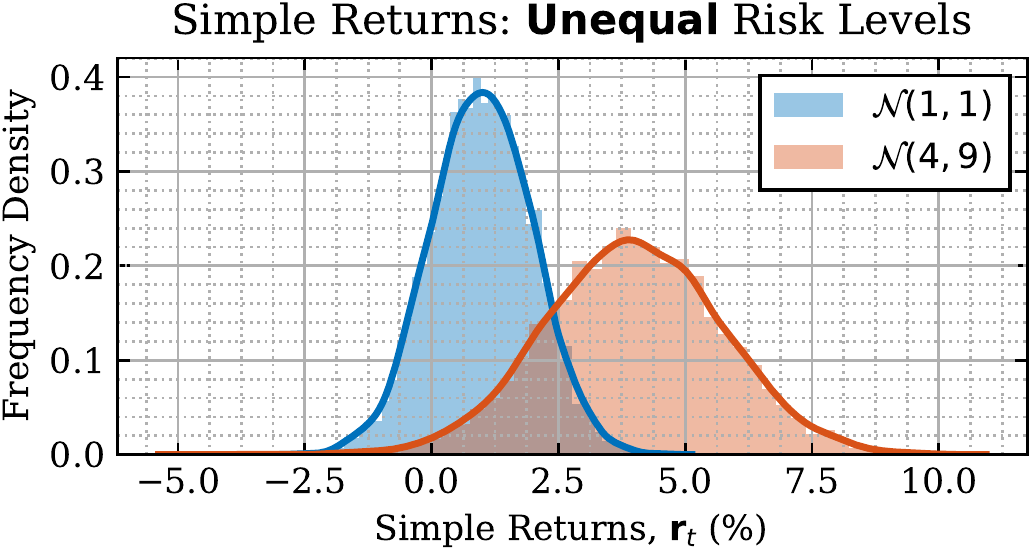}
    \end{subfigure}
    \caption{Greedy criterion for equally risky assets (left) and unequally risky assets (right).}
    \label{fig:greedy-criterion}
\end{figure}

Figure \ref{fig:greedy-criterion} illustrates to cases where the \textbf{Greedy Criterion}
\ref{adv:greedy-criterion} is applied. In case of assets with equal risk levels (i.e., left sub-figure)
we prefer the one that maximizes expected returns, thus the red ($\mu_{\text{blue}} = 1 < \mu_{\text{red}} = 4$).
On the other hand, when the assets have unequal risk levels (i.e., right sub-figure)
the criterion does not apply and we cannot draw any conclusions without employing other metrics as well.

The definition of the mean value is extended to the multivariate case as the juxtaposition of
the mean value (\ref{def:univariate-mean}) of the marginal distribution of each entry:

\begin{equation}
    \E[\rmX] = \vmu_{\rmX} \triangleq
    \renewcommand\arraystretch{1.5}
    \begin{bmatrix}
    \E[X_{1}] \\ \E[X_{2}] \\ \vdots \\ \E[X_{M}]
    \end{bmatrix}
    \in \sR^{M}
\label{def:multivariate-mean}
\end{equation}

A portfolio with vector $\vw_{t}$ and single asset mean simple returns $\vmu_{r}$ has expected simple returns:

\begin{equation}
    \mu_{r} = \vw_{t}^{T} \vmu_{r}
\label{eq:portfolio-expected-simple-returns}
\end{equation}

An alternative choice for the location parameter is the \textbf{median}, which is the quantile relative
to the specific cumulative probability $p=1/2$:

\begin{equation}
    \text{Med}[X] \triangleq Q_{X} \bigg( \frac{1}{2} \bigg) \in \sR
\label{def:median}
\end{equation}

The juxtaposition of the median, or any other quantile, of each entry of a random variable
does not satisfy the affine equivariance property\footnote{$\text{Med}[\va + \mB\rmX] \neq \va + \mB \text{Med}[\rmX]$.}
\citep{meucci2009risk} and therefore it does not define a suitable location parameter.

A third parameter of location is the mode, which refers to the shape of the probability density function $f_{X}$.
Indeed, the mode is defined as the point that corresponds to the highest peak of the density function:

\begin{equation}
    \text{Mod}[X] \triangleq \argmax_{\rx \in \sR} f_{X}(\rx) \in \sR
\label{def:univariate-mode}
\end{equation}

Intuitively, the mode is the most frequently occurring data point in the distribution.
It is trivially extended to multivariate distributions, namely as the highest peak of the
joint probability density function:

\begin{equation}
    \text{Mod}[\rmX] \triangleq \argmax_{\rvx \in \sR^{M}} f_{\rmX}(\rvx) \in \sR^{M}
\label{def:multivariate-mode}
\end{equation}

Note that the relative position of the location parameters provide qualitative information about
the symmetry, the tails and the concentration of the distribution. Higher-order moments quantify these properties.

\begin{figure}[h]
    \centering
    \begin{subfigure}[t]{0.48\textwidth}
        \includegraphics[width=\textwidth]{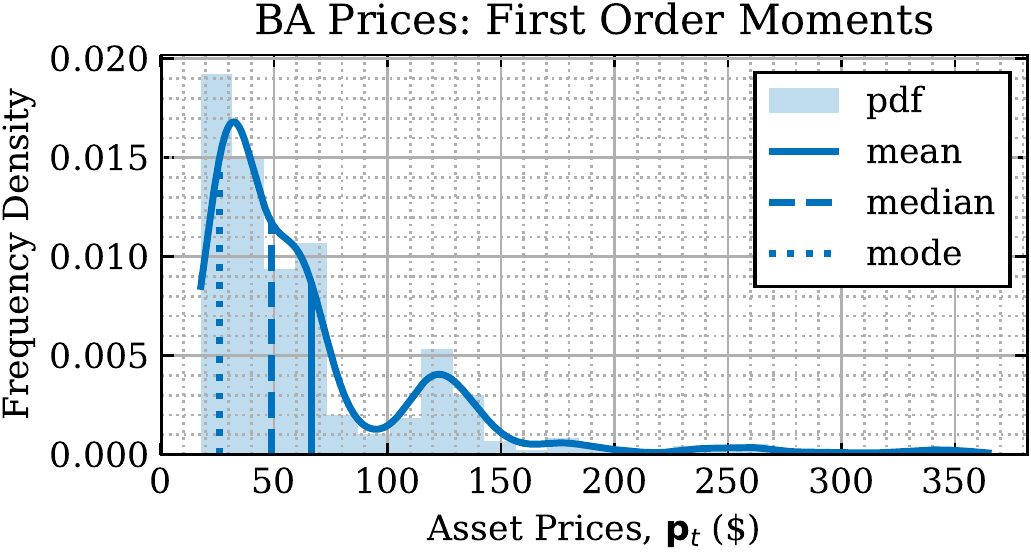}
    \end{subfigure}
    ~ 
    \begin{subfigure}[t]{0.48\textwidth}
        \includegraphics[width=\textwidth]{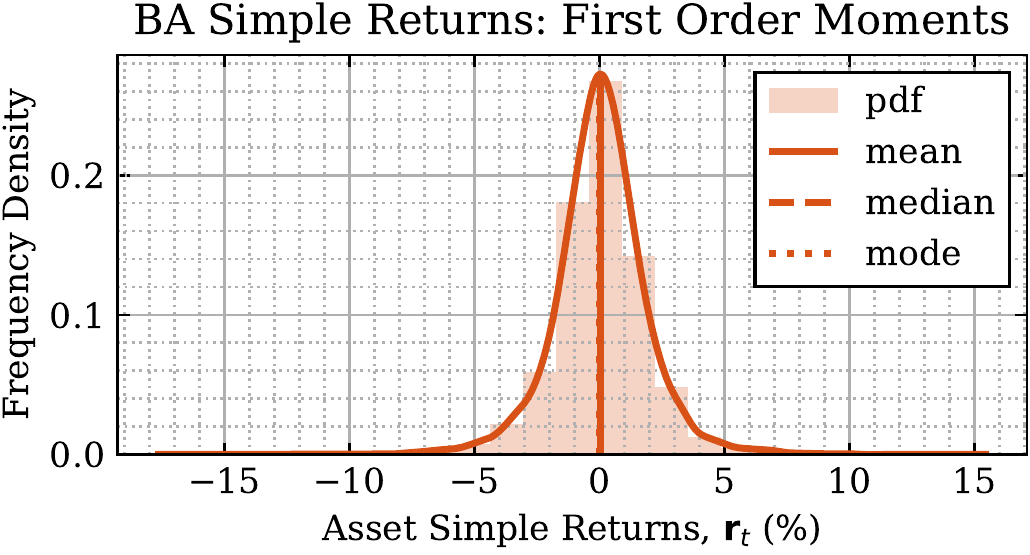}
    \end{subfigure}
    \caption{First order moments for \texttt{BA} (Boeing Company) prices (left) and simple returns (right).}
    \label{fig:first-order-moments}
\end{figure}

Figure \ref{fig:first-order-moments} illustrates the distribution of
the prices and the corresponding simple returns of the asset \texttt{BA} (Boeing Company),
along with their location parameters. In case of the simple returns, we highlight that
the mean, the median and the mode are very close to each other,
reflecting the symmetry and the concentration of the distribution, properties that motivated the
selection of returns over raw asset prices.

%%%
\subsubsection{Volatility \& Covariance}

The dilemma we faced in selecting between assets in Figure \ref{fig:greedy-criterion} motivates
the introduction of a metric that quantifies risk level. On other words, we are looking for a
dispersion parameter that yields an indication of the extent to which the location parameter (i.e., mean, median)
might be wrong in guessing the outcome of the random variable $X$.

The \textbf{variance} is the benchmark dispersion parameter, measuring how far the random variable
$X$ is spread out of its mean, given by:

\begin{equation}
    \Var[X] = \sigma_{X}^{2} \triangleq \E[(X-\E[X])^{2}] \in \sR
\label{def:variance}
\end{equation}

The square root of the variance, $\sigma_{X}$, namely the \textbf{standard deviation} 
or \textbf{volatility} in finance, is a more physically interpretable parameter, since it has
the same units as the random variable under consideration (i.e., prices, simple returns).

Note that the variance is also the second order statistical central moment of the distribution $f_{X}$.
When a finite number of observations $T$ is available, and there is no closed form expression
for the probability density function, $f_{X}$, the Bessel's correction formula \citep{sp:time-series}
is used as an unbiased estimate of the variance, according to:

\begin{equation}
    \Var[X] \approx \frac{1}{T-1} \sum_{t=1}^{T} (\rx_{t} - \mu_{X})^{2}
\label{def:bessels-correction-formula-variance}
\end{equation}

\begin{advice}[Risk-Aversion Criterion]
    For the same expected returns, choose the portfolio that minimizes the volatility \citep{finance:quant-finance}.
\label{adv:risk-aversion-criterion}
\end{advice}

\begin{figure}[h]
    \centering
    \begin{subfigure}[t]{0.48\textwidth}
        \includegraphics[width=\textwidth]{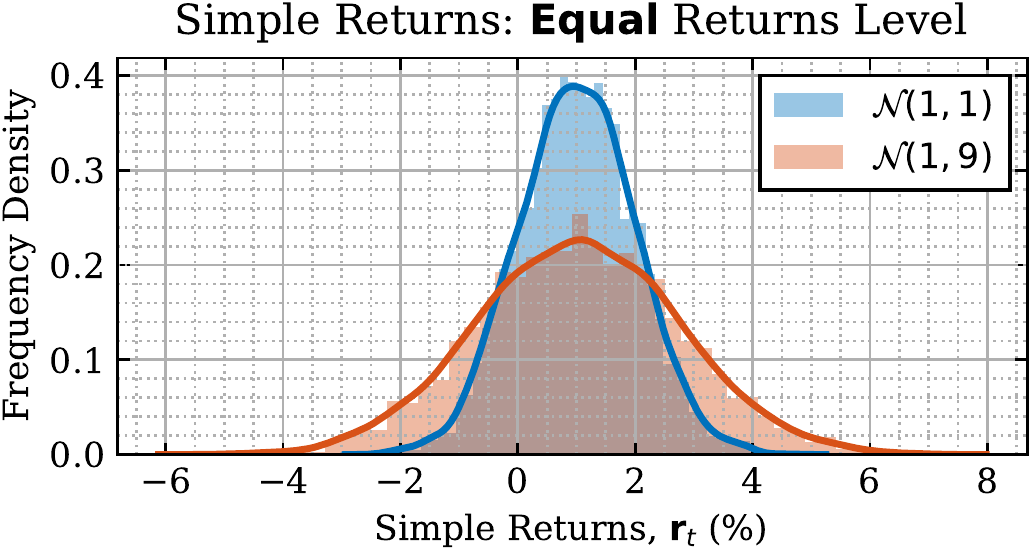}
    \end{subfigure}
    ~ 
    \begin{subfigure}[t]{0.48\textwidth}
        \includegraphics[width=\textwidth]{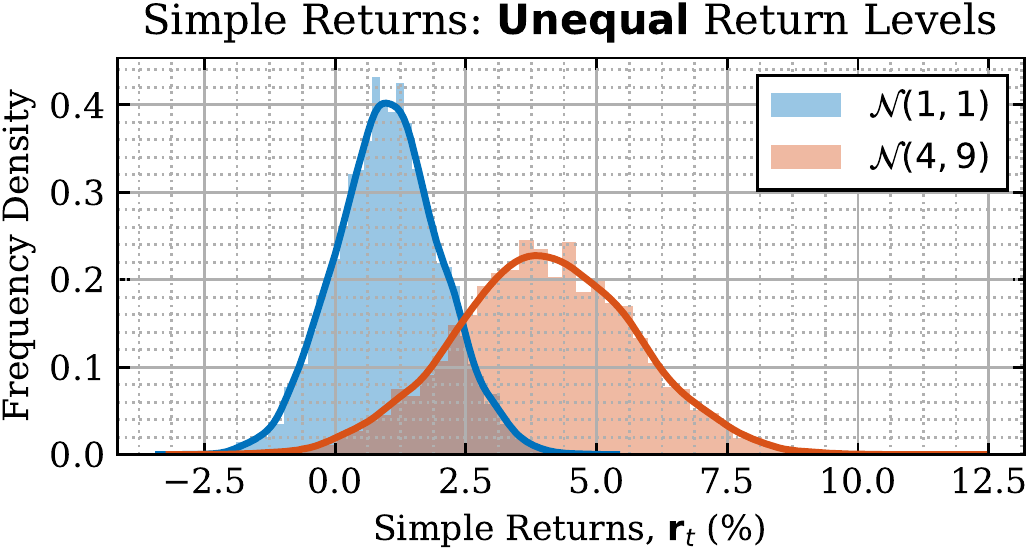}
    \end{subfigure}
    \caption{Risk-aversion criterion for equal returns (left) and unequal returns (right).}
    \label{fig:risk-aversion-criterion}
\end{figure}

According to the \textbf{Risk-Aversion Criterion}, in Figure \ref{fig:risk-aversion-criterion},
for the same returns level (i.e., left sub-figure) we choose the less risky asset, the blue,
since the red is more spread out ($\sigma_{\text{blue}}^{2} = 1 < \sigma_{\text{red}}^{2} = 9$).
However, in case of unequal return levels (i.e., right sub-figure) the criterion is inconclusive.

The definition of variance is extended to the multivariate case by introducing \textbf{covariance},
which measures the joint variability of two variables, given by:

\begin{equation}
    \Cov[\rmX] = \mSigma_{\rmX} \triangleq \E[(\rmX-\E[\rmX]) (\rmX-\E[\rmX])^{T}] \in \sR^{M \times M}
\label{def:covariance}
\end{equation}

or component-wise:

\begin{equation}
    \Cov[X_{m}, X_{n}] = [\Cov[\rmX]]_{mn} = \Sigma_{m n} \triangleq \E[(X_{m}-\E[X_{m}]) (X_{n}-\E[X_{n}])] \in \sR
\label{def:covariance-component-wise}
\end{equation}

By direct comparison of (\ref{def:variance}) and (\ref{def:covariance-component-wise}), we note that:

\begin{equation}
    \Var[X_{m}] = \Cov[X_{m}, X_{m}] = [\Cov[\rmX]]_{mm} = \Sigma_{m m}
\label{eq:variance-covariance}
\end{equation}

hence the $m$-th diagonal element of the covariance matrix $\Sigma_{m m}$ is the variance of the
$m$-th component of the multivariate random variable $\rmX$, while the non-diagonal terms
$\Sigma_{m n}$ represent the joint variability of the $m$-th with the $n$-th component of $\rmX$.
Note that, by definition (\ref{def:covariance}), the covariance is a symmetric and real matrix,
thus it is semi-positive definite \citep{mandic:asp-estimators}.

Empirically, we estimate the covariance matrix entries using again the Bessel's correction formula
\citep{sp:time-series}, in order to obtain an unbiased estimate:

\begin{equation}
    \Cov[X_{m}, X_{n}] \approx \frac{1}{T-1} \sum_{t=1}^{T} (\rx_{m, t} - \mu_{X_{m}})(\rx_{n, t} - \mu_{X_{n}})
\label{def:bessels-correction-formula-covariance}
\end{equation}

A portfolio with vector $\vw_{t}$ and covariance matrix of assets simple returns $\mS$ has variance:

\begin{equation}
    \sigma_{r}^{2} = \vw_{t}^{T} \mSigma \vw_{t}
\label{eq:portfolio-variance}
\end{equation}

The \textbf{correlation coefficient} is also frequently used to quantify the linear dependency between
random variables. It takes values in the range $[-1, 1]$ and hence it is a normalized way to compare
dependencies, while covariances are highly influenced by the scale of the random variables' variance.
The correlation coefficient is given by:

\begin{equation}
    \corr[X_{m}, X_{n}] = [\corr[\rmX]]_{mn} = \rho_{m n} \triangleq \frac{\Cov[X_{m}, X_{n}]}{\sigma_{X_{m}} \sigma_{X_{n}}}
        \in [-1, 1] \subset \sR
\label{def:correlation-component-wise}
\end{equation}

\begin{figure}[h]
    \centering
    \begin{subfigure}[t]{0.48\textwidth}
        \includegraphics[width=\textwidth]{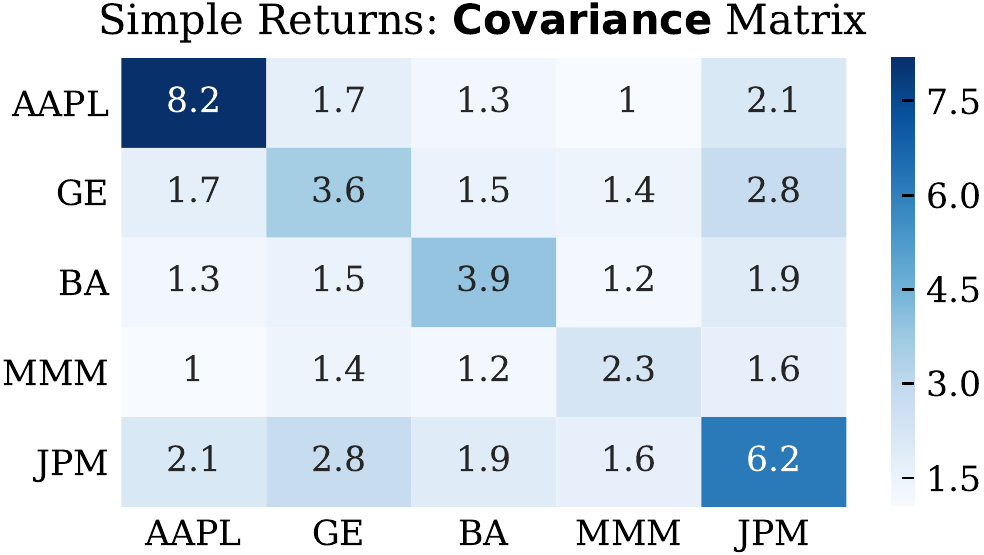}
    \end{subfigure}
    ~ 
    \begin{subfigure}[t]{0.48\textwidth}
        \includegraphics[width=\textwidth]{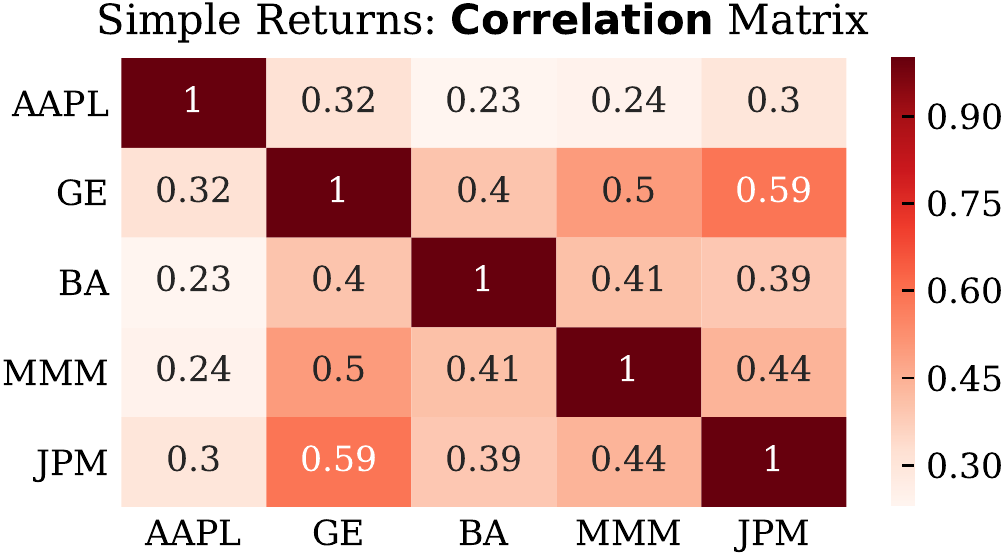}
    \end{subfigure}
    \caption{Covariance and correlation matrices for assets simple returns.}
    \label{fig:second-order-moments}
\end{figure}

%%%
\subsubsection{Skewness}

The standard measure of symmetry of a distribution is the \textbf{skewness},
which is the third central moment normalized by the standard deviation, in such a way to make it scale-independent:

\begin{equation}
    \skewness[X] \triangleq \frac{\E \big[(X - \E[X])^{3}\big] }{\sigma_{X}^{3}}
\label{def:skewness}
\end{equation}

In particular, a distribution whose probability density function is symmetric around its expected value has null skewness.
If the skewness is positive (negative), occurrences larger than the expected value are less (more) likely
than occurrences smaller than the expected value.

\begin{advice}[Negatively Skewed Criterion]
    Choose negatively skewed returns, rather than positively skewed. \citep{finance:quant-finance}.
\label{adv:negatively-skewed-criterion}
\end{advice}

%%%
\subsubsection{Kurtosis}

The fourth moment provides a measure of the relative weight of the tails with respect to the central body of a distribution.
The standard quantity to evaluate this balance is the \textbf{kurtosis}, defined as the normalized fourth central moment:

\begin{equation}
    \kurtosis[X] \triangleq \frac{\E \big[(X - \E[X])^{4} \big] }{\sigma_{X}^{4}}
\label{def:kurtosis}
\end{equation}

The kurtosis gives an indication of how likely it is to observe a measurement far in the tails of the distribution:
a large kurtosis implies that the distribution displays "fat tails".

\subsection{Financial Risk and Performance Metrics} \label{sub:financial-risk-and-performance-metrics}

Despite the insight into the statistical properties we gain by studying moments of returns distribution,
we can combine them in such ways to fully capture the behaviour of our strategies and better assess them.
Inspired by standard metrics used in signal processes (e.g. signal-to-noise ratio) and sequential decision making
we introduce the following performance evaluators: cumulative returns, sharpe ratio, drawdown
and value at risk.

%%%
\subsubsection{Cumulative Returns}

In subsetion \ref{sub:returns} we defined returns relative to the change in asset prices in one time period.
Nonetheless, we usually get involved into a multi-period investment, hence we are extending the definition
of vanilla returns to the \textbf{cumulative returns}, which represent the change in asset prices over
larger time horizons.

Based on (\ref{def:gross-returns}), the \textbf{cumulative gross return} $R_{t \rightarrow T}$
between time indexes $t$ and $T$ is given by:

\begin{equation}
    R_{t \rightarrow T} \triangleq \frac{p_{T}}{p_{t}}
    = \bigg( \frac{p_{T}}{p_{T-1}} \bigg) \bigg( \frac{p_{T-1}}{p_{T-2}} \bigg) \cdots \bigg( \frac{p_{t+1}}{p_{t}} \bigg)
    \textoversign{=}{(\ref{def:gross-returns})} R_{T} R_{T-1} \cdots R_{t+1}
    = \prod_{i=t+1}^{T} R_{i}
    \in \sR
\label{def:cumulative-gross-returns}
\end{equation}

The cumulative gross return is usually also termed \textbf{Profit \& Loss} (PnL), since it represents
the wealth level of the investment. If $R_{t \rightarrow T} > 1$ ($< 1$) the investment was profitable
(lossy).

\begin{advice}[Profitability Criterion]
    Aim to maximize profitability of investment.
\label{adv:profitability-criterion}
\end{advice}

Moreover, the \textbf{cumulative simple return} $r_{t \rightarrow T}$ is given by:

\begin{equation}
    r_{t \rightarrow T} \triangleq \frac{p_{T}}{p_{t}} - 1
    \ \textoversign{=}{(\ref{def:cumulative-gross-returns})}\ \bigg[ \prod_{i=t+1}^{T} R_{i} - 1 \bigg]
    \textoversign{=}{(\ref{def:simple-returns})} \bigg[ \prod_{i=t+1}^{T} (1 + r_{i}) - 1 \bigg] \in \sR
\label{def:cumulative-simple-returns}
\end{equation}

while the \textbf{cumulative log return} $\rho_{t \rightarrow T}$ is:

\begin{equation}
    \rho_{t \rightarrow T} \triangleq ln(\frac{p_{T}}{p_{t}}) 
    \ \textoversign{=}{(\ref{def:cumulative-gross-returns})}\ ln \bigg( \prod_{i=t+1}^{T} R_{i} \bigg)
    = \sum_{i=t+1}^{T} ln(R_{i})
    \textoversign{=}{(\ref{def:log-returns})} \sum_{i=t+1}^{T} \rho_{i}
    \in \sR
\label{def:cumulative-log-returns}
\end{equation}

\begin{figure}[h]
    \centering
    \begin{subfigure}[t]{0.48\textwidth}
        \includegraphics[width=\textwidth]{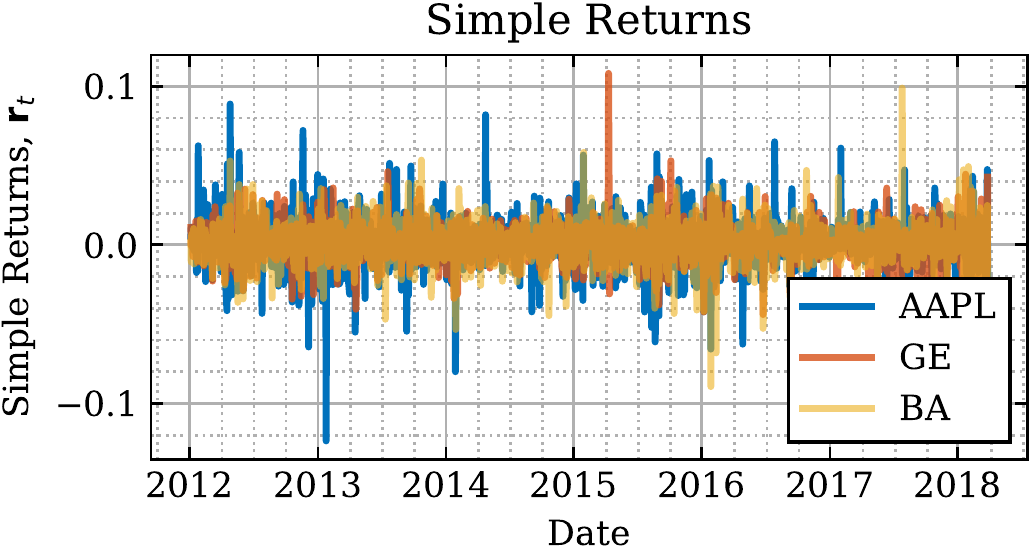}
    \end{subfigure}
    ~ 
    \begin{subfigure}[t]{0.48\textwidth}
        \includegraphics[width=\textwidth]{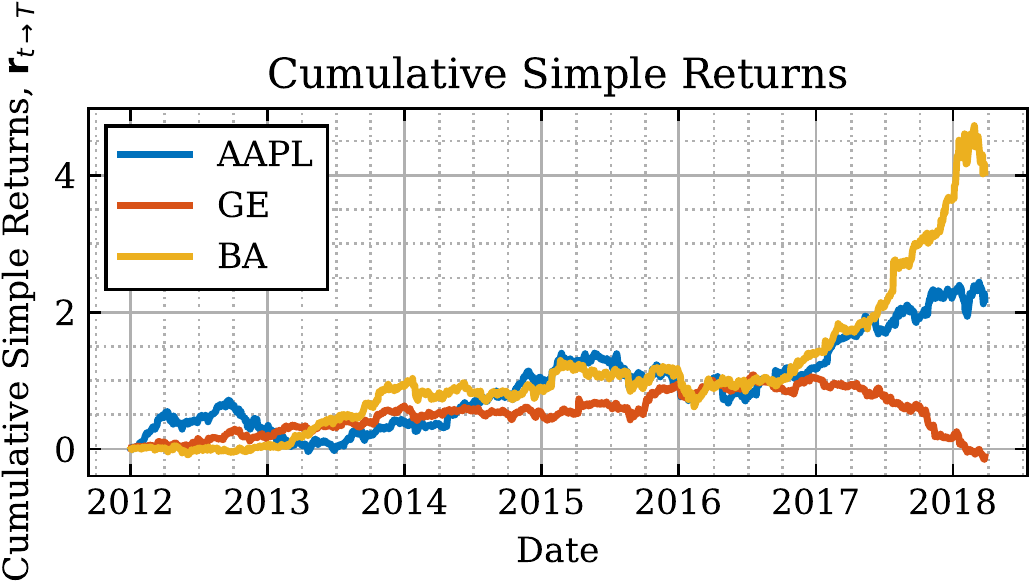}
    \end{subfigure}
    \caption{Assets cumulative simple returns.}
    \label{fig:cumulative-returns}
\end{figure}

Figure \ref{fig:cumulative-returns} demonstrates the interpretation power of cumulative returns over simple returns.
Simple visual inspection of simple returns is inadequate for comparing the performance of the different assets.
On the other hand, cumulative simple returns exhibit that \texttt{BA}'s (Boeing Company) price increased by $\approx 400\%$,
while \texttt{GE}'s (General Electric) price declines by $\approx 11\%$, in the time period $2012$ to $2018$.

%%% 
\subsubsection{Sharpe Ratio}

\begin{remark}
The criteria \ref{adv:greedy-criterion} and \ref{adv:risk-aversion-criterion} can sufficiently distinguish and
prioritize investments which either have the same risk level or returns level, respectively.
Nonetheless, they fail in all other cases, when risk or return levels are unequal.
\end{remark}

The failure of greedy criterion and risk-aversion criterion is demonstrated in
both examples in Figures \ref{fig:greedy-criterion} and \ref{fig:risk-aversion-criterion},
where it can be observed that the more risky asset,
the red one, has a higher expected returns (i.e., the red
distribution is wider, hence has larger variance, but it is
centered around a larger value, compared to the blue distribution).
Consequently, none of the criteria applies and the comparison is inconclusive.

In order to address this issue and motivated by \textbf{Signal-to-Noise Ratio} (SNR) \citep{sp:finance, sp:financial-engineering},
we define \textbf{Sharpe Ratio} (SR) as the ratio of expected returns (i.e., signal power)
to their standard deviation (i.e., noise power\footnote{The variance of the noise is equal to the noise power.
Standard deviation is used in the definition of SR to provide a unit-less metric.}),
adjusted by a scaling factor:

\begin{equation}
    \textbf{SR}_{1:T} \triangleq \sqrt{T} \frac{\E[\rvr_{1:T}]}{\sqrt{\Var[\rvr_{1:T}]}} \in \sR
\label{def:sharpe-ratio}
\end{equation}

where $T$ is the number of samples considered in the calculation of the empirical mean and standard deviation.

\begin{advice}[Sharpe Ratio Criterion]
    Aim to maximize Sharpe Ratio of investment.
\label{adv:sharpe-ratio-criterion}
\end{advice}

Considering now the example in Figures \ref{fig:greedy-criterion}, \ref{fig:risk-aversion-criterion},
we can quantitatively compare the two returns streams and select the one that maximizes the
Sharpe Ratio:

\begin{align}
    \textbf{SR}_{blue}  &= \sqrt{T} \frac{\mu_{blue}}{\sigma_{blue}} = \sqrt{T} \frac{1}{1} = \sqrt{T} \\
    \textbf{SR}_{red} &= \sqrt{T} \frac{\mu_{red}}{\sigma_{red}} = \sqrt{T} \frac{4}{3} \\
    \textbf{SR}_{blue}  &< \textbf{SR}_{red} \Rightarrow \text{choose }\textbf{red}
\end{align}

\subsubsection{Drawdown}

The \textbf{drawdown} (DD) is a measure of the decline from a historical peak in cumulative returns \citep{finance:investment-science}.
A drawdown is usually quoted as the percentage between the peak and the subsequent trough and is defined as:

\begin{equation}
    \textbf{DD}(t) = - \max \{0, \big[ \max_{\tau \in (0, t)} r_{0 \rightarrow \tau} \big] - r_{0 \rightarrow  t} \}
\label{def:drawdown}
\end{equation}

The \textbf{maximum drawdown} (MDD) up to time $t$ is the maximum of the drawdown over the history of the
cumulative returns, such that:

\begin{equation}
    \textbf{MDD}(t) = - \max_{x \in (0, t)} \{ \big[ \max_{\tau \in (0, T)} r_{0 \rightarrow  \tau} \big] - r_{0 \rightarrow  T} \}
\label{def:max-drawdown}
\end{equation}

\begin{figure}[h]
    \centering
    \begin{subfigure}[t]{0.48\textwidth}
        \includegraphics[width=\textwidth]{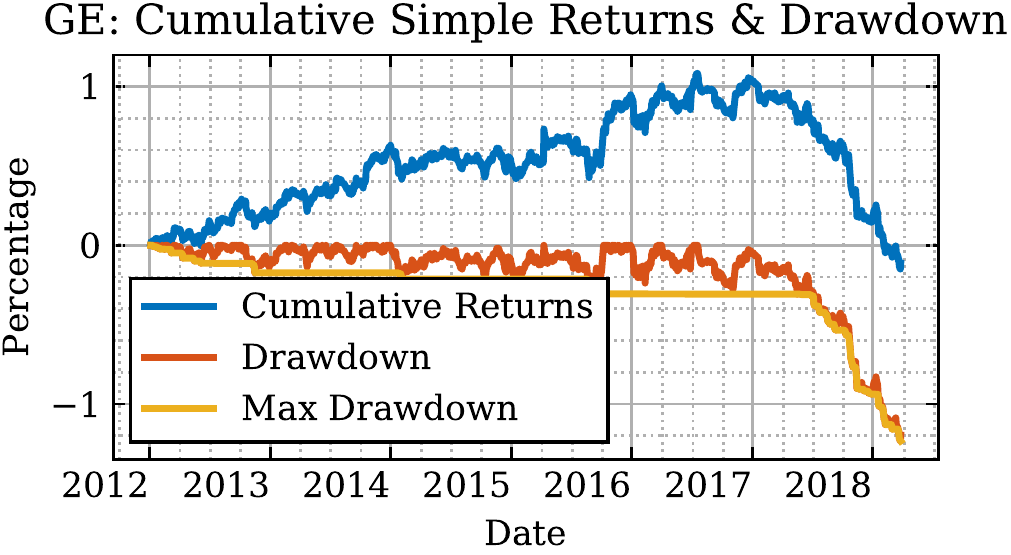}
    \end{subfigure}
    ~ 
    \begin{subfigure}[t]{0.48\textwidth}
        \includegraphics[width=\textwidth]{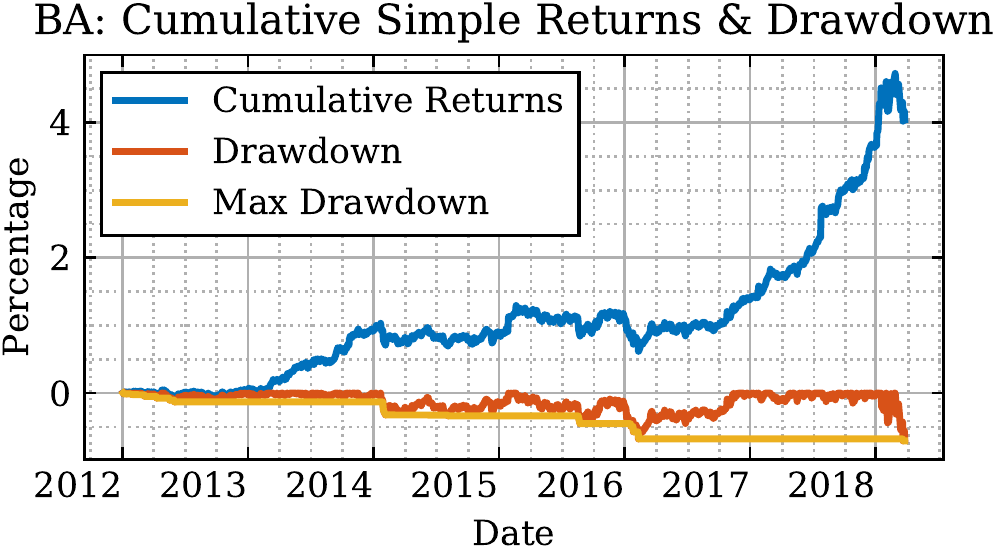}
    \end{subfigure}
    \caption{(Maximum) drawdown and cumulativer returns for \texttt{GE} and \texttt{BA}.}
    \label{fig:drawdown}
\end{figure}

The drawdown and maximum drawdown plots are provided in Figure \ref{fig:drawdown} along with the cumulative returns
of assets \texttt{GE} and \texttt{BA}. Interestingly, the decline of \texttt{GE}'s cumulative returns starting in
early $2017$ is perfectly reflected by the (maximum) drawdown curve.

% %%%
% \subsubsection{Average Profitability Per Trade}

% Another frequently used risk metric is \textbf{average profitability per trade} (APPT),
% summarizing in a single scalar number (point estimator) the percentage win (or loss)
% per trade, given by:

% \begin{equation}
%     \textbf{APPT} \triangleq (\text{probability\_win}) (\text{average\_win}) - (\text{probability\_loss}) (\text{average\_loss})
% \label{def:appt}
% \end{equation}

% where \text{probability\_win} and \text{average\_win} are empirical estimates obtained only from the positive simple returns samples,
% while \text{probability\_loss} and \text{average\_loss} from the negative samples:

% \begin{align}
%     \text{probability\_*} &= \frac{\text{number\_of\_*}}{\text{total\_number\_of\_trades}} \\
%     \text{average\_*} &= \frac{\sum *}{\text{total\_number\_of\_trades}}
% \end{align}

% For the example assets in Figure \ref{fig:cumulative-returns}, by application of (\ref{def:appt}), we obtain:

% \begin{equation}
%     \textbf{AAPT}_{\text{AAPL}} = 1.10\%, \quad
%     \textbf{AAPT}_{\text{GE}}   = 0.86\%, \quad
%     \textbf{AAPT}_{\text{BA}}   = 0.96\%
% \end{equation}

% which we interpret as any trade in \texttt{AAPL} (Apple) is on average $1.10\%$ profitable.

%%%
\subsubsection{Value at Risk}

The \textbf{value at risk} (VaR) is another commonly used metric to assess the performance of a returns
time-series (i.e., stream). Given daily simple returns $r_{t}$ and cut-off $c \in (0, 1)$,
the value at risk is defined as the $c$ quantile of their distribution, representing
the worst $100c\%$ case scenario:

\begin{equation}
    \textbf{VaR}(c) \triangleq Q_{\rr} (c) \in \sR
\end{equation}

\begin{figure}[h]
    \centering
    \begin{subfigure}[t]{0.48\textwidth}
        \includegraphics[width=\textwidth]{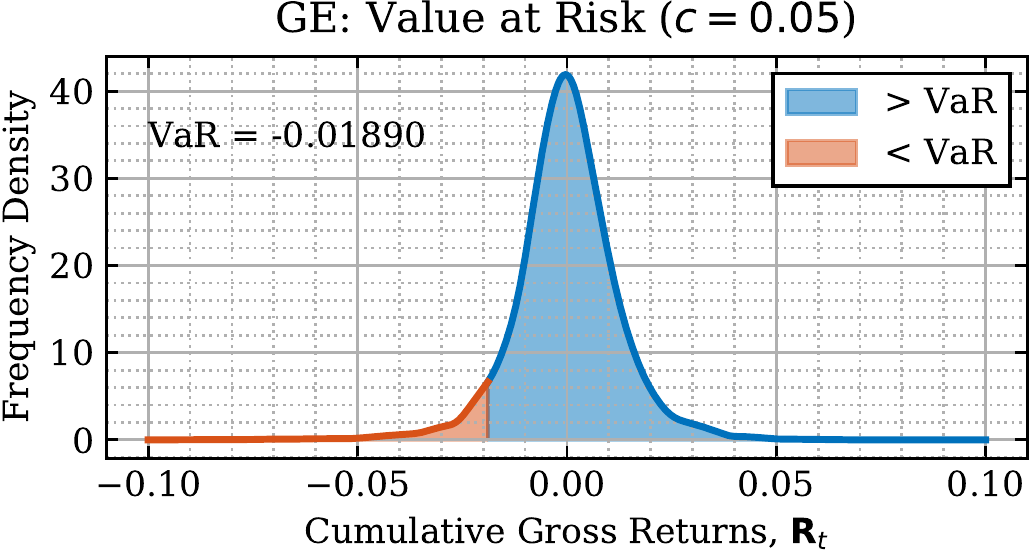}
    \end{subfigure}
    ~ 
    \begin{subfigure}[t]{0.48\textwidth}
        \includegraphics[width=\textwidth]{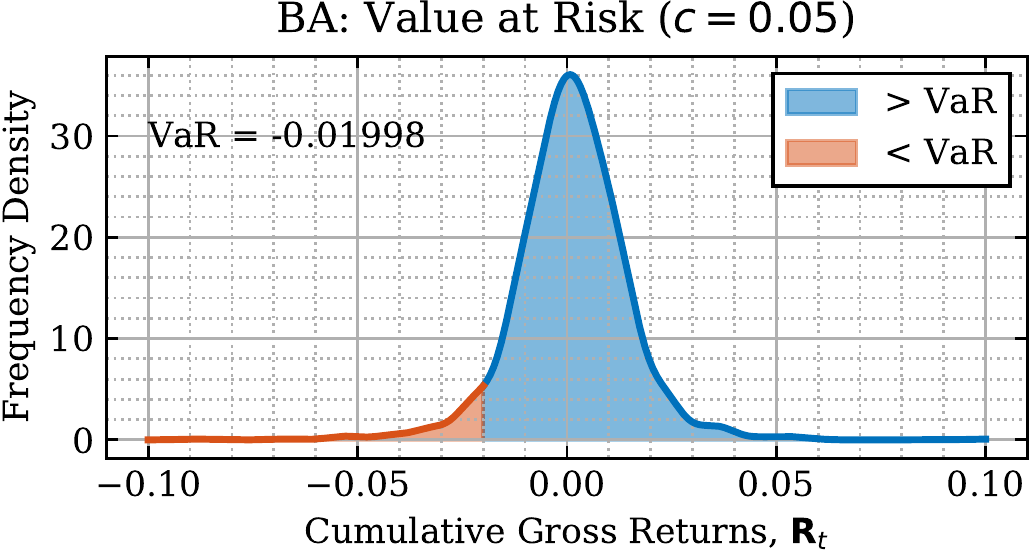}
    \end{subfigure}
    \caption{Illustration of the $5\%$ value at risk (VaR) of \texttt{GE} and \texttt{BA} stocks.}
    \label{fig:value-at-risk}
\end{figure}

Figure \ref{fig:value-at-risk} depicts \texttt{GE}'s value at risk at $-1.89 \%$ for cut-off parameter $c=0.05$.
We interpret this as "$5 \%$ of the trading days, General Electric's stock declines more than $1.89 \%$".

\section{Time-Series Analysis} \label{sec:time-series-analysis}

Time-series analysis is of major importance in a vast range of research topics,
and many engineering applications. This relates to analyzing time-series data for
estimating meaningful statistics and identifying patterns of sequential data.
Financial time-series analysis deals with the extraction of underlying features
to analyze and predict the temporal dynamics of financial assets
\citep{deep-learning:time-series}. Due to the inherent uncertainty and
non-analytic structure of financial markets \citep{sp:time-series},
the task is proven challenging, where classical linear statistical methods
such as the VAR model, and statistical machine learning models have been
widely applied \citep{machine-learning:time-series}.
In order to efficiently capture the non-linear nature of the financial time-series,
advanced non-linear function approximators, such as RNN models
\citep{mandic:rnn} and Gaussian Processes
\citep{gaussian-process:time-series} are also extensively used.

In this section, we introduce the VAR and RNN models, which comprise the basis
for the model-based approach developed in Section \ref{sec:model-based-reinforcement-learning}.

\subsection{Vector Autoregression (VAR)} \label{sub:var}

Autoregressive (AR) processes can model univariate time-series and specify that
future values of the series depend linearly on the past realizations of the series \citep{mandic:asp-arma}.
In particular, a $p$-order autoregressive process AR($p$) satisfies:

\begin{align}
    \rx_{t} &= a_{1} \rx_{t-1} + a_{2} \rx_{t-2} + \cdots + a_{p} \rx_{t-p} + \epsilon_{t} \nonumber\\
            &= \sum_{i=1}^{p} a_{i} \rx_{t-i} + \epsilon_{t} \nonumber\\
            &= \va^{T} \rvvx_{t-p:t-1} + \epsilon_{t}
            \in \sR
\label{def:ar}
\end{align}

where $\epsilon_{t}$ is a stochastic term (an imperfectly predictable term),
which is usually treated as white noise and $\va = [a_{1}, a_{2}, \cdots, a_{p}]^{T}$
the $p$ model parameters/coefficients.

Extending the AR model for multivariate time-series, we obtain the \textbf{vector autoregressive} (VAR) process,
which enables us to capture the cross-dependencies between series. For the general case of a
$M$-dimensional $p$-order vector autoregressive process VAR$_{M}$($p$), it follows that:

\begin{align}
    \renewcommand\arraystretch{1.5}
    \begin{bmatrix}
        \rx_{1, t} \\ \rx_{2, t} \\ \vdots \\ \rx_{M, t}
    \end{bmatrix}
    =
    \renewcommand\arraystretch{1.5}
    \begin{bmatrix}
        c_{1} \\ c_{2} \\ \vdots \\ c_{M}
    \end{bmatrix}
    +
    &\renewcommand\arraystretch{1.5}
    \begin{bmatrix}
        a_{1, 1}^{(1)} & a_{1,2}^{(1)} & \cdots & a_{1, M}^{(1)} \\
        a_{2, 1}^{(1)} & a_{2,2}^{(1)} & \cdots & a_{2, M}^{(1)} \\
        \vdots & \vdots & \ddots & \vdots \\
        a_{M, 1}^{(1)} & a_{M,2}^{(1)} & \cdots & a_{M, M}^{(1)} \\
    \end{bmatrix}
    \begin{bmatrix}
        \rx_{1, t-1} \\ \rx_{2, t-1} \\ \vdots \\ \rx_{M, t-1}
    \end{bmatrix}
    +
    \nonumber\\
    &\renewcommand\arraystretch{1.5}
    \begin{bmatrix}
        a_{1, 1}^{(2)} & a_{1,2}^{(2)} & \cdots & a_{1, M}^{(2)} \\
        a_{2, 1}^{(2)} & a_{2,2}^{(2)} & \cdots & a_{2, M}^{(2)} \\
        \vdots & \vdots & \ddots & \vdots \\
        a_{M, 1}^{(2)} & a_{M,2}^{(2)} & \cdots & a_{M, M}^{(2)} \\
    \end{bmatrix}
    \begin{bmatrix}
        \rx_{1, t-2} \\ \rx_{2, t-2} \\ \vdots \\ \rx_{M, t-2}
    \end{bmatrix}
    + \cdots +
    \nonumber\\
    &\renewcommand\arraystretch{1.5}
    \begin{bmatrix}
        a_{1, 1}^{(p)} & a_{1,2}^{(p)} & \cdots & a_{1, M}^{(p)} \\
        a_{2, 1}^{(p)} & a_{2,2}^{(p)} & \cdots & a_{2, M}^{(p)} \\
        \vdots & \vdots & \ddots & \vdots \\
        a_{M, 1}^{(p)} & a_{M,2}^{(p)} & \cdots & a_{M, M}^{(p)} \\
    \end{bmatrix}
    \begin{bmatrix}
        \rx_{1, t-p} \\ \rx_{2, t-p} \\ \vdots \\ \rx_{M, t-p}
    \end{bmatrix}
    +
    % \nonumber\\
    % &\renewcommand\arraystretch{1.5}
    \begin{bmatrix}
        \re_{1, t} \\ \re_{2, t} \\ \vdots \\ \re_{M, t}
    \end{bmatrix}
\label{def:var-complete}
\end{align}

or equivalently in compact a form:

\begin{equation}
    \rvx_{t} = \vc + \mA_{1} \rvx_{t-1} + \mA_{2} \rvx_{t-2} + \cdots + \mA_{p} \rvx_{t-p} + \rve_{t}
    = \vc + \sum_{i=1}^{p} \mA_{i} \rvx_{t-i} + \rve_{t}
    \in \sR^{M}
\label{def:var}
\end{equation}

where $\vc \in \sR^{M}$ a vector of constants (intercepts), $\mA_{i} \in \sR^{M \times M}$ for $i=1,2,\ldots,p$,
the $p$ parameter matrices and $\rve_{t} \in \sR^{M}$ a stochastic term, noise. Hence,
VAR processes can adequately capture the dynamics of linear systems,
under the assumption that they follow a Markov process of
finite order, at most $p$ \citep{Murphy:2012:MLP:2380985}. In other words, the effectiveness of a $p$-th
order VAR process relies on the assumption that the last
$p$ observations have all the sufficient statistics and information to
predict and describe the future realizations of the process. As
a result, we \textit{enforce} a memory mechanism, keeping the
$p$ last values, $\mvX_{t-p:t-1}$, of the multivariate time-series and making predictions
according to (\ref{def:var}). Increasing the order of the model $p$,
results in increased computational and memory complexity as well as
a tendency to overfit the noise of the observed data. A
VAR$_{M}$($p$) process has:

\begin{equation}
    |P|_{\text{VAR}_{M}(p)} = M \times M \times p + M
\label{eq:var-number-of-parameters}
\end{equation}

parameters, hence they increase linearly with
the model order. The systematic selection of the model order $p$
can be achieved by minimizing an information criterion, such as the
\textbf{Akaike Information Criterion} (AIC) \citep{mandic:asp-arma}, given by:

\begin{equation}
    p_{\text{AIC}} = \min_{p \in \sN} \bigg[ ln(\text{MSE}) + \frac{2p}{N} \bigg]
\label{def:aic}
\end{equation}

where MSE the mean squared error of the model and
$N$ the number of samples.

After careful investigation of equation (\ref{def:var}), we note that the target $\rvvx_{t}$ is
given by an ensemble (i.e., linear combination) of $p$ linear regressions, where
the $i$-th regressor\footnote{Let one of the regressors to has
a bias vector that corresponds to $\vc$.} has (trainable) weights $\mA_{i}$
and features $\rvx_{t-i}$. This interpretation of a VAR model allows us to interpret its
strengths and weaknesses on a common basis with the neural
network architectures, covered in subsequent parts. Moreover, this enables adaptive training
(e.g., via Least-Mean-Square \citep{mandic2004generalized} filter), which will prove useful in online learning,
covered in Section \ref{sec:model-based-reinforcement-learning}.

\begin{figure}[h]
    \centering
    \begin{subfigure}[t]{0.48\textwidth}
        \includegraphics[width=\textwidth]{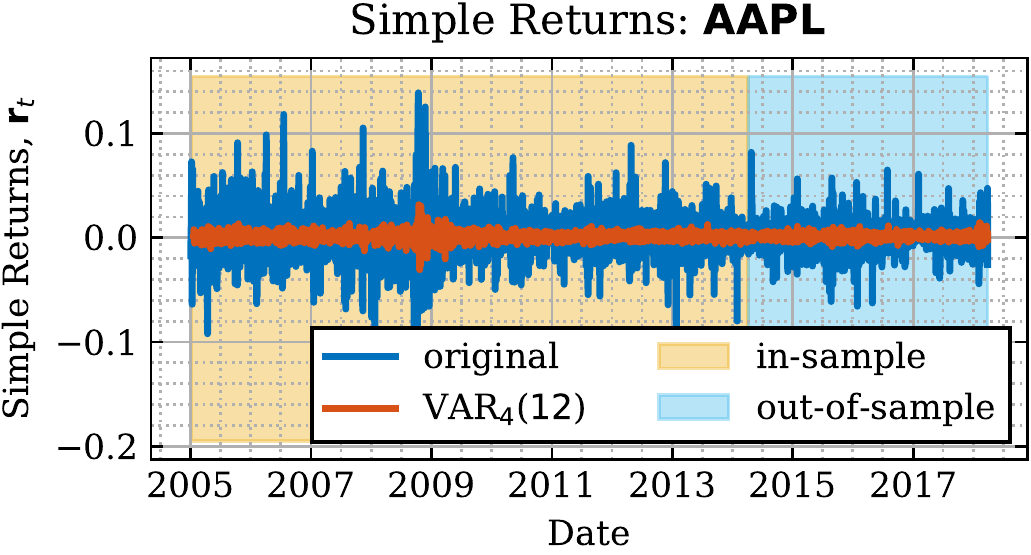}
    \end{subfigure}
    ~ 
    \begin{subfigure}[t]{0.48\textwidth}
        \includegraphics[width=\textwidth]{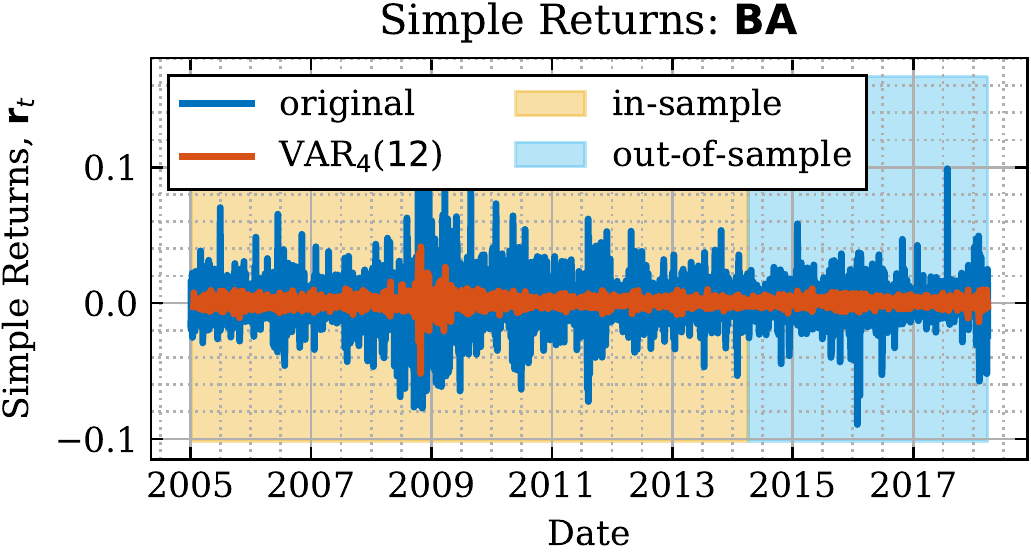}
    \end{subfigure}
    
    \vspace{0.5cm}
    
    \begin{subfigure}[t]{0.48\textwidth}
        \includegraphics[width=\textwidth]{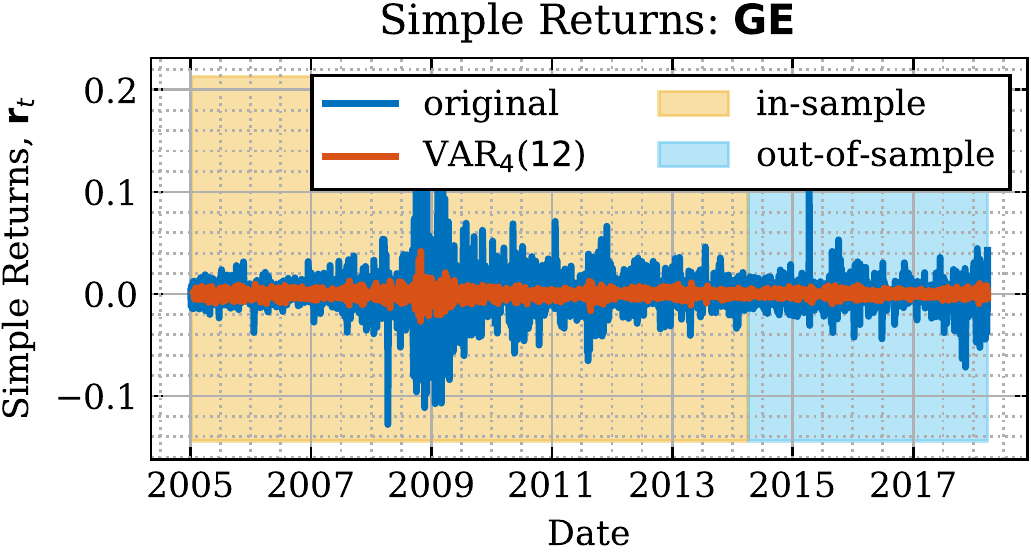}
    \end{subfigure}
    ~    
    \begin{subfigure}[t]{0.48\textwidth}
        \includegraphics[width=\textwidth]{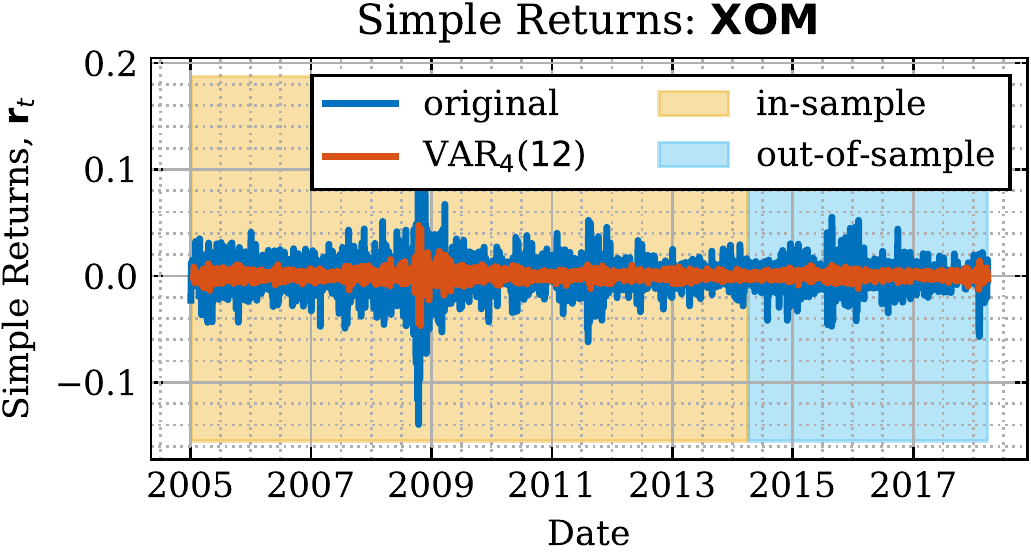}
    \end{subfigure}
    \caption{Vector autoregressive (VAR) time-series predictive model for assets simple returns.
    One step prediction is performed, where the realized observations are
    used as they come.}
    \label{fig:var-model-time-series}
\end{figure}

Figure \ref{fig:var-model-time-series} illustrates a fitted VAR$_{4}$($12$) process,
where the $p = p_{\text{AIC}} = 12$. We note that both in-sample (i.e., training) performance and
out-of-sample (i.e., testing) performance are poor (see Table \ref{tab:var-rnn-mse}), despite the large
number of parameters  $|P|_{\text{VAR}_{4}(12)} = 196$.

\subsection{Recurrent Neural Networks (RNN)} \label{sub:rnn}

\textbf{Recurrent neural networks} (RNN) \citep{mandic:rnn}, are a family of neural networks with feedback loops
which are very successful in processing sequential data. Most recurrent networks can also process
sequences of variable length \citep{deep-learning:goodfellow}.

Consider the classical form of a dynamical system:

\begin{equation}
    \vs_{t} = f(\vs_{t-1}, \vx_{t}; \vtheta)
\label{def:dynamical-system}
\end{equation}

where $\vs_{t}$ and $\vx_{t}$ the system state and input signal at time step $t$,
respectively, while $f$ a function parametrized by $\vtheta$ that maps the
previous state and the input signal to the new state.
Unfolding the recursive definition in (\ref{def:dynamical-system}) for a finite value of $t$:

\begin{align}
    \vs_{t} &= f(\vs_{t-1}, \vx_{t};\ \vtheta) \nonumber\\
    \vs_{t} &= f(f(\vs_{t-2}, \vx_{t-1};\ \vtheta), \vx_{t};\ \vtheta) \nonumber\\
    \vs_{t} &= f(f(f(\cdots(f(\cdots), \vx_{t-1};\ \vtheta), \vx_{t};\ \vtheta)))
\label{def:dynamical-system-unrolled}
\end{align}

In general, $f$ can be a highly non-linear function. Interestingly, a composite
function of nested applications of $f$ is responsible for generating the
next state.

Many recurrent neural networks use equation (\ref{def:rnn-hidden-state}) or a similar equation
to define the values of their hidden units. To indicate
that the state is the hidden units of the network,
we now rewrite equation (\ref{def:dynamical-system}) using the variable $\vh$ to represent the state:

\begin{equation}
    \vh_{t} = f(\vh_{t-1}, \vx_{t}; \vtheta)
\label{def:rnn-hidden-state}
\end{equation}

Then the hidden state $\vh_{t}$ can be used to obtain the
output signal $\vy_{t}$ (i.e., observation) at time index $t$, assuming a non-linear relationship,
described by function $g$ that is parametrized by $\vphi$:

\begin{equation}
    \hat{\vy}_{t} = g(\vh_{t};\ \vphi)
\label{def:rnn-output}
\end{equation}

\begin{figure}[h]
    \centering
    \includegraphics[width=0.75\textwidth]{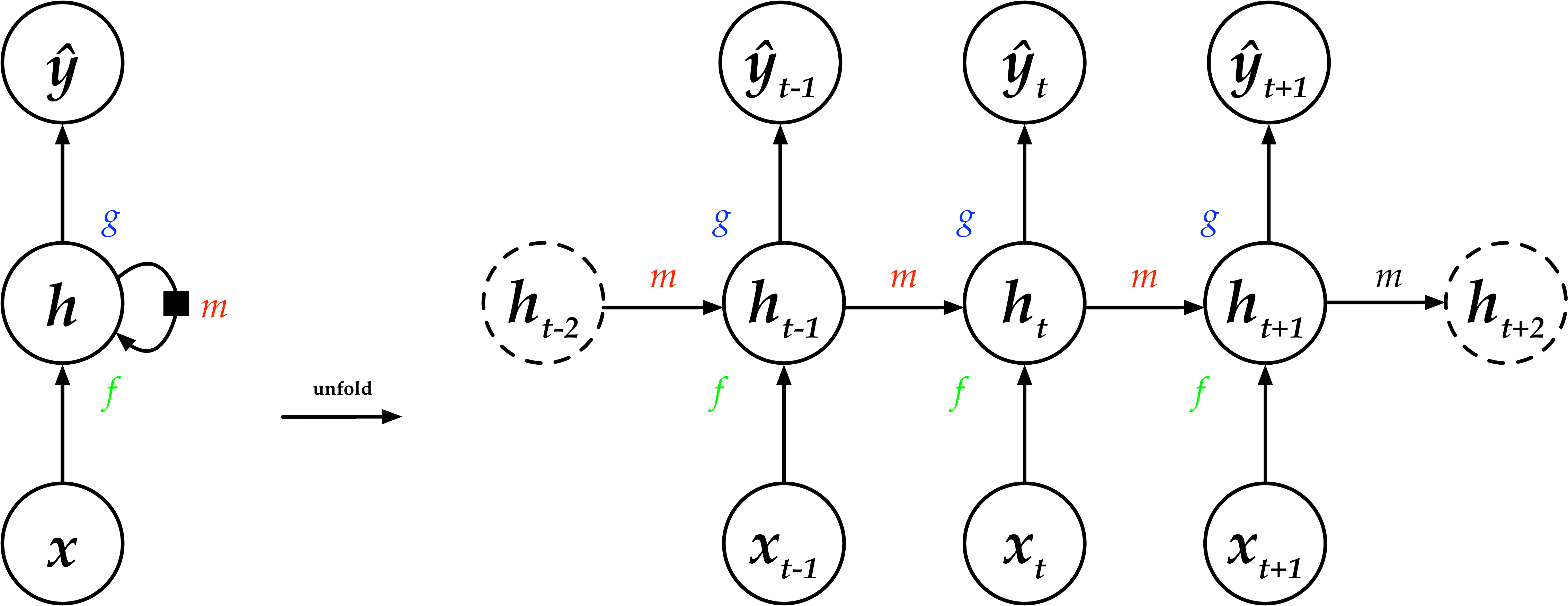}
    \caption{A generic recurrent network computational graph. This recurrent network
    processes information from the input $\vx$ by incorporating it into the
    state $\vh$ that is passed forward through time, which in turn
    is used to predict the target variable $\hat{\vy}$. (\textit{Left}) Circuit diagram.
    The black square indicates a delay of a single time step. (\textit{Right})
    The same network seen as an unfolded computational graph, where
    each node is now associated with one particular time instance \citep{deep-learning:goodfellow}.}
    \label{fig:rnn}
\end{figure}

The computational graph corresponding to (\ref{def:rnn-hidden-state}) and (\ref{def:rnn-output}) is provided in Figure
\ref{fig:rnn}. It can be shown that recurrent neural networks\footnote{And any neural
network with certain non-linear activation functions, in general.} are universal
function approximators \citep{cybenko1989approximation}, which means that if there is a relationship between
past states and current input with next states, RNNs have
the capacity to model it.

Another important aspect of RNNs is \textit{parameter sharing}. Note in (\ref{def:dynamical-system-unrolled})
that $\vtheta$ are the only parameters, shared between time steps. Consequently, the
number of parameters of the model decreases significantly, enabling faster
training and limiting model overfitting \citep{deep-learning:graves}, compared to feedforward neural networks
(i.e., multi-layer-perceptrons), which do not allow loops or any recursive connection.
Feedforward networks can be also used with sequential data  when
memory is brute-forced\footnote{Similar to VAR process memory mechanics.}, leading
to very large and deep architectures, and requiring a lot
more time to train and effort to avoid overfitting, in
order to achieve similar results with smaller RNNs \citep{mandic:rnn}.

% The basic RNN cell in Figure \ref{fig:rnn} can be configured in
% various ways to simulate $f$ from (\ref{def:rnn-hidden-state}). The most popular implementation includes
% two affine transformations and a non-linear activation:

% \begin{align}
%     \vh_{t} &= \sigma( \vb + \mW \vh_{t-1} + \mU \vx_{t} ) \\
%     \vy_{t} &= \vc + \mV \vh_{t}
% \end{align}

% where $\sigma$ a non-linearity, such as the sigmoid or the rectified
% linear unit (ReLU) \citep{deep-learning:goodfellow}, $\vb$ and $\vc$ bias vectors of the two affine
% transformations and $\mW$, $\mU$ and $\mV$ the coefficient matrices of the affine transformations.

Simple RNNs, such that the one implementing equation (\ref{def:dynamical-system-unrolled}),
are not used because of the vanishing gradient problem
\citep{hochreiter1998vanishing, deep-learning:vanishing-gradient}, but instead variants, such as the
\textbf{Gated Rectified Unit}s (GRU),
are preferred because of their low computational complexity\footnote{Compared to LSTM
\citep{deep-learning:rnn-comparison}.} and simple architecture. The most variants introduce some
filter/forget mechanism, responsible for selectively filtering out (or "forgetting")
past states, shaping the hidden state $\vh$ in a highly non-linear
fashion, but without accumulating the effects from all the history
of observations, alleviating the vanishing gradients problem. Consulting the schematic
in Figure \ref{fig:rnn}, this filtering operation is captured by $m$ (in red),
which stands for selective "memory".

Given a loss function $\mathcal{L}$ and historic data $\mathcal{D}$, then the process
of training involves minimization of $\mathcal{L}$ conditioned on $\mathcal{D}$, where the parameters
$\vtheta$ and $\vphi$ are the decision variables of the optimization problem:

\begin{equation}
    \underset{\vtheta,\ \vphi}{\text{minimize}} \quad \mathcal{L}(\vtheta, \vphi;\ \mathcal{D})
\label{obj:rnn}
\end{equation}

which is usually addressed by adaptive variants of Stochastic Gradient
Descent (SGD), such as \textbf{Adam} \citep{kingma2014adam}, to ensure faster convergence and
avoid saddle points. All these optimization algorithms
rely on (estimates of) descent directions, obtained by the gradients of the
loss function $\mathcal{L}$ with respect to the network parameters ($\vtheta,\ \vphi$), namely $\nabla_{\vtheta}(\mathcal{L})$
and $\nabla_{\vphi}(\mathcal{L})$. Due to the parameter sharing mechanics of RNNs, obtaining
the gradients is non-trivial and thus \textbf{Backpropagation Through Time} (BPTT) \citep{deep-learning:bptt}
algorithm is used, which efficiently calculates the contribution of each
parameter to the loss function across all time steps.

The selection of the hyperparameters, such as the size of
the hidden state $\vh$ and the activation functions, can be performed
using cross-validation or other empirical methods. Nonetheless, we choose
to allow excess degrees of freedom to our model but
regularize it using weight decay in the form of $\normlone$ and $\normltwo$ norms,
as well as dropout, according to the \citet{gal2016theoretically} guidelines.

As any neural network layer, recurrent layers can be stack
together or connected with other layers (i.e., affine or convolutional layers)
forming deep architectures, capable of dealing with complex datasets.

\begin{figure}[h]
    \centering
    \begin{subfigure}[t]{0.48\textwidth}
        \includegraphics[width=\textwidth]{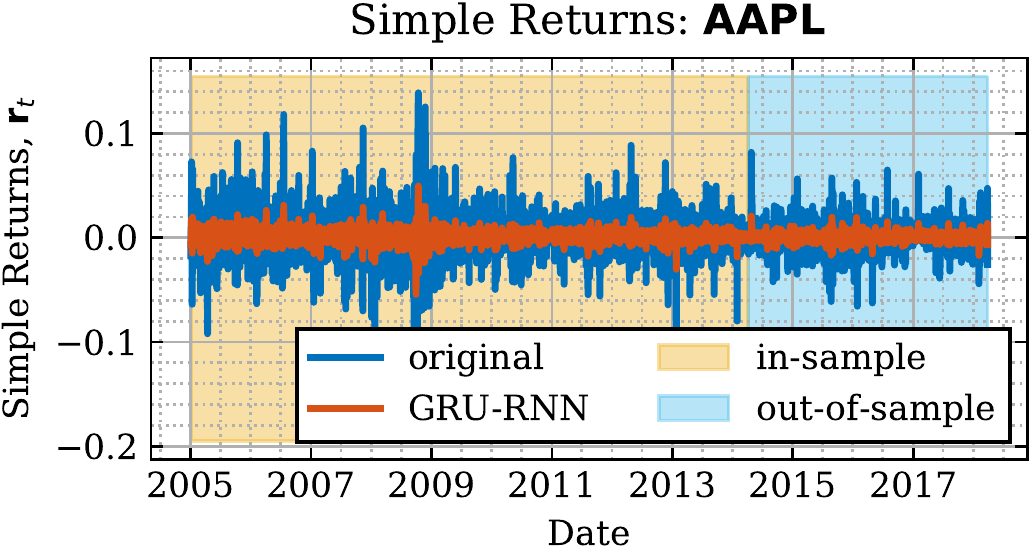}
    \end{subfigure}
    ~ 
    \begin{subfigure}[t]{0.48\textwidth}
        \includegraphics[width=\textwidth]{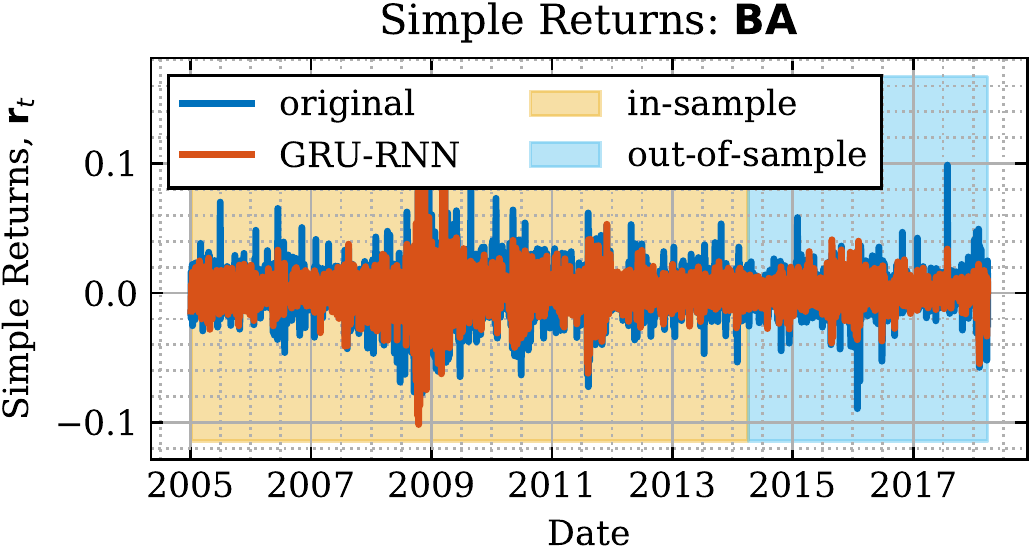}
    \end{subfigure}
    
    \vspace{0.5cm}
    
    \begin{subfigure}[t]{0.48\textwidth}
        \includegraphics[width=\textwidth]{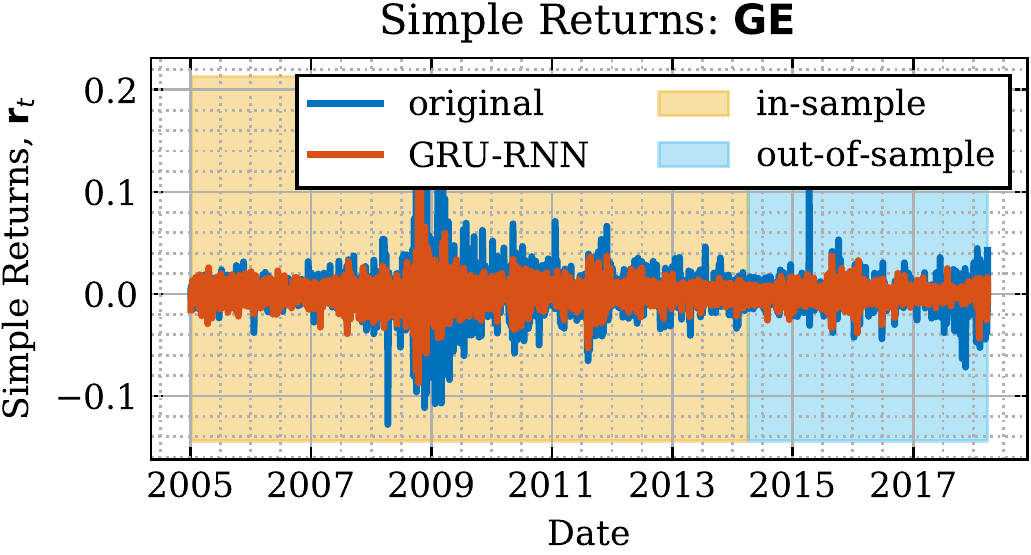}
    \end{subfigure}
    ~    
    \begin{subfigure}[t]{0.48\textwidth}
        \includegraphics[width=\textwidth]{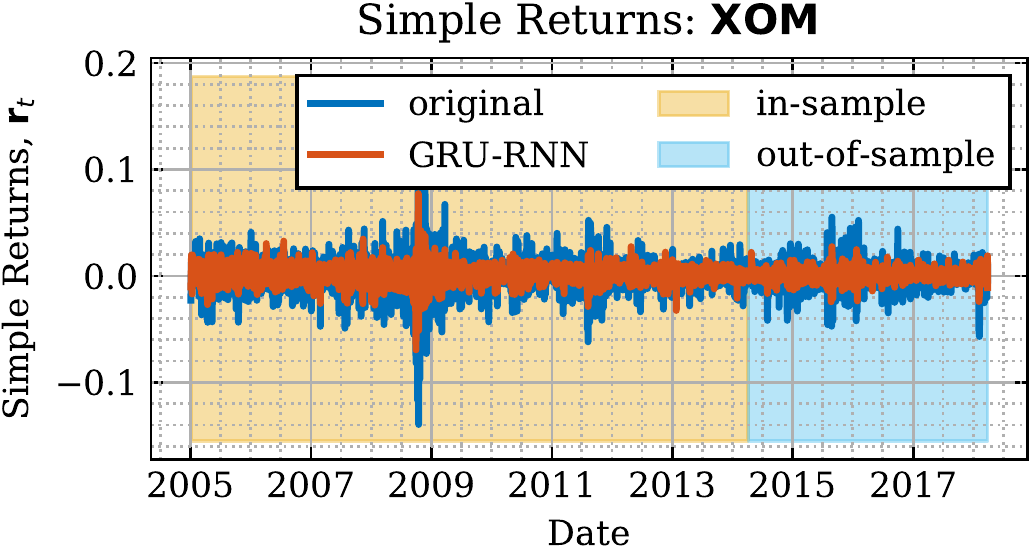}
    \end{subfigure}
    \caption{Gated recurrent unit recurrent neural network (GRU-RNN) time-series predictive model
    for assets simple returns. One step prediction is performed, where
    the realized observations are used as they come (i.e., set observation
    $\vy_{t-1}$ equal to $\vx_{t}$, rather than predicted value $\hat{\vy_{t-1}}$).}
    \label{fig:rnn-model-time-series}
\end{figure}

For comparison with the VAR$_{4}$($12$) process in Section \ref{sub:var}, we train
an RNN, comprised of two layers, one GRU layer followed by an affine layer, where
the size of the hidden state is 3 or $\vh \in \sR^{3}$. The
number of model parameters is $|P|_{\text{GRU-RNN}(4\rightarrow3\rightarrow4)} = 88$, but it significantly outperforms the
VAR model, as suggested by Figure \ref{fig:rnn-model-time-series} and summary in Table \ref{tab:var-rnn-mse}.

\begin{table}[h]
\centering
\begin{tabular}{|l||c|c||c|c|} 
\hline
& \multicolumn{2}{c||}{\textbf{Training Error}} & \multicolumn{2}{c|}{\textbf{Testing Error}}\\
\hline
                & \textbf{VAR}  & \textbf{GRU-RNN}  & \textbf{VAR}  & \textbf{GRU-RNN}  \\
\hline\hline
\texttt{AAPL}   & \cred0.000411 & \cgreen0.000334   & \cred0.000520 & \cgreen0.000403   \\ \hline
\texttt{BA}     & \cred0.000315 & \cgreen0.000090   & \cred0.000415 & \cgreen0.000124   \\ \hline
\texttt{GE}     & \cred0.000284 & \cgreen0.000132   & \cred0.000392 & \cgreen0.000172   \\ \hline
\texttt{XOM}    & \cred0.000210 & \cgreen0.000140   & \cred0.000341 & \cgreen0.000198   \\ \hline
\end{tabular}
\caption{Training and testing Mean Square Error (MSE) of VAR and GRU-RNN for Figures \ref{fig:var-model-time-series},
\ref{fig:rnn-model-time-series}. Despite the lower number of parameters, the GRU-RNN model can capture
the non-linear dependencies and efficiently construct a "memory" (hidden state)
to better model temporal dynamics.}
\label{tab:var-rnn-mse}
\end{table}

%% assets folder to reference
\renewcommand{\assets}{report/background/portfolio-optimization/assets}

\chapter{Portfolio Optimization} \label{ch:portfolio-optimization}

The notion of a portfolio has already been introduced in
subsection \ref{sub:portfolio}, as a master asset, highlighting its representation advantage over
single assets. Nonetheless, portfolios allow also investors to combine properties
of individual assets in order to "amplify" the positive aspects
of the market, while "attenuating" its negative impacts on the
investment.

\begin{figure}[h]
    \centering
    \begin{subfigure}[t]{0.48\textwidth}
        \includegraphics[width=\textwidth]{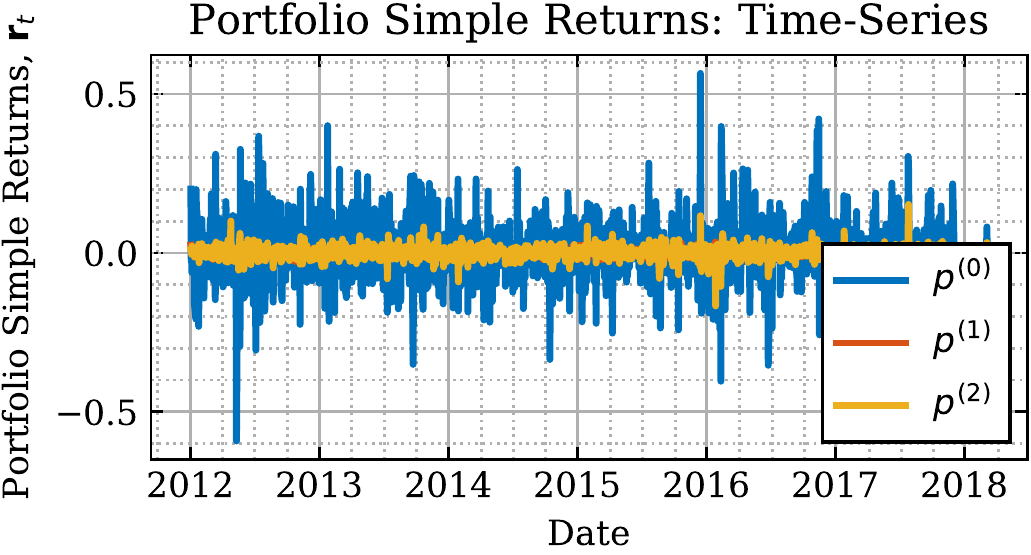}
    \end{subfigure}
    ~ 
    \begin{subfigure}[t]{0.48\textwidth}
        \includegraphics[width=\textwidth]{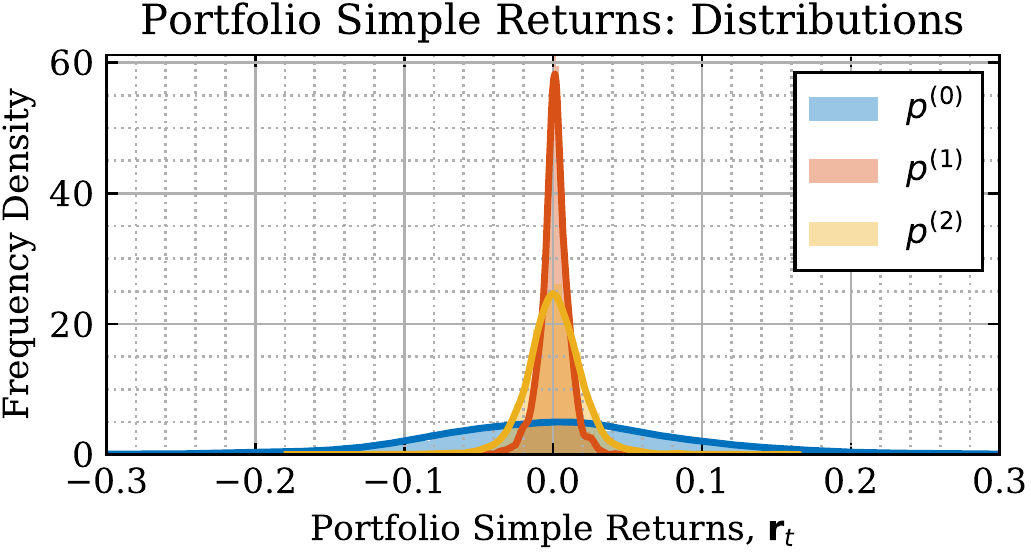}
    \end{subfigure}
    \caption{Simple returns of randomly allocated portfolios.}
    \label{fig:random-portfolio}
\end{figure}

Figure \ref{fig:random-portfolio} illustrates three randomly generated portfolios (with fixed portfolio weights
over time given in Table \ref{tab:random-portfolio-weights}), while Table \ref{tab:random-portfolio-summary} summarizes their
performance. Importantly, we note the significant differences between the random
portfolios, highlighting the importance that portfolio construction and asset allocation
plays in the success of an investment. As Table \ref{tab:random-portfolio-weights} implies,
short-selling is allowed in the generation of the random portfolios
owing to the negative portfolio weights, such as $\vp_{AAPL}^{(0)}$ and $\vp_{MMM}^{(2)}$.
Regardless, portfolio vector definition (\ref{def:portfolio-vector}) is satisfied in all cases, since
the portfolio vectors' elements sum to one (column-wise addition).

\begin{table}[h]
\centering
\begin{tabular}{|l||c|c|c|}
\hline
\multicolumn{4}{|c|}{\textbf{Randomly Allocated Portfolios Performance Summary}}                 \\ \hline\hline
                    Performance Metrics &    $\vp^{(0)}$ &       $\vp^{(1)}$ &       $\vp^{(2)}$ \\ \hline\hline
\textbf{Mean Returns (\%)}              &  \cred0.334685 &   \cgreen0.097516 &         0.0869129 \\ \hline
\textbf{Cumulative Returns (\%)}        &  \cred-83.8688 &    \cgreen297.869 &           167.605 \\ \hline
\textbf{Volatility (\%)}                &   \cred9.49456 &    \cgreen0.94646 &           2.02911 \\ \hline
\textbf{Sharpe Ratio}                   &   \cred1.35885 &    \cgreen3.97176 &           1.65115 \\ \hline
\textbf{Max Drawdown (\%)}              &   \cred173.656 &    \cgreen44.8894 &           101.782 \\ \hline
\textbf{Average Drawdown Time (days)}   &        \cred70 &          \cgreen7 &                24 \\ \hline
\textbf{Skewness}                       &  \cred0.162362 &  \cgreen-0.242675 &        -0.0217292 \\ \hline
\textbf{Kurtosis}                       &        3.27544 &           1.90174 &           7.83347 \\ \hline
\textbf{Value at Risk, $c=0.05$ (\%)}   & \cred-14.0138  &   \cgreen-1.52798 &          -2.88773 \\ \hline
\textbf{Conditional Value at Risk (\%)} &   \cred-20.371 &   \cgreen-2.18944 &          -4.44684 \\ \hline
\textbf{Hit Ratio (\%)}                 &\cyellow50.9091 &     \cgreen56.633 &           51.4478 \\ \hline
\textbf{Average Win to Average Loss}    &        1.06156 &           1.01662 &    \cgreen1.06534 \\ \hline
\end{tabular}
\caption{Performance summary of Figure \ref{fig:random-portfolio}, illustrating a tremendous
impact of portfolio construction on performance. Portfolio 0 ($\vp^{(0)}$) is outperformed by portfolio 1 ($\vp^{(1)}$)
in all metrics, motivating the introduction of portfolio optimization.}
\label{tab:random-portfolio-summary}
\end{table}

\begin{table}[h]
\centering
\begin{tabular}{|c||c|c|c|}
\hline
              &  $p^{(0)}$ &  $p^{(1)}$ &  $p^{(2)}$ \\
\hline\hline
\texttt{AAPL} &  -2.833049 &   0.172436 &   0.329105 \\
\hline
\texttt{GE}   &  -2.604941 &  -0.061177 &   0.467233 \\
\hline
\texttt{BA}   &   3.622328 &   0.313936 &   1.484903 \\
\hline
\texttt{JPM}  &   6.764848 &   0.233765 &   0.092537 \\
\hline
\texttt{MMM}  &  -3.949186 &   0.341040 &  -1.373778 \\
\hline
\end{tabular}
\caption{Random Portfolio Vectors of Figure \ref{fig:random-portfolio}.}
\label{tab:random-portfolio-weights}
\end{table}

\textbf{Portfolio Optimization} aims to address the allocation problem in a
systematic way, where an objective function reflecting the investor's preferences
is constructed and optimized with respect to the portfolio vector.

In this chapter, we introduce the Markowitz Model (section \ref{sec:markowitz-model}), the first
attempt to mathematically formalize and suggest an optimization method to
address portfolio management. Moreover, we extend this framework to generic
utility and objective functions (section \ref{sec:generic-objective-functions}), by taking transaction costs into
account (section \ref{sec:transaction-cost-optimization}). The shortcomings of the methods discussed here encourage
the development of context-agnostic agents, which is the focus of
this thesis (section \ref{sec:model-free-reinforcement-learning}). However, the simplicity and robustness of the
traditional portfolio optimization methods have motivated the supervised pre-training (Chapter \ref{ch:pre-training})
of the agents with Markowitz-like models as ground truths.

\section{Markowitz Model} \label{sec:markowitz-model}

The \textbf{Markowitz model} \citep{finance:markowitz-2,finance:markowitz-1} mathematically formulates the portfolio allocation
problem, namely finding a portfolio vector $\vw$ in a universe of $M$ assets, according
to the investment Greedy (Investment Advice \ref{adv:greedy-criterion}) and Risk-Aversion (Investment
Advice \ref{adv:risk-aversion-criterion}) criteria, constrained on the portfolio vector definition (\ref{def:portfolio-vector}).
Hence, we the Markowitz model gives the optimal portfolio vector
$\vw_{*}$ which minimizes volatility for a given returns level, such that:

\begin{equation}
    \sum_{i=1}^{M} \evw_{*, i} = 1, \quad \vw_{*} \in \sR^{M}
\end{equation}

\subsection{Mean-Variance Optimization} \label{sub:mean-variance-optimization}

For a trading universe of $M$ assets, provided historical data,
we obtain empirical estimates of the expected returns
$
\vmu =
\begin{bmatrix}
    \mu_{1}, & \mu_{2}, & \ldots & \mu_{M}
\end{bmatrix}^{T} \in \sR^{M}
$,
where $\mu_{i}$ the sample mean (\ref{def:univariate-empirical-mean}) of the $i$-th asset and the
covariance
$
\mSigma \in \sR^{M \times M}
$,
such that $\Sigma_{ij}$ the empirical covariance (\ref{def:bessels-correction-formula-covariance}) of the $i$-th and
the $j$-th assets. 

For given target expected return $\bar{\mu}_{target}$, determine the portfolio vector $\vw \in \sR^{M}$ such that:

\begin{align}
\underset{\vw}{\text{minimize}} \quad &\frac{1}{2}\vw^{T} \mSigma \vw
\label{obj:markowitz-model-1}\\
\text{subject to} \quad &\vw^{T} \vmu = \bar{\mu}_{target}
\label{con:markowitz-model-2}\\
\text{and} \quad &\boldsymbol{1}_{M}^{T}\vw = 1
\label{con:markowitz-model-3}
\end{align}

where $\sigma^{2} = \vw^{T} \mSigma \vw$ is the portfolio variance, $\mu = \vw^{T} \vmu$ the portfolio expected return and the
$M$-dimensional column vector of ones is denoted by $\boldsymbol{1}_{M}$.

For Lagrangian multipliers $\lambda, \kappa \in \sR$, we form the Lagrangian function $\mathcal{L}$
\citep{optimization:papadimitriou} such that:
\begin{equation}
\mathcal{L}(\vw, \lambda, \kappa) =
\frac{1}{2}\vw^{T} \mSigma \vw
- \lambda (\vw^{T}\vmu - \bar{\mu}_{target})
- \kappa (\boldsymbol{1}_{M}^{T}\vw - 1)
\label{def:markowitz-model-lagrangian}
\end{equation}

We differentiating the Lagrangian function $\mathcal{L}$
\footnote{The covariance matrix $\mSigma$ is by definition (\ref{def:covariance}) symmetric, so
$\frac{\partial(\vw^{T} \mSigma \vw)}{\partial \vw} = \mSigma \vw$.}
and apply the first order necessary condition of optimality:

\begin{equation}
\frac{\partial \mathcal{L}}{\partial \vw} =
\boldsymbol{\Sigma} \vw
- \lambda \boldsymbol{\mu}
- \kappa \boldsymbol{1}_{M}
= 0
\label{def:markowitz-model-lagrangian-derivative}
\end{equation}

Upon combining the optimality condition equations (\ref{con:markowitz-model-2}),
(\ref{con:markowitz-model-3}) and (\ref{def:markowitz-model-lagrangian-derivative})
into a matrix form, we have:
\begin{equation}
\begin{bmatrix}
\mSigma & \vmu & \boldsymbol{1}_{M}\\
\vmu^{T} & 0 & 0\\
\boldsymbol{1}_{M}^{T} & 0 & 0\\
\end{bmatrix}
\begin{bmatrix}
\vw\\
-\lambda\\
-\kappa\\
\end{bmatrix}
=
\begin{bmatrix}
0\\
\bar{\mu}_{target}\\
\boldsymbol{1}_{M}\\
\end{bmatrix}
\label{eq:markowitz-model-optimality-conditions}
\end{equation}

Under the assumption that $\mSigma$ is full rank and $\vmu$ is not
a multiple of $\boldsymbol{1}_{M}$, then equation (\ref{eq:markowitz-model-optimality-conditions}) is solvable by matrix inversion
\citep{optimization:boyd}. The resulting portfolio vector $\vw_{MVP}$ defines the \textbf{mean-variance optimal portfolio}.

\begin{equation}
\begin{bmatrix}
\vw_{MVP}\\
-\lambda\\
-\kappa\\
\end{bmatrix}
=
\begin{bmatrix}
\boldsymbol{\Sigma} & \boldsymbol{\mu} & \boldsymbol{1}\\
\boldsymbol{\mu}^{T} & 0 & 0\\
\boldsymbol{1}^{T} & 0 & 0\\
\end{bmatrix}^{-1}
\begin{bmatrix}
0\\
\bar{\mu}_{target}\\
\boldsymbol{1}\\
\end{bmatrix}
\label{eq:markowitz-model-solution}
\end{equation}

Note that mean-variance optimization is also used in signal processing and
wireless communications in order to determine the optimal beamformer \citep{sp:beamforming},
using, for example, Minimum Variance Distortion Response filters \citep{mandic:mvdr}.

\subsection{Quadratic Programming} \label{sub:quadratic-programming}

Notice that the constraint (\ref{con:markowitz-model-3}) suggests that short-selling is allowed. This simplifies
the formulation of the problem by relaxing conditions, enabling a
closed form solution (\ref{eq:markowitz-model-solution}) as a set of linear equations.
If short sales are prohibited, then an additional constraint should
be added, and in this case, the optimization problem becomes:

% \optim{
\begin{align}
\underset{\vw}{\text{minimize}} \quad &\vw^{T} \mSigma \vw\nonumber\\
\text{subject to} \quad &\vw^{T}\vmu = \bar{\mu}_{target}\nonumber\\
\text{and} \quad &\boldsymbol{1}_{M}^{T}\vw = 1\nonumber\\
\text{and} \quad & \vw \succeq 0
\label{con:markowitz-model-4}
\end{align}
% }

where $\succeq$ designates an element-wise inequality operator.
This problem cannot be reduced to the solution of a set of linear equations.
It is termed a \textbf{quadratic program} and it is solved
numerically using gradient-based algorithms \citep{optimization:ricketts}.
Optimization problems with quadratic objective functions and linear constraints
fall into this framework.

\begin{figure}[h]
    \centering
    \begin{subfigure}[t]{0.48\textwidth}
        \includegraphics[width=\textwidth]{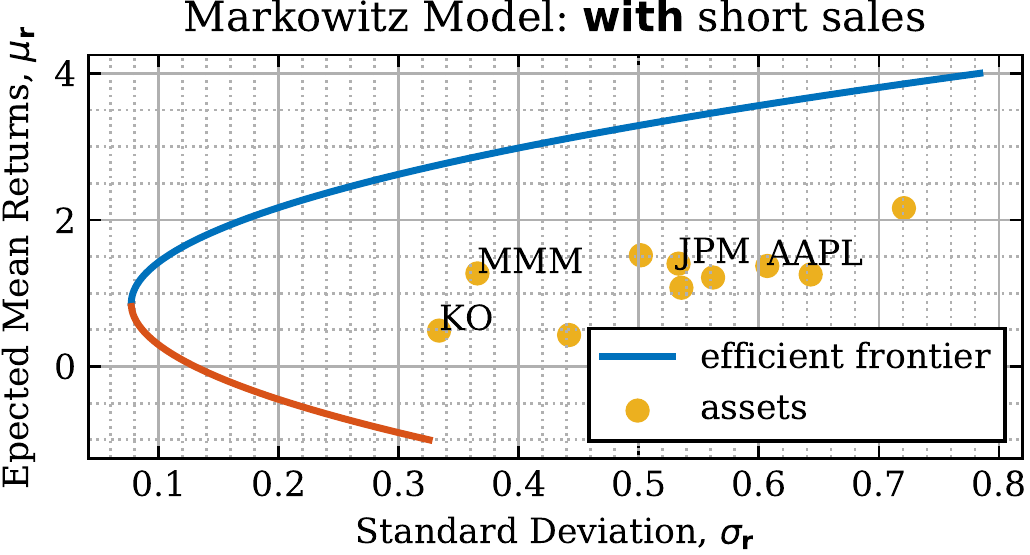}
    \end{subfigure}
    ~ 
    \begin{subfigure}[t]{0.48\textwidth}
        \includegraphics[width=\textwidth]{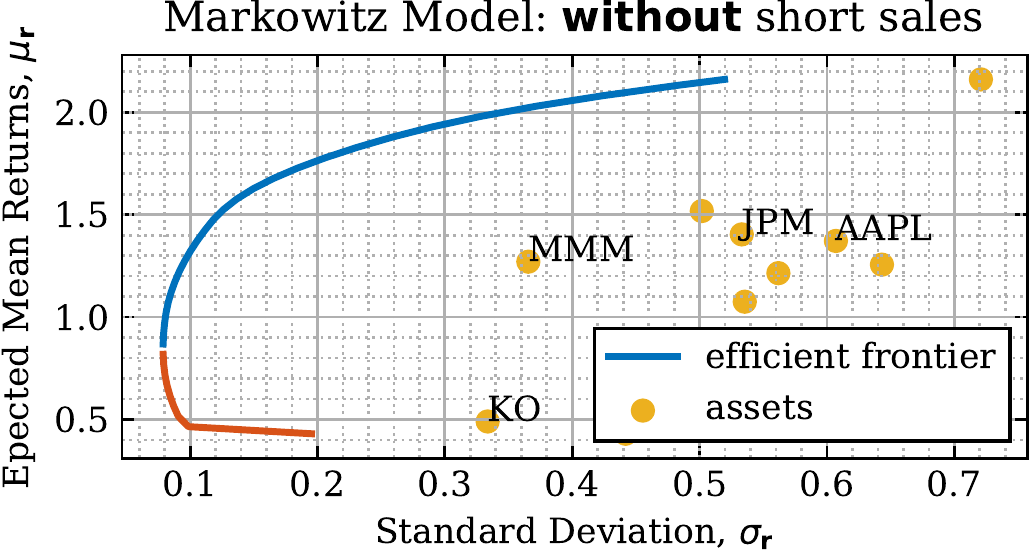}
    \end{subfigure}
    \caption{Efficient frontier for Markowitz model the with (\textit{Left}) and without (\textit{Right}) short-selling.
    The \textit{yellow} points are the projections of single assets historic performance on
    the $\sigma-\mu$ (i.e., volatility-returns) plane. The solid lines are the portfolios, obtained by
    solving the Markowitz model optimization problem (\ref{obj:markowitz-model-1}) for different
    values of $\bar{\mu}_{target}$. Note that the red points are rejected and only the \textit{blue} loci is efficient.}
    \label{fig:markowitz}
\end{figure}

\begin{remark}
Figure \ref{fig:markowitz} illustrates the volatility to expected returns
dependency of an example trading universe (i.e., yellow scatter points),
along with all optimal portfolio solutions with and
without short selling, left and right subfigures, respectively.
The blue part of the solid curve is termed \textbf{Efficient Frontier} \citep{finance:investment-science},
and the corresponding portfolios are called efficient. These portfolios are
obtained by the solution of (\ref{obj:markowitz-model-1}).
\end{remark}

Interestingly, despite the inferior performance
of the individual assets' performance, appropriate linear combinations of them
results in less volatile and more profitable master assets, demonstrating
once again the power of portfolio optimization. Moreover, we note that the red part of the solid curve
is inefficient, since for the same risk level (i.e., standard
deviation) there are portfolios with higher expected returns, which aligns
with the Greedy Criterion \ref{adv:greedy-criterion}. Finally, we highlight that in case
of short-selling (left subfigure) there are feasible portfolios, which have
higher expected returns than any asset from the universe. This
is possible since the low-performing assets can be shorted and the
inventory from them can be invested in high-performance assets, amplifying
their high returns. On the other hand, when short-selling is
not allowed, the expected returns of any portfolio is restricted in the
interval defined by the lowest and the highest empirical returns
of the assets in the universe.

\section{Generic Objective Functions} \label{sec:generic-objective-functions}

In Section \ref{sec:evaluation-criteria} we introduced various evaluation metrics that reflect our
investment preferences. Extending the vanilla Markowitz model which minimizes risk
(i.e., variance), constraint on a predetermined profitability level (i.e., expected returns), we
can select any metric as the optimization objective function, constrained
on the budget (\ref{con:markowitz-model-3}) and any other criteria we favor, as
long as the fit in the quadratic programming framework. More complex objectives
may be solvable with special non-linear optimizers, but there is no
general principle. In this Section we exhibit how to translate evaluation
metrics to objective functions, suitable for quadratic programming.
Any of the functions presented can be used with and without
short-selling, so in order to address the more difficult of the two cases,
we will consider that long positions are only allowed. The
other case is trivially obtained by ignoring the relevant constraint
in the sign of weights.

%%%
\subsection{Risk Aversion} \label{sub:risk-aversion}

Motivated by the Lagrangian formulation of the Markowitz model (\ref{def:markowitz-model-lagrangian}, we
define the \textbf{Risk Aversion} portfolio, named after the risk aversion coefficient
$\alpha \in \sR_{+}$, given by the solution of the program:

\begin{align}
\underset{\vw}{\text{maximize}} \quad &\vw^{T} \vmu - \alpha \vw^{T} \mSigma \vw \label{obj:risk-aversion}\\
\text{subject to} \quad &\boldsymbol{1}_{M}^{T}\vw = 1\nonumber\\
\text{and} \quad & \vw \succeq 0\nonumber
\end{align}

\begin{figure}[h]
    \centering
    \begin{subfigure}[t]{0.48\textwidth}
        \includegraphics[width=\textwidth]{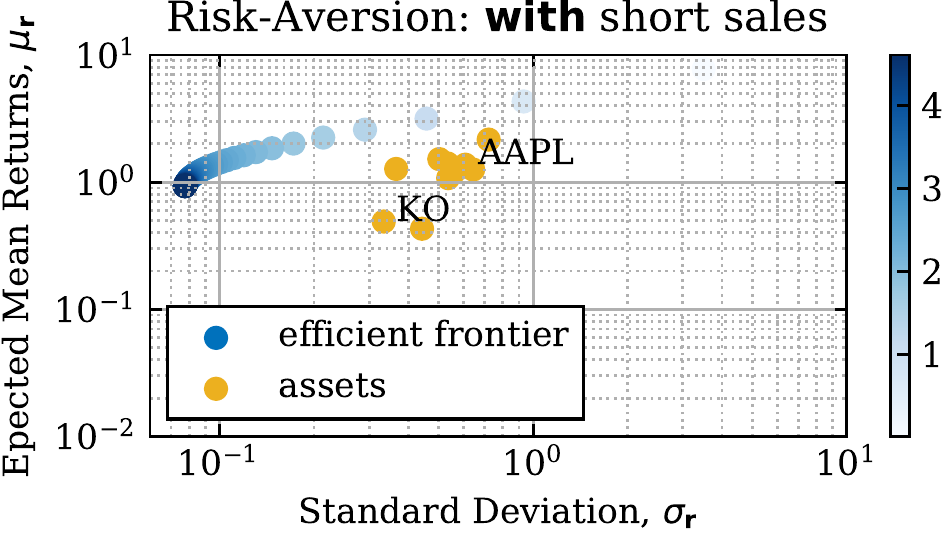}
    \end{subfigure}
    ~ 
    \begin{subfigure}[t]{0.48\textwidth}
        \includegraphics[width=\textwidth]{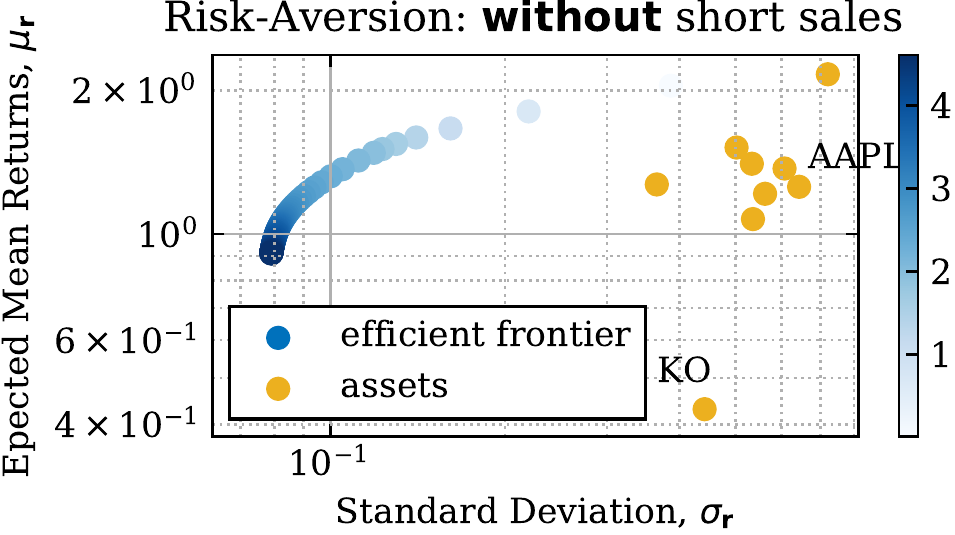}
    \end{subfigure}
    \caption{Risk-Aversion optimization efficient frontier with (\textit{Right}) and without (\textit{left}) short-selling.
    The \textit{yellow} points are the projections of single assets historic performance on
    the $\sigma-\mu$ (i.e., volatility-returns) plane. The \textit{blue} points are the efficient portfolios, or
    equivalently the solutions of the risk-aversion optimization problem (\ref{obj:risk-aversion}) for different
    values of $\alpha$, designated by the opacity of the blue color (see colorbar).}
    \label{fig:risk-aversion-optimization}
\end{figure}

The risk aversion coefficient $\alpha$ is model hyperparameter, which reflect the
trade-off between portfolio expected returns ($\vw^{T} \vmu$) and risk level ($\vw^{T} \mSigma \vw$) \citep{finance:quant-finance}.
For $\alpha \rightarrow 0$ the investor is infinitely greedy, they do not consider volatility
and aim to maximize only returns. On the other hand,
for $\alpha \rightarrow \infty$, the investor is infinitely risk-averse and selects the least
risky portfolio, regardless its returns performance. Any positive value for $\alpha$
results in a portfolio which balances the two objectives weighted by the risk aversion coefficient.
Figure \ref{fig:risk-aversion-optimization} illustrates the volatility-returns plane for the same trading universe
as in Figure \ref{fig:markowitz}, but the curve is obtained by the solution of (\ref{obj:risk-aversion}).
As expected, high values of $\alpha$ result in less volatile portfolios,
while low risk aversion coefficient leads to higher returns.

%%%
\subsection{Sharpe Ratio} \label{sub:sharpe-ratio-optimization}

Both objective functions so far require hyperparameter tuning ($\bar{\mu}_{target}$ or $\alpha$),
hence either cross-validation or hand-picked selection is required \citep{finance:stochastic-financial-models}. On the
other hand, in Section \ref{sec:evaluation-criteria} we motivated the use of Sharpe Ratio (\ref{def:sharpe-ratio})
as a Signal-to-Noise Ratio equivalent for finance, which is not
parametric. Considering the Sharpe Ratio as the objective function of
the program:

\begin{align}
\underset{\vw}{\text{maximize}} \quad &\frac{\vw^{T} \vmu}{\sqrt{\vw^{T} \mSigma \vw}} \label{obj:sharpe-ratio}\\
\text{subject to} \quad &\boldsymbol{1}_{M}^{T}\vw = 1\nonumber\\
\text{and} \quad & \vw \succeq 0\nonumber
\end{align}

Unlike the aforementioned methods, by solving the optimization problem in (\ref{obj:sharpe-ratio}) we obtain
a single portfolio which is guaranteed to be optimal for the provided universe.

\section{Transaction Costs} \label{sec:transaction-cost-optimization}

In real stock exchanges, such as NYSE \citep{wikipedia:nyse},
NASDAQ \citep{wikipedia:nasdaq} and LSE \citep{wikipedia:lse},
trading activities (buying or selling) are accompanied with
expenses, including brokers' commissions and spreads
\citep{investopedia:transaction-costs,quantopian:commission},
they are usually referred to as \textbf{transaction costs}.
Therefore every time a new portfolio vector is determined
(portfolio re-balancing), the corresponding transaction costs
should be subtracted from the budget.

In order to simplify the analysis around transaction costs
we will use the rule of thumb \citep{quantopian:commission},
charging $0.2\%$ for every activity.
For example, if three stocks A are bought with price $100\$$ each, then the
transaction costs will be $0.6\$$. Let the price of the stock A raising at price $107\$$,
when we decide to sell all three stocks, then the transaction costs will be $0.642\$$.

%%%
\subsection{Mathematical Formalization}

Given any objective function, $\mathcal{J}$, the transaction costs are subtracted
from the returns term in order to adjust the profit \& loss, accounting for the expenses of trading activities.
Therefore, we solve the optimization program:

\begin{align}
\underset{\vw}{\text{maximize}} \quad & \mathcal{J} - \boldsymbol{1}_{M}^{T} \beta \|\vw_{0} - \vw\|_{1}\label{obj:transaction-costs}\\
\text{subject to} \quad &\boldsymbol{1}_{M}^{T}\vw = 1\nonumber\\
\text{and} \quad & \vw \succeq 0\nonumber
\end{align}

where $\beta \in \sR$ the transactions cost (i.e., 0.002 for standard $0.2\%$ commissions),
and $\vw_{0} \in \sR^{M}$ the initial portfolio, since the last re-balancing. All the
parameters of the model\footnote{Not
considering the objective function's $\mathcal{J}$ parameters.} (\ref{obj:transaction-costs})
are given, since $\beta$ is market-specific,
and $\vw_{0}$ the current position. Additionally, the transactions cost term can be seen as
a regularization term which penalizes excessive trading and restricts large trades (i.e., large $\|\vw_{0} - \vw\|$).

The objective function $\mathcal{J}$ can be any function that can be optimized
according to the framework developed in Section \ref{sec:generic-objective-functions}.
For example the risk-aversion with transaction costs optimization program is given by:

\begin{equation}
    \underset{\vw}{\text{maximize}} \quad
    \vw^{T} \vmu - \alpha \vw^{T} \mSigma \vw - \boldsymbol{1}_{M}^{T} \beta \|\vw_{0} - \vw\|_{1}
\label{opt:risk-aversion-transaction-costs}
\end{equation}

while the Sharpe Ratio optimization with transaction costs is:

\begin{equation}
    \underset{\vw}{\text{maximize}} \quad
    \frac{\vw^{T} \vmu - \boldsymbol{1}_{M}^{T} \beta \|\vw_{0} - \vw\|_{1}} {\sqrt{\vw^{T} \mSigma \vw}}
\label{opt:sharpe-ratio-transaction-costs}
\end{equation}

We highlight that in (\ref{opt:risk-aversion-transaction-costs}) is subtracted from $\mathcal{J}$ directly
since all terms have the same units\footnote{The parameter $\alpha$ is not dimensionless.}, while in
(\ref{opt:sharpe-ratio-transaction-costs}) the transaction cost term is subtracted directly from the expected returns,
since Sharpe Ratio is unitless.

%%%
\subsection{Multi-Stage Decision Problem} \label{sub:multi-stage-dicesion-problem}

The involvement of transaction costs make Portfolio Management a \textbf{Multi-Stage Decision Problem}
\citep{paper:adaptive-dynamic-programming}, which in simple terms means that two sequences
of states with the same start and end state will have
different value. For instance, imagine two investors, both of which have an initial budget of $100\$$.
On Day 1, the first investor uses all of his budget
to construct a portfolio according to his preferred objective function,
paying $0.2\$$ for transaction costs, according to \citet{quantopian:commission}. By Day 3, the
market prices have changed but the portfolio of the first investor has not changed in value
and decided to liquidate all of his investment, paying another $0.2\$$ for selling his portfolio.
On Day 5 both of the investors (re-)enter the market make identical investments
and hence pay the same commission fees. Obviously, the two investors have
the same start and end states but their intermediate trajectories lead to different reward streams.

From (\ref{obj:transaction-costs}) it is obvious that $\vw_{0}$ affects the optimal allocation $\vw$.
In a sequential portfolio optimization setting, the past decisions (i.e., $\vw_{0}$)
will have a direct impact on the optimality of the future decisions
(i.e., $\vw$), therefore apart from the maximization of immediate rewards, we should also
focus on eliminating the negative effects on the future decisions.
As a consequence, sequential asset allocation is a multi-stage decision problem
where myopic optimal actions can lead to sub-optimal cumulative rewards.
This setting encourages the use of Reinforcement Learning agents which aim to maximize
long-term rewards, even if that means acting sub-optimally in the near-future
from the traditional portfolio optimization point of view.

%% assets folder to reference
\renewcommand{\assets}{report/background/reinforcement-learning/assets}

\chapter{Reinforcement Learning} \label{ch:reinforcement-learning}

% Having already motivated reinforcement learning to address the sequential portfolio
% management task (Section \ref{sub:multi-stage-dicesion-problem}), now, we introduce its fundamental
% principles  and components.

\textbf{Reinforcement learning} (RL) refers to both a learning problem and
a subfield of machine learning \citep{deep-learning:goodfellow}. As a learning problem
\citep{reinforcement-learning:ualberta}, it refers to learning to control a system (\textit{environment}) so
as to maximize some numerical value, which represents a long-term
objective (\textit{discounted cumulative reward signal}). Recalling the analysis in Section \ref{sub:multi-stage-dicesion-problem},
sequential portfolio management  

In this chapter we introduce the necessary tools to analyze
stochastic dynamical systems (Section \ref{sec:dynamical-systems}). Moreover, we review the major
components of a reinforcement learning algorithm (Section \ref{sec:major-components-of-reinforcement-learning}, as well as
extensions of the formalization of dynamical systems (Section \ref{sec:mdp-extensions}), enabling us
to reuse some of those tools to more general and
intractable otherwise problems.

\section{Dynamical Systems} \label{sec:dynamical-systems}

\begin{figure}[h]
    \centering
    \includegraphics[width=0.85\textwidth]{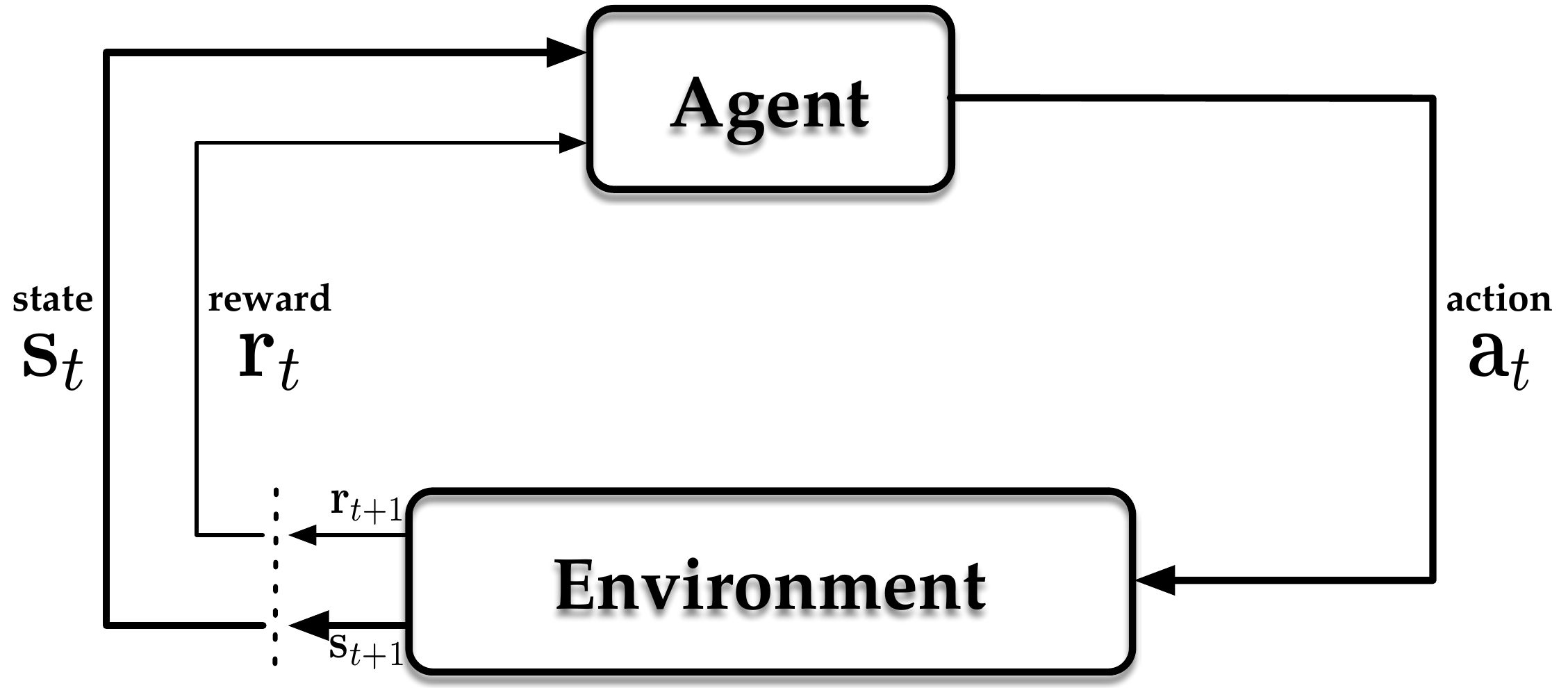}
    \caption{High-level stochastic dynamical system schematic \citep{reinforcement-learning:sutton}.}
    \label{fig:dynamical-system}
\end{figure}

Reinforcement learning is suitable in optimally controlling dynamical systems, such
as the general one illustrated in Figure \ref{fig:dynamical-system}: A controller (\textit{agent})
receives the controlled \textit{state} of the system and a \textit{reward} associated with
the last \textit{state transition}. It then calculates a control signal
(\textit{action}) which is sent back to the system. In response,
the system makes a transition to a \textit{new state} and
the cycle is repeated. The goal is to learn a way
of controlling the system (\textit{policy}) so as to maximize the total reward.
The focus of this report is on \textbf{discrete-time} dynamical systems, thought most of the notions
developed extend to continuous-time systems.

%%%
\subsection{Agent \& Environment} \label{sub:agent-and-environment}

The term \textbf{agent} is used to refer to the controller, while \textbf{environment}
is used interchangeably with the term system.
The goal of a reinforcement learning algorithm is the development (training) of an
agent capable of successfully interacting with the environment, such that it maximizes
some scalar objective over time.

% At each time step $t$, the agent (controller)
% \begin{itemize}
% \item executes action $A_{t}$
% \item receives observation $O_{t}$
% \item receives scalar reward $R_{t}$
% \end{itemize}
% while the environment (system)
% \begin{itemize}
% \item receives action $A_{t}$
% \item emits observation $O_{t+1}$
% \item emits scalar reward $R_{t+1}$
% \end{itemize}

%%%
\subsection{Action} \label{sub:action}

\textbf{Action} $\va_{t} \in \sA$ is the control signal that the agent sends back to the system
at time index $t$. It is the only way that the agent can influence the environment
state and as a result, lead to different reward signal sequences.
The \textbf{action space} $\sA$ refers to the set of actions that
the agent is allowed to take and it can be:
\begin{itemize}
    \item Discrete: $\sA = \{ a_{1}, a_{2}, \ldots, a_{M} \}$;
    \item Continuous: $\sA \subseteq [c, d]^{M}$.
\end{itemize}

%%%
\subsection{Reward} \label{sub:reward}

A \textbf{reward} $r_{t} \in \sB \subseteq \sR$ is a scalar feedback signal, which indicates how
well the agent is doing at discrete time step $t$. The agent
aims to maximize cumulative reward, over a sequence of steps.

Reinforcement learning addresses \textbf{sequential decision making} tasks
\citep{reinforcement-learning:silver-intro}, by training agents
that optimize delayed rewards and can evaluate the long-term consequences
of their actions, being able to sacrifice immediate reward to
gain more long-term reward. This special property of reinforcement learning
agents is very attractive for financial applications, where investment horizons
range from few days and weeks to years or decades.
In the latter cases, myopic agents can perform very poorly
since evaluation of long-term rewards is essential in order to succeed \citep{reinforcement-learning:a3c}.
However, the applicability of reinforcement learning depends vitally on the hypothesis:

\begin{hypothesis}[Reward Hypothesis]
    All goals can be described by the maximization of expected cumulative reward.
\label{hyp:reward}
\end{hypothesis}

Consequently, the selection of the appropriate reward signal for each application
is very crucial. It influences the agent learned strategies since it reflects its goals.
In Section \ref{sec:reward-signal}, a justification for the
selected reward signal is provided, along with an empirical comparison between
other metrics mentioned in Section \ref{sec:evaluation-criteria}.

%%%
\subsection{State \& Observation} \label{sub:state-and-observation}

The \textbf{state}, $\vs_{t} \in \sS$, is also a fundamental element of reinforcement learning, but it is usually
used to refer to both the environment state and the agent state. 

The agent does not always have direct access to the state, but at every time step, $t$,
it receives an observation, $\vo_{t} \in \sO$.

%%%%
\subsubsection{Environment State} \label{sub:environment-state}

The \textbf{environment state} $\vs_{t}^{e}$ is the internal representation of the system,
used in order to determine the next observation $\vo_{t+1}$ and reward $r_{t+1}$.
The environment state is usually invisible to the agent and even if it visible,
it may contain irrelevant information \citep{reinforcement-learning:sutton}.

%%%%
\subsubsection{Agent State}

The \textbf{history} $\vvh_{t}$ at time $t$ is the sequence of observations,
actions and rewards up to time step $t$, such that:
\begin{equation}
    \vvh_{t} = (\vo_{1}, \va_{1}, r_{1}, \vo_{2}, \va_{2}, r_{2}, \ldots, \vo_{t}, \va_{t}, r_{t})
\label{def:history}
\end{equation}

The \textbf{agent state} (a.k.a \textbf{state}) $\vs_{t}^{a}$ is the internal
representation of the agent about the environment, used in order to select the next action $\va_{t+1}$
and it can be any function of the history:
\begin{equation}
    \vs_{t}^{a} = f(\vvh_{t})
\label{def:agent-state}
\end{equation}
The term \textbf{state space} $\sS$ is used to refer to the set of possible states
the agents can observe or construct. Similar to the action space, it can be:
\begin{itemize}
    \item Discrete: $\sS = \{ s_{1}, s_{2}, \ldots, s_{n} \}$;
    \item Continuous: $\sS \subseteq \sR^{N}$.
\end{itemize}

%%%
\subsubsection{Observability} \label{sub:observability}

\textbf{Fully observable} environments allow the agent to directly observe the environment
state, hence:
\begin{equation}
\vo_{t} = \vs_{t}^{e} = \vs_{t}^{a}
\label{def:full-observability}
\end{equation}

\textbf{Partially observable} environments offer indirect access to the environment state,
therefore the agent has to construct its own state representation $\vs_{t}^{a}$ \citep{deep-learning:graves}, using:
\begin{itemize}
    \item Complete history: $\vs_{t}^{a} \equiv \vvh_{t}$;
    \item Recurrent neural network: $\vs_{t}^{a} \equiv f(\vs_{t-1}^{a}, \vo_{t};\ \vtheta)$.
\end{itemize}

\begin{figure}[h]
    \centering
    \includegraphics[width=0.85\textwidth]{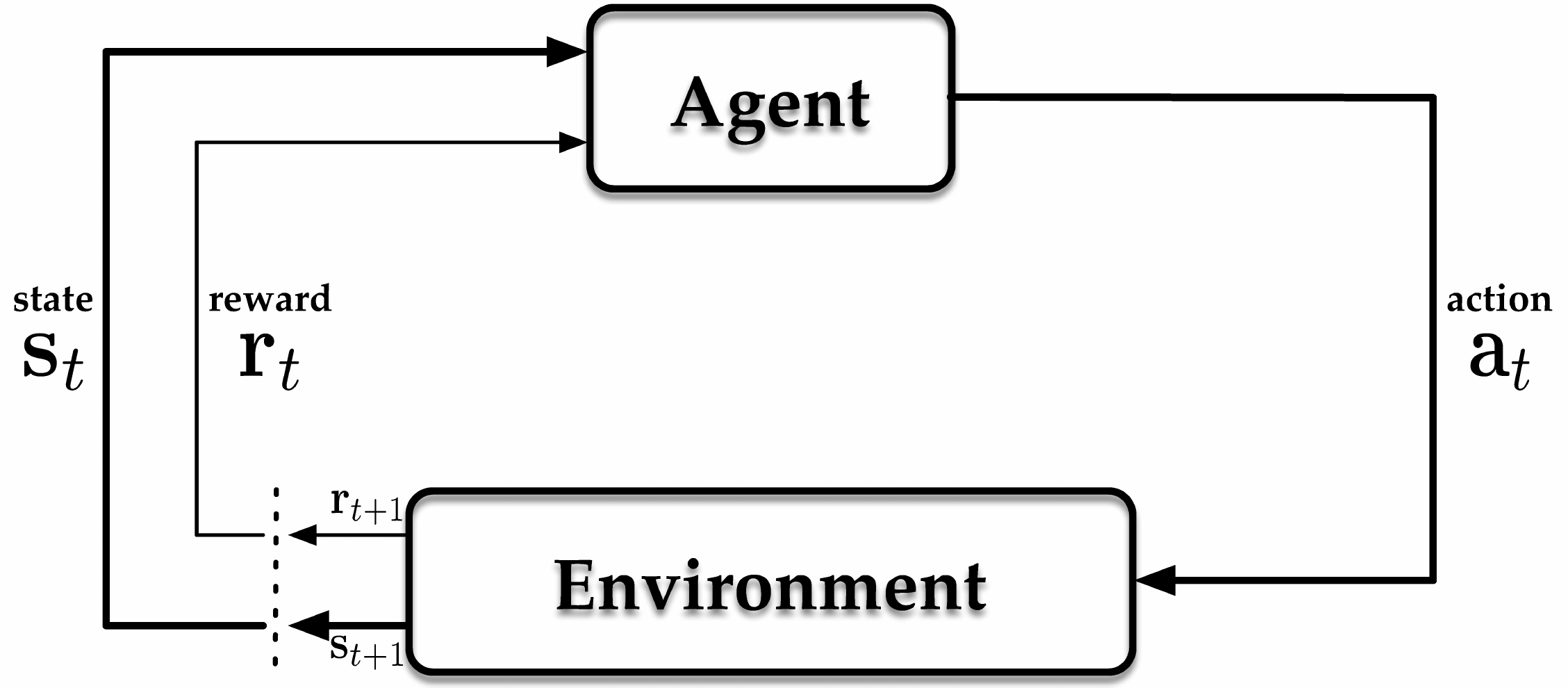}
    \caption{High-level stochastic partially observable dynamical system schematic.}
    \label{fig:dynamical-system-partial}
\end{figure}

Upon modifying the basic dynamical system in Figure \ref{fig:dynamical-system} in order to
take partial observability into account, we obtain the schematic in Figure \ref{fig:dynamical-system-partial}.
Note that $f$ is function unknown to the agent, which has access to the observation $\vo_{t}$
but not to the environment state $\vs_{t}$.
Moreover, $\mathcal{R}_{\vs}^{\va}$ and $\mathcal{P}_{\vs \vs'}^{\va}$ are the reward generating function and the
transition probability matrix (function) of the MDP, respectively.
Treating the system as a probabilistic graphical model, the state $\vs_{t}$ is a latent variable that
either deterministically or stochastically (depending on the nature of $f$) determines the observation $\vo_{t}$.
In a partially observable environment, the agent needs to reconstruct the environment state,
either by using the complete history $\vh_{t}$ or a stateful sequential
model (i.e., recurrent neural network, see (\ref{def:rnn-hidden-state})).

%%%%%%% ALGORITHM BLOCK %%%%%%%%%%%%

%%
\section{Major Components of Reinforcement Learning} \label{sec:major-components-of-reinforcement-learning}

Reinforcement Learning agents may include one or more of the following components
\citep{reinforcement-learning:silver-intro}:
\begin{itemize}
\item \textbf{Policy}: agent’s behavior function;
\item \textbf{Value function}: how good is each state, or state-action pair;
\item \textbf{Model}: agent’s representation of the environment.
\end{itemize}

In this Section, we discuss these components and highlight their importance and impact on algorithm design.

%%%
\subsection{Return} \label{sub:return}

Let $\gamma$ be the \textbf{discount factor} of future rewards,
then the \textbf{return} $G_{t}$
(also known as \textbf{future discounted reward}) at time index $t$ is given by:
\begin{align}
G_{t}     = r_{t+1} + \gamma r_{t+2} + \gamma^{2} r_{t+3} + \ldots
        = \sum_{k=0}^{\infty} \gamma^{k} r_{t+k+1}, \quad \gamma \in [0, 1]
\label{def:discounted-future-reward}
\end{align}

%%%
\subsection{Policy} \label{sub:policy}

\textbf{Policy}, $\pi$, refers to the behavior of an agent. It is a mapping function
from "state to action" \citep{reinforcement-learning:witten}, such that:
\begin{equation}
\pi: \sS \rightarrow \sA
\end{equation}
where $\sS$ and $\sA$ are respectively the state space and the action space. A policy function
can be:
\begin{itemize}
\item Deterministic: $A_{t+1} = \pi(\vs_{t})$;
\item Stochastic: $\pi(a|s) = \sP[\va_{t}=a|\vs_{t}=s]$.
\end{itemize}

%%%
\subsection{Value Function} \label{sub:value-function}

\textbf{State-value function}, $v_{\pi}$,
is the expected return, $G_{t}$, starting from state $s$, which then follows a policy $\pi$
\citep{reinforcement-learning:ualberta}, that is:
\begin{equation}
   v_{\pi} : \sS \rightarrow \sB, \quad
v_{\pi}(s) = \E_{\pi}[G_{t} | \vs_{t} = s]
\label{def:state-value-function}
\end{equation}
where $\sS$ and $\sB$ are respectively the state space and the rewards set ($\sB \subseteq \sR$).

\textbf{Action-value function}, $q_{\pi}$, is the expected return, $G_{t}$,
starting from state $s$, upon taking action $a$, which then follows a policy $\pi$
\citep{reinforcement-learning:silver-intro}:
\begin{equation}
   q_{\pi} : \sS \times \sA \rightarrow \sB, \quad
q_{\pi}(s, a) = \E_{\pi}[G_{t} | \vs_{t} = s, \va_{t} = a]
\label{def:action-value-function}
\end{equation}
where $\sS, \sA, \sB$ the state space, the action space and and the reward set, 
respectively.

%%%
\subsection{Model} \label{sub:model}

A \textbf{model} predicts the next state of the environment, $\vs_{t+1}$, and the corresponding reward signal, $r_{t+1}$,
given the current state, $\vs_{t}$, and the action taken, $\va_{t}$, at time step, $t$.
It can be represented by a \textbf{state transition probability matrix} $\gP$ given by:
\begin{equation}
\gP_{ss'}^{a}: \sS \times \sA \rightarrow \sS, \quad \gP_{ss'}^{a} = \sP[\vs_{t+1} = s' | \vs_{t} = s, \va_{t} = a]
\label{def:transition-probability-matrix}
\end{equation}
and a \textbf{reward generating function} $\gR$:
\begin{equation}
\gR: \sS \times \sA \rightarrow \sB, \quad \gR_{s}^{a} = \E[r_{t+1} | \vs_{t} = s, \va_{t} = a]
\label{def:reward-function}
\end{equation}
where $\sS, \sA, \sB$ the state space, the action space and and the reward set, 
respectively.

\section{Markov Decision Process} \label{sec:mdp}

A special type of discrete-time stochastic dynamical systems are Markov Decision Processes (MDP).
They posses strong properties that guarantee converge to the global optimum policy (i.e., strategy),
while by relaxing some of the assumption, they can describe any dynamical system, providing a
powerful representation framework and a common way of controlling dynamical systems.

%%%
\subsection{Markov Property}

A state $S_{t}$ \citep{reinforcement-learning:silver-mdp}
satisfies the \textbf{Markov property} if and only if (iff):
\begin{equation}
\sP[\vs_{t+1} | \vs_{t}, \vs_{t-1}, \ldots, \vs_{1}] = \sP[\vs_{t+1} | \vs_{t}]
\label{def:markov-property}
\end{equation}
This implies that the previous state, $\vs_{t}$, is a sufficient statistic for predicting the future,
therefore the longer-term history, $\vvh_{t}$, can be discarded.

%%%
\subsection{Definition} \label{sub:mdp-definition}

Any fully observable environment, which satisfies equation (\ref{def:full-observability}),
can be modeled as a \textbf{Markov Decision Process} (MDP).
A Markov Decision Process \citep{reinforcement-learning:poole}
is an object (i.e., 5-tuple) $\langle \sS, \sA, \gP, \gR, \gamma \rangle$ where:
\begin{itemize}
\item $\sS$ is a \textit{finite} set of states (state space),
    such that they satisfy the Markov property, as in definition
    (\ref{def:markov-property})
\item $\sA$ is a \textit{finite} set of actions (action space);
\item $\gP$ is a state transition probability matrix;;
\item $\gR$ is a reward generating function;
\item $\gamma$ is a discount factor.
\end{itemize}

\subsection{Optimality} \label{sub:mdp-optimality}

Apart from the expressiveness of MDPs, they can be optimally solved, making them very attractive.

\subsubsection{Value Function}

The \textbf{optimal state-value function}, $v_{*}$, is the maximum state-value function
over all policies:
\begin{equation}
v_{*}(s) = \max_{\pi}\  v_{\pi}(s), \quad \forall s \in \sS
\label{def:optimal-state-value-function}
\end{equation}

The \textbf{optimal action-value function}, $q_{*}$, is the maximum action-value function
over all policies:
\begin{equation}
q_{*}(s, a) = \max_{\pi}\  q_{\pi}(s, a),
    \quad \forall s \in \sS, a \in \sA
\label{def:optimal-action-value-function}
\end{equation}

\subsubsection{Policy}

Define a partial ordering over policies \citep{reinforcement-learning:silver-mdp}
\begin{eqnarray}
\pi \leq \pi' \quad \Leftarrow \quad v_{\pi}(s) \leq v_{\pi}'(s), \ \forall s \in \sS
\label{def:policy-partial-ordering}
\end{eqnarray}

For an MDP the following theorems
\footnote{The proofs are based on the contraction property of Bellman operator
\citep{reinforcement-learning:poole}.} are true:

\begin{theorem}[Policy Optimality]
There exists an optimal policy, $\pi_{*}$, that is better than or equal
to all other policies, such that $\pi_{*} \geq \pi, \ \forall \pi$.
\label{th:optimal-policy-mdp}
\end{theorem}

\begin{theorem}[State-Value Function Optimality]
All optimal policies achieve the optimal state-value function, such that
$v_{\pi_{*}}(s) = v_{*}(s), \  \forall s \in \sS$.
\label{th:optimal-state-value-function-mdp}
\end{theorem}

\begin{theorem}[Action-Value Function Optimality]
All optimal policies achieve the optimal action-value function, such that
$q_{\pi_{*}}(s, a) = q_{*}(s, a), \  \forall s \in \sS,\ a \in \sA$.
\label{th:optimal-action-value-function-mdp}
\end{theorem}

\subsection{Bellman Equation} \label{sub:bellman-equation}

Given a Markov Decision Process $\langle \sS, \sA, \gP, \gR, \gamma \rangle$,
because of the Markov property (\ref{def:markov-property}) that states in $\sS$ satisfy:
\begin{itemize}
\item The policy $\pi$ is a distribution over actions given states
\begin{equation}
\pi(s|a) = \sP[\va_{t} = a | \vs_{t} = s]
\label{eq:policy-mdp}
\end{equation}
Without loss of generality we assume that the policy $\pi$ is stochastic because of the
state transition probability matrix $\gP$ \citep{reinforcement-learning:sutton}.
Owing to the Markov property, MDP policies depend only on the current state and are
time-independent, stationary \citep{reinforcement-learning:silver-mdp}, such that
\begin{equation}
\va_{t} \sim \pi(\cdot | \vs_{t}), \quad \forall t > 0
\label{eq:policy-stationarity-mdp}
\end{equation}
\item The state-value function $v_{\pi}$ can be decomposed into two parts:
the immediate reward and the discounted reward of successor state $\gamma r_{t+1}$:
\begin{align}
v_{\pi}(s) &= \E_{\pi}[G_{t} | \vs_{t} = s]\nonumber\\
     &\textoversign{=}{(\ref{def:discounted-future-reward})}
         \E_{\pi}[r_{t+1} + \gamma r_{t+2} + \gamma^{2} r_{t+3} + \ldots | \vs_{t} = s]\nonumber\\
     &= \E_{\pi}[r_{t+1} + \gamma (r_{t+2} + \gamma r_{t+3} + \ldots) | \vs_{t} = s]\nonumber\\
     &\textoversign{=}{(\ref{def:discounted-future-reward})}
         \E_{\pi}[r_{t+1} + \gamma G_{t+1} | \vs_{t} = s]\nonumber\\
     &\textoversign{=}{(\ref{def:markov-property})}
         \E_{\pi}[r_{t+1} + \gamma v_{\pi}(\vs_{t+1}) | \vs_{t} = s]
\label{eq:state-value-function-mdp}
\end{align}
\item The action-value function $q_{\pi}$ can be similarly decomposed to
\begin{equation}
q_{\pi}(s) = \E_{\pi}[r_{t+1} + \gamma q_{\pi}(\vs_{t+1}, \vs_{t+1}) | \vs_{t} = s, \va_{t} = a]
\label{eq:action-value-function-mdp}
\end{equation}
\end{itemize}

Equations (\ref{eq:state-value-function-mdp}) and (\ref{eq:action-value-function-mdp})
are the \textbf{Bellman Expectation Equations} for Markov Decision Processes formulated
by \citet{reinforcement-learning:bellman}.

\subsection{Exploration-Exploitation Dilemma} \label{sub:exploration-exploitation-dilemma}

Search, or seeking a goal under uncertainty, is a ubiquitous requirement
of life\footnote{Metaphorically speaking, an agent "lives" in the environment.} \citep{Hills2015}.
Not only machines but also humans and animals usually face
the trade-off between \textit{exploiting} known opportunities and \textit{exploring} for better
opportunities elsewhere. This is a fundamental dilemma in reinforcement learning,
where the agent may need to act "sub-optimally" in order
to explore new possibilities, which may lead it to better
strategies. Every reinforcement learning algorithm takes into account this trade-off,
trying to balance search for new opportunities (exploration) with secure\footnote{This does not reflect any risk-sensitive metric or strategy,
"secure" is used here to describe actions that have been tried in the past and their outcomes are predictable to some extend.} actions (exploitation).
From an optimization point of view, if an algorithm is
greedy and only exploits, it may converges fast, but it
runs the risk of sticking to a local minimum. Exploration,
may at first slow down convergence, but it can lead
to previously unexplored regions of the search space, resulting in
an improved solution. Most algorithms perform exploration either by artificially
adding noise to the actions, which is attenuated while the
agent gains experience, or modelling the uncertainty of each action \citep{gal2016uncertainty}
in the Bayesian optimization framework.

\section{Extensions} \label{sec:mdp-extensions}

Markov Decisions Processes can be exploited by reinforcement learning agents,
who can optimally solve them
\citep{reinforcement-learning:sutton,reinforcement-learning:ualberta}. Nonetheless,
most real-life applications are not satisfying one or more of the conditions stated
in Section \ref{sub:mdp-definition}. As a consequence, modifications of them
lead to other types of processes, such as Infinite MDP and Partially Observable MDP,
which in turn can realistically fit a lot of application domains.

\subsection{Infinite Markov Decision Process} \label{sub:imdp}

In the case of either the state space $\sS$, or the action space $\sA$, or both
being infinite \footnote{Countably infinite (discrete) or continuous.} then
the environment can be modelled as an \textbf{Infinite Markov Decision Process} (IMDP).
Therefore, in order to implement the policy $\pi$ or/and the action-value function
$q_{\pi}$ in a computer, a differentiable function approximation method must be used
\citep{reinforcement-learning:policy-gradient-function-approximation}, such as
a least squares function approximation or a neural network
\citep{machine-learning:mitchell}. An IMDP action-state dynamics are described by a \textbf{transition
probability function} $\mathcal{P}_{\vs \vs'}^{\va}$ and not a matrix, since the state or/and the action
spaces are continuous.

\subsection{Partially Observable Markov Decision Process} \label{sub:pomdp}

If $\vs_{t}^{e} \neq \vs_{t}^{a}$ then the environment is partially observable and it can be
modeled as a \textbf{Partially Observable Markov Decision Process} (\textbf{POMDP}).
POMDP is a tuple
$\langle \sS, \sA, {\color{red}{\sO}}, \gP, \gR, {\color{red}{\gZ}}, \gamma \rangle$ where:
\begin{itemize}
\item ${\color{red}{\sO}}$ is a \textit{finite} set of observations (observation space)
\item ${\color{red}{\gZ}}$ is an observation function,
    $\gZ_{s'o}^{a} = \sP[O_{t+1} | \vs_{t+1} = s', \va_{t} = a]$
\end{itemize}

It is important to notice that, any dynamical system can be viewed as a POMDP and all the algorithms
used for MDPs are applicable, \textit{without convergence guarantees} though.

\part{Innovation} \label{part:innovation}

%% assets folder to reference
\renewcommand{\assets}{report/innovation/financial-market-as-discrete-time-stochastic-dynamical-system/assets}

\chapter{Financial Market as Discrete-Time Stochastic Dynamical System}
\label{ch:financial-market-as-discrete-time-stochastic-dynamical-system}

In Chapter \ref{ch:portfolio-optimization}, the task of static asset allocation as well
as traditional methods of its assessment were introduced. Our interest
in dynamically (i.e., sequentially) constructing portfolios led to studying Reinforcement Learning
basic components and concepts in Chapter \ref{ch:reinforcement-learning}, which suggest a framework to deal
with sequential decision making tasks. However, in order to leverage
the reinforcement learning tools, it is necessary to translate the problem
(i.e., asset allocation) into a discrete-time stochastic dynamical system and,
in particular, into a Markov Decision Process (MDP). Note that not all of
the strong assumptions of an MDP (Section \ref{sub:mdp-definition}) can be satisfied,
hence we resort to the relaxation of some of the
assumptions and consideration of the MDP extensions, discussed in Section \ref{sec:mdp-extensions}.
However, the convergence and optimality guarantees are obviously not applicable
under this formalization.

In this chapter, we mathematically formalize financial markets as discrete-time
stochastic dynamical systems. Firstly, we consider the necessary assumptions for
this formalization (Section \ref{sec:assumptions}), followed by the framework
(Sections \ref{sec:action-space}, \ref{sec:state-observation-space}, \ref{sec:reward-signal}) which enables
reinforcement learning agents to interact with the financial market in order
to optimality address portfolio management.

\section{Assumptions} \label{sec:assumptions}

Back-test tradings are only considered, where the trading agent pretends to be
back in time at a point in the market history, not knowing any ”future” market information,
and does paper trading from then onward \citep{paper:drl-pm}.
As a requirement for the back-test experiments, the following three assumptions must apply:
\textit{sufficient liquidity}, \textit{zero slippage} and \textit{zero market impact}, all of which are realistic if the
traded assets' volume in a market is high enough \citep{finance:quant-finance}.

\subsection{Sufficient Liquidity} \label{sub:liquidity}

An asset is termed \textbf{liquid} if it can be converted
into cash quickly, with little or no loss in value \citep{investopedia:liquidity}.

\begin{assumption}[Sufficient Liquidity]
    All market assets are liquid and every transaction can be
    executed under the same conditions.
\label{ass:sufficient-liquidity}
\end{assumption}

\subsection{Zero Slippage} \label{sub:zero-slippage}

\textbf{Slippage} refers to the difference between the expected price of a trade
and the price at which the trade is actually executed \citep{investopedia:slippage}.

\begin{assumption}[Zero Slippage]
    The liquidity of all market assets is high enough that,
    each trade can be carried out immediately at the last price when an order is placed.
\label{ass:zero-slippage}
\end{assumption}

\subsection{Zero Market Impact} \label{sub:zero-market-impact}

Asset prices are determined by the \textbf{Law of Supply and Demand}
\citep{investopedia:law-of-supply-and-demand}, therefore any trade impacts the balance between them, hence
affects the price of the asset being traded.

\begin{assumption}[Zero Market Impact]
    The capital invested by the trading agent is so insignificant
    that is has no influence on the market.
\label{ass:zero-market-impact}
\end{assumption}

\section{Action Space} \label{sec:action-space}

In order to solve the asset allocation task, the trading
agent should be able to determine the portfolio vector $\vw_{t}$
at every time step $t$, therefore the action $\va_{t}$ at time $t$ is the
portfolio vector $\vw_{t+1}$ at time $t+1$:
\begin{equation}
    \va_{t} \equiv \vw_{t+1} \textoversign{\triangleq}{(\ref{def:portfolio-vector})}
        \renewcommand\arraystretch{1.5}
        \begin{bmatrix}
            \evw_{1, t+1}, & \evw_{1, t+1}, & \ldots, & \evw_{M, t+1}
        \end{bmatrix}
\label{def:financial-market-action}
\end{equation}

hence the action space $\sA$ is a subset of the continuous
$M$-dimensional real space $\sR^{M}$:

\begin{equation}
    \va_{t} \in \sA \subseteq \sR^{M}, \quad \forall t \geq 0
    \quad \text{subject to} \quad
    \sum_{i=1}^{M} a_{i, t} = 1
\label{def:financial-market-action-space}
\end{equation}

If short-selling is prohibited, the portfolio weights are strictly
non-negative, or:

\begin{equation}
    \va_{t} \in \sA \subseteq [0, 1]^{M}, \quad \forall t \geq 0
    \quad \text{subject to} \quad
    \sum_{i=1}^{M} a_{i, t} = 1
\label{def:financial-market-action-space-short-selling}
\end{equation}

In both cases, nonetheless, the action space is infinite (continuous) and hence the financial market needs to
be treated as an Infinite Markov Decision Process (IMDP), as described in Section \ref{sub:imdp}.

\section{State \& Observation Space} \label{sec:state-observation-space}

\subsection{Observation} \label{sub:financial-market-observation}

At any time step $t$, we can only observe asset prices, thus
the price vector $\vp_{t}$ (\ref{def:price-vector}) is the observation $\vo_{t}$, or equivalently:

\begin{equation}
    \vo_{t} \equiv \vp_{t} \textoversign{\triangleq}{(\ref{def:price-vector})}
        \renewcommand\arraystretch{1.5}
        \begin{bmatrix}
            p_{1, t} & p_{2, t} & \cdots & p_{M, t}
        \end{bmatrix}
\label{def:financial-market-observation}
\end{equation}

hence the observation space $\sO$ is a subset of the continuous
$M$-dimensional positive real space $\sR_{+}^{M}$, since prices are non-negative real values:

\begin{equation}
    \vo_{t} \in \sO \subseteq \sR_{+}^{M}, \quad \forall t \geq 0
\label{def:financial-market-observation-space}
\end{equation}

Since one-period prices do not fully capture the market state\footnote{If
prices were a VAR(1) process \citep{mandic:asp-arma}, then financial markets are pure MDPs.},
\textit{financial markets are partially observable} \citep{reinforcement-learning:silver-intro}.
As a consequence, equation (\ref{def:full-observability}) is not satisfied and we should
construct the agent's state $\vs_{t}^{a}$ by processing the observations $\vo_{t} \in \sO$.
In Section \ref{sub:observability}, two alternatives to deal with partial
observability were suggested, considering:

\begin{enumerate}
    \item Complete history: $\vs_{t}^{a} \equiv \vvh_{t} \textoversign{\triangleq}{(\ref{def:history})}
        (\vo_{1}, \va_{1}, r_{1}, \ldots, \vo_{t}, \va_{t}, r_{t})$;
    \item Recurrent neural network: $\vs_{t}^{a} \equiv f(\vs_{t-1}^{a}, \vo_{t};\ \vtheta)$;
\end{enumerate}

where in both cases we assume that the agent state
approximates the environment state $\vs_{t}^{a} = \hat{\vs}_{t}^{e} \approx \vs_{t}^{e}$.
While the first option may contain all the environment information
by time $t$, it does not scale well, since the memory and
computational load grow linearly with time $t$. A GRU-RNN
(see Section \ref{sub:rnn}), on the other hand, can store and process efficiently
the historic observation in an adaptive manner as they arrive,
filtering out any uninformative observations out. We will be referring to
this recurrent layer as the \textbf{state manager}, since it is responsible
for constructing (i.e., managing) the agent state. This layer can be
part of any neural network architecture, enabling end-to-end differentiability and training.

\begin{figure}[h]
    \centering
    \begin{subfigure}[t]{0.48\textwidth}
        \includegraphics[width=\textwidth]{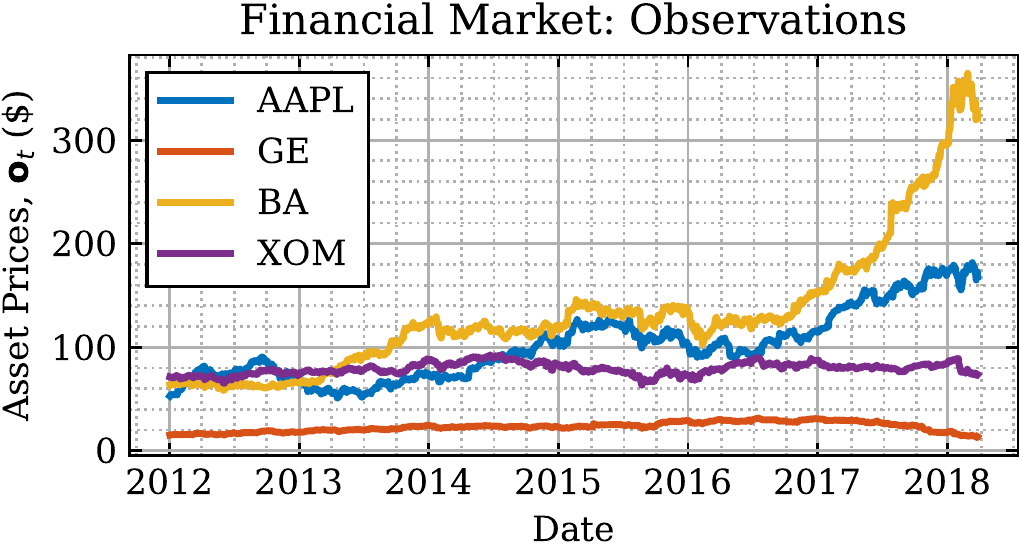}
    \end{subfigure}
    ~ 
    \begin{subfigure}[t]{0.48\textwidth}
        \includegraphics[width=\textwidth]{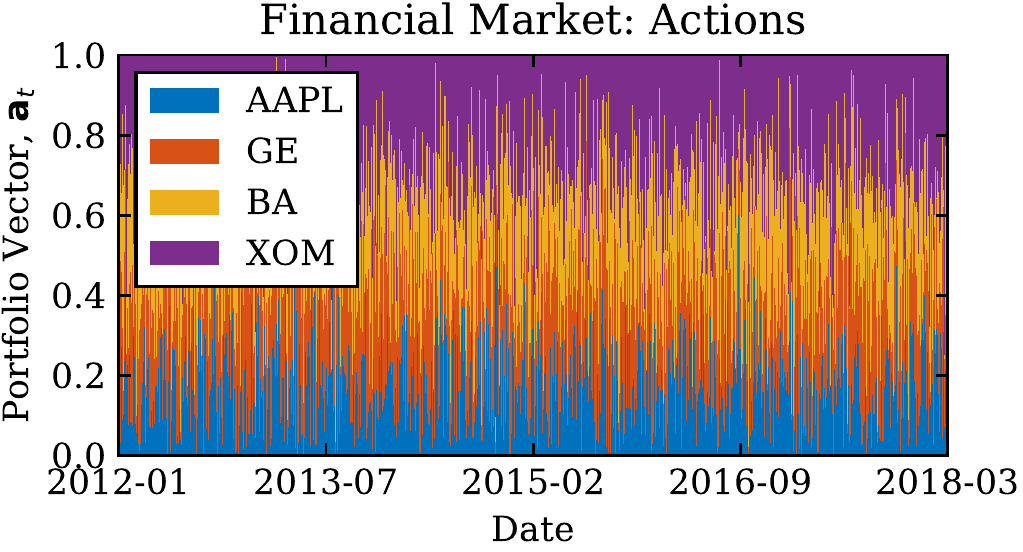}
    \end{subfigure}
    \caption{Example universe of assets as dynamical system, including \texttt{AAPL} (Apple), \texttt{GE} (General Electric),
    \texttt{BA} (Boeing Company) and \texttt{XOM} (Exxon Mobil Corporation).
    (\textit{Left}) Financial market asset prices, observations $\vo_{t}$.
    (\textit{Right}) Portfolio manager, agent, portfolio vectors, actions $\va_{t}$; the portfolio coefficients
    are illustrated in a stacked bar chat, where at each
    time step, they sum to unity according to equation (\ref{def:portfolio-vector}).}
    \label{fig:financial-market-actions-observations}
\end{figure}

Figure \ref{fig:financial-market-actions-observations} illustrates examples of a financial market observations $\vo_{t}$ and the
corresponding actions $\va_{t}$ of a random agent.

\subsection{State} \label{sub:financial-market-state}

In order to assist and speed-up the training of the
state manager, we process the raw observations $\vo_{t}$, obtaining $\hat{\vs}_{t}$. In particular,
thanks to the representation and statistical superiority of log returns
over asset prices and simple returns (see \ref{sub:returns}), we use log
returns matrix $\vvrho_{t-T:t}$ (\ref{def:simple-return-matrix}), of fixed window size\footnote{Expanding windows are
more appropriate in case the RNN state manager is replaced
by complete the history $\vvh_{t}$.} $T$. We also demonstrate another important property
of log returns, which suits the nature of operations performed
by neural networks, the function approximators used for building agents.
Neural network layers apply non-linearities to weighted \textit{sums} of input features,
hence the features do not multiply \citep{nasrabadi2007pattern} with each other\footnote{This
is the issue that multiplicative neural networks \citep{salinas1996model} try to address.
Consider two scalar feature variables $x_{1}$ and $x_{2}$ and the target scalar
variable $y$ such that:
\begin{equation}
    f(x_{1}, x_{2}) \triangleq y = x_{1} * x_{2}, \quad x_{1}, x_{2} \in \sR
\end{equation}
It is very hard for a neural network to learn this
function, but a logarithmic transformation of the features transforms the
problem to a very simple sum of logarithms using the property:
\begin{equation}
    log(x_{1}) + log(x_{2}) = log(x_{1} * x_{2}) = log(y)
\end{equation}
},
but only with the layer weights (i.e., parameters). Nonetheless, by summing
the log returns, we equivalently multiply the gross returns, hence
the networks are learning non-linear functions of the products of
returns (i.e., asset-wise and cross-asset) which are the building blocks of
the covariances between assets. Therefore, by simply using the log
returns we enable cross-asset dependencies to be easily captures.

Moreover, transaction costs are taken into account, and since (\ref{obj:transaction-costs}) suggests that
the previous time step portfolio vector $\vw_{t-1}$ affects transactions costs, we
also append the $\vw_{t}$, or equivalently $\va_{t-1}$ by (\ref{def:financial-market-action}), to the agent state,
obtaining the 2-tuple:

\begin{equation}
    \hat{\vs}_{t} \triangleq \langle \vw_{t}, \vvrho_{t-T:t} \rangle =
        \bigg\langle
        \renewcommand\arraystretch{1.5}
        \begin{bmatrix}
            \evw_{1, t} \\ \evw_{2, t} \\ \vdots \\ \evw_{M, t}
        \end{bmatrix}
        ,
        \renewcommand\arraystretch{1.5}
        \begin{bmatrix}
            \rho_{1, t-T} & \rho_{1, t-T+1} & \cdots & \rho_{1, t} \\
            \rho_{2, t-T} & \rho_{2, t-T+1} & \cdots & \rho_{2, t} \\
            \vdots & \vdots & \ddots & \vdots \\
            \rho_{M, t-T} & \rho_{M, t-T+1} & \cdots & \rho_{M, t}
        \end{bmatrix}
        \bigg\rangle
\label{def:financial-market-processed-observation}
\end{equation}

where $\rho_{i, (t-\tau) \rightarrow t}$ the log cumulative returns of asset $i$ between the time
interval $[t-\tau, t]$. Overall, the agent state is given by:

\begin{equation}
    \vs_{t}^{a} \equiv f(\vs_{t-1}^{a}, {\color{red}\hat{\vs}_{t}};\ \vtheta)
\label{def:financial-market-state}
\end{equation}

where $f$ the state manager non-linear mapping function. When an observation arrives,
we calculate\footnote{Most terms can be stored or pre-calculated.} $\hat{\vs}_{t}$
and feed it to the state manager (GRU-RNN).

Therefore the state space $\sS$ is a subset of the continuous
$K$-dimensional real space $\sR^{K}$, where $K$ the size of the hidden state in
the GRU-RNN state manager:

\begin{equation}
    \vs_{t}^{a} \in \sS \subseteq \sR^{K}, \quad \forall t \geq 0
\label{def:financial-market-state-space}
\end{equation}

\begin{figure}[h]
    \centering
    \begin{subfigure}[t]{0.48\textwidth}
        \includegraphics[width=\textwidth]{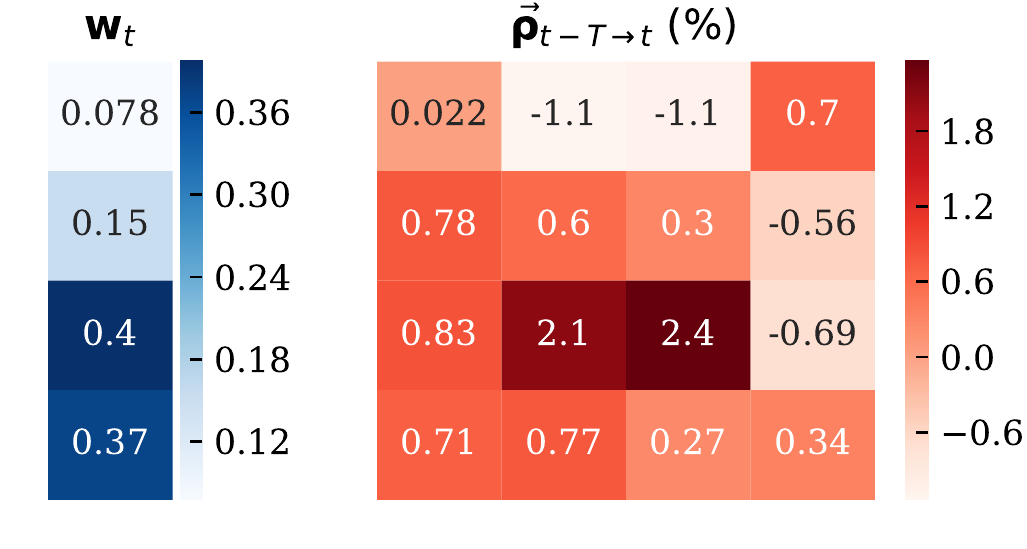}
    \end{subfigure}
    ~ 
    \begin{subfigure}[t]{0.48\textwidth}
        \includegraphics[width=\textwidth]{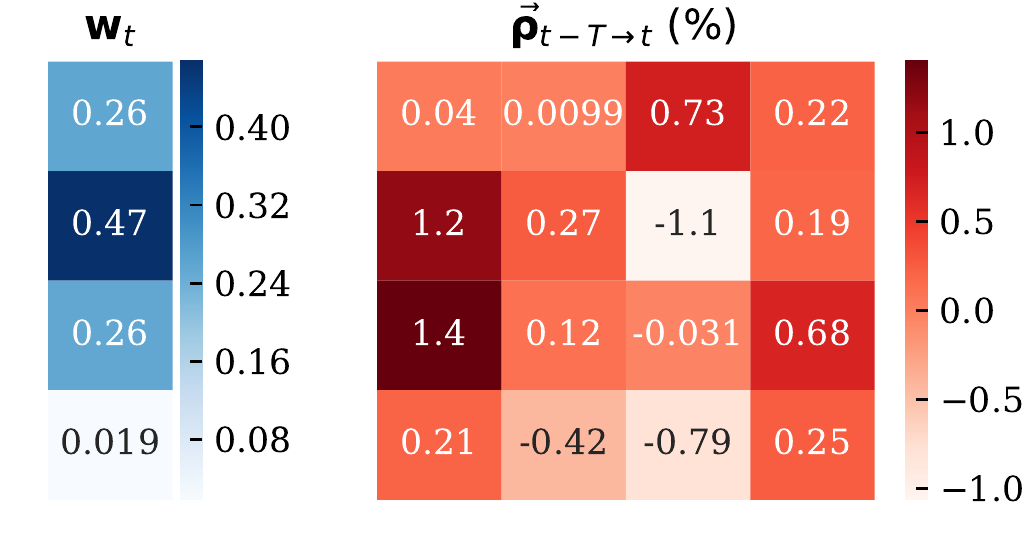}
    \end{subfigure}
    \caption{Examples of processed observation 2-tuples for two randomly selected time steps.}
    \label{fig:financial-market-processed-observations}
\end{figure}

Figure \ref{fig:financial-market-processed-observations} illustrates two examples of the processed observation $\hat{\vs}_{t}$.
The agent uses this as input to determine its internal state, which in term drives its policy.
They look meaningless and impossible to generalize from, with the naked eye,
nonetheless, in Chapter \ref{ch:trading-agents}, we demonstrated the effectiveness of this,
representation, especially thanks to the combination of the convolutional
and recurrent layers combination.

Overall, the financial market should be modelled as an Infinite
Partially Observable Markov Decision Process (IPOMDP), since:

\begin{itemize}
    \item The action space is continuous (infinite), $\sA \subseteq \sR^{M}$;
    \item The observations $\vo_{t}$ are not sufficient statistics (partially observable)
    of the environment state;
    \item The state space is continuous (infinite), $\sS \subseteq \sR^{K}$.
\end{itemize}

\section{Reward Signal} \label{sec:reward-signal}

The determination of the reward signal is usually the most
challenging step in the design of a reinforcement learning problem.
According to the Reward Hypothesis \ref{hyp:reward}, the reward is a scalar
value, which fully specifies the goals of the agent, and
the maximization of the expected cumulative reward leads to the
optimal solution of the task. Specifying the optimal reward
generating function is the field of study of Inverse Reinforcement
Learning \citep{ng2000algorithms} and Inverse Optimal Control \citep{moylan1973nonlinear}.

In our case, we develop a generic, modular framework, which enables comparison of various reward
generating functions\footnote{Transaction costs are included in all case.}, including
log returns, (negative) volatility and Sharpe Ratio.
Section \ref{sec:evaluation-criteria} motivates a few reward function candidates, most of which
are implemented and tested in Chapter \ref{ch:market-data}. It is worth highlighting that
the reinforcement learning methods, by default, aim to maximize the
\textit{expected cumulative reward signal}, hence the optimization problem
that the agent (parametrized by $\vtheta$) solves is given by:

\begin{equation}
    \underset{\vtheta}{\text{maximize}} \quad
    \sum_{t=1}^{T} \E[\gamma^{t} r_{t}]
\label{opt:reward-signal}
\end{equation}

For instance, when we refer to log returns (with transaction
costs) as the reward generating function, the agent solves the
optimization problem:

\begin{equation}
    \underset{\vtheta}{\text{maximize}} \quad
    \sum_{t=1}^{T} \E[\gamma^{t} ln(1 + \vw_{t}^{T} \vr_{t} - \beta \|\vw_{t-1} - \vw_{t}\|_{1})]
\label{opt:reward-signal-log-returns}
\end{equation}

where the argument of the logarithm is the adjusted by
the transaction costs gross returns at time index $t$ (see (\ref{def:log-returns}) and (\ref{sec:transaction-cost-optimization})).

\renewcommand{\assets}{report/innovation/trading-agents/assets}

\chapter{Trading Agents} \label{ch:trading-agents}

Current state-of-the-art algorithmic portfolio management methods:

\begin{itemize}
    \item Address the decision making task of asset allocation by solving a prediction problem, heavily relying on
    the accuracy of predictive models for financial time-series \citep{aldridge2013high, heaton2017deep}, which are
    usually unsuccessful, due to the stochasticity of the financial markets;
    \item Make unrealistic assumptions about the second and higher-order statistical moments
    of the financial signals \citep{paper:pg-asset-allocation, paper:drl-pm};
    \item Deal with binary trading signals (i.e., BUY, SELL, HOLD) \citep{paper:adaptive-dynamic-programming, deng2017deep},
    instead of assigning portfolio weights to each asset, and hence limiting the scope of their applications.
\end{itemize}

On the other hand, the representation of the financial market as a discrete-time stochastic
dynamical system, as derived in Chapter \ref{ch:financial-market-as-discrete-time-stochastic-dynamical-system}
enables the development of a unified framework for
training reinforcement learning trading agents. In this chapter, this framework is exploited by:

\begin{itemize}
    \item Model-based Reinforcement Learning agents, as in Section \ref{sec:model-based-reinforcement-learning}, where vector
    autoregressive processes (VAR) and recurrent neural networks (RNN) are fitted
    to environment dynamics, while the derived agents perform planning and
    control \citep{reinforcement-learning:silver-model-based}. Similar to \citep{aldridge2013high, heaton2017deep}, these agents
    are based on a predictive model of the environment, which
    is in turn used for decision making. Their performance is
    similar to known algorithms and thus they are used as baseline models for comparison;
    \item Model-free Reinforcement Learning agents, as in Section \ref{sec:model-free-reinforcement-learning}, which directly
    address the decision making task of sequential and multi-step optimization. Modifications to the 
    state-of-the-art reinforcement learning algorithms, such as Deep Q-Network (DQN) \citep{deepmind:atari}
    and Deep Deterministic Policy Gradient (DDPG) \citep{lillicrap2015continuous}, enable their incorporation to
    the trading agents training framework.
\end{itemize}

Algorithm \ref{algo:general-setup} provides the general setup for
reinforcement learning algorithms discussed in this chapter, based on which, experiments
on a small universe of real market data are carried
out, for testing their efficacy and illustration purposes. In Part \ref{part:experiments},
all the different agents are compared on a larger universe
of assets  with different reward functions, a more realistic and
practical setting.

\begin{algorithm}
	\SetKwInOut{Args}{inputs}  
    \SetKwInOut{Output}{output}
    \DontPrintSemicolon
    \Args{%
    	trading universe of $M$-assets\\
    	initial portfolio vector $\vw_{1}=\va_{0}$\\
    	initial asset prices $\vp_{0}=\vo_{0}$\\
        objective function $\mathcal{J}$
    }
    \Output{optimal agent parameters $\vtheta_{*}, \vphi_{*}$}
    \vspace{0.25cm}
    \SetAlgoLined\SetArgSty{}
    \Repeat{convergence}{
        \For{$t=1, 2, \ldots T$}{
            observe 2-tuple $\langle \vo_{t}, r_{t} \rangle$ \\
            calculate gradients $\nabla_{\vtheta} \mathcal{J}(r_{t})$ and $\nabla_{\vphi} \mathcal{J}(r_{t})$ \tcp*{BPTT}
            update agent parameters $\vtheta, \vphi$\;
                \Indp using adaptive gradient optimizers \tcp*{ADAM} \Indm
            get estimate of agent state: $\vs_{t} \approx f(\cdot \ ,\ o_{t})$ \tcp*{(\ref{def:financial-market-state})}
            sample and take action: $\va_{t} \sim \pi(\cdot | \vs_{t} ; \vphi )$ \tcp*{portfolio rebalance}
        }
    }
    set $\vtheta_{*}, \vphi_{*} \leftarrow \vtheta, \vphi$
\caption{General setup for trading agents.}
\label{algo:general-setup}
\end{algorithm}

\section{Model-Based Reinforcement Learning} \label{sec:model-based-reinforcement-learning}

Upon a revision of the schematic of a generic partially observable environment (i.e.,
dynamical system) as in Figure \ref{fig:dynamical-system-partial}, it is noted that given the
transition probability function $\mathcal{P}_{\vs \vs'}^{\va}$ of the system, the reinforcement learning task
reduces to \textit{planning} \citep{atkeson1997comparison}; simulate future states by recursively calling
$\mathcal{P}_{\vs \vs'}^{\va}$ $L$ times and choose the roll-outs (i.e., trajectories) which maximize cumulative reward,
via dynamic programming \citep{bertsekas1995dynamic}:

\begin{gather}
    \vs_{t}     \textoversign{\rightarrow}{$\mathcal{P}_{\vs \vs'}^{\va}$}
    \vs_{t+1}   \textoversign{\rightarrow}{$\mathcal{P}_{\vs \vs'}^{\va}$}
    \cdots  \textoversign{\rightarrow}{$\mathcal{P}_{\vs \vs'}^{\va}$}
    \vs_{t+L} \\
    \va_{t+1} \equiv \max_{\va \in \sA} \mathcal{J}(\va | \va_{t}, \vs_{t}, \ldots, \vs_{t+L})
\label{def:planning}
\end{gather}

Note that due to the assumptions made in Section \ref{sec:assumptions}, and especially the Zero Market Impact assumption
\ref{ass:zero-market-impact}, the agent actions do not affect the environment state transitions,
or equivalently the financial market is an \textit{open loop system} \citep{sp:financial-engineering},
where the agent actions do not modify the system state,
but only the received reward:

\begin{equation}
    p(\vs_{t+1} | \vs, \va) = p(\vs_{t+1} | \vs)
    \Rightarrow
    \mathcal{P}_{\vs \vs'}^{\va} = \mathcal{P}_{\vs \vs'}
\label{def:zero-market-impact}
\end{equation}

Moreover, the reward generating function is known, as explained in
section \ref{sec:reward-signal}, hence \textit{a model of the environment is obtained by learning
only the transition probability function} $\mathcal{P}_{\vs \vs'}$.

\subsection{System Identification} \label{sub:system-identification}

In the area of Signal Processing and Control Theory, the task under consideration is usually termed
as \textbf{System Identification} (SI), where an approximation of the environment, the "model",
is fitted such that it captures the environment dynamics:

\begin{equation}
    \underbrace{\hat{\mathcal{P}}_{\vs \vs'}}_\text{model} \approx \underbrace{\mathcal{P}_{\vs \vs'}}_\text{environment}
\label{def:system-identification}
\end{equation}

\begin{figure}
    \centering
    \includegraphics[width=0.85\textwidth]{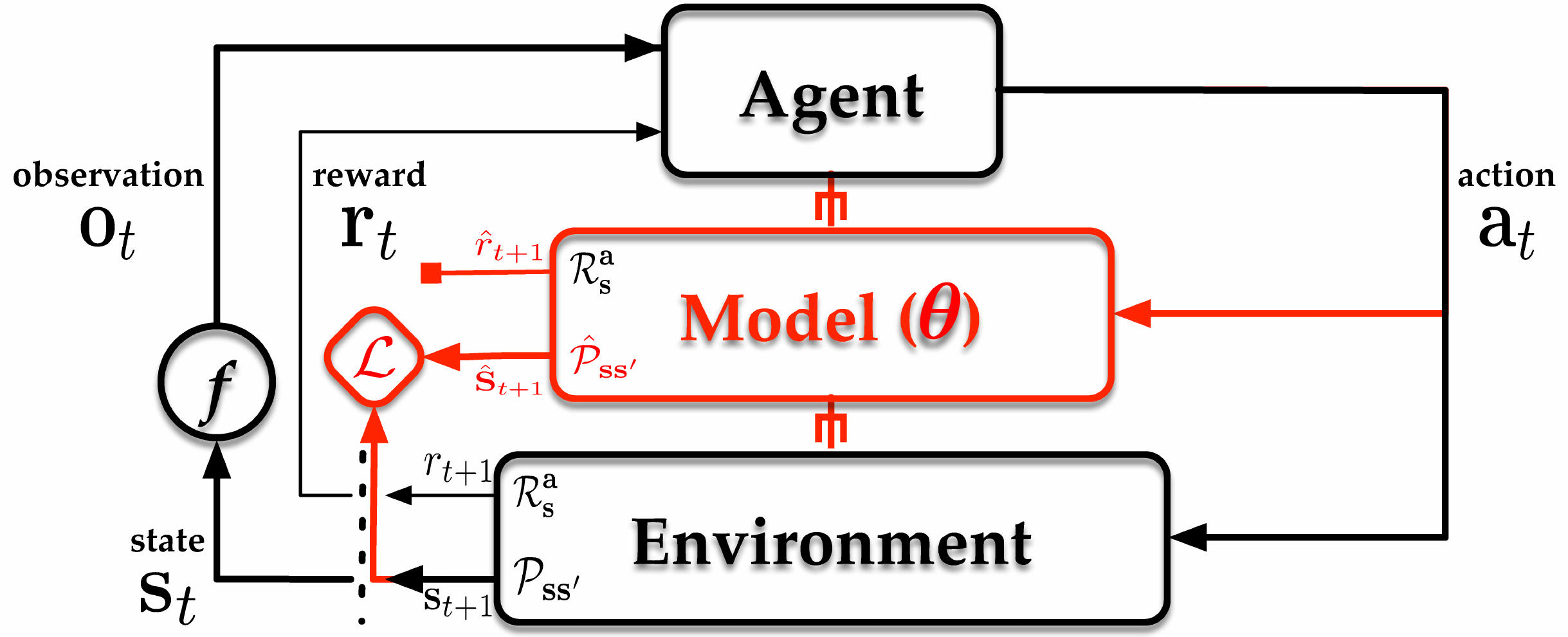}
    \caption{General setup for System Identification (SI) (i.e., model-based reinforcement learning) for solving
    a discrete-time stochastic partially observable dynamical system.}
    \label{fig:model-based}
\end{figure}

Figure \ref{fig:model-based} illustrates schematically the system identification wiring of the circuit,
where the model, represented by {\color{red}$\hat{\mathcal{P}}_{\vs \vs'}$}, is compared against the true
transition probability function $\mathcal{P}_{\vs \vs'}$ and the loss function $\mathcal{L}$ (i.e., mean squared error)
is minimized by optimizing with respect to the model parameters $\vtheta$.
It is worth highlighting that the transition probability function is by definition stochastic (\ref{def:transition-probability-matrix})
hence the candidate fitted models should ideally be able to
capture and incorporate this uncertainty. As a result, model-based reinforcement
learning usually \citep{deisenroth2011pilco, levine2016end, gal2016improving} relies on probabilistic graphical models, such as
Gaussian Processes \citep{rasmussen2004gaussian} or Bayesian Networks \citep{ghahramani2001introduction}, which are
non-parametric models that do not output point estimates, but learn
the generating process of the data $\pdata$, and hence enable sampling
from the posterior distribution. Sampling from the posterior distribution allows
us to have stochastic predictions that respect model dynamics.

In this section we shall will focus, nonetheless, only on vector
autoregressive processes (VAR) and recurrent neural networks (RNN) for modelling $\mathcal{P}_{\vs \vs'}$,
trained on an adaptive fashion \citep{mandic:rnn}, given by Algorithm \ref{algo:model-based}.
An extension of vanilla RNNs to bayesian RNNs could be also
tried using the MC-dropout trick from \citep{gal2016theoretically}.

\begin{algorithm}
	\SetKwInOut{Args}{inputs}  
    \SetKwInOut{Output}{output}
    \DontPrintSemicolon
    \Args{%
    	trading universe of $M$-assets\\
    	initial portfolio vector $\vw_{1}=\va_{0}$\\
    	initial asset prices $\vp_{0}=\vo_{0}$\\
        loss function $\mathcal{L}$\\
        historic dataset $\mathcal{D}$
    }
    \Output{optimal model parameters $\vtheta_{*}$}
    \vspace{0.25cm}
    \SetAlgoLined\SetArgSty{}
    batch training on $\mathcal{D}$\;
        \Indp $\vtheta \leftarrow \argmax_{\vtheta} p(\vtheta|\mathcal{D})$ \tcp*{MLE} \Indm
    \Repeat{convergence}{
        \For{$t=1, 2, \ldots T$}{
            predict next state $\hat{\vs}_{t}$ \tcp*{via $\hat{\mathcal{P}}_{\vs \vs'}$}
            observe tuple $\langle \vo_{t}, r_{t} \rangle$ \\
            get estimate of agent state: $\vs_{t} \approx f(\cdot \ ,\ o_{t})$ \tcp*{(\ref{def:financial-market-state})}
            calculate gradients: $\nabla_{\vtheta} \mathcal{L}(\hat{\vs}_{t}, \vs_{t})$ \tcp*{backprop}
            update model parameters $\vtheta$\;
                \Indp using adaptive gradient optimizers \tcp*{ADAM} \Indm
            plan and take action $\va_{t}$ \tcp*{portfolio rebalance}
        }
    }
    set $\vtheta_{*}\leftarrow \vtheta$
\caption{General setup for adaptive model-based trading agents.}
\label{algo:model-based}
\end{algorithm}

\subsection{Vector Autoregression (VAR)} \label{sub:var-agent}

Following on the introduction of the vector autoregressive processes (VAR) in
Section \ref{sub:var}, and using the fact that the transition probability model $\hat{\mathcal{P}}_{\vs \vs'}$ is a
one-step time-series predictive model, we investigate the effectiveness
of VAR processes as time-series predictors.

%%%
\subsubsection{Agent Model}

The vector autoregressive processes (VAR) regress past values of multivariate time-series with the future values
(see equation (\ref{def:var})). In order to satisfy the covariance stationarity assumption \citep{mandic:asp-arma},
we fit a VAR process on the log-returns $\vrho_{t}$, and not the raw
observations $\vo_{t}$ (i.e., price vectors), since the latter is known to
be highly non-stationary\footnote{In the wide-sense \citep{bollerslev1986generalized}.}. The model is
pre-trained on historic data (i.e., batch supervised learning training \citep{Murphy:2012:MLP:2380985})
and it is updated online, following the gradient $\nabla_{\vtheta} \mathcal{L}(\hat{\vrho_{t}}_{t}, \vrho_{t})$,
as described in Algorithm \ref{algo:model-based}. The model takes the form:

\begin{align}
    \mathcal{P}_{\vs \vs'}&: \quad
    \vrho_{t} \textoversign{\approx}{(\ref{def:var})} \vc + \sum_{i=1}^{p} \mA_{i} \vrho_{t-i}\\
    \text{planning}&: \quad \bigg( \vs_{t} \textoversign{\rightarrow}{$\mathcal{P}_{\vs \vs'}^{\va}$}
    \cdots \textoversign{\rightarrow}{$\mathcal{P}_{\vs \vs'}^{\va}$}
    \vs_{t+L} \bigg) \textoversign{\Rightarrow}{(\ref{def:planning})}
    \va_{t+1} \equiv \max_{\va \in \sA} \mathcal{J}(\va | \va_{t}, \vs_{t}, \ldots, \vs_{t+L})
\label{eq:trading-agent-var}
\end{align}

%%%
\subsubsection{Related Work}

Vector autoregressive processes have been widely used for modelling financial
time-series, especially returns, due to their pseudo-stationary nature \citep{sp:time-series}.
In the context of model-based reinforcement learning there is no results in the open literature
on using VAR processes in this context, nonetheless, control engineering applications \citep{akaike1998markovian}
have extensively used autoregressive models to deal with dynamical systems.

%%%
\subsubsection{Evaluation}

\begin{figure}[h]
    \centering
    \begin{subfigure}[t]{0.48\textwidth}
        \includegraphics[width=\textwidth]{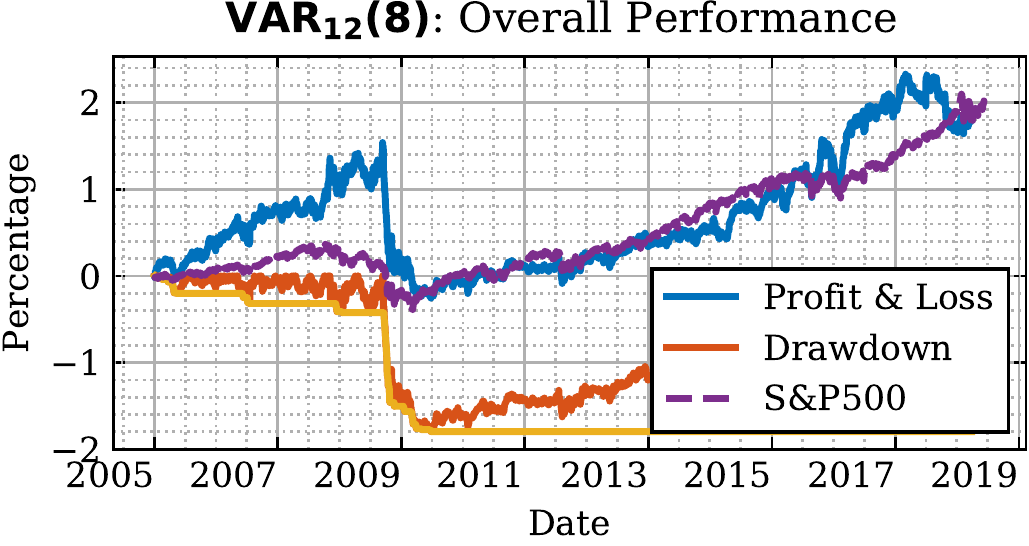}
    \end{subfigure}
    ~ 
    \begin{subfigure}[t]{0.48\textwidth}
        \includegraphics[width=\textwidth]{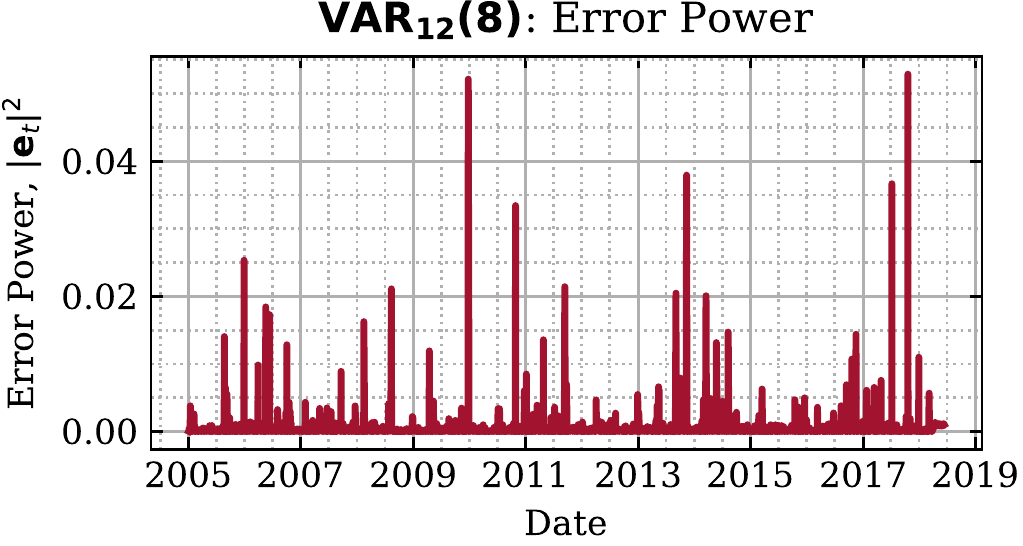}
    \end{subfigure}
    \caption{Order-eight vector autoregressive model-based reinforcement learning agent on a
    $12$-asset universe (i.e., VAR$_{12}$($8$)), pre-trained on historic data between 2000-2005 and trained
    online onward. (\textit{Left}) Cumulative rewards and (maximum) drawdown of the
    learned strategy, against the S\&P 500 index (traded as \texttt{SPY}). (\textit{Right}) Mean squared prediction
    error for single-step predictions.}
    \label{fig:model-based-var}
\end{figure}

Figure \ref{fig:model-based-var} illustrates the cumulative rewards and the prediction error power are illustrated
Observe that the agent is highly correlated with the market (i.e., S\&P 500)
and overall collects lower cumulative returns. Moreover, note that the
market crash in 2009 \citep{farmer2012stock}, affects the agent significantly, leading to
a decline by $179.6 \%$ (drawdown), taking as many as \textit{2596 days to recover}, from 2008-09-12 to 2015-10-22.

%%%
\subsubsection{Weaknesses}

The order-$p$ VAR model, VAR($p$), assumes that the underlying generating process:
\begin{enumerate}
    \item Is covariance stationary;
    \item Satisfies the order-$p$ Markov property;
    \item Is linear, conditioned on past samples.
\end{enumerate}

Unsurprisingly, most of these assumptions are not realistic for real
market data, which reflects the poor performance of the
method illustrated in Figure \ref{fig:model-based-var}.

\begin{property}[Non-Stationary Dynamics]
    The agent should be able to capture non-stationary dynamics.
\label{pro:non-stationarity}
\end{property}

\begin{property}[Long-Term Memory]
    The agent should be able to selectively remember past events
    without brute force memory mechanisms (e.g., using lagged values as features).
\label{pro:long-term-memory}
\end{property}

\begin{property}[Non-Linear Model]
    The agent should be able to learn non-linear dependencies between features.
\label{pro:non-linearity}
\end{property}

\subsection{Recurrent Neural Network (RNN)} \label{sub:rnn-agent}

The limitations of the vector autoregressive processes regarding stationarity, linearity and
finite memory assumptions are overcome by the
recurrent neural network (RNN) environment model. Inspired by the effectiveness
of recurrent neural networks in time-series prediction \citep{gers1999learning, mandic:rnn, henaff2011real} and the encouraging
results obtained from the initial one-step predictive GRU-RNN model in Figure \ref{fig:rnn-model-time-series},
we investigate the suitability of RNNs in the context of
model-based reinforcement learning, used as environment predictors.

%%%
\subsubsection{Agent Model}

\begin{figure}[h]
    \centering
    \includegraphics[width=0.85\textwidth]{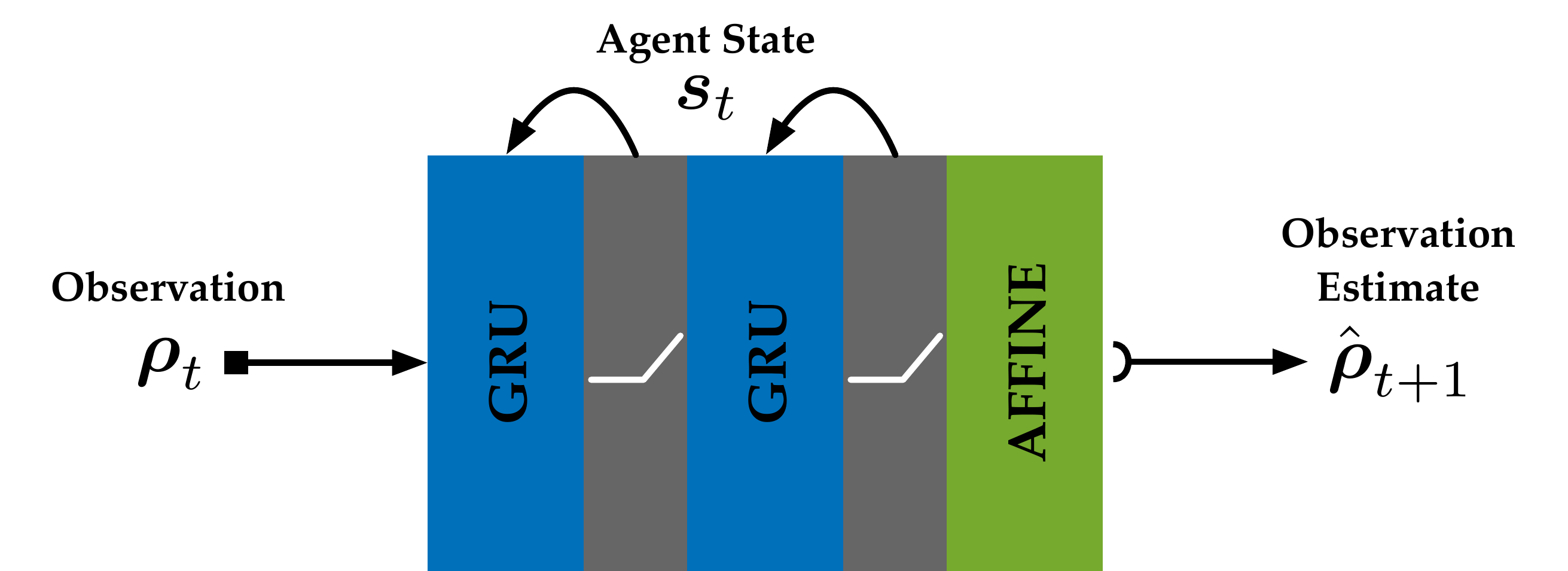}
    \caption{Two layer gated recurrent unit recurrent neural network (GRU-RNN) model-based
    reinforcement learning agent; receives log returns $\vrho_{t}$ as input, builds internal
    state $\vs_{t}$ and estimates future log returns $\hat{\vrho}_{t+1}$. Regularized mean squared error
    is used as the loss function, optimized with the ADAM \citep{kingma2014adam}
    adaptive optimizer.}
    \label{fig:model-based-rnn-schematic}
\end{figure}

Revisiting Algorithm \ref{algo:model-based} along with the formulation of RNNs in Section \ref{sub:rnn},
we highlight the steps:

\begin{align}
    \text{state manager}&:\quad
        \vs_{t} \textoversign{=}{(\ref{def:agent-state})} f(\vs_{t-1}, \vrho_{t}) \\
    \text{prediction}&: \quad
        \hat{\vrho}_{t+1} \approx \mV \sigma(\vs_{t}) + \vb \\
    \text{planning}&: \quad \bigg( \vs_{t} \textoversign{\rightarrow}{$\mathcal{P}_{\vs \vs'}^{\va}$}
    \cdots \textoversign{\rightarrow}{$\mathcal{P}_{\vs \vs'}^{\va}$}
    \vs_{t+L} \bigg) \textoversign{\Rightarrow}{(\ref{def:planning})}
    \va_{t+1} \equiv \max_{\va \in \sA} \mathcal{J}(\va | \va_{t}, \vs_{t}, \ldots, \vs_{t+L})
\label{eq:trading-agent-rnn}
\end{align}

where $\mV$ and $\vb$ the weights matrix and bias vector of the
output affine layer\footnote{In Deep Learning literature \citep{deep-learning:goodfellow}, the term
\textbf{affine} refers to a neural network layer with parameters $\mW$ and $\vb$ that performs a mapping from an
input matrix $\mX$ to an output vector $\vy$ according to
\begin{equation}
    f_{\textnormal{affine}}(\mX; \mW, \vb) = \hat{\vy} \triangleq \mX \mW + \vb
\label{def:affine-layer}
\end{equation}
The terms \textit{affine}, \textit{fully-connected} (FC) and \textit{dense} refer to the same
layer configuration.} of the network and $\sigma$ a non-linearity (i.e., rectified linear unit \citep{nair2010rectified},
hyperbolic tangent, sigmoid function.). Schematically the network is depicted in Figure \ref{fig:model-based-rnn-schematic}.

%%%
\subsubsection{Related Work}

Despite the fact that RNNs were first used
decades ago \citep{hopfield1982neural, hochreiter1997long, mandic:rnn}, recent advances in adaptive optimizers
(e.g., RMSProp \citep{tieleman2012lecture}, ADAM \citep{kingma2014adam}) and deep learning
have enabled the development of \textit{deep} recurrent neural networks for
sequential data modelling (e.g., time-series, text). Since financial markets are dominated
by dynamic structures and time-series, RNNs have been extensively used for modelling dynamic financial systems
\citep{tino2001financial, chen2015lstm, heaton2016deep, bao2017deep}. In most cases, feature engineering prior to training is
performed so that meaningful financial signals are combined, instead of
raw series. Our approach was rather context-free, performing pure technical
analysis of the series, without involving manual extraction and validation
of high-order features.

%%%
\subsubsection{Evaluation}

\begin{figure}[h]
    \centering
    \begin{subfigure}[t]{0.48\textwidth}
        \includegraphics[width=\textwidth]{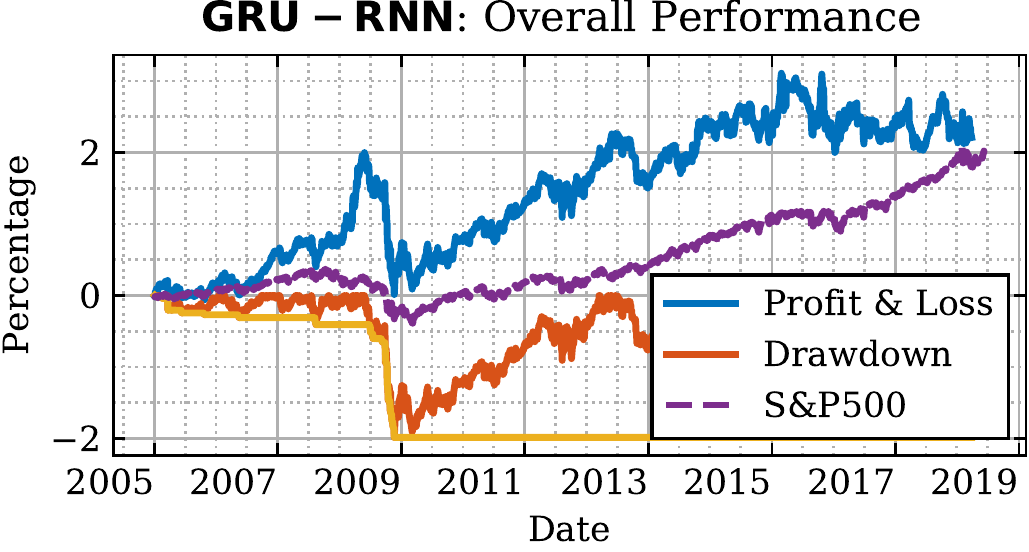}
    \end{subfigure}
    ~ 
    \begin{subfigure}[t]{0.48\textwidth}
        \includegraphics[width=\textwidth]{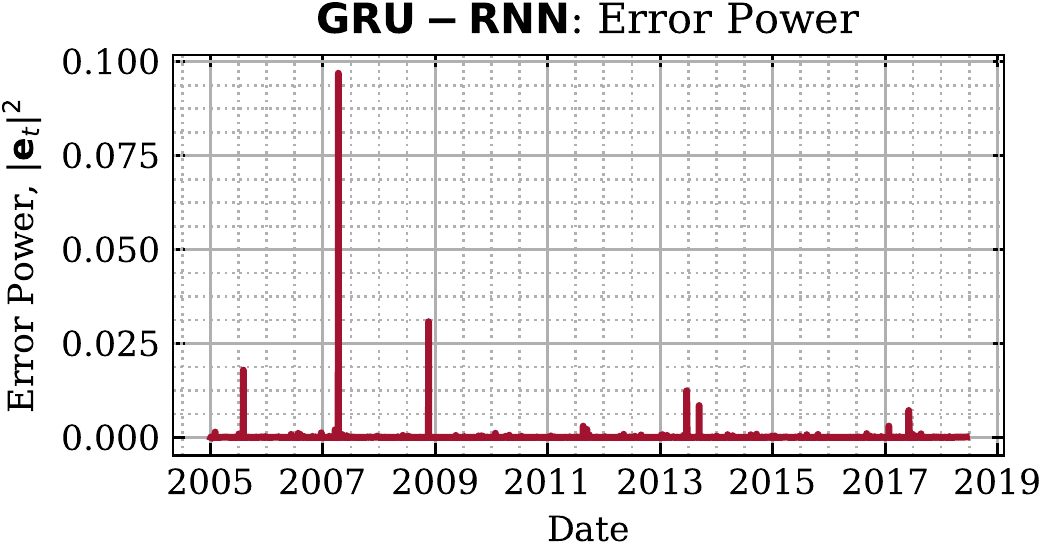}
    \end{subfigure}
    \caption{Two-layer gated recurrent unit recurrent neural network (GRU-RNN) model-based
    reinforcement learning agent on a $12$-asset universe, pre-trained on historic data between 2000-2005 and trained
    online onward (\textit{Left}) Cumulative rewards and (maximum) drawdown of the
    learned strategy, against the S\&P 500 index (traded as \texttt{SPY}). (\textit{Right}) Mean square prediction
    error for single-step predictions.}
    \label{fig:model-based-rnn}
\end{figure}

Figure \ref{fig:model-based-rnn} illustrates the performance of a two-layer
gated recurrent unit recurrent neural network, which is not outperforming the
vector autoregressive predictor as much as it was expected. Again, we note the strong correlation
with the market (i.e., S\&P 500). The 2008 market crash affects the RNN
agent as well, which manages to recover faster than the VAR
agent, leading to an overall $221.1 \%$ cumulative return, slightly higher than
the market index.

\subsection{Weaknesses} \label{sub:model-based-weaknesses}

Having developed both vector autoregressive and recurrent neural network
model-based reinforcement learning agents, we conclude that despite the architectural
simplicity of system identification, who are under-performing. The inherent randomness
(i.e., due to uncertainty) of the financial time-series (e.g., prices, returns) affects the model
training and degrades predictability.

In spite of the promising results in Figures \ref{fig:var-model-time-series}, \ref{fig:rnn-model-time-series}, where one-step
predictions are considered, control and planning (\ref{def:planning}) are only effective when
accurate multi-step predictions are available. Therefore, the process of first
fitting a model (e.g., VAR, RNN or Gaussian Process) and then use an external optimization step,
results in two sources of approximation error, where the
error propagates over time and reduces performance.

A potential improvement of these methods would be \textit{manually feature engineering the state space},
such as extracting meaningful econometric signals \citep{greene2003econometric} (e.g., volatility regime shifts, earning
or dividends announcements, fundamentals) which in turn are used for
predicting the returns. This is, in a nutshell, the traditional
approach that quantitative analysts \citep{lebaron2001builder} have been using for the past
decades. The computational power has been radically improved over the years,
which permits larger (i.e., deeper and wider) models to be fitted,
while elaborate algorithms, such as variational inference \citep{archer2015black}, have made previously intractable tasks possible.
Nonetheless, this approach involves a lot of tweaks and human
intervention, which are the main aspects we aim to attenuate.

\section{Model-Free Reinforcement Learning} \label{sec:model-free-reinforcement-learning}

In the final Section, we assumes that solving a system
identification problem (i.e., explicitly inferring environment dynamics) is easier than addressing
directly the initial objective; the maximization of a cumulative reward
signal (e.g., log returns, negative volatility, sharpe ratio). Nonetheless, predicting accurately
the evolution of the market was proven challenging, resulting in ill-performing agents.

In this section, we adapt an orthogonal approach, where we do
not rely on an explicit model of the environment, but
we parametrize the agent value function or/and policy directly. At
first glance, it may seem counter-intuitive how skipping the modelling
of the environment can lead to a meaningful agent at
all, but consider the following example from daily life. Humans
are able to easily handle objects or move them around.
Unarguably, this is a consequence of experience that we have
gained over time, however, if we are asked to explain
the environment model that justifies our actions, it is much
more challenging, especially for an one year old kid,
who can successfully play with toys but fails to explain
this task using Newtonian physics.

Another motivating example from the portfolio management and trading field:
\textit{pairs trading} is a simple trading strategy \citep{gatev2006pairs}, which relies on
the assumption that historically correlated assets will preserve this relationship
over time\footnote{Pairs trading is selected for educational purposes only, we are not claiming that it is optimal in any sense.}.
Hence when the two assets start deviating from
one another, this is considered an arbitrage opportunity \citep{finance:quant-finance}, since they
are expected to return to a correlated state.
This opportunity is exploited by taking a long position for
the rising stock and a short position for the falling.
If we would like to train a model-based reinforcement learning
agent to perform pairs trading, it would be almost impossible
or too unstable, regardless the algorithm simplicity. On the other
hand, a value-based or policy gradient agent could 
perform this task with minimal effort, replicating the strategy, because the pairs trading strategy
does not rely on future value prediction, but much simpler
statistical analysis (i.e., cross-correlation), which, in case of model-based approaches, should
be translated into an optimization problem of an unrelated objective - the prediction error.
Overall, using model-free reinforcement learning improves efficiency (i.e., only one episode
fitting) and also allows finer control over the policy, but
it also limits the policy to only be as good as
the learned model. More importantly, for the task under consideration (i.e.,
asset allocation) it is shown to be easier to represent
a good policy than to learn an accurate model.

The model-free reinforcement learning agents are summarized in Algorithm \ref{algo:general-setup}, where different
objective functions and agent parametrizations lead to different approaches and
hence strategies. We classify these algorithms as:

\begin{itemize}
    \item \textbf{Value-based}: learn a state value function $v$ (\ref{eq:state-value-function-mdp}),
    or an action-value function $q$ (\ref{eq:action-value-function-mdp}) and use it with an \textit{implicit}
    policy (e.g., $\epsilon$-greedy \citep{reinforcement-learning:sutton});
    \item \textbf{Policy-based}: learn a policy directly by using the reward signal to guide adaptation.
    % \item \textbf{actor-critic}: learn both a state value function and a policy, where
    % the value function provides a baseline to the policy gradient
    % estimation, reducing in this way the variance of the estimator
\end{itemize}

% The model-free reinforcement learning algorithms can be categorized by their components
% \citep{reinforcement-learning:silver-intro}, as in table \ref{tab:reinforcement-learning-algorithms-policy-value}:

% \begin{table}[h]
% \centering
% \begin{tabular}{|c||c|c|c|} 
% \hline
%                         & \textbf{Value-Based}  & \textbf{Policy-Based} & \textbf{Actor-Critic} \\ \hline\hline
% \textbf{Policy}         & \cyellow implicit        & \cgreen yes           & \cgreen yes\\ \hline
% \textbf{Value Function} & \cgreen yes           & \cred no              & \cgreen yes\\ \hline
% \end{tabular}
% \caption{Reinforcement learning algorithms classification, with respect to Policy \& Value Function.}
% \label{tab:reinforcement-learning-algorithms-policy-value}
% \end{table}

In this chapter, we will, first, focus on value-based algorithms, which exploit
the state (action) value function, as defined in equations (\ref{def:state-value-function}), (\ref{def:action-value-function})
as an estimate for expected cumulative rewards. Then policy gradient methods will be covered,
which parametrize directly the policy of the agent, and perform gradient ascent to optimize performance.
Finally, a universal agent will be introduced, which reduces
complexity (i.e., computational and memory) and generalizes strategies across assets, regardless
the trained universe, based on parameter sharing \citep{bengio2003neural} and transfer
learning \citep{pan2010survey} principles. 

\subsection{Deep Soft Recurrent Q-Network (DSRQN)} \label{sub:dsrqn-agent}

A wide range of value-based reinforcement learning algorithms
\citep{reinforcement-learning:sutton, reinforcement-learning:silver-model-free-control, reinforcement-learning:ualberta}
have been suggested and used over time. The Q-Learning is one
of the simplest and best performing ones \citep{reinforcement-learning:overview}, which motivates us
to extend it to continuous action spaces to fit our system formulation.

%%%
\subsubsection{Q-Learning}

\textbf{Q-Learning} is a simple but very effective value-based model-free reinforcement
learning algorithm \citep{reinforcement-learning:q-learning}. It works by successively improving its
evaluations of the action-value function $q$, and hence the name. Let $\hat{q}$, be the estimate
of the true action-value function, then $\hat{q}$ is updated online
(every time step) according to:
\begin{equation}
\hat{q}(\vs_{t}, \va_{t}) \leftarrow \hat{q}(\vs_{t}, \va_{t}) +
    \alpha \bigg[
    \underbrace{
        r_{t} + \gamma \max_{\va' \in \sA}\  \hat{q}(\vs_{t+1}, a') - \hat{q}(\vs_{t}, \va_{t})
    }_{\text{TD error, } \delta_{t+1}}
    \bigg]
\label{eq:q-learning}
\end{equation}
where $\alpha \geq 0$ the learning rate and $\gamma \in [0, 1]$ the discount factor.
In the literature, the term in the square brackets is usually
called \textbf{Temporal Difference Error} (TD error), or $\delta_{t+1}$ \citep{reinforcement-learning:sutton}.

\begin{theorem}
For a Markov Decision Process, Q-learning converges to the optimum action-values
with probability 1, as long as all actions are repeatedly sampled in all states and
the action-values are represented discretely.
\label{th:q-learning-convergence}
\end{theorem}
The proof of Theorem \ref{th:q-learning-convergence} is provided by
\citet{reinforcement-learning:q-learning} and relies on the contraction property of
the Bellman Operator\footnote{The Bellman Operator, $\mathfrak{B}$ is \textbf{a-contraction} with
respect to some norm $\|\cdot\|$ since it can be shown that \citep{rust1997using}:
\begin{equation}
    \| \mathfrak{B}s - \mathfrak{B}\bar{s} \| \leq a \| s - \bar{s} \|
\end{equation}
Therefore it follows that:
\begin{enumerate}
    \item The sequence $s, \mathfrak{B}s, \mathfrak{B}^{2}s, \ldots$ converges for every $s$;
    \item $\mathfrak{B}$ has a unique fixed point $s^{*}$, which satisfies $\mathfrak{B}s^{*} = s^{*}$ and 
    all sequencues $s, \mathfrak{B}s, \mathfrak{B}^{2}s, \ldots$ converge to this unique fixed point $s^{*}$.
\end{enumerate}
}
\citep{reinforcement-learning:sutton}, showing that:
\begin{equation}
\hat{q}(\vs, \va) \rightarrow q_{*}(\vs, \va) 
\label{eq:q-learning-optimality}
\end{equation}

Note that equation (\ref{eq:q-learning}) is practical only in the cases that:

\begin{enumerate}
    \item The state space is discrete, and hence the action-value function
    can be stored in a digital computer, as a grid of scalars;
    \item The action space is discrete, and hence at each iteration, the
    maximization over actions $\va \in \sA$ is tractable.
\end{enumerate}

The Q-Learning steps are summarized in Algorithm \ref{algo:q-learning}.

\begin{algorithm}
	\SetKwInOut{Args}{inputs}  
    \SetKwInOut{Output}{output}
    \DontPrintSemicolon
    \Args{%
        trading universe of $M$-assets\\
    	initial portfolio vector $\vw_{1}=\va_{0}$\\
    	initial asset prices $\vp_{0}=\vo_{0}$\\
    	initial 
    }
    \Output{optimal action-value function $q_{*}$}
    \vspace{0.25cm}
    \SetEndCharOfAlgoLine{}
    \SetAlgoLined\SetArgSty{}
    initialize q-table: $\hat{q}(\vs, \va) \leftarrow 0, \ \forall \vs \in \sS, \va \in \sA$ \\
    \While{convergence}{
        \For{$t=0, 1, \ldots T$}{
            select greedy action: $\va_{t} = \max_{\va' \in \sA} \hat{q}(\vs_{t}, \va')$ \\
            observe tuple $\langle \vs_{t+1}, r_{t} \rangle$ \\
            update q-table: $\hat{q}(\vs_{t}, \va_{t}) \leftarrow \hat{q}(\vs_{t}, \va_{t}) +
                \alpha \big[ r_{t} + \gamma \max_{\va' \in \sA}\ \hat{q}(\vs_{t+1}, \va') - \hat{q}(\vs_{t}, \va_{t}) \big]$
        }
    }
\caption{Q-Learning with greedy policy.}
\label{algo:q-learning}
\end{algorithm}

%%%
\subsubsection{Related Work}

Due to the success of the Q-Learning algorithm, early attempts to
modify it to fit the asset allocation task were made
by \citet{paper:adaptive-dynamic-programming}, who attempted to used a differentiable function approximator (i.e., neural network)
to represent the action-value function, and hence enabled the
use of Q-Learning in \textit{continuous state spaces}. Nonetheless, he was
restricted to discrete action spaces and thus was acting on \textit{buy} and \textit{sell} signals only.
In the same year, \citet{paper:rl-trading-portfolios} used a similar approach but introduced
new reward signals, namely general utility functions and the differential
Sharpe Ratio, which are considered in Chapter \ref{ch:market-data}.

Recent advances in deep learning \citep{deep-learning:goodfellow} and stochastic optimization methods \citep{optimization:boyd}
led to the first practical use case of Q-Learning in high-dimensional
state spaces \citep{deepmind:atari}. Earlier work was limited to low-dimensional applications, where
shallow neural network architectures were effective. \citet{deepmind:atari} used a few algorithmic tricks and
heuristics, and managed to stabilize the training process of the
\textbf{Deep Q-Network} (DQN). The first demonstration was performed on Atari
video games, where trained agents outperformed human players in most games.

Later, \citet{hausknecht2015deep} published a modified version of the
DQN for partially observable environments, using a recurrent layer
to construct the agent state, giving rise to the \textbf{Deep Recurrent Q-Network} (DRQN).
Nonetheless, similar to the vanilla Q-Learning and enhanced DQN algorithms,
the action space was always discrete.

%%%
\subsubsection{Agent Model}

Inspired by the breakthroughs in DQN and DRQN, we suggest
a modification to the last layers to handle pseudo-continuous action spaces,
as required for the portfolio management task. The
current implementation, termed the \textbf{Deep Soft Recurrent Q-Network} (DSRQN) relies
on a fixed, implicit policy (i.e., exponential normalization or \textit{softmax} \citep{mccullagh1984generalized}) while
the action-value function $q$ is adaptively fitted.

\begin{figure}[h]
    \centering
    \includegraphics[width=\textwidth]{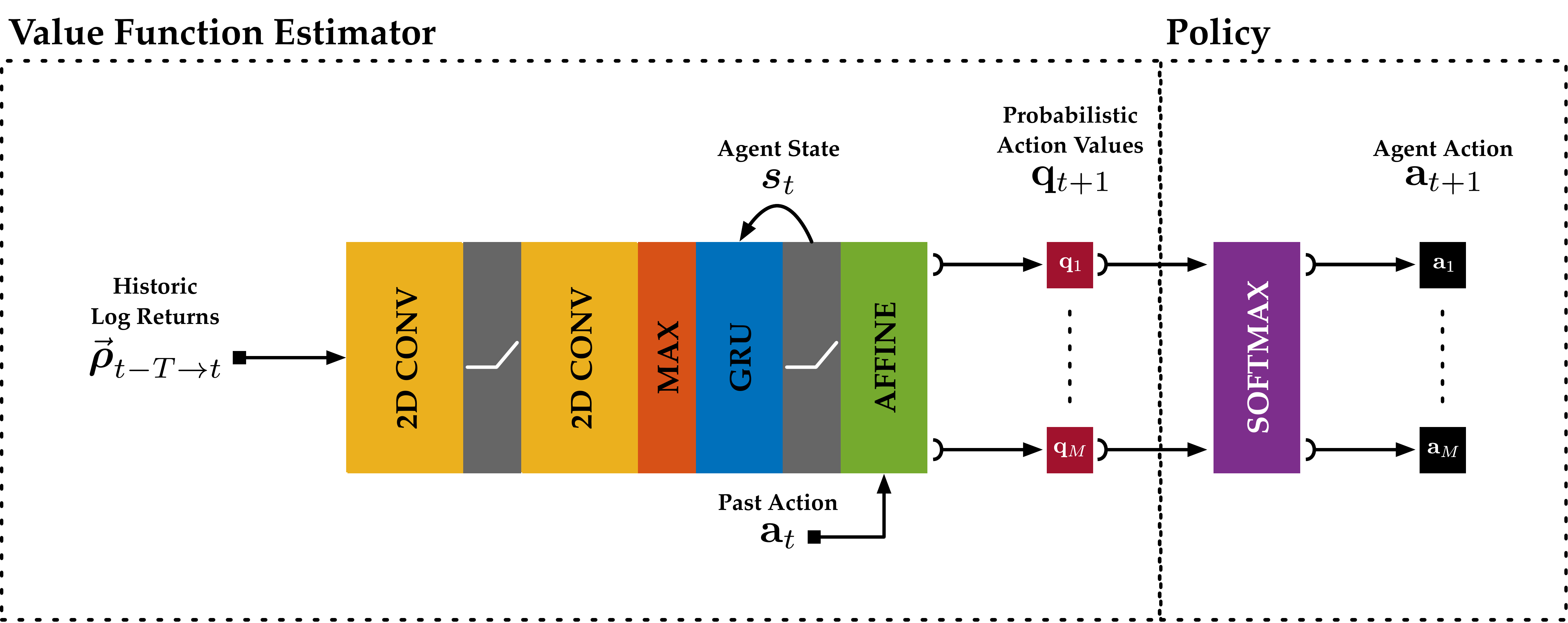}
    \caption{Deep Soft Recurrent Q-Network (DSRQN) architecture. The historic log returns
    $\vvrho_{t-T \rightarrow t} \in \sR^{M \times T}$ are passed throw two 2D-convolution layers, which generate a feature
    map, which is, in turn, processed by the GRU state
    manager. The agent state produced is combined (via matrix flattening and vector concatenation) with the past
    action (i.e., current portfolio positions) to estimate action-values $q_{1}, q_{2}, \ldots, q_{M}$. The action
    values are used both for calculating the TD error (\ref{eq:q-learning}), showing
    up in the gradient calculation, as well as for determining
    the agents actions, after passed throw a softmax activation layer.}
    \label{fig:dsqrn-schematic}
\end{figure}

A neural network architecture as in Figure \ref{fig:dsqrn-schematic} is used to estimate the action-value function.
The two 2D-convolution (2D-CONV) layers are followed by a max-pooling (MAX) layer, which
aim to extract non-linear features from the raw historic log returns \citep{lecun1995convolutional}.
The feature map is then fed to the gated recurrent
unit (GRU), which is the state manager (see Section \ref{sub:financial-market-state}), responsible
for reconstructing a meaningful agent state from a partially
observable environment. The generated agent state, $\vs_{t}$, is then regressed along
with the past action, $\va_{t}$ (i.e., current portfolio vector), in order to produce
the action-value function estimates. Those estimates are used with
the realized reward $r_{t+1}$ to calculate the TD error $\delta_{t+1}$ (\ref{eq:q-learning}) and train
the DSRQN as in Algorithm \ref{algo:dsqrn}. The action-values estimates $\hat{q}_{t+1}$ are passed to
a \textbf{softmax} layer, which produces the agent action $\va_{t+1}$. We select
the softmax function because it provides the favourable property of
forcing all the components (i.e., portfolio weights) to sum to unity
(see Section \ref{sec:terms-concepts}). Analytically the actions are given by:

\begin{equation}
    \forall i \in \{1, 2, \ldots, M\}: \quad a_{i} = \frac{e^{a_{i}}}{\sum_{j=1}^{M} e^{a_{j}}}
    \Longrightarrow
    \sum_{i=1}^{M} a_{i} = 1
\label{def:softmax}
\end{equation}

Note that the last layer (i.e, softmax) is not trainable \citep{deep-learning:goodfellow}, which
means that it can be replaced by any function (i.e., deterministic or stochastic), even
by a quadratic programming step, since differentiability is not required.
For this experiment we did not consider more advanced policies,
but anything is accepting as long as constraint (\ref{def:portfolio-vector}) is satisfied.
Moreover, comparing our implementation with the original DQN \citep{deepmind:atari}, no experience
replay is performed, in order to avoid reseting the GRU
hidden state for each batch, which will lead to an
unused latent state, and hence poor state manager.

\begin{algorithm}
	\SetKwInOut{Args}{inputs}  
    \SetKwInOut{Output}{output}
    \DontPrintSemicolon
    \Args{%
        trading universe of $M$-assets\\
    	initial portfolio vector $\vw_{1}=\va_{0}$\\
    	initial asset prices $\vp_{0}=\vo_{0}$\\
        objective function $\mathcal{J}$\\
        initial agent weights $\vtheta_{0}$
    }
    \Output{optimal agent parameters $\vtheta_{*}$}
    \vspace{0.25cm}
    \SetEndCharOfAlgoLine{}
    \SetAlgoLined\SetArgSty{}
    \Repeat{convergence}{
        \For{$t=1, 2, \ldots T$}{
            observe tuple $\langle \vo_{t}, r_{t} \rangle$ \\
            calculate TD error $\delta_{t+1}$ \tcp*{(\ref{eq:q-learning})}
            calculate gradients $\nabla_{\theta_{i}}\mathcal{L}(\theta_{i}) =
                \delta_{t+1} \nabla_{\theta_{i}} q(\vs, \va; \vtheta)$ \tcp*{BPTT}
            update agent parameters $\vtheta$\;
                \Indp using adaptive gradient optimizers \tcp*{ADAM} \Indm
            get estimate of value function $\vq_{t} \approx \textbf{NN}(\vvrho_{t-T \rightarrow t})$ \tcp*{(\ref{eq:q-learning})}
            take action $\va_{t} \text{softmax}( \vq_{t} )$ \tcp*{portfolio rebalance}
        }
    }
    set $\vtheta_{*} \leftarrow \vtheta$
\caption{Deep Soft Recurrent Q-Learning.}
\label{algo:dsqrn}
\end{algorithm}

%%%
\subsubsection{Evaluation}

\begin{figure}
    \centering
    \begin{subfigure}[t]{0.48\textwidth}
        \includegraphics[width=\textwidth]{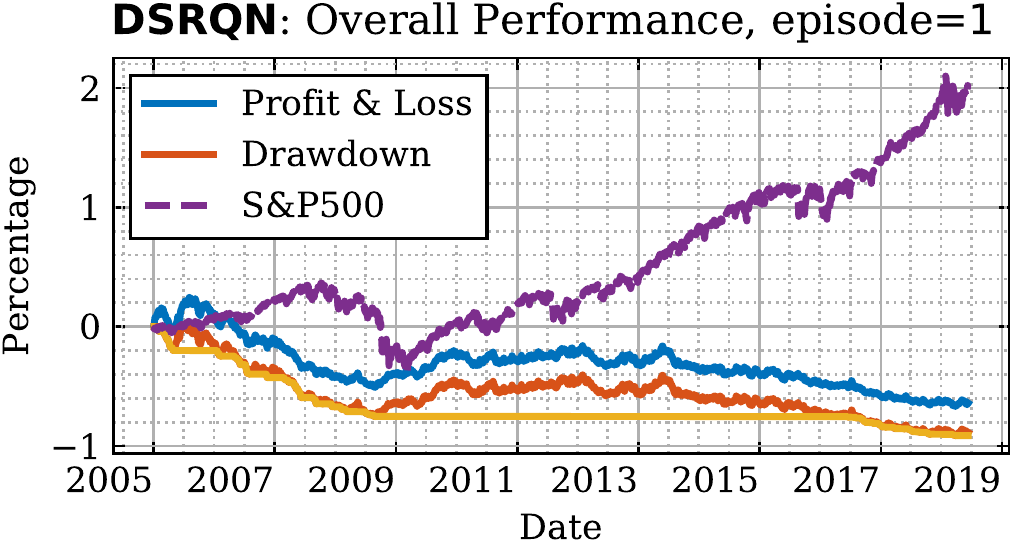}
    \end{subfigure}
    ~ 
    \begin{subfigure}[t]{0.48\textwidth}
        \includegraphics[width=\textwidth]{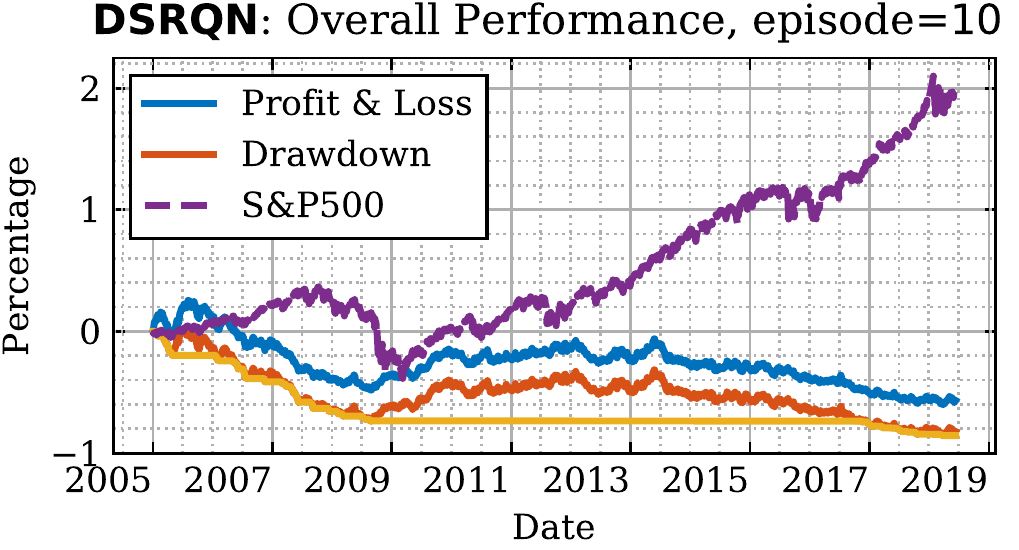}
    \end{subfigure}
    
    \vspace{0.5cm}
    
    \begin{subfigure}[t]{0.48\textwidth}
        \includegraphics[width=\textwidth]{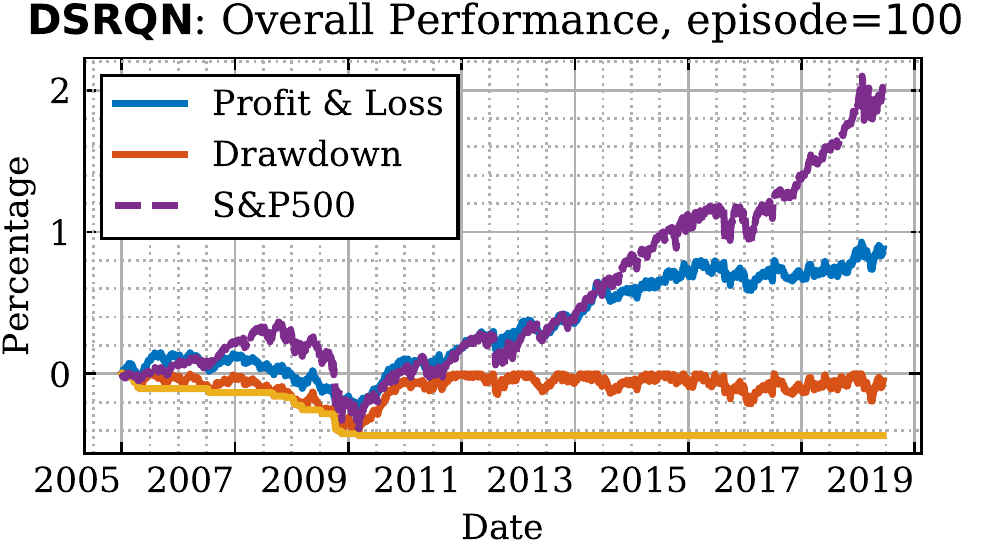}
    \end{subfigure}
    ~    
    \begin{subfigure}[t]{0.48\textwidth}
        \includegraphics[width=\textwidth]{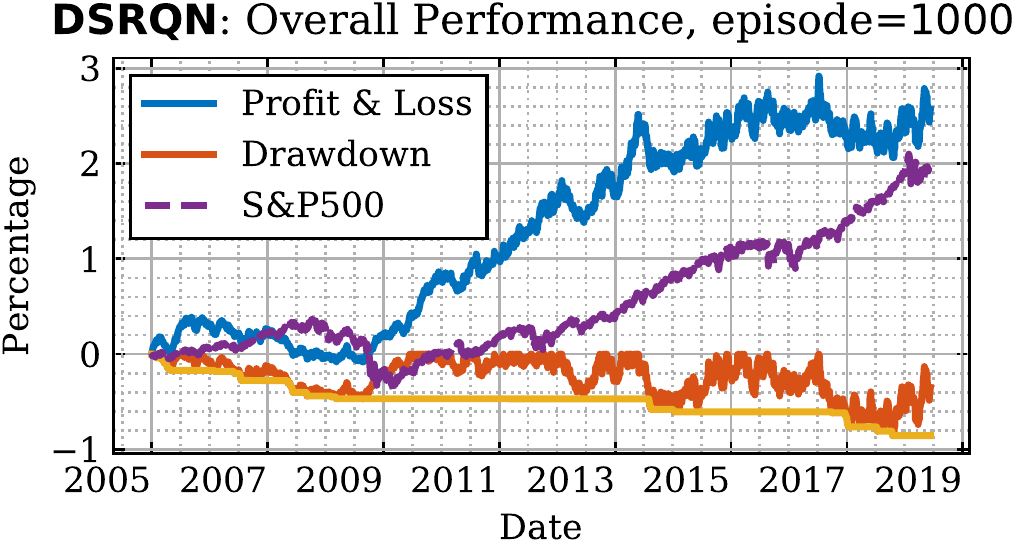}
    \end{subfigure}
    \caption{Deep Soft Recurrent Q-Network (DSRQN) model-free reinforcement learning agent on a $12$-asset universe,
    trained on historic data between 2000-2005 and tested onward,
    for different number of episodes $e = \{1, 10, 100, 1000 \}$.
    Visualization of cumulative rewards and (maximum) drawdown of the
    learned strategy, against the S\&P 500 index (traded as \texttt{SPY}).}
    \label{fig:dsqrn-report}
\end{figure}

Figure \ref{fig:dsqrn-report} illustrates the results obtained on a small scale experiment
with $12$ assets from S\&P 500 market using the DSRQN. The agent is
trained on historic data between 2000-2005 for 5000 episodes, and
tested on 2005-2018. The performance evaluation of the agent for different episodes $e$ is highlighted.
For $e=1$ and $e=10$, the agent acted completely randomly, leading to poor performance.
For $e=100$, the agent did not beat the market (i.e., S\&P 500) but it
learned how to follow it (hence high correlation) and to yiel profit out of it (since the market is increasing in this case).
Finally, for $e=1000$, we note that the the agent is both profitable and less correlated with the market, compared to
previous model-based agents (i.e., VAR and GRU-RNN), which were highly impacted
by 2008 market crash, while DSRQN was almost unaffected (e.g., 85 \% drowdawn).

\begin{figure}
    \centering
    \begin{subfigure}[t]{0.48\textwidth}
        \includegraphics[width=\textwidth]{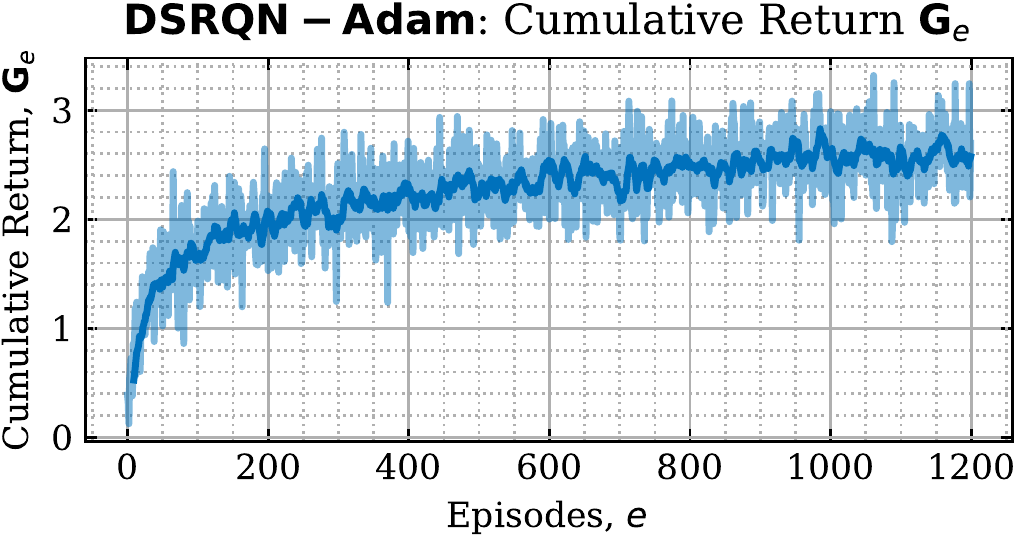}
    \end{subfigure}
    ~ 
    \begin{subfigure}[t]{0.48\textwidth}
        \includegraphics[width=\textwidth]{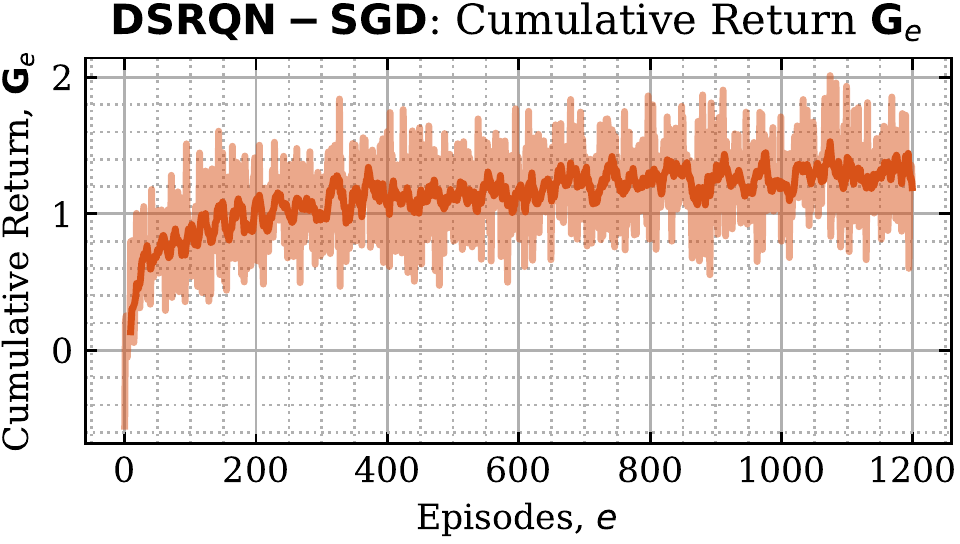}
    \end{subfigure}
    \caption{Out-of-sample cumulative returns per episode during training phase for DSRQN. Performance improvement
    saturates after $e \gtrapprox 1000$. (\textit{Left}) Adaptive Neural network optimization algorithm \textbf{ADAM} \citep{kingma2014adam}. (\textit{Right}) Neural network optimized with Stochastic Gradient Descent (SGD)
    \citep{mandic2004generalized}.}
    \label{fig:dsqrn-learning-curve}
\end{figure}

In Figure \ref{fig:dsqrn-learning-curve}, we illustrate the out-of-sample cumulative return of the
DSRQN agent, which flattens for $e \gtrapprox 1000$, and hence the neural network optimizer converges
to a (local) minimum, thus we terminate training.

Figure \ref{fig:dsqrn-learning-curve} also
highlights the importance of the optimization algorithm used in training
the neural network, since ADAM (i.e., the adaptive optimizer) did not
only converge faster than Stochastic Gradient Descent (SGD), but it also
found a better (local) minimum.

%%%
\subsubsection{Weaknesses}

Despite the improved performance of DSRQN compared to the model-based agents,
its architecture has severe weaknesses.

Firstly, the selection of the policy (e.g., softmax layer) is a
manual process that can be only verifies via empirical means,
for example, cross-validation. This complicates the training process, without guaranteeing
any global optimality of the selected policy.

\begin{property}[End-to-End Differentiable Architecture]
    Agent policy should be part of the trainable architecture
    so that it adapts to (locally) optimal strategy via gradient
    optimization during training.
\label{pro:end-to-end-differentiable-architecture}
\end{property}

Secondly, DSRQN is a Many-Input-Many-Output (MIMO) model, whose number of
parameters grows polynomially as a function of the universe size
(i.e., number of assets $M$), and hence its training complexity. Moreover, under this
setting, the learned strategy is universe-specific, which means that the same
trained network does not generalize to other universes. It even fails
to work on permutations of the original universe; for example,
if we interchange the order assets in the processed observation $\hat{\vs}_{t}$
after training, then DSRQN will break down. 

\begin{property}[Linear Scaling]
    Model should scale linearly (i.e, computation and memory) with respect
    to the universe size.
\label{pro:linear-scaling}
\end{property}

\begin{property}[Universal Architecture]
    Model should be universe-agnostic and replicate its strategy regardless the
    underlying assets.
\label{pro:universal-architecture}
\end{property}

\subsection{Monte-Carlo Policy Gradient (REINFORCE)} \label{sub:reinforce-agent}

In order to address the first weakness of the DSRQN
(i.e., manual selection of policy), we consider policy gradient algorithms, which
directly address the learning of an agent policy, without intermediate
action-value approximations, resulting in an end-to-end differentiable model.

%%%
\subsubsection{Policy Gradient Theorem}

In Section \ref{sub:policy} we defined policy of an agent, $\pi$, as:

\begin{equation}
    \pi: \sS \rightarrow \sA
\tag{\ref{eq:policy-mdp}}
\end{equation}

In policy gradient algorithms, we parametrize the policy with parameters
$\vtheta$ as $\pi_{\vtheta}$ and optimize them according to a long-term objective function $\mathcal{J}$,
such as average reward per time-step, given by:

\begin{equation}
    \mathcal{J}(\vtheta) \triangleq \sum_{\vs \in \sS} \mathcal{P}^{\pi_{\vtheta}}(s)
        \sum_{\va \in \sA} \pi_{\vtheta}(\vs, \va) \mathcal{R}_{\vs}^{\va}
\label{def:average-reward-per-time-step}
\end{equation}

Note, that any differentiable parametrization of the policy is valid (e.g., neural network, linear model).
Moreover, the freedom of choosing the reward generating function, $\mathcal{R}_{\vs}^{\va}$, is still available.

In order to optimize the parameters, $\vtheta$, the gradient, $\nabla_{\vtheta} \mathcal{J}$,
should be calculated at each iteration. Firstly, we consider an
one-step Markov Decision Process:

\begin{align}
    \mathcal{J}(\vtheta)    &= \E_{\pi_{\vtheta}} [ \mathcal{R}_{\vs}^{\va} ] \tag{\ref{def:average-reward-per-time-step}}\\
                            &= \sum_{\vs \in \sS} \mathcal{P}^{\pi_{\vtheta}}(s)
                            \sum_{\va \in \sA} \pi_{\vtheta}(\vs, \va) \mathcal{R}_{\vs}^{\va} \\
    \nabla_{\vtheta} \mathcal{J}(\vtheta)
                            &= \sum_{\vs \in \sS} \mathcal{P}^{\pi_{\vtheta}}(s)
                            \sum_{\va \in \sA} \nabla_{\vtheta} \pi_{\vtheta}(\vs, \va) \mathcal{R}_{\vs}^{\va} \nonumber\\
                            &= \sum_{\vs \in \sS} \mathcal{P}^{\pi_{\vtheta}}(s) \sum_{\va \in \sA}
                            \pi_{\vtheta}(\vs, \va) \frac{\nabla_{\vtheta} \pi_{\vtheta}(\vs, \va)}{\pi_{\vtheta}(\vs, \va)}
                            \mathcal{R}_{\vs}^{\va} \nonumber\\
                            &= \sum_{\vs \in \sS} \mathcal{P}^{\pi_{\vtheta}}(s) \sum_{\va \in \sA}
                            \pi_{\vtheta}(\vs, \va) \nabla_{\vtheta} log[\pi_{\vtheta}(\vs, \va)] \mathcal{R}_{\vs}^{\va}
                            \nonumber\\
                            &= \E_{\pi_{\vtheta}} \bigg[ \nabla_{\vtheta} log[\pi_{\vtheta}(\vs, \va)] \mathcal{R}_{\vs}^{\va} \bigg]
\label{def:policy-gradient-one-step-mdp}
\end{align}

The policy gradient calculation is extended to multi-step MDPs by replacing the
instantaneous reward $\mathcal{R}_{\vs}^{\va}$ with the long-term (action) value $q^{\pi}(\vs, \va)$.

\begin{theorem}[Policy Gradient Theorem]
    For any differentiable policy $\pi_{\vtheta}(\vs, \va)$ and for $\mathcal{J}$ the average discounted future
    rewards per step, the policy gradient is:
    \begin{equation}
        \nabla_{\vtheta} \mathcal{J}(\vtheta) =
        \E_{\pi_{\vtheta}} \bigg[ \nabla_{\vtheta} log[\pi_{\vtheta}(\vs, \va)] q^{\pi}(\vs, \va) \bigg]
    \label{def:policy-gradient}
    \end{equation}
\end{theorem}

where the proof is provided by \citet{sutton2000policy}. The theorem applies to continuous
settings (i.e., Infinite MDPs) \citep{reinforcement-learning:sutton}, where the summations are replaced by integrals.

%%%
\subsubsection{Related Work}

Policy gradient methods have gained momentum the past years due
to their better covergence policites \citep{reinforcement-learning:sutton}, their efffective in high-dimensional or
continuous action spaces and natural fit to stochastic environments \citep{reinforcement-learning:silver-policy-gradient}.
Apart from the their extensive application in robotics \citep{smart2002effective, kohl2004policy, kober2009policy}, policy gradient
methods have been used also in financial markets. \citet{paper:pg-asset-allocation} develops a
general framework for policy gradient agents to be trained to solve
the asset allocation task, but only successful back-tests for synthetic market data are provided.

%%%
\subsubsection{Agent Model}

From equation (\ref{def:policy-gradient}), we note two challenges:

\begin{enumerate}
    \item Calculation of an expectation over the (stochastic) policy, $\E_{\pi_{\vtheta}}$, leading to
    integration over unknown quantities;
    \item Estimation of the unknown action-value, $q^{\pi}(\vs, \va)$
\end{enumerate}

The simplest, successful algorithm to address both of the challenges
is \textbf{Monte-Carlo Policy Gradient}, also know as \textbf{REINFORCE}. In particular,
different trajectories are generated following policy $\pi_{\vtheta}$ which are then used
to estimate the expectation\footnote{The empirical mean is an unbiased estimate
of expected value \citep{mandic:asp-estimators}.} and the discounted future rewards, which are
an unbiased estimate of $q^{\pi}(\vs, \va)$. Hence we obtain Monte-Carlo estimates:

\begin{align}
    q^{\pi}(\vs, \va)   &\approx G_{t} \triangleq \sum_{i=1}^{t} r_{i}
\label{eq:monte-carlo-returns} \\
    \E_{\pi_{\vtheta}} \bigg[ \nabla_{\vtheta} log[\pi_{\vtheta}(\vs, \va)] q^{\pi}(\vs, \va) \bigg]
                        &\approx \frac{1}{T} \bigg[ 
                            \nabla_{\vtheta} log[\pi_{\vtheta}(\vs, \va)] \sum_{i=1}^{T} G_{i} \bigg]
\label{eq:monte-carlo-log-gradient}
\end{align}

where the gradient of the log term $\nabla_{\vtheta} log[\pi_{\vtheta}(\vs, \va)]$, is obtained by Backpropagation Through Time
\citep{deep-learning:bptt}.

\begin{figure}[h]
    \centering
    \includegraphics[width=\textwidth]{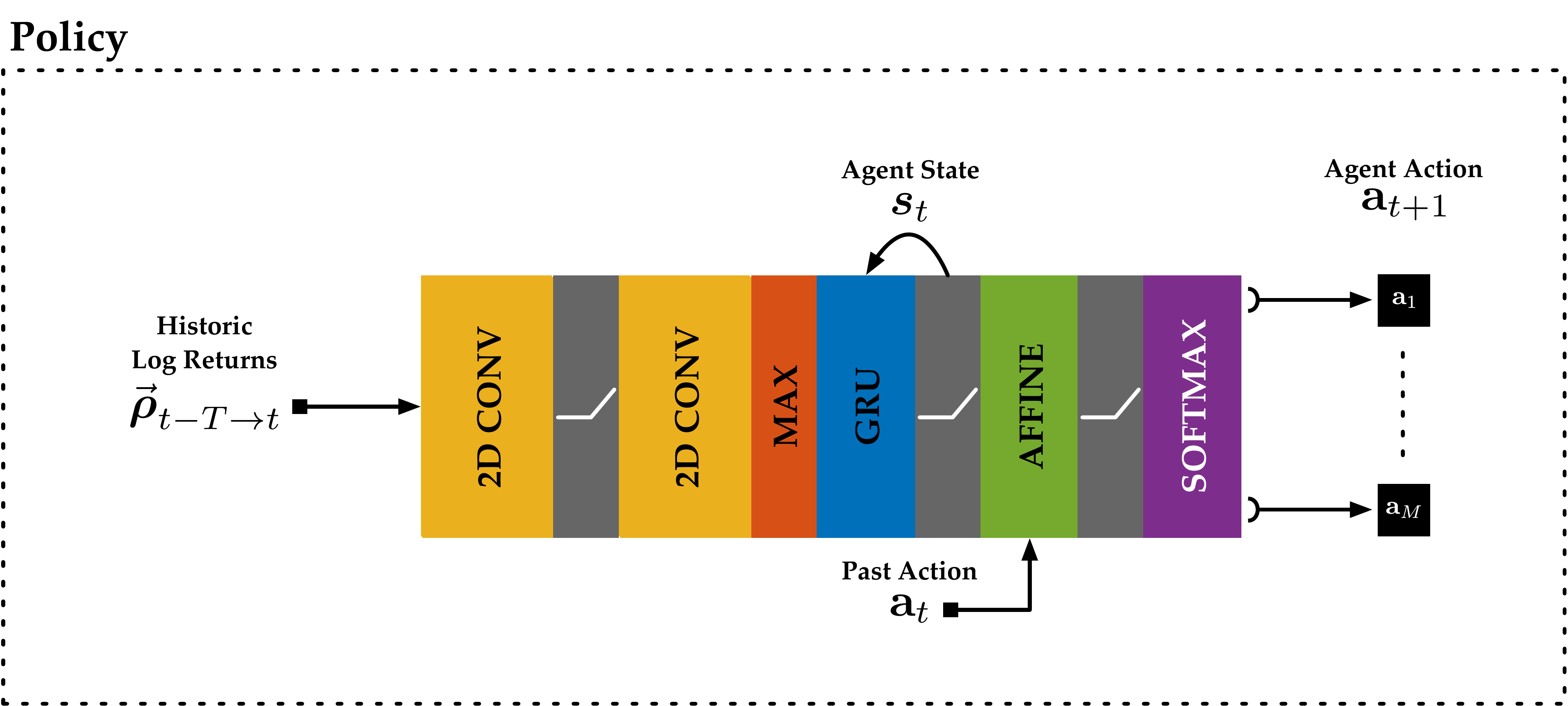}
    \caption{Monte-Carlo Policy Gradient (REINFORCE) architecture. The historic log returns
    $\vvrho_{t-T \rightarrow t} \in \sR^{M \times T}$ are passed throw two 2D-convolution layers, which generate a feature
    map, which is, in turn, processed by the GRU state
    manager. The agent state produced and the past
    action (i.e., current portfolio positions) are non-linearly regressed and exponentially normalized by
    the affine and the  softmax layer, respectively, to generate the agent actions.}
    \label{fig:reinforce-schematic}
\end{figure}

We choose to parametrize the policy using a neural network
architecture, illustrated in Figure \ref{fig:reinforce-schematic}. The configuration looks very similar to
the DSRQN, but the important difference is in the last two layers, where the REINFORCE network does not estimate the
state action-values, but directly the agent actions. Algorithm \ref{algo:reinforce} describes
the steps for training a REINFORCE agent.

\begin{algorithm}
	\SetKwInOut{Args}{inputs}  
    \SetKwInOut{Output}{output}
    \DontPrintSemicolon
    \Args{%
    	trading universe of $M$-assets\\
    	initial portfolio vector $\vw_{1}=\va_{0}$\\
    	initial asset prices $\vp_{0}=\vo_{0}$\\
        objective function $\mathcal{J}$\\
        initial agent weights $\vtheta_{0}$
    }
    \Output{optimal agent policy parameters $\vtheta_{*}$}
    \vspace{0.25cm}
    \SetAlgoLined\SetArgSty{}
    initialize buffers: $G,\ {\Delta \vtheta}_{c} \leftarrow 0$
    \Repeat{convergence}{
        \For{$t=1, 2, \ldots T$}{
            observe tuple $\langle \vo_{t}, r_{t} \rangle$ \\
            sample and take action: $\va_{t} \sim \pi_{\vtheta}(\cdot | \vs_{t} ; \vtheta )$ \tcp*{portfolio rebalance}
            cache rewards: $G \leftarrow G + r_{t}$ \tcp*{(\ref{eq:monte-carlo-returns})}
            cache log gradients: ${\Delta \vtheta}_{c} \leftarrow {\Delta \vtheta}_{c} +
                \nabla_{\vtheta} log[\pi_{\vtheta}(\vs, \va)] G$
            \tcp*{(\ref{eq:monte-carlo-log-gradient})}
        }
        update policy parameters $\vtheta$ using buffered\;
            \Indp Monte-Carlo estimates via adaptive optimization \tcp*{(\ref{def:policy-gradient}), ADAM} \Indm
        empty buffers: $G,\ {\Delta \vtheta}_{c} \leftarrow 0$
    }
    set $\vtheta_{*} \leftarrow \vtheta$
\caption{Model-Carlo Policy Gradient (REINFORCE).}
\label{algo:reinforce}
\end{algorithm}

%%%
\subsubsection{Evaluation}

\begin{figure}
    \centering
    \begin{subfigure}[t]{0.48\textwidth}
        \includegraphics[width=\textwidth]{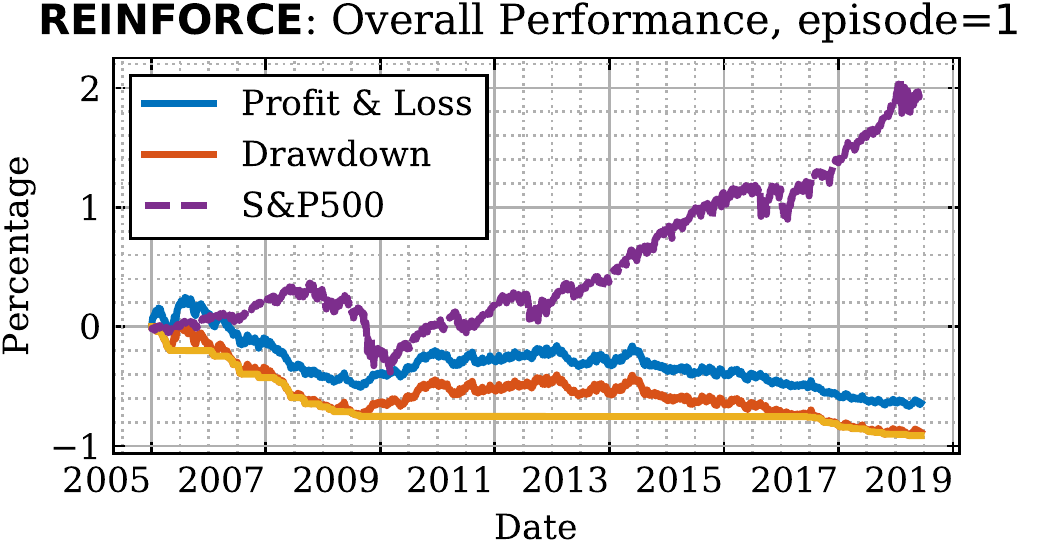}
    \end{subfigure}
    ~ 
    \begin{subfigure}[t]{0.48\textwidth}
        \includegraphics[width=\textwidth]{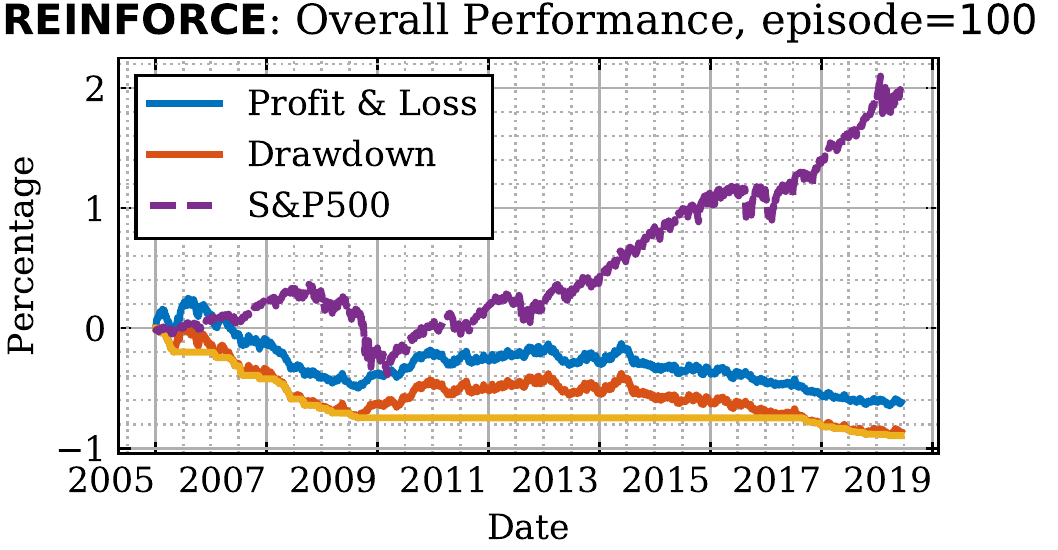}
    \end{subfigure}
    
    \vspace{0.5cm}
    
    \begin{subfigure}[t]{0.48\textwidth}
        \includegraphics[width=\textwidth]{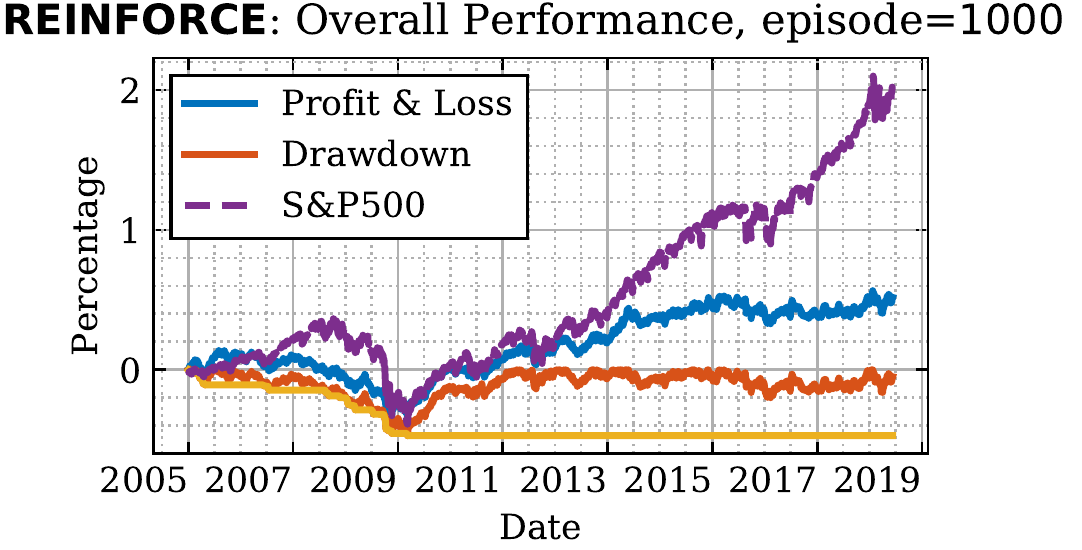}
    \end{subfigure}
    ~    
    \begin{subfigure}[t]{0.48\textwidth}
        \includegraphics[width=\textwidth]{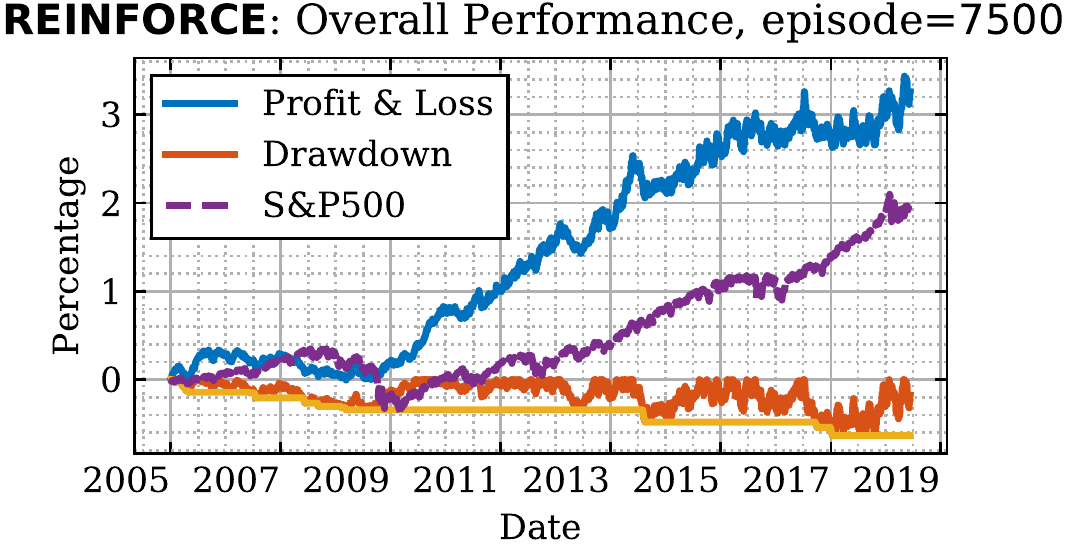}
    \end{subfigure}
    \caption{Model-Carlo Policy Gradient (REINFORCE) model-free reinforcement learning agent on a $12$-asset universe,
    trained on historic data between 2000-2005 and tested onward,
    for different number of episodes $e = \{1, 100, 1000, 7500 \}$.
    Visualization of cumulative rewards and (maximum) drawdown of the
    learned strategy, against the S\&P 500 index (traded as \texttt{SPY}).}
    \label{fig:reinforce-report}
\end{figure}

In Figure \ref{fig:reinforce-report}, we present the results from an experiment performed
on a small universe comprising of $12$ assets from S\&P 500 market using
the REINFORCE agent. The agent is trained on historic data
between 2000-2005 for 5000 episodes, and tested on 2005-2018.
Similar to the DSRQN, at early episodes the REINFORCE agent
performs poorly, but it learns a profitable strategy after a few thousands of episodes ($e \approx 7500$).
Note that almost $8$ times more episodes are required to train
the REINFORCE agent compared to the DSRQN, however, the latter performs a parameters update at every step,
while the former only once every episode (i.e., approximately steps $1260$ in an episodes).
The performance of the REINFORCE agent is significantly improved, with total cumulative returns $325.9 \%$
and $63.5 \%$ maximum drawdown.

In Figure \ref{fig:dsqrn-learning-curve}, we illustrate the out-of-sample cumulative return of the
DSRQN agent, which flattens for $e \gtrapprox 7500$, and hence the neural network optimizer converges
to a (local) minimum, thus we terminate training.

\begin{figure}
    \centering
    \begin{subfigure}[t]{0.48\textwidth}
        \includegraphics[width=\textwidth]{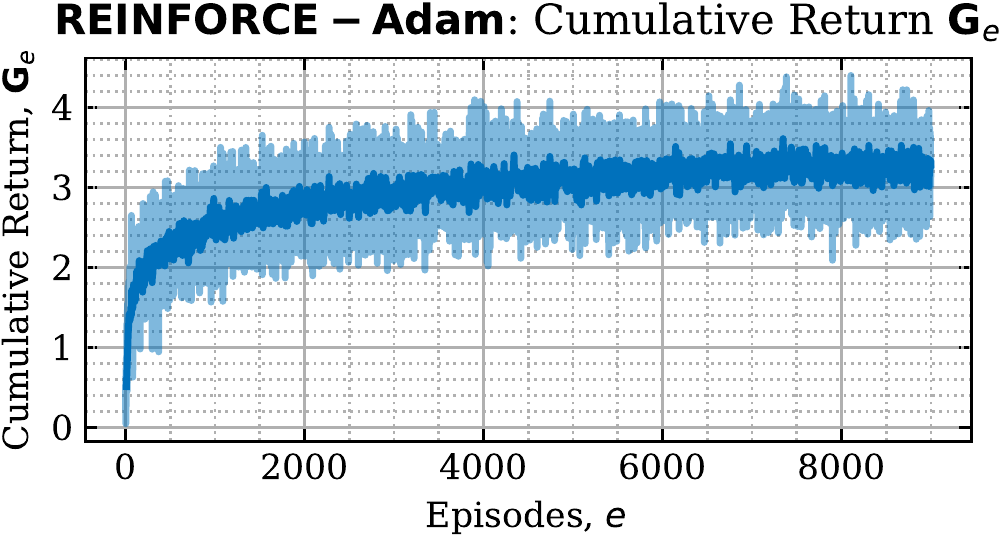}
    \end{subfigure}
    ~ 
    \begin{subfigure}[t]{0.48\textwidth}
        \includegraphics[width=\textwidth]{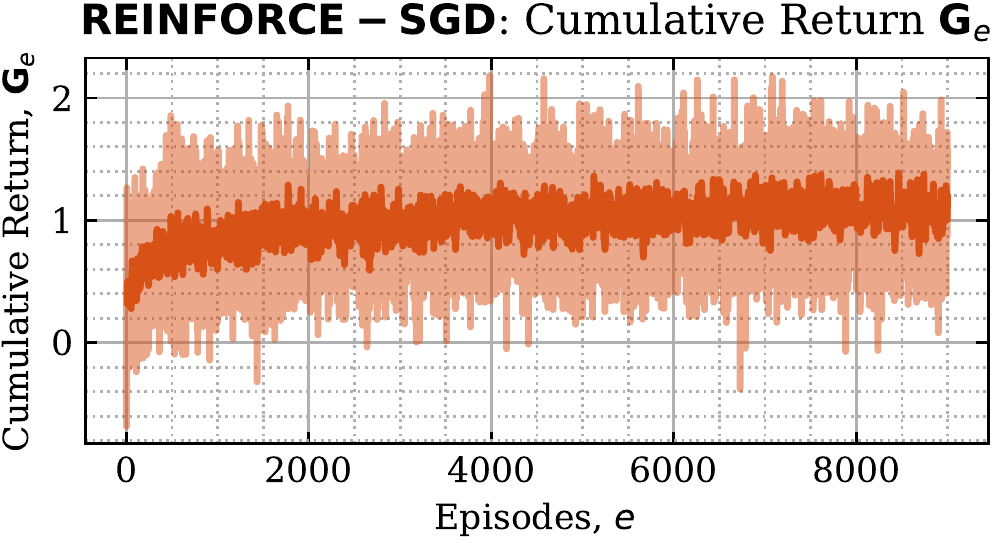}
    \end{subfigure}
    \caption{Out-of-sample cumulative returns per episode during training phase for REINFORCE. Performance improvement
    saturates after $e \gtrapprox 7500$. (\textit{Left}) Adaptive Neural network optimization algorithm \textbf{ADAM}.
    (\textit{Right}) Neural network optimized with Stochastic Gradient Descent (SGD).}
    \label{fig:reinforce-learning-curve}
\end{figure}

%%%
\subsubsection{Weaknesses}

Policy gradient addressed only the end-to-end differentiability weakness of the
DSRQN architecture, leading to siginificant improvements. Nonetheless, the polynomial scaling
and universe-specific nature of the model are still restricting the
applicability and generalization of the learned strategies. Moreover, the intractability
of the policy gradient calculation given by (\ref{def:policy-gradient}) lead to the comprising
solution of using Monte-Carlo estimates by running numerous simulations, leading
to increased number of episodes required for convergence. Most importantly,
the empirical estimation of the state action-value $q^{\pi_{\vtheta}}(\vs, \va)$ in (\ref{eq:monte-carlo-returns})
has high-variance (see returns in Figure \ref{fig:reinforce-learning-curve}) \citep{reinforcement-learning:sutton}.

\begin{property}[Low Variance Estimators]
    Model should rely on low variance estimates.
\label{pro:low-variance-etimators}
\end{property}

\subsection{Mixture of Score Machines (MSM)} \label{sub:mixture-of-score-machines}

% In this subsection, we introduce the \textbf{Advantage Actor-Critic} (A2C) model \citep{mnih2016asynchronous},
% which reduces the high variance of the state-action-value estimate of REINFORCE
% by \textit{learning a state-value function} along with the policy. Usually,
% the policy estimator is called \textbf{actor}, while the \textbf{critic} is
% the state-action-value estimator. The critic is trained by minimizing
% the TD error (\ref{eq:q-learning}), similar to the DSRQN except for the
% last softmax layer. The actor uses the state-action-value estimates
% of the critic to approximate the policy gradient (\ref{def:policy-gradient}), enabling training
% without costly and volatile Monte-Carlo estimates. The variance of policy
% gradient estimate can be further reduce by means of a baseline, or the \textbf{advantage function}.
% Finally, we reduce the model complexity and provide the architecture
% of a \textbf{universal} model, which without retraining can replicate learned
% strategies across assets and markets.

In this subsection, we introduce the \textbf{Mixture of Score Machines} (MSM) model
with the aim to provide a universal model that reduces the agent model
complexity and generalizes strategies across assets, regardless of the trained universe.
These properties are obtained by virtue of principles of parameter
sharing \citep{bengio2003neural} and transfer learning \citep{pan2010survey}.

%%%
\subsubsection{Related Work}

\citet{paper:drl-pm} suggested a universal policy gradient architecture, the \textit{Ensemble of Identical
Independent Evaluators} (EIIE), which reduces significantly the model complexity, and
hence enables larger-scale applications. Nonetheless, it operates only on independent
(e.g., uncorrelated) time-series, which is a highly unrealistic assumption for real
financial markets.

%%%
\subsubsection{Agent Model}

As a generalization to the universal model of \citet{paper:drl-pm}, we introduce
the \textbf{Score Machine} (SM), an estimator of statistical moments of
stochastic multivariate time-series:

\begin{itemize}
    \item A \textbf{First-Order Score Machine} SM($1$) operates on univariate time-series,
        generating a score that summarizes the characteristics of the \textit{location} 
        parameters of the time-series (e.g., mode, median, mean). An $M$-components multivariate
        series will have $\binom{M}{1} = M$ first-order scores, one for each component;
    \item A \textbf{Second-Order Score Machine} SM($2$) operates on bivariate time-series,
        generating a score that summarizes the characteristics of the \textit{dispersion} 
        parameters of the joint series (e.g., covariance, modal dispersion \citep{meucci2009risk}).
        An $M$-components multivariate series will have $\binom{M}{2} = \frac{M!}{2!(M-2)!}$ second-order scores, one
        for each distinct pair;
    \item An \textbf{$N$-Order Score Machine} SM($N$) operates on $N$-component multivariate series
        and extracts information about the $N$\textit{-order statistics} (i.e., statistical moments) of
        the joint series. An $M$-components multivariate series, for $M \geq N$, will have $\binom{M}{N} = \frac{M!}{N!(M-N)!}$
        $N$-order scores, one for each distinct $N$ combination of components.
\end{itemize}

Note that the extracted scores are \textit{not} necessarily equal to
the statistical (central) moments of the series, but a compressed and
informative representation of the statistics of the time-series. The
universality of the score machine is based on parameter sharing \citep{bengio2003neural} across assets.

\begin{figure}
    \centering
    \begin{subfigure}[t]{0.48\textwidth}
        \includegraphics[width=\textwidth]{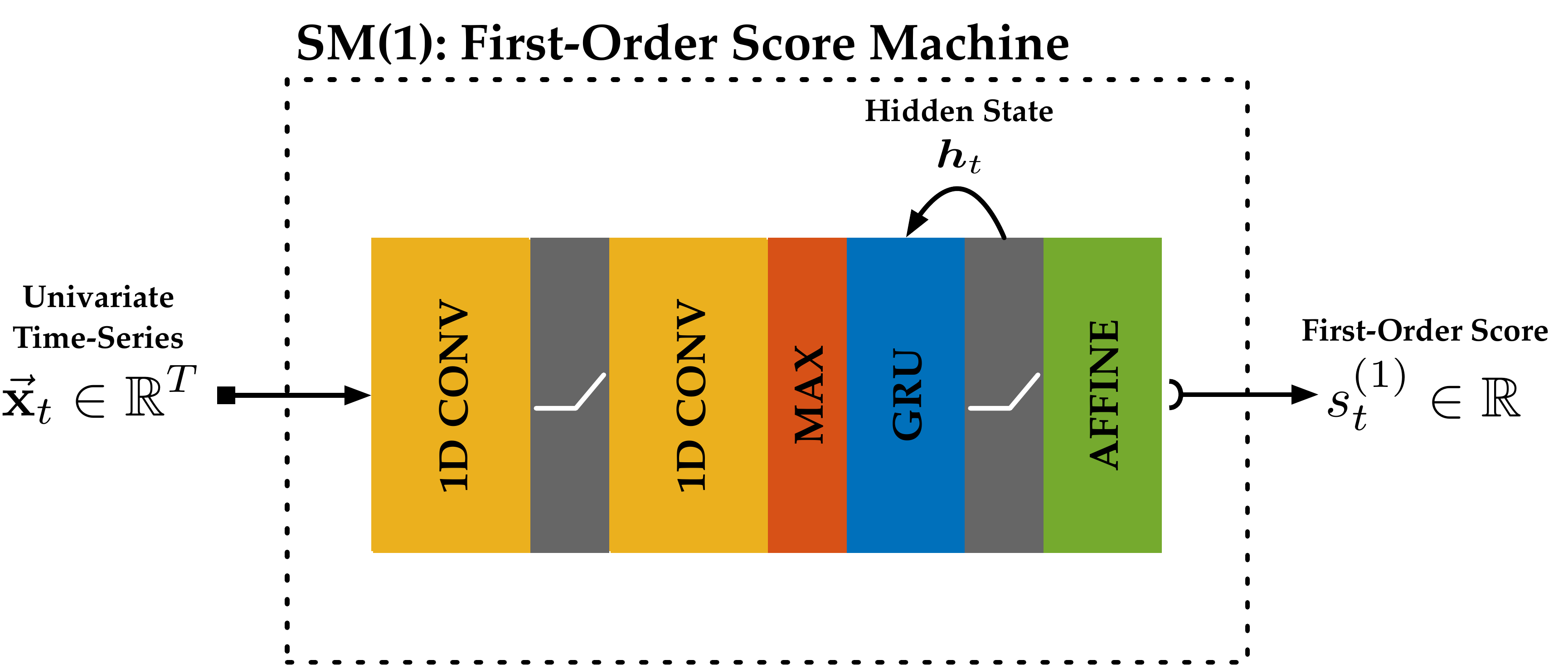}
    \end{subfigure}
    ~ 
    \begin{subfigure}[t]{0.48\textwidth}
        \includegraphics[width=\textwidth]{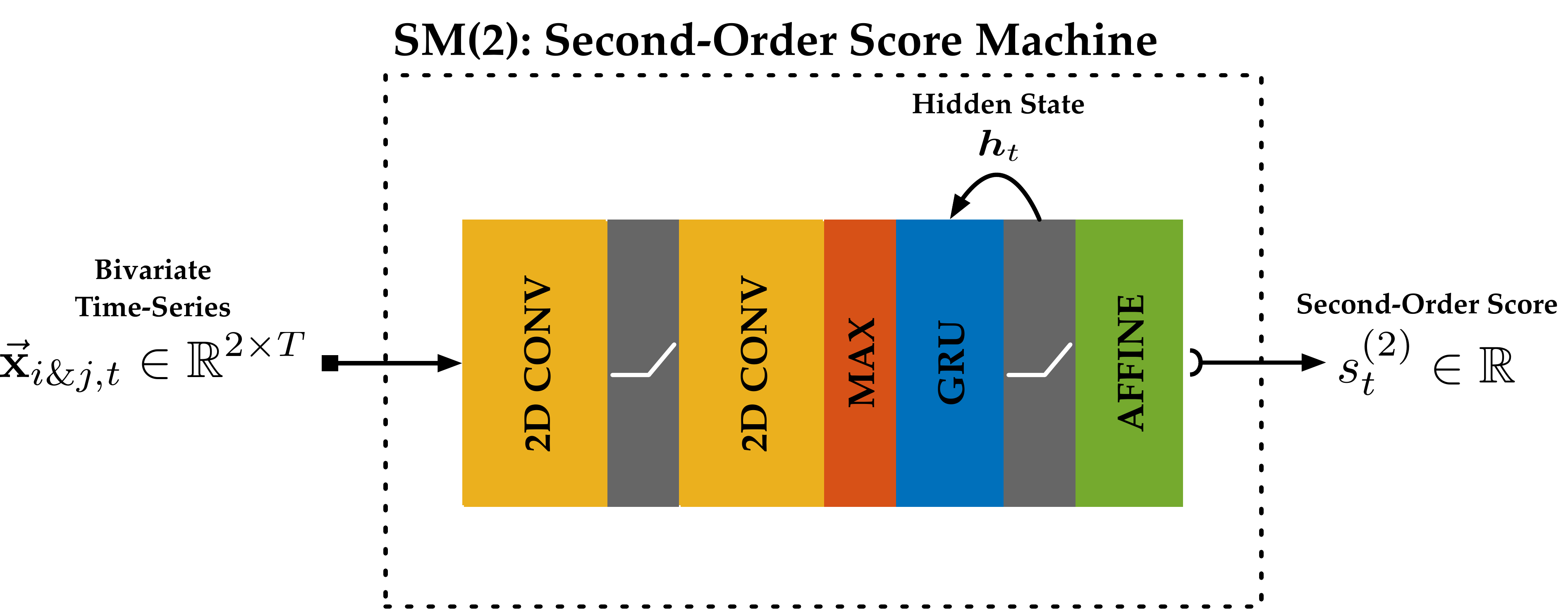}
    \end{subfigure}
    \caption{Score Machine (SM) neural netowrk architecture. Convolutional layers followed by
    non-linearities (e.g., ReLU) and Max-Pooling \citep{giusti2013fast} construct a feature map, which is
    selectively stored and filtered by the Gate Recurrent Unit (GRU)
    layer. Finally, a linear layer combines the GRU output components
    to a single scalar value, the score. (\textit{Left}) First-Order Score Machine SM($1$);
    given a univariate time-series $\vvx_{t}$ of $T$ samples, it produces a scalar score
    value $s_{t}^{(1)}$. (\textit{Right}) Second-Order Score Machine SM($2$); given a bivariate time-series
    $\vvx_{i\&j, t}$, with components $\vvx_{i, t}$ and $\vvx_{j, t}$, of $T$ samples each,
    it generates a scalar value $s_{t}^{(2)}$.}
    \label{fig:score-machine}
\end{figure}

Figure \ref{fig:score-machine} illustrates the first and second order score machines,
where the transformations are approximated by neural networks.
Higher order score machines can be used in order to captures
higher-order moments.

\begin{figure}[h]
    \centering
    \includegraphics[width=\textwidth]{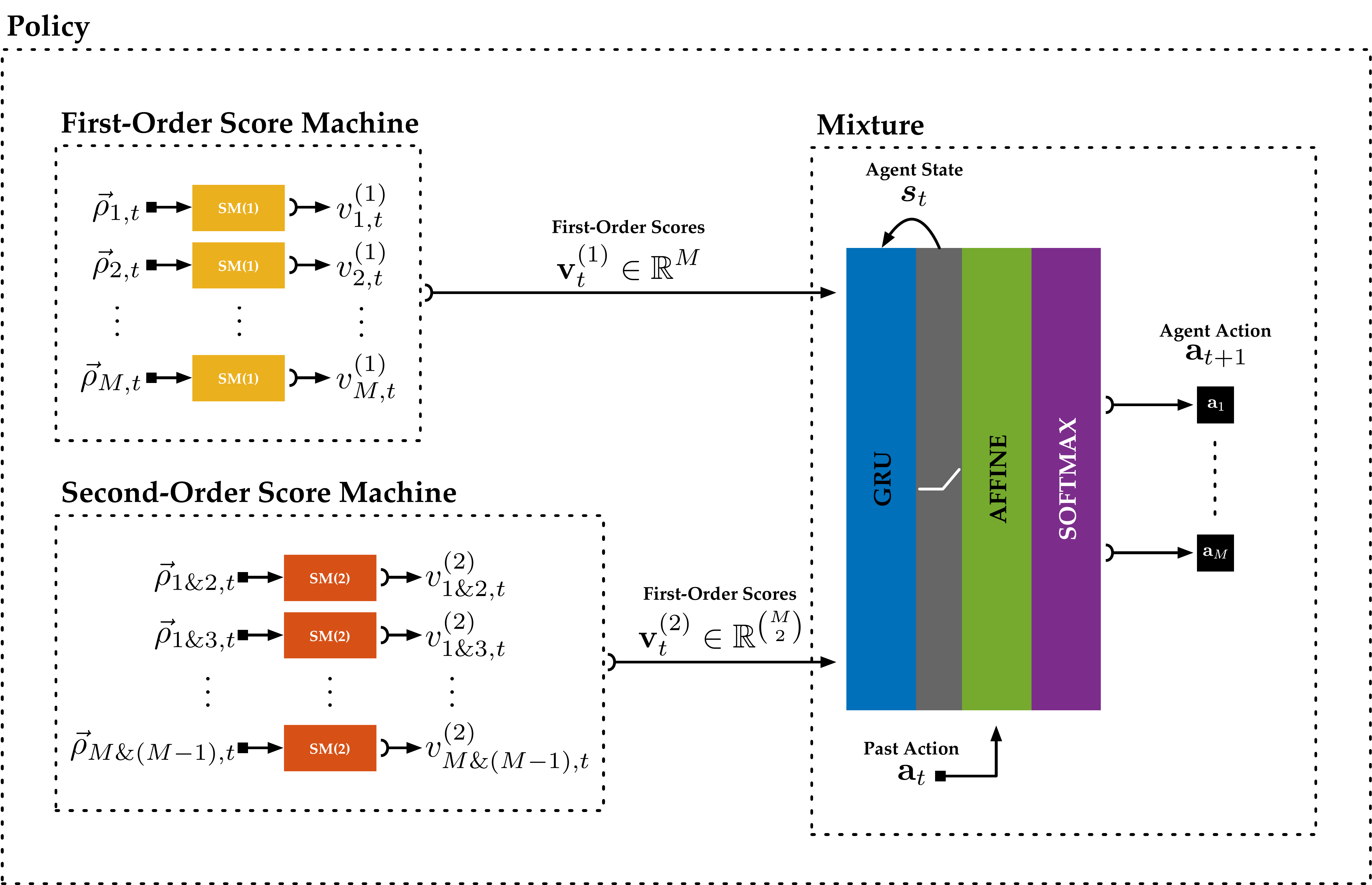}
    \caption{Mixture of Score Machines (MSM) architecture. The historic log returns
    $\vvrho_{t-T \rightarrow t} \in \sR^{M \times T}$ processes by the score machines SM($1$) and SM($2$), which assign
    scores to each asset ($\vv_{t}^{(1)}$) and pair of assets ($\vv_{t}^{(2)}$), respectively. The scores
    concatenated and passed to the mixture network, which combines them with
    the past action (i.e., current portfolio vector) and the generated agent
    state (i.e., state manager hidden state) to determine the next optimal action.}
    \label{fig:msm-schematic}
\end{figure}

By combining the score machines, we construct the \textbf{Mixture of Score
Machines} (MSM), whose architecture is illustrated in Figure \ref{fig:msm-schematic}. We identify
three main building blocks:

\begin{enumerate}
    \item \textbf{SM(1)}: a first-order score machine that processes all single-asset log returns,
        generating the first-order scores;
    \item \textbf{SM(2)}: a second-order score machine that processes all pairs of
        assets log-returns, generating the second-order scores;
    \item \textbf{Mixture Network}: an aggregation mechanism that accesses the scores from
        SM($1$) and SM($2$) and infers the action-values.
\end{enumerate}

We emphasize that there is only one SM for each
order, shared across the network, hence during backpropagation the gradient
of the loss function with respect to the parameters of
each SM is given by the sum of all the
paths that contribute to the loss \citep{deep-learning:goodfellow} that pass through that particular SM. 
The mixture network, inspired by the Mixtures of Expert Networks
by \citet{jacobs1991adaptive}, gathers all the extracted information from first and second
order statistical moments and combines them with the past action
(i.e., current portfolio vector) and the generated agent state (i.e., state manager
hidden state) to determine the next optimal action.

Neural networks, and especially deep architectures, are very \textit{data
hungry} \citep{Murphy:2012:MLP:2380985}, requiring thousands (or even millions) of data points to
converge to meaningful strategies. Using daily market data (i.e., daily prices),
almost 252 data points are only collected every year, which means
that very few samples are available for training. Thanks to
the parameter sharing, nonetheless, the SM networks are trained on
orders of magnitude more data. For example, a $12$-assets universe
with $5$ years history is given by the dataset $\mathcal{D} \in \sR{12 \times 1260}$.
The asset-specific agents (i.e., VAR, RNN, DSRQN, REINFORCE) have
$1260$ samples of a multivariate time series (i.e., with $12$ components)
from which they have to generalize. On the other hand,
the MSM agent has $12 \cdot 1260 = 15120$ samples for training the SM($1$)
network and $\binom{12}{2} \cdot 1260 = 66 \cdot 1260 = 83160$ samples for training the SM($2$) network.

The architecture is end-to-end differentiable and can be trained by backpropagating
the policy gradient to the composite network. While the score
machines (i.e., SM($1$) and SM($2$)) can be extended to any number
of assets $M$ without modification, by stacking more copies of the same
machine with shared parameters, the \textit{mixture network is universe-specific}.
As a result, a different mixture network is required to
be trained for different universes. Consider the case of an
$M$-asset market, then the the mixture network has the interface:

\begin{equation}
    N_\text{mixture-network}^{\text{inputs}}    = M + \binom{M}{2}, \quad
    N_\text{mixture-network}^{\text{outputs}}   = M
\end{equation}

Consequently, selecting a different number of assets would break the
interface of the mixture network. Nonetheless, the score machines can
be trained with different mixture networks hence when a new
universe of assets is given, we freeze the training of
the score machines and train only the mixture network. This
operation is \textit{cheap} since the mixture network 
comprises of only a small fraction of the total number of
trainable parameters of the MSM.

Practically, the score machines are trained on large historic datasets
and kept fixed, while transfer learning is performed on the mixture network.
Therefore, the score machines can be viewed as rich, universal feature
extractors, while the mixture network is the small (i.e., in size
and capacity) mechanism that enables mapping from the abstract space
of scores to the feasible action space, capable of preserving
asset-specific information as well.

% The aforementioned architectures (i.e., VAR, LSTM, DSRQN and REINFORCE) were
% treating each asset separately, adhering to the specific structure and
% dynamics of each asset and hence could not generalize
% to global "truths" and strategies. On the other hand,
% score machines treat all assets in the same way, since the
% parameters of each order of SMs are shared across assets,
% converging to globally applicable strategies; the estimators are asset-agnostic.

% For instance, the 2015 Volkswagen (VW) Emissions Scandal \citep{jung2017uncovering} shaved \$20 billion
% off of the company’s market cap and the stock dropped
% nearly 30\% virtually overnight. In fact, VW was down more
% than 40\% from May 2015 highs \citep{investopedia:vw-scandal}. The asset-specific models would
% capture this event and all future actions would indirectly (i.e., via
% the hidden state of the GRU state manager) take into
% account possible effects of such an event happening again. Nonetheless,
% the agents assume that only VW has this volatile behaviour
% and it will not propagate the information to the other
% assets, which is unrealistic, because scandals, big events and announcements,
% in general, are very infrequently observed and hardly ever repeat
% for the same asset. Consequently, harnessing the power of the score machines 

%%%
\subsubsection{Evaluation}

\begin{figure}
    \centering
    \begin{subfigure}[t]{0.48\textwidth}
        \includegraphics[width=\textwidth]{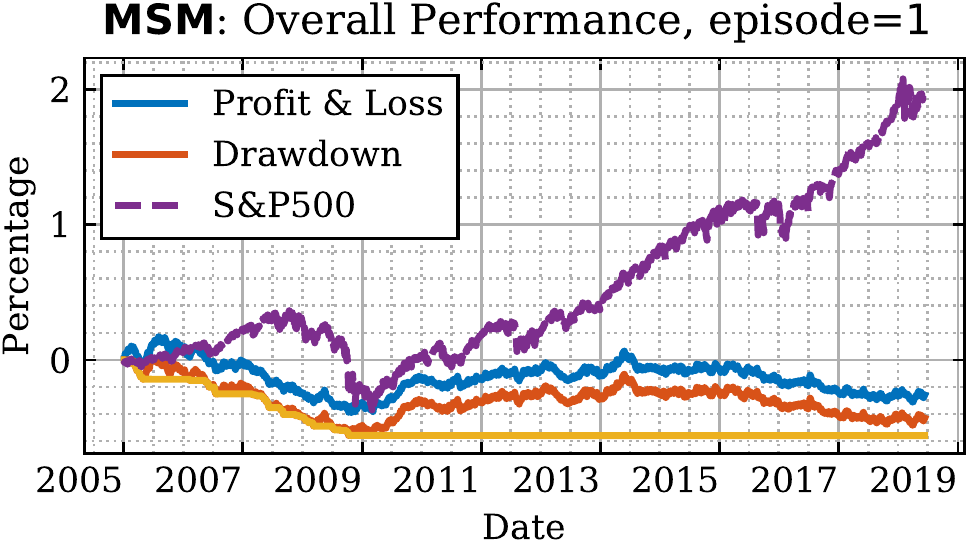}
    \end{subfigure}
    ~ 
    \begin{subfigure}[t]{0.48\textwidth}
        \includegraphics[width=\textwidth]{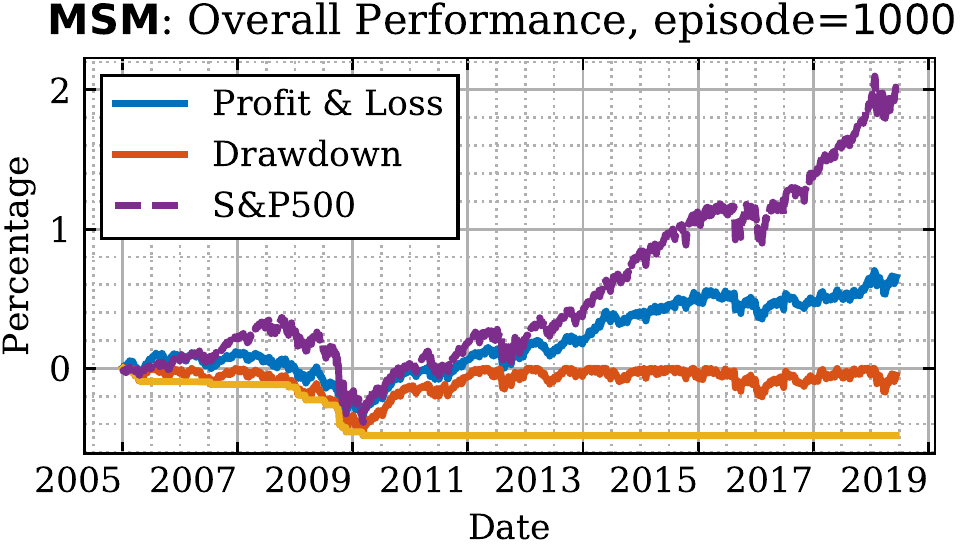}
    \end{subfigure}
    
    \vspace{0.5cm}
    
    \begin{subfigure}[t]{0.48\textwidth}
        \includegraphics[width=\textwidth]{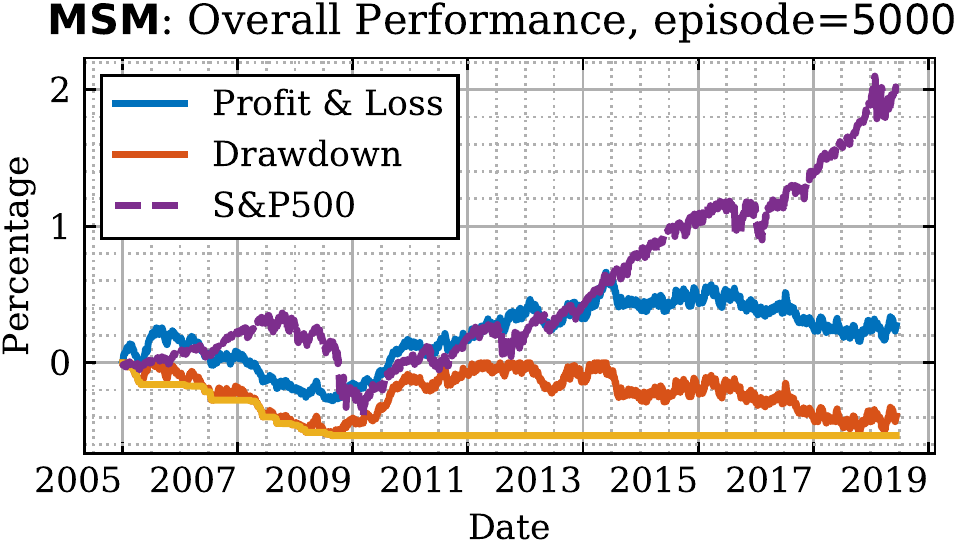}
    \end{subfigure}
    ~    
    \begin{subfigure}[t]{0.48\textwidth}
        \includegraphics[width=\textwidth]{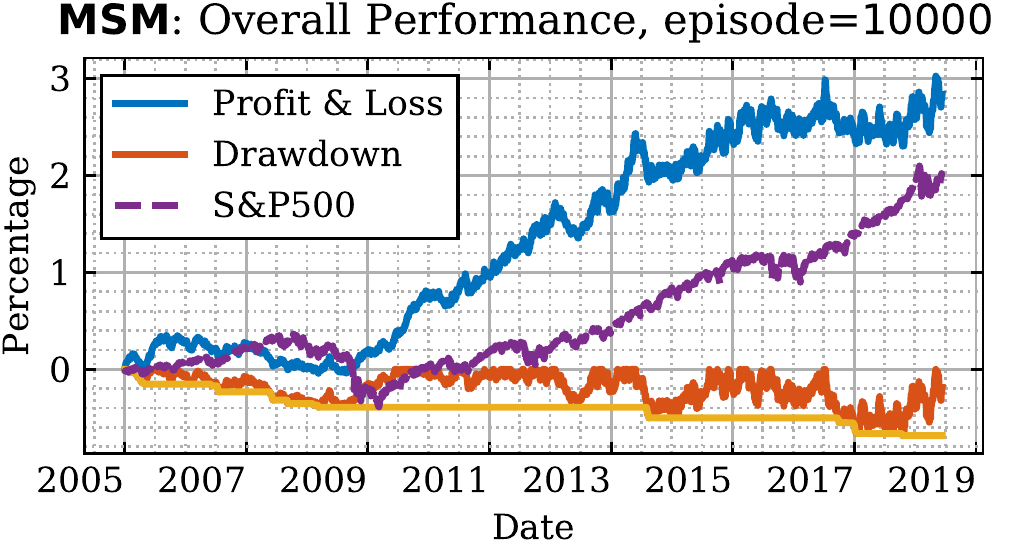}
    \end{subfigure}
    \caption{Mixture of Score Machines (MSM) model-free reinforcement learning agent on a $12$-asset universe,
    trained on historic data between 2000-2005 and tested onward,
    for different number of episodes $e = \{1, 1000, 5000, 10000 \}$.
    Visualization of cumulative rewards and (maximum) drawdown of the
    learned strategy, against the S\&P 500 index (traded as \texttt{SPY}).}
    \label{fig:msm-report}
\end{figure}

Figure \ref{fig:msm-report} shows the results from an experiment performed
on a small universe comprising of $12$ assets from S\&P 500 market using
the MSM agent. The agent is trained on historic data
between 2000-2005 for 10000 episodes, and tested on 2005-2018.
Conforming with our analysis, the agent underperformed early in the training, but after $9000$ episodes it became profitable
and its performance saturated after $10000$ episodes, with total cumulative returns of $283.9 \%$
and $68.5 \%$ maximum drawdown. It scored slightly worse than the REINFORCE agent, but in Part \ref{part:experiments}
it is shown that in a larger scale experiments the MSM is both more effective and efficient (i.e., computationally and memory-wise).

\begin{figure}
    \centering
    \begin{subfigure}[t]{0.48\textwidth}
        \includegraphics[width=\textwidth]{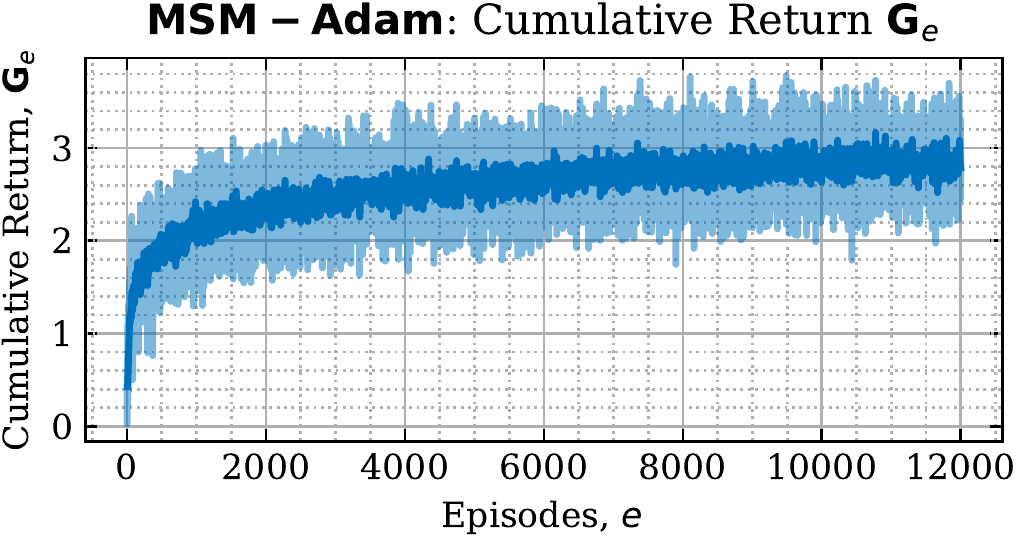}
    \end{subfigure}
    ~ 
    \begin{subfigure}[t]{0.48\textwidth}
        \includegraphics[width=\textwidth]{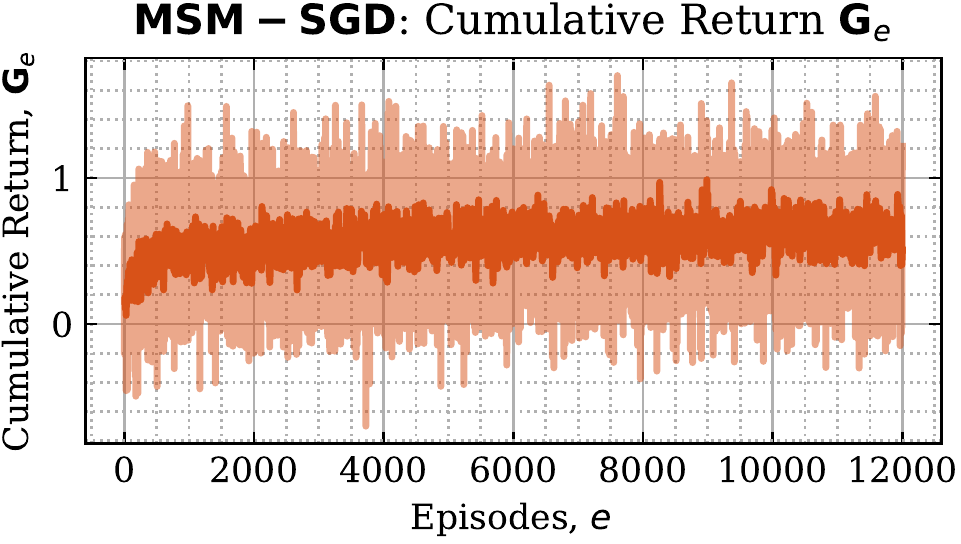}
    \end{subfigure}
    \caption{Out-of-sample cumulative returns per episode during training phase for MSM. Performance improvement
    saturates after $e \gtrapprox 10000$. (\textit{Left}) Adaptive Neural network optimization algorithm \textbf{ADAM}.
    (\textit{Right}) Neural network optimized with Stochastic Gradient Descent (SGD).}
    \label{fig:msm-learning-curve}
\end{figure}

%%%
\subsubsection{Weaknesses}

The Mixture of Score Machines (MSM) is an architecture that
addresses the weaknesses of all the aforementioned approaches (i.e., VAR, LSTM, DSRQN and REINFORCE).
However, it could be improved by incorporating the expected properties \ref{pro:short-sales} and \ref{pro:optimality-guarantees}:

\begin{property}[Short Sales]
    Model should output negative portfolio weights, corresponding to short selling, as well.
\label{pro:short-sales}
\end{property}

\begin{property}[Optimality Guarantees]
    Explore possible re-interpretations of the framework, which would allow proof
    of optimality (if applicable).
\label{pro:optimality-guarantees}
\end{property}

%%% SUMMARY TABLE %%%

\begin{table}
\begin{center}
\begin{tabular}{ |c||l||c|c||c|c|c| }
	\hline
    \multicolumn{7}{|c|}{\textbf{Trading Agents Comparison Matrix: $12$-assets of S\&P 500}} \\
    \hline
    \hline
    \multicolumn{2}{|c||}{} & \multicolumn{2}{c||}{\textbf{Model-Based}} & \multicolumn{3}{c||}{\textbf{Model-Free}} \\ 
	\hline
    \hline
    \multicolumn{2}{|c||}{} & \textbf{VAR} & \textbf{RNN} & \textbf{DSRQN} & \textbf{REINFORCE} & \textbf{MSM} \\ 
    \hline
    \hline
    \multirow{3}{*}{\rotatebox[origin=c]{90}{\textbf{Metrics}}}
    	& Cumulative Returns (\%) & \cred 185.1 & \cred 221.0 & \cyellow 256.7 & \cgreen 325.9 & \cyellow 283.9 \\
    \cline{2-7}
    	& Sharpe Ratio (SNR) & \cred 1.53 & \cred 1.62 & \cyellow 2.40 & \cgreen 3.02 & \cyellow 2.72 \\
    \cline{2-7}
        & Max Drawdown (\%) & \cred 179.6 & \cred 198.4 & \cyellow 85.6 & \cgreen 63.5 & \cgreen 68.5 \\
    \hline
    \hline
    \multirow{8}{*}{\rotatebox[origin=c]{90}{\textbf{Expected Properties}}}
        & Non-Stationary Dynamics   & \cred $\times$
                                    & \cgreen $\checkmark$
                                    & \cgreen $\checkmark$
                                    & \cgreen $\checkmark$
                                    & \cgreen $\checkmark$ \\
    \cline{2-7}
        & Long-Term Memory          & \cred $\times$
                                    & \cgreen $\checkmark$
                                    & \cgreen $\checkmark$
                                    & \cgreen $\checkmark$
                                    & \cgreen $\checkmark$ \\
    \cline{2-7}
        & Non-Linear Model          & \cred $\times$
                                    & \cgreen $\checkmark$
                                    & \cgreen $\checkmark$
                                    & \cgreen $\checkmark$
                                    & \cgreen $\checkmark$ \\
    \cline{2-7}
        & End-to-End                & \cred $\times$
                                    & \cred $\times$
                                    & \cred $\times$
                                    & \cgreen $\checkmark$
                                    & \cgreen $\checkmark$ \\
    \cline{2-7}
        & Linear Scaling            & \cred $\times$
                                    & \cred $\times$
                                    & \cred $\times$
                                    & \cred $\times$
                                    & \cgreen $\checkmark$ \\
    \cline{2-7}
        & Universality              & \cred $\times$
                                    & \cred $\times$
                                    & \cred $\times$
                                    & \cred $\times$
                                    & \cgreen $\checkmark$ \\
    \cline{2-7}
        & Low Variance Estimators   & \cred $\times$
                                    & \cred $\times$
                                    & \cred $\times$
                                    & \cred $\times$
                                    & \cred $\times$ \\
    \cline{2-7}
        & Short Sales               & \cgreen $\checkmark$
                                    & \cgreen $\checkmark$
                                    & \cred $\times$
                                    & \cred $\times$
                                    & \cred $\times$ \\
    \hline
\end{tabular}
\end{center}
\caption{Comprehensive comparison of evaluation metrics and their weaknesses (i.e., expected properties)
of trading algorithms addressed in this chapter. Model-based agents (i.e., VAR
and RNN) underperform, while the best performing agent is the
REINFORCE. As desired, the MSM agent scores also well above the index
(i.e., baseline) and satisfies most wanted properties.}
\label{tab:trading-agents-12-assets-comparison}
\end{table}

%% assets folder to reference
\renewcommand{\assets}{report/innovation/pre-training/assets}

\chapter{Pre-Training} \label{ch:pre-training}

In Chapter \ref{ch:trading-agents}, model-based and model-free reinforcement learning agents were introduce,
which address the asset allocation task. It was demonstrated (see
comparison table \ref{tab:trading-agents-12-assets-comparison}) that model-based (i.e., VAR and RNN) and value-based model-free
agents (i.e., DSRQN) are outperformed by the policy gradient agents (i.e., REINFORCE
and MSM). However, policy gradient algorithms usually converge to local
optima \citep{reinforcement-learning:sutton}.

Inspired by the approach taken by the authors of the original
DeepMind AlphaGo paper \citep{silver2016alphago}, the local optimality of policy gradient agents is
addressed via pre-training the policies in order to replicate the
strategies of baseline models. It is shown that any one-step optimization
method, discussed in Chapter \ref{ch:portfolio-optimization} that reduces to a quadratic program, can
be reproduced by the policy gradient networks (i.e., REINFORCE and MSM), when
the networks are trained to approximate the quadratic program solution.

Because of the highly non-convex policy search space \citep{reinforcement-learning:ualberta}, the randomly
initialised agents (i.e., agnostic agents) tend to either get stuck to vastly
sub-optimal local minima or need a lot more episodes and samples
to converge to meaningful strategies. Therefore, the limited number of available
samples (e.g., 10 years of market data is equivalent to approximately
2500 samples), motivates pre-training, which is expected to improve convergence speed
and performance, assuming that the baseline model is sub-optimal but
a proxy to the optimal strategy. Moreover, the pre-trained models can be
viewed as \textit{priors}\footnote{As in the context of Bayesian Inference.} to
the policies and episodic training with reinforcement learning steers them
to the updated strategies, in a data-driven and data-efficient manner.

In Section \ref{sec:pre-training:baseline-models}, a few candidate baseline models are introduced,
including the algorithm for generating synthetic (supervised) datasets according
to these baseline models as well as the corresponding steps for
pre-training the agents using the generated data. In addition, in
Section \ref{sec:pre-training:model-evaluation} the pre-training process is assessed as well as the performance gain
compared to randomly initialized agents is quantified.

\section{Baseline Models} \label{sec:pre-training:baseline-models}

\begin{figure}[h]
    \centering
    \includegraphics[width=0.75\textwidth]{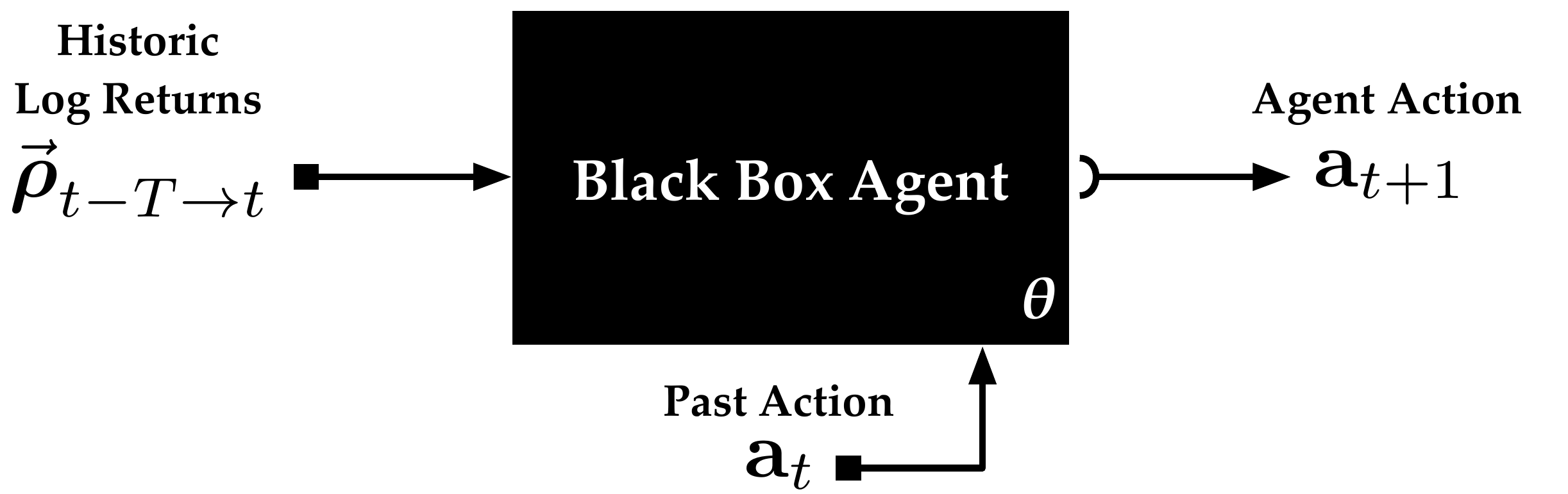}
    \caption{Interfacing with policy gradient agents as black boxes, with inputs
    (1) historic log returns $\vvrho_{t-T \rightarrow t}$ and (2) past action (i.e., current portfolio vector) $\va_{t}$
    and output the next agent actions $\va_{t+1}$. The black-box is parametrized
    by $\vtheta$ which can be updated and optimized.}
    \label{fig:black-box-agents}
\end{figure}

In Chapter \ref{ch:portfolio-optimization}, the traditional one-step (i.e., static) portfolio optimization methods were described,
derived from the Markowitz model \citep{finance:markowitz-2}. Despite the assumptions about covariance
stationarity (i.e., time-invariance of first and second statistical moments) and myopic
approach of those methods, they usually form the basis of
other more complicated and effective strategies. As a result, it
is attempted to replicate those strategies with the REINFORCE (see subsection \ref{sub:reinforce-agent})
and the Mixture of Score Machines (MSM) (see subsection \ref{sub:mixture-of-score-machines}) agents. In both cases, the
architecture of the agents (i.e., underlying neural networks) are treated as
black boxes, represented by a set of parameters, which thanks
to their end-to-end differentiability, can be trained via backpropagation.
Figure \ref{fig:black-box-agents} summarizes the pipeline used to (pre-)train the policy
gradient agents.

\subsection{Quadratic Programming with Transaction Costs} \label{sub:quadratic-programming-with-transaction-costs}

The one-step optimal portfolio for given commission rates (i.e., transaction costs coefficient $\beta$)
and hyperparameters (e.g., risk-aversion coefficient) is obtained, by solving the optimization
task in (\ref{opt:risk-aversion-transaction-costs}) or (\ref{opt:sharpe-ratio-transaction-costs}) via quadratic programming.
The Sharpe Ratio with transaction costs objective function is selected as
the baseline for pre-training, since it has no hyperparameter to
tune and inherently balances profit-risk trade-off.

Without being explicitly given the mean vector $\vmu$, the covariance matrix $\mSigma$
and the transaction coefficient $\beta$, the black-box agents should be able
to solve the optimization task (\ref{opt:sharpe-ratio-transaction-costs}), or equivalently:

\begin{align*}
\underset{\va_{t+1} \in \sA}{\text{maximize}} \quad
    &\frac{\va_{t+1}^{T} \vmu - \boldsymbol{1}_{M}^{T} \beta \| \va_{t} - \va_{t+1} \|_{1}}{\sqrt{\va_{t+1}^{T} \mSigma \va_{t+1}}} \\
\text{and} \quad &\boldsymbol{1}_{M}^{T} \va_{t+1} = 1 \\
\text{and} \quad & \va_{t+1} \succeq 0
\end{align*}

\subsection{Data Generation} \label{sub:data-generation}

Since there is a closed form formula that connects the
black-box agents' inputs $\vvrho_{t-T \rightarrow t}$ and $\va_{t}$ with the terms in the optimization problem
(\ref{opt:sharpe-ratio-transaction-costs}), $N$ supervised pairs $\{ (\mX_{i}, \vy_{i}) \}_{i=1}^{N}$ are generated
by solving the optimization for $N$ distinct cases, such that:

\begin{align}
    \mX_{i} &= \big[ \vvrho_{t_{i}-T \rightarrow t_{i}},\ \va_{t_{i}} \big] \\
    \vy_{i} &= \va_{t_{i}+1}
\end{align}

Interestingly, myriad of examples (i.e., $\mX_{i},\ \vy_{i}$ pairs) can be produced to enrich
the dataset and allow convergence. This is a very rare
situation where the generating process of the data is known
and can be used to produce valid samples, which respect
the dynamics of the target model. The data generation process is given in algorithm \ref{algo:data-generation}.

\begin{algorithm}
	\SetKwInOut{Args}{inputs}
    \SetKwInOut{Output}{output}
    \DontPrintSemicolon
    \Args{%
        number of pairs to generate $N$\\
    	number of assets in portfolio $M$\\
    	look back window size $T$\\
    	transaction costs coefficient $\beta$
    }
    \Output{dataset $\{ (\mX_{i}, \vy_{i}) \}_{i=1}^{N}$}
    \vspace{0.25cm}
    \SetAlgoLined\SetArgSty{}
    \For{$i=1, 2, \ldots N$}{
        sample valid random initial portfolio vector $\vw_{t_{i}}$ \\
        sample random lower triangular matrix $\mL \in \sR^{M \times M}$ \tcp*{Cholesky decomposition}
        sample randomly distributed log returns:
            $\vvrho_{t_{i}-T \rightarrow t_{i}} \sim \mathcal{N}(\mathbf{1}, \mL \mL^{T})$ \\
        calculate empirical mean vector of log returns: $\vmu = \E[ \vvrho_{t_{i}-T \rightarrow t_{i}} ]$ \\
        calculate empirical covariance matrix of log returns: $\mSigma = \Cov[ \vvrho_{t_{i}-T \rightarrow t_{i}} ]$ \\
        determine $\va_{t_{i}+1}$ by solving quadratic program (\ref{opt:sharpe-ratio-transaction-costs}) \\
        set $\mX_{i} = [\vvrho_{t_{i}-T \rightarrow t_{i}},\ \va_{t_{i}}]$ and $\vy_{i} = \va_{t_{i}+1}$
    }
\caption{Pre-training supervised dataset generation.}
\label{algo:data-generation}
\end{algorithm}

\section{Model Evaluation} \label{sec:pre-training:model-evaluation}

The parameters of the black-box agents are steered in the gradient
direction that minimizes the \textbf{Mean Square Error} between the predicted portfolio
weights, $\hat{\vy}_{t_{i}}$, and the baseline model target portfolio weights, $\vy_{t_{i}}$.
An $L^{2}$-norm weight decaying, \textbf{regularization}, term is also considered to avoid
overfitting, obtaining the loss function:

\begin{equation}
    \mathcal{L}(\vtheta) = \| \vy_{t_{i}} - \hat{\vy}_{t_{i}; \vtheta} \|_{2}^{2} + \lambda \| \vtheta \|_{2}^{2}
\label{def:pre-training:loss-function}
\end{equation}

The parameters are adaptively optimized by Adam \citep{kingma2014adam}, while the network
parameters gradients are obtained via Backpropagation Through Time \citep{deep-learning:bptt}.

\subsection{Convergence to Quadratic Programming} \label{sec:convergence-to-quadratic-programming}

\begin{figure}[h]
    \centering
    \begin{subfigure}[t]{0.48\textwidth}
        \includegraphics[width=\textwidth]{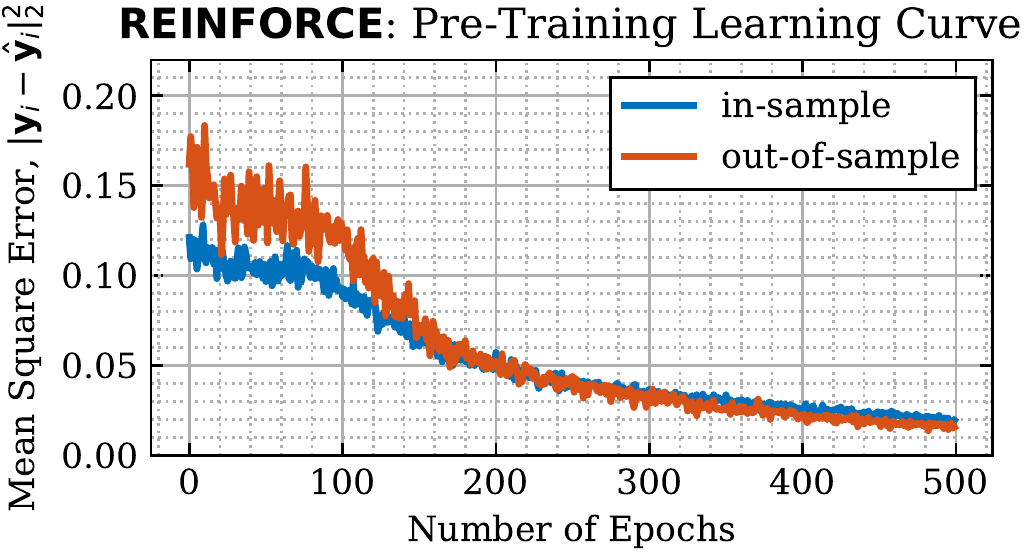}
    \end{subfigure}
    ~ 
    \begin{subfigure}[t]{0.48\textwidth}
        \includegraphics[width=\textwidth]{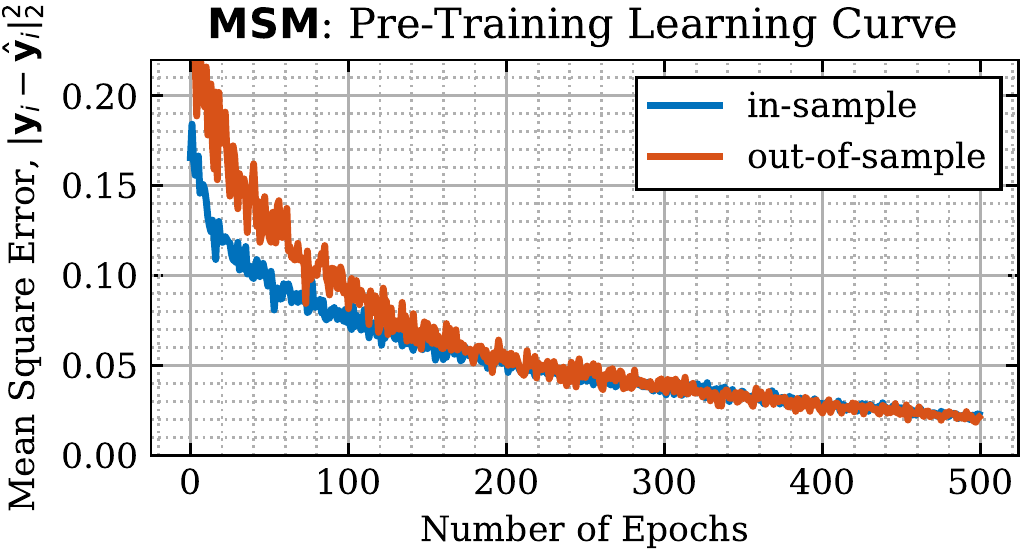}
    \end{subfigure}
    \caption{Mean square error (MSE) of Monte-Carlo Policy Gradient (REINFORCE) and
    Mixture of Score Machines (MSM) during pre-training. After $\approx 150$ epochs the
    gap between the training (in-sample) and the testing (out-of-sample) errors
    is eliminated and error curve plateaus after $\approx 400$ epochs, when training is terminated.}
    \label{fig:pre-training:learning-curves}
\end{figure}

Figure \ref{fig:pre-training:learning-curves} depicts the the learning curves, in-sample and out-of sample,
of the supervised learning training process. Both the REINFORCE and
the MSM converge after $\approx 400$ epochs (i.e., iterations). The 

\subsection{Performance Gain} \label{sec:pre-training:performance-gain}

\begin{figure}[h]
    \centering
    \begin{subfigure}[t]{0.48\textwidth}
        \includegraphics[width=\textwidth]{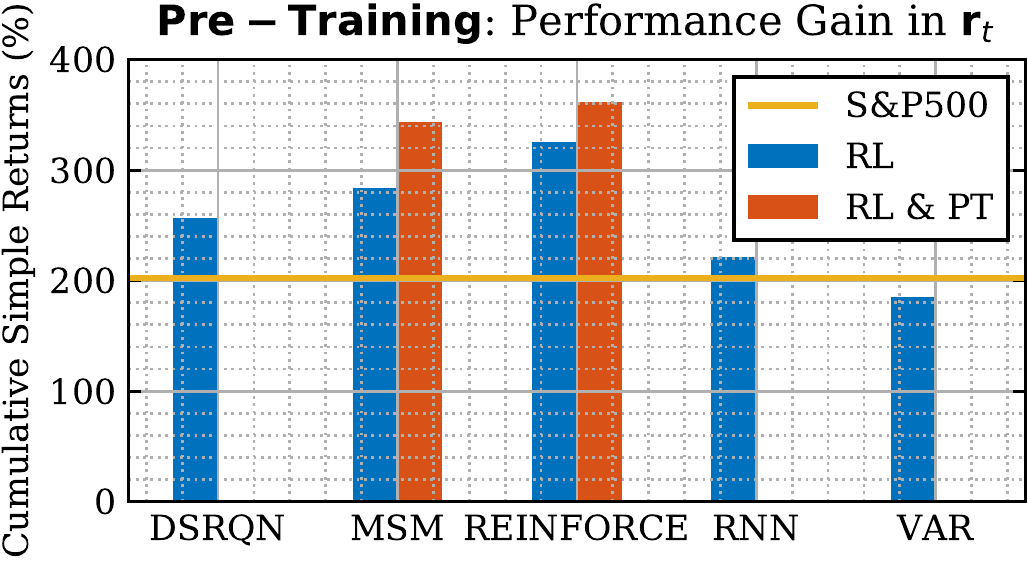}
    \end{subfigure}
    ~ 
    \begin{subfigure}[t]{0.48\textwidth}
        \includegraphics[width=\textwidth]{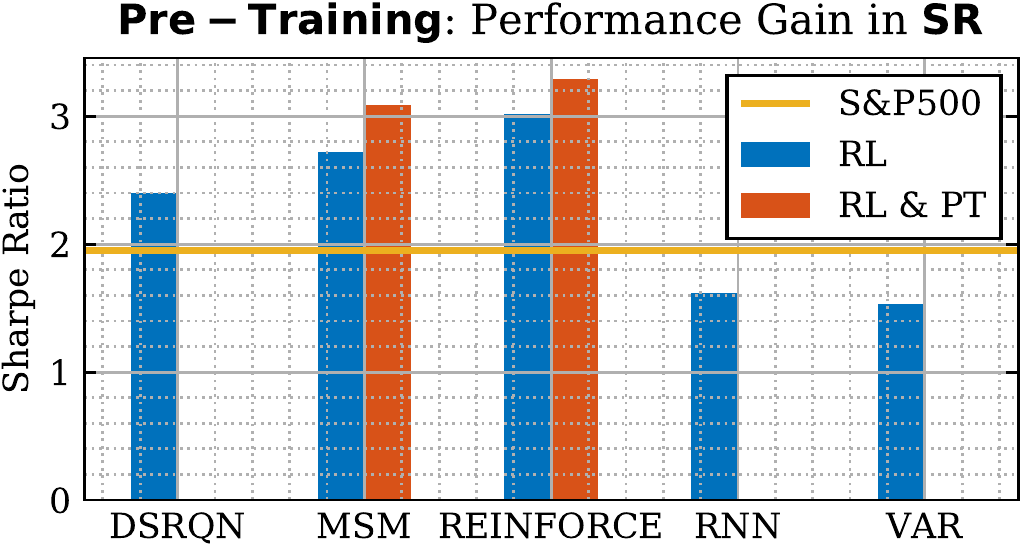}
    \end{subfigure}
    \caption{Performance evaluation of trading with reinforcement learning (RL) and reinforcement
    learning and pre-training (RL \& PT). The Mixture of Score Machines (MSM)
    improves cumulative returns by $21.02\%$ and Sharpe Ratio by $13.61\%$. The model-based
    (i.e., RNN and VAR) and the model-free value-based (i.e., DSRQN) agents are \textit{not}
    end-to-end differentiable and hence cannot be pre-trained.}
    \label{fig:pre-training:performance-gain}
\end{figure}

As suggested by Figure \ref{fig:pre-training:performance-gain}, the pre-training improves the
cumulative returns and Sharpe Ratio of the policy gradient agents up to $21.02\%$ and $13.61\%$, respectively.

\part{Experiments} \label{part:experiments}

%% assets folder to reference
\renewcommand{\assets}{report/experiments/synthetic-data/assets}

\chapter{Synthetic Data} \label{ch:synthetic-data}

It has been shown that the trading agents of Chapter \ref{ch:trading-agents},
and especially REINFORCE and MSM, outperform the market index
(i.e., S\&P500) when tested in a small universe of $12$-assets, see
Table \ref{tab:trading-agents-12-assets-comparison}.
For rigour, the validity and effectiveness of the developed reinforcement
agents is investigated via a series of experiments on:

\begin{itemize}
    \item \textbf{Deterministic} series, including sine, sawtooth and chirp waves,
    as in Section \ref{sec:deterministic-processes};
    \item \textbf{Simulated} series, using data surrogate methods, such as AAFT, as in Section \ref{sec:simulated-data}.
\end{itemize}

As expected, it is demonstrated that model-based agents (i.e., VAR and RNN)
excel in deterministic environments. This is attributed to the fact
that given enough capacity they have the predictive power to
accurately forecast the future values, based on which they can
act optimally via planning.

On the other hand, on simulated (i.e., surrogate)
time-series, it is shown that model-free agents score higher,
especially after the pre-training process of Chapter \ref{ch:pre-training}, which contributes to
up to $21\%$ improvement in Sharpe Ratio and up to $40\%$
reduction in the number of  episodic runs.

\section{Deterministic Processes} \label{sec:deterministic-processes}

To begin with, via interaction with the environment (i.e., paper trading),
the agents construct either an explicit (i.e., model-based reinforcement learning) or
implicit model (i.e., model-free reinforcement learning) of the environment. In Section \ref{sec:model-based-reinforcement-learning},
It has been demonstrated that explicit modelling of financial time
series is very challenging due to the stochasticity of the involved
time-series, and, as a result, model-based methods underperform.
On the other hand, should the market series were sufficiently predictable,
these methods would be expected to optimally allocate assets of the
portfolio via dynamic programming and planning. In this section, we
investigate the correctness of this hypothesis by generating a universe
of deterministic time-series.

\subsection{Sinusoidal Waves}

\begin{figure}[h]
    \centering
    \begin{subfigure}[t]{0.48\textwidth}
        \includegraphics[width=\textwidth]{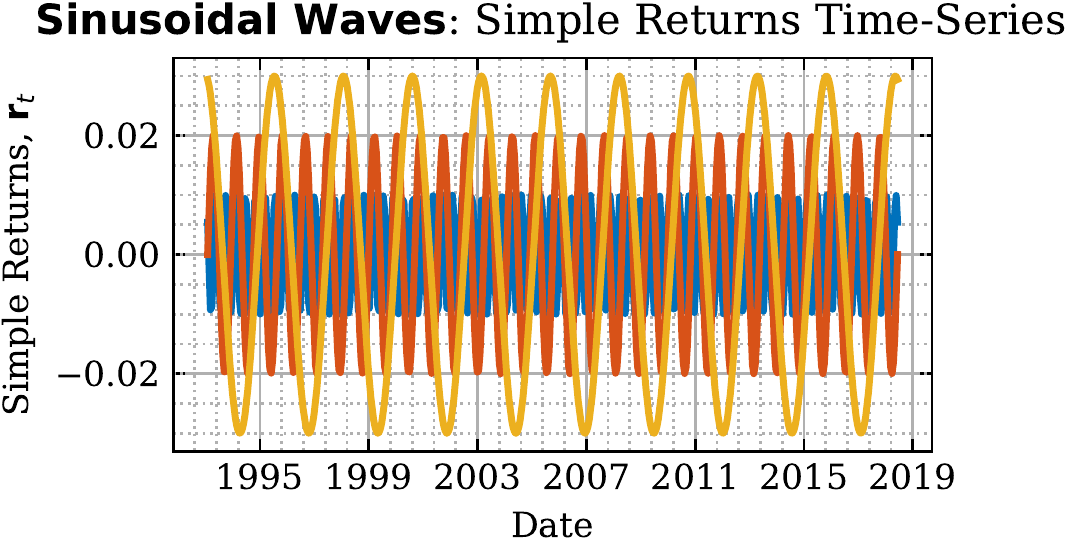}
    \end{subfigure}
    ~ 
    \begin{subfigure}[t]{0.48\textwidth}
        \includegraphics[width=\textwidth]{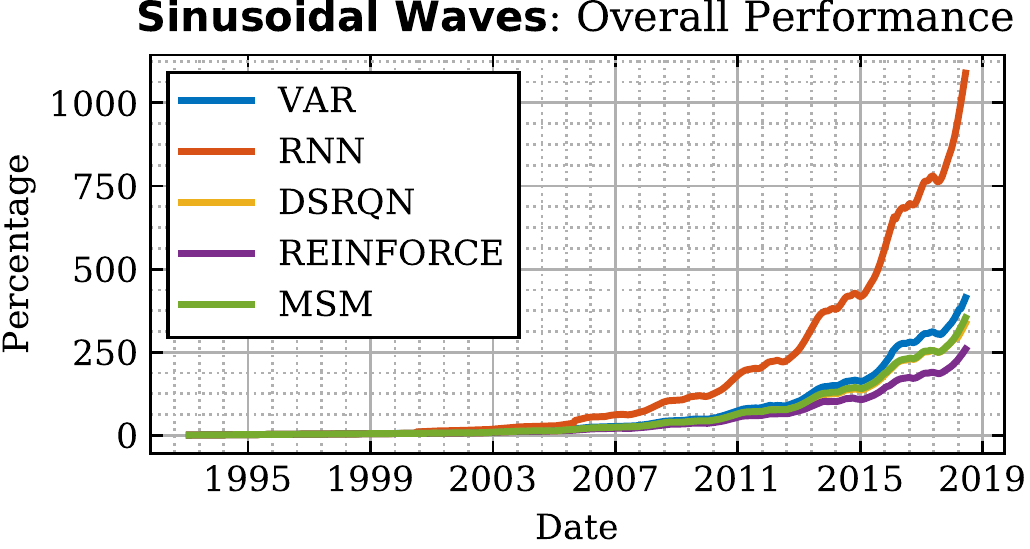}
    \end{subfigure}
    \caption{Synthetic universe of deterministic sinusoidal waves. (\textit{Left}) Example series from universe.
    (\textit{Right}) Cumulative returns of reinforcement learning trading agents.}
    \label{fig:datasets:sinusoidal-waves}
\end{figure}

A set of $100$ sinusoidal waves of constant parameters (i.e., amplitude, circular frequency
and initial phase) is generated, while example series are
provided in Figure \ref{fig:datasets:sinusoidal-waves}. Note the dominant performance of the model-based
recurrent neural network (RNN) agent, which exploits its accurate predictions
of future realizations and scores over three times better
than the best-scoring model-free agent, the Mixture of Score Machines (MSM).

\begin{figure}[h]
    \centering
    \includegraphics[width=\textwidth]{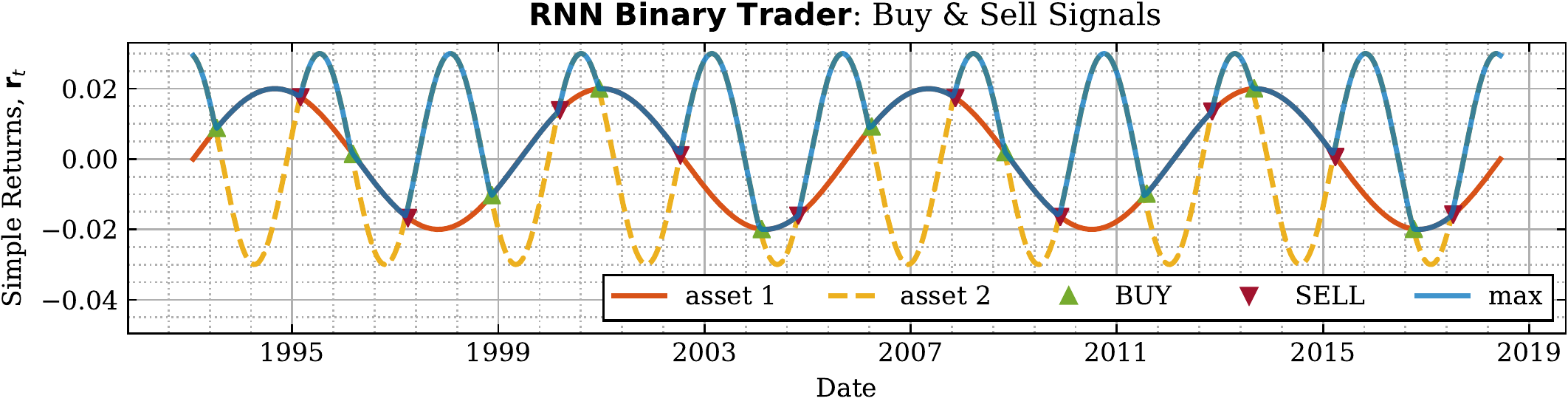}
    \caption{Recurrent neural network (RNN) model-based reinforcement learning agent trained on
    binary trading between two sinusoidal waves. The triangle trading signals
    (i.e., BUY or SELL) refer to \textit{asset 1} (i.e., red), while opposite actions
    are taken for asset 2, but not illustrated.}
    \label{fig:datasets:sinusoidal-waves-binary-trading}
\end{figure}

For illustration purposes and in order to gain a finer
insight into the learned trading strategies, a universe of only
two sinusoids is generated the RNN agent is trained on
binary trading the two assets; at each time step the
agent puts all its budget on a single asset. As
shown in Figure \ref{fig:datasets:sinusoidal-waves-binary-trading}, the RNN agent learns
the theoretically optimal strategy\footnote{Note that transaction costs are not
considered in this experiment, in which case we would expect
a time-shifted version of the current strategy so that it
offsets the fees.}:

\begin{equation}
\vw_{t} =
\left\{
\begin{array}{ll}
    \evw_{i, t} = 1, & \text{if } i = \argmax\{\vr_{t}\} \\
    \evw_{i, t} = 0, & \text{otherwise}
\end{array} 
\right. 
\end{equation}

or equivalently, the returns of the constructed portfolio is the
max of the single asset returns at each time step.

\subsection{Sawtooth Waves}

\begin{figure}[h]
    \centering
    \begin{subfigure}[t]{0.48\textwidth}
        \includegraphics[width=\textwidth]{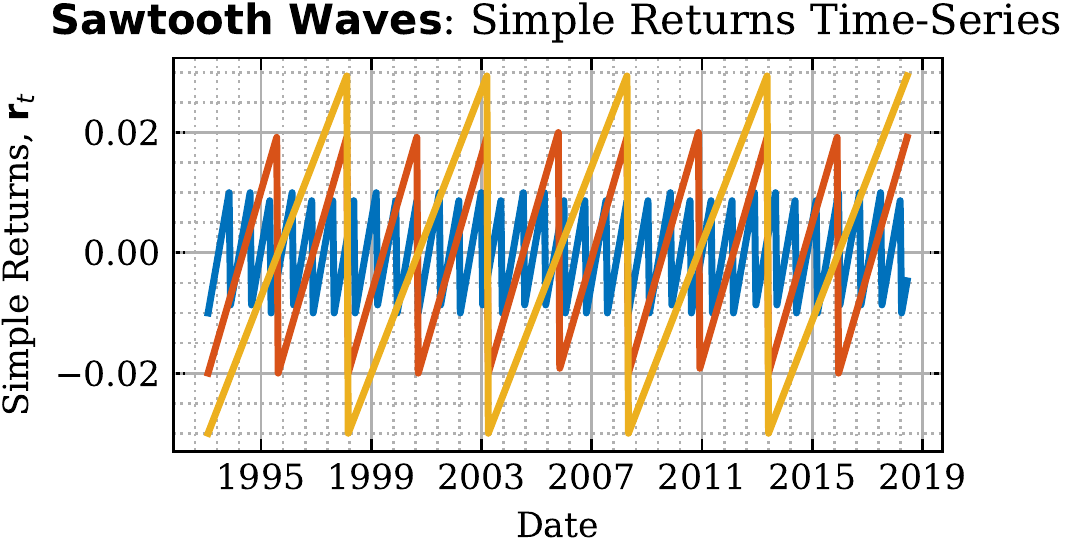}
    \end{subfigure}
    ~ 
    \begin{subfigure}[t]{0.48\textwidth}
        \includegraphics[width=\textwidth]{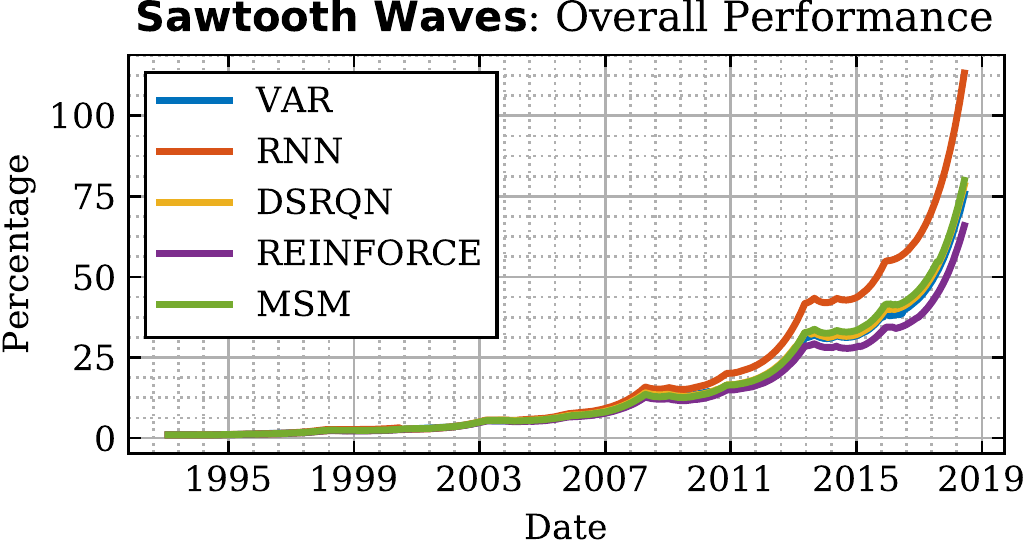}
    \end{subfigure}
    \caption{Synthetic universe of deterministic sawtooth waves. (\textit{Left}) Example series from universe.
    (\textit{Right}) Cumulative returns of reinforcement learning trading agents.}
    \label{fig:datasets:sawtooth-waves}
\end{figure}

A set of $100$ deterministic sawtooth waves is generated next
and examples are illustrated in Figure \ref{fig:datasets:sawtooth-waves}. Similar to the sinusoidal
waves universe, the RNN agent outperforms the rest of the
agents. Interestingly, it can be observed in the cumulative returns
time series, right Figure \ref{fig:datasets:sawtooth-waves}, that all strategies have a low-frequency
component, which corresponds to the highest amplitude sawtooth wave (i.e., yellow).

\subsection{Chirp Waves}

\begin{figure}[h]
    \centering
    \begin{subfigure}[t]{0.48\textwidth}
        \includegraphics[width=\textwidth]{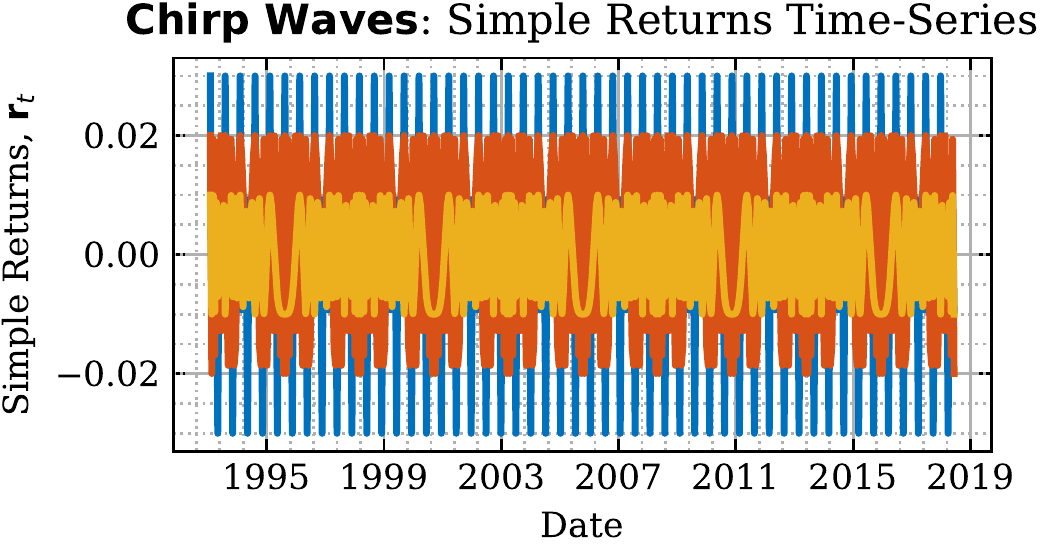}
    \end{subfigure}
    ~ 
    \begin{subfigure}[t]{0.48\textwidth}
        \includegraphics[width=\textwidth]{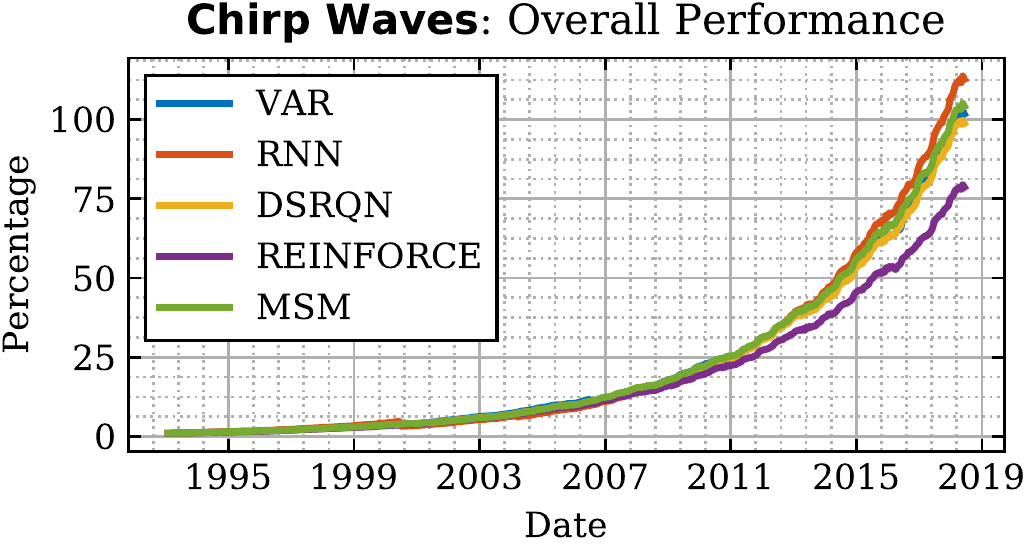}
    \end{subfigure}
    \caption{Synthetic universe of deterministic chirp waves. (\textit{Left}) Example series from universe.
    (\textit{Right}) Cumulative returns of reinforcement learning trading agents.}
    \label{fig:datasets:chirp-waves}
\end{figure}

Last but not least, the experiment is repeated with a set
of $100$ deterministic chirp waves (i.e., sinusoidal wave with linearly modulated frequency).
Three example series are plotted in \ref{fig:datasets:chirp-waves}, along with the cumulative
returns of each trading agent. Note that the RNN agent
is only $8.28\%$ better than the second, the MSM, agent, compared to
the $> 300\%$ marginal benefit in case of the sinusoidal waves. This is
explained by the imperfect predictions of the RNN due to
the increased difficulty to learn the chirp signals.

\begin{remark}
Overall, in a deterministic financial market, all trading agents learn
profitable strategies and solve the asset allocation task. As expected,
model-based agents, and especially the RNN, are significantly outperforming in case of 
well-behaving, easy-to-model and deterministic series (e.g., sinusoidal, sawtooth). On the
other hand, in more complicated settings (e.g., chirp waves universe) the
model-free agents perform almost as good as model-based agents.
\end{remark}

\section{Simulated Data} \label{sec:simulated-data}

Having asserted the successfulness of reinforcement learning trading agents in deterministic
universes, their effectiveness is challenged in stochastic universes, in this section.
Instead of randomly selecting families of stochastic processes and corresponding parameters for
them, real market data is used to learn the parameters of
candidate generating processes that explain the data. The purpose of this approach is two-fold:

\begin{enumerate}
    \item There is \textbf{no} need for \textbf{hyperparameter tuning};
    \item The training dataset is expanded, via \textbf{data augmentation}, giving the opportunity
    to the agents to gain more experience and further explore the
    joint state-action space.
\end{enumerate}

It is worth highlighting that data augmentation improves overall performance,
especially when strategies learned in the simulated environment are transferred
and polished on real market data, via \textbf{Transfer Learning} \citep{deep-learning:transfer-learning}.

\subsection{Amplitude Adjusted Fourier Transform (AAFT)} \label{sub:aaft}

The simulated universe is generated using surrogates with random Fourier
phases \citep{2009tcsp.conf..274R}. In particular the \textbf{Amplitude Adjusted Fourier Transform} (AAFT)
method \citep{prichard1994generating} is used, explained in Algorithm \ref{algo:aaft}.
Given a real univariate time-series, the AAFT algorithm operates in Fourier
(i.e., frequency) domain, where it preserves the amplitude spectrum of the series,
but randomizes the phase, leading to a new realized signal.

AAFT can be explained by the \textbf{Wiener–Khinchin–Einstein Theorem} \citep{cohen1998generalization}, which
states that the autocorrelation function of a wide-sense-stationary random process
has a spectral decomposition given by the power spectrum of that process.
In other words, first and second order statistical moments (i.e., due
to autocorrelation) of the signal are encoded in its power spectrum,
which is purely dependent on the amplitude spectrum. Consequently,
the randomization of the phase does not impact the first and
second order moments of the series, hence the surrogates share
statistical properties of the original signal.

Since the original time-series (i.e., asset returns) are real-valued signals, their
Fourier Transform after randomization of the phase should preserve \textit{conjugate
symmetry}, or equivalently, the randomly generated phase component should be
an odd function of frequency. Then the Inverse Fourier Transform
(IFT) returns real-valued surrogates. 

\begin{algorithm}
	\SetKwInOut{Args}{inputs}
    \SetKwInOut{Output}{output}
    \DontPrintSemicolon
    \Args{%
        $M$-variate original time-series $\mvX$
    }
    \Output{$M$-variate synthetic time-series $\hat{\mvX}$}
    \vspace{0.25cm}
    \SetAlgoLined\SetArgSty{}
    \For{$i=1, 2, \ldots M$}{
        calculate Fourier Transform of\;
            \Indp univariate series $\mathfrak{F}[\mvX_{:i}]$ \\ \Indm
        randomize phase component \tcp*{preserve odd symmetry of phase}
        calculate Inverse Fourier Transform of\;
            \Indp unchanged amplitude and randomized phase $\hat{\mvX}_{:i}$
    }
\caption{Amplitude Adjusted Fourier Transform (AAFT).}
\label{algo:aaft}
\end{algorithm}

\begin{remark}
Importantly, the AAFT algorithm works on univariate series, therefore the
first two statistical moments of the \textit{single asset} are preserved
but the cross-asset dependencies (i.e., cross-correlation, covariance) are modified due to
the data augmentation.
\end{remark}
\subsection{Results}

\begin{figure}[h]
    \centering
    \begin{subfigure}[t]{0.48\textwidth}
        \includegraphics[width=\textwidth]{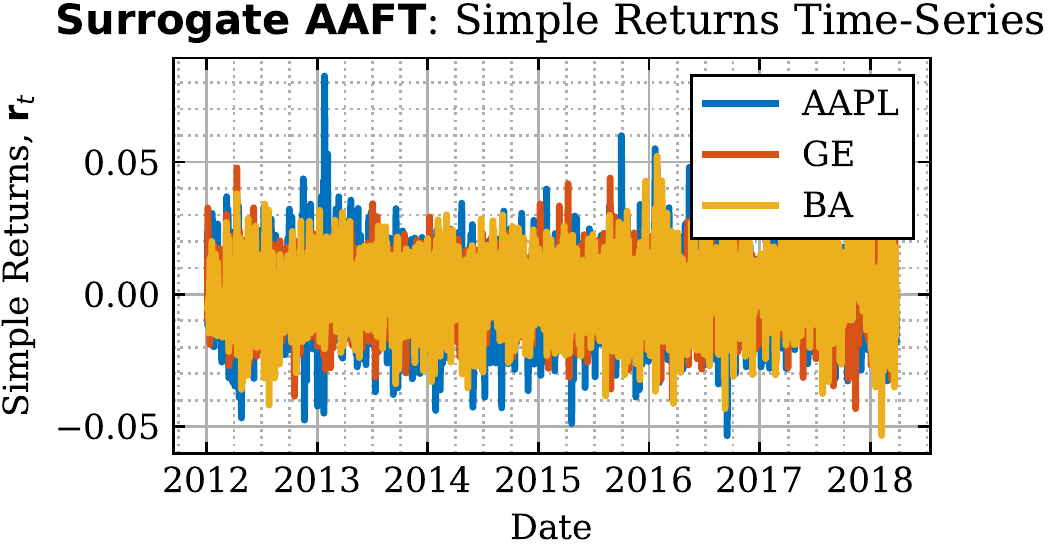}
    \end{subfigure}
    ~ 
    \begin{subfigure}[t]{0.48\textwidth}
        \includegraphics[width=\textwidth]{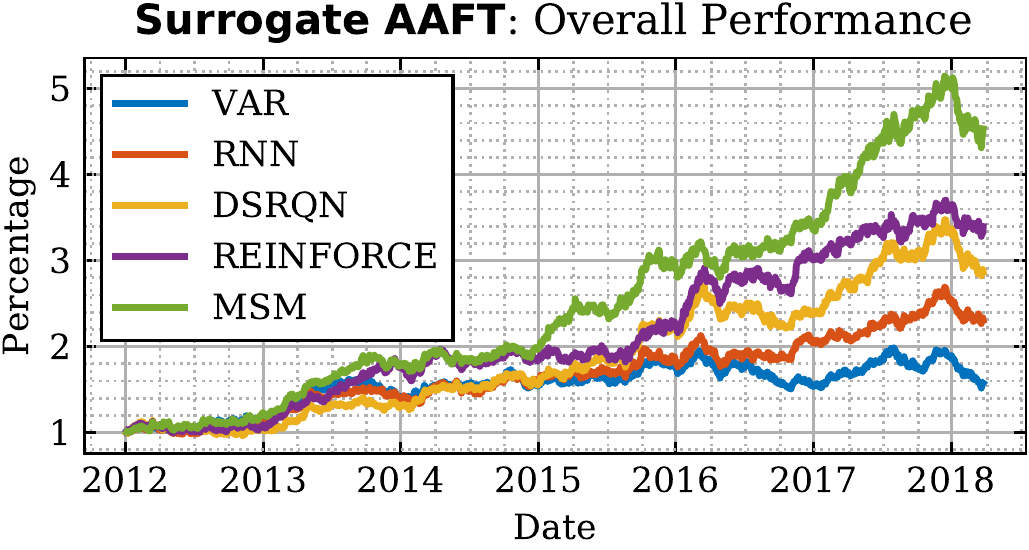}
    \end{subfigure}
    \caption{Synthetic, simulated universe of $12$-assets from S\&P500 via Amplitude Adjusted
    Fourier Transform (AAFT). (\textit{Left}) Example series from universe. (\textit{Right}) Cumulative
    returns of reinforcement learning trading agents.}
    \label{fig:datasets:aaft}
\end{figure}

Operating on the same $12$-assets universe used in experiments of
Chapter \ref{ch:trading-agents}, examples of AAFT surrogates are given in Figure \ref{fig:datasets:aaft}, along
with the cumulative returns of each trading agent on this
simulated universe. As expected, the model-free agents
outperform the model-based agents, corroborating the results obtained in Chapter \ref{ch:trading-agents}.

    % \item \textbf{Real Market} data, in Section \ref{sec:market-data}, selected such that sufficient liquidity
    %     and zero slippage Assumptions \ref{ass:sufficient-liquidity}, \ref{ass:zero-slippage} are satisfied
    
%% assets folder to reference
\renewcommand{\assets}{report/experiments/market-data/assets}

\chapter{Market Data} \label{ch:market-data}

Having verified the applicability of the trading agents in synthetic environments
(i.e., deterministic and stochastic) in Section \ref{ch:synthetic-data}, their effectiveness is challenged
in real financial markets, namely the underlying stocks of the Standard
\& Poor's 500 \citep{investopedia:sp500} and the EURO STOXX 50 \citep{investopedia:eurostoxx50} indices.
In detail, in this chapter:

\begin{itemize}
    \item Candidate reward generating functions are explored, in Section \ref{sec:reward-generating-functions};
    \item Paper trading experiments are carried out on U.S. and European most liquid assets
    (see Sufficient Liquidity Assumption \ref{ass:sufficient-liquidity}), as in Sections \ref{sec:s&p500} and \ref{sec:euro-stoxx-50},
    respectively;
    \item Comparison matrices and insights into the learned agent strategies are obtained.
\end{itemize}

\section{Reward Generating Functions} \label{sec:reward-generating-functions}

Reinforcement learning relies fundamentally on the hypothesis that the goal
of the agent can be fully described by the maximization
of the cumulative reward over time, as suggested by the
Reward Hypothesis \ref{hyp:reward}. Consequently, the selection of the reward generating
function can significantly affect the learned strategies and hence the performance of the agents.
Motivated by the returns-risk trade-off arising in investments (see Chapter \ref{ch:portfolio-optimization}),
two reward functions are implemented and tested: the log returns and
the Differential Sharpe Ratio \citep{paper:rl-trading-portfolios}.

\subsection{Log Rewards} \label{sub:log-rewards}

The agent at time step $t$ observes asset prices $\vo_{t} \equiv \vp_{t}$
and computes the log returns, given by:

\begin{equation}
    \vrho_{t} = log(\vp_{t} \oslash {\vp_{t-1}})
\tag{\ref{def:log-returns}}
\end{equation}

where $\oslash$ designates element-wise division and the $log$ function is also applied element-wise, or equivalently:

\begin{equation}
    \vrho_{t} \triangleq
    \renewcommand\arraystretch{1.5}
    \begin{bmatrix}
        \rho_{1, t} \\
        \rho_{2, t} \\
        \vdots \\
        \rho_{M, t}
    \end{bmatrix}
    =
    \renewcommand\arraystretch{1.5}
    \begin{bmatrix}
        log(\frac{p_{1, t}}{p_{1, t-1}}) \\
        log(\frac{p_{2, t}}{p_{1, t-1}}) \\
        \vdots \\
        log(\frac{p_{M, t}}{p_{M, t-1}})
    \end{bmatrix} \in \sR^{M} 
\end{equation}

Using one-step log returns for reward, results in the multi-step maximization
of \textbf{cumulative log returns}, which focuses only on the profit,
without considering any risk (i.e., variance) metric. Therefore, agents are expected
to be highly volatile when trained with this reward function.

\subsection{Differential Sharpe Ratio} \label{sub:differential-sharpe-ratio}

In Section \ref{sec:evaluation-criteria}, the Sharpe Ratio \citep{finance:sharpe} was introduced, motivated by Signal-to-Noise
Ratio (SNR), given by:

\begin{equation}
    \textbf{SR}_{t} \triangleq \sqrt{t} \frac{\E[\vr_{t}]}{\sqrt{\Var[\vr_{t}]}} \in \sR
\tag{\ref{def:sharpe-ratio}}
\end{equation}

where $T$ is the number of samples considered in the calculation
of the empirical mean and standard deviation. Therefore, empirical
estimates of the mean and the variance of the portfolio
are used in the calculation, making Sharpe Ratio an inappropriate
metric for online (i.e., adaptive) episodic learning. Nonetheless, the \textbf{Differential
Sharpe Ratio} (DSR), introduced by \citet{paper:rl-trading-portfolios}, is a suitable reward function.
DSR is obtained by:
\begin{enumerate}
    \item Considering exponential moving averages of the
    returns and standard deviation of returns in \ref{def:sharpe-ratio};
    \item Expanding to first order in the decay rate:
    \begin{equation}
        \textbf{SR}_{t} \approx \textbf{SR}_{t-1} +
        \eta \left. \frac{\partial \textbf{SR}_{t}}{\partial \eta}\right\vert_{\eta = 0} + O(\eta^{2})
    \label{def:sharpe-ratio-first-order-taylor}
    \end{equation}
\end{enumerate}

Noting that only the first order term in expansion (\ref{def:sharpe-ratio-first-order-taylor})
depends upon the return, $r_{t}$, at time step, $t$, the differential Sharpe Ratio, $\textbf{D}_{t}$, is defined as:

\begin{equation}
    \textbf{D}_{t} \triangleq \frac{\partial \textbf{SR}_{t}}{\partial \eta}
    = \frac{B_{t-1} \Delta A_{t} + \frac{1}{2} A_{t-1} \Delta B_{t}}{(B_{t-1} - A_{t-1}^{2} )^{\frac{3}{2}}}
\label{def:differential-sharpe-ratio}
\end{equation}

where $A_{t}$ and $B_{t}$ are exponential moving estimates of the first and second
moments of $r_{t}$, respectively, given by:

\begin{align}
    A_{t} &= A_{t-1} + \eta \Delta A_{t} = A_{t-1} + \eta (r_{t} - A_{t-1}) \\
    B_{t} &= B_{t-1} + \eta \Delta B_{t} = B_{t-1} + \eta (r_{t}^{2} - B_{t-1})
\end{align}

Using differential Sharpe Ratio for reward, results in the multi-step maximization
of \textbf{Sharpe Ratio}, which balances risk and profit, and hence
it is expected to lead to better strategies, compared
to log returns.

\section{Standard \& Poor's 500} \label{sec:s&p500}

\subsection{Market Value} \label{sub:market-value}

Publicly traded companies are usually also compared in terms of their
\textbf{Market Value} or \textbf{Market Capitalization} (Market Cap), given by multiplying the
number of their outstanding shares by the current share price \citep{investopedia:market-value}, or equivalently:

\begin{equation}
    \textbf{Market Cap}_{\text{asset i}} = \textbf{Volume}_{\text{asset i}} \times \textbf{Share Price}_{\text{asset i}}
\label{def:market-cap}
\end{equation}

The Standard \& Poor's 500 Index (S\&P 500) is a market capitalization
weighted index of the 500 largest U.S. publicly traded companies by
market value \citep{investopedia:market-value}. According to the Capital Asset Pricing Model (CAPM) \citep{finance:investment-science}
and the Efficient Market Hypothesis (EMH) \citep{fama1970efficient}, the market index, S\&P 500,
is efficient and portfolio derived by its constituent assets \textit{cannot}
perform better (as in the context of Section \ref{sub:quadratic-programming}). Nonetheless, CAPM
and EMH are not exactly satisfied and trading opportunities can be exploited via proper strategies.

\subsection{Evaluation}

\begin{figure}[h]
    \centering
    \begin{subfigure}[t]{0.48\textwidth}
        \includegraphics[width=\textwidth]{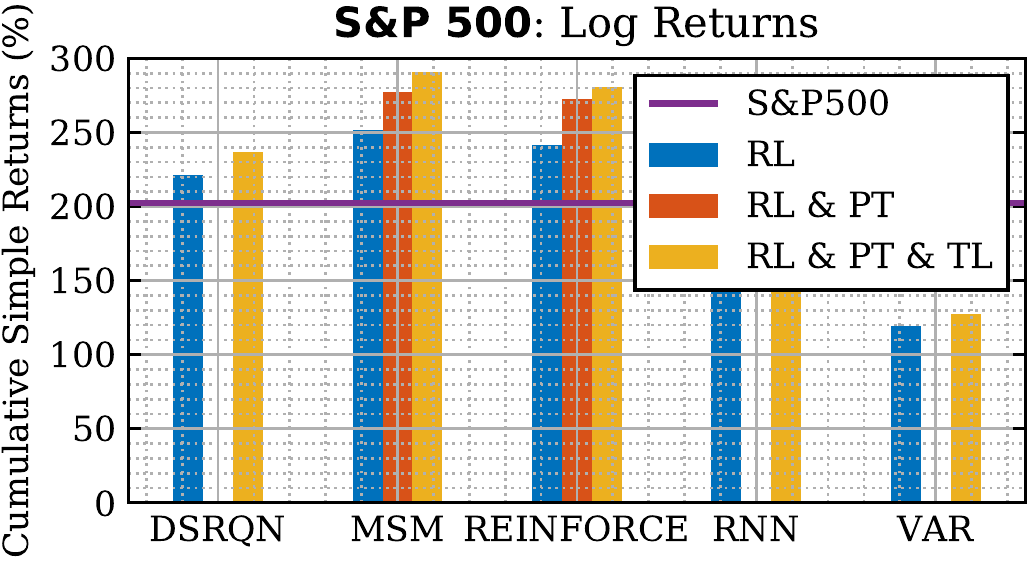}
    \end{subfigure}
    ~ 
    \begin{subfigure}[t]{0.48\textwidth}
        \includegraphics[width=\textwidth]{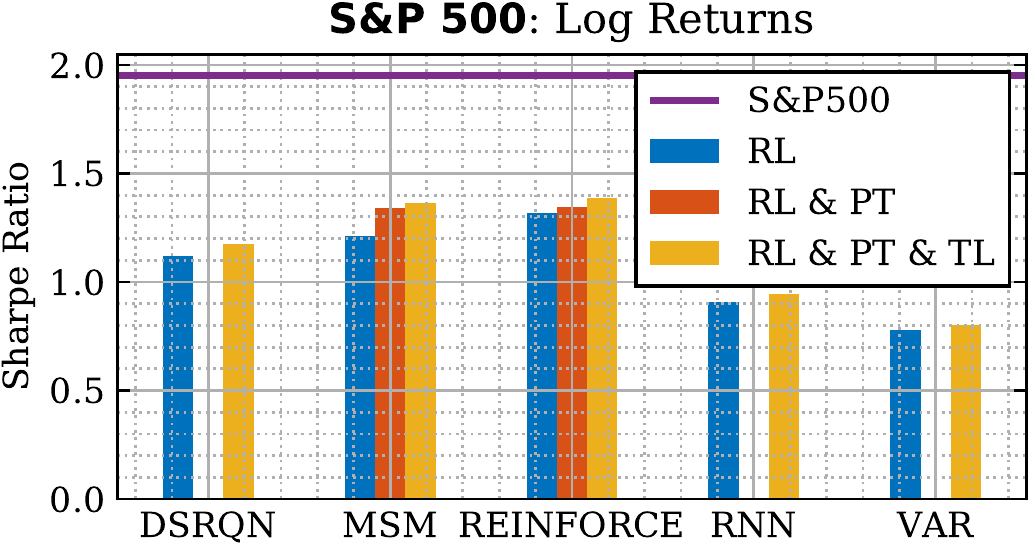}
    \end{subfigure}
    
    \vspace{0.5cm}
    
    \begin{subfigure}[t]{0.48\textwidth}
        \includegraphics[width=\textwidth]{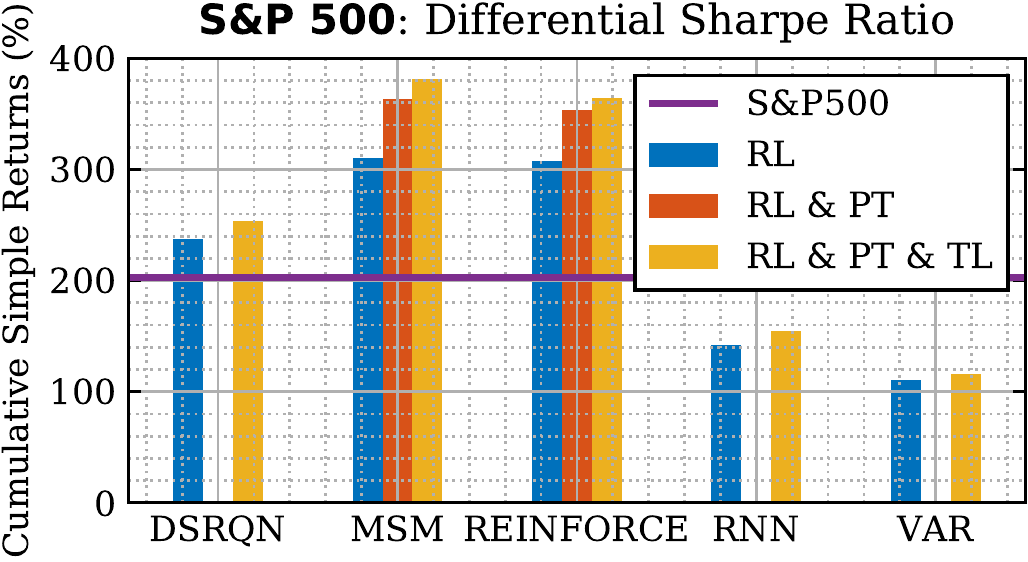}
    \end{subfigure}
    ~    
    \begin{subfigure}[t]{0.48\textwidth}
        \includegraphics[width=\textwidth]{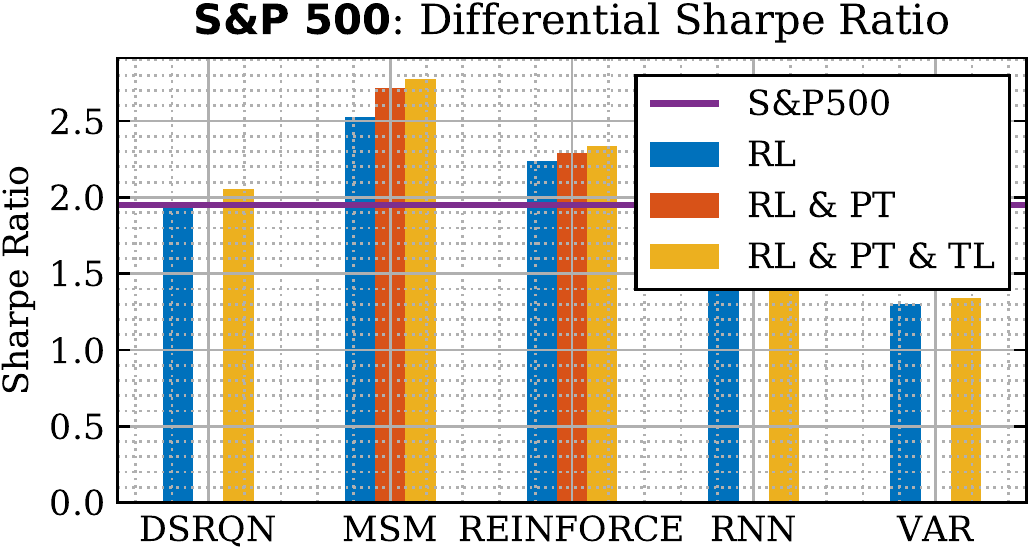}
    \end{subfigure}
    \caption[a b c]{Comparison of reinforcement learning trading agents on cumulative returns and
    Sharpe Ratio, trained with:
    (\textit{RL}) Reinforcement Learning;
    (\textit{RL \& PT}) Reinforcement Learning and Pre-Training;
    (\textit{RL \& PT \& TL}) Reinforcement Learning, Pre-Training and Transfer Learning from simulated data.}
    \label{fig:s&p500}
\end{figure}

In order to compare the different trading agents introduced in Chapter \ref{ch:trading-agents}, as well
as variants in Chapter \ref{ch:pre-training} and Section \ref{sec:simulated-data}, all agents are trained
on the constituents of S\&P 500 (i.e., $500$ U.S. assets) and the
results of their performance are provided in Figure \ref{fig:s&p500} and Table \ref{tab:s&p500}.
As expected, the differential Sharpe Ration (DSR) is more stable than
log returns, yielding higher Sharpe Ratio strategies, up to $2.77$ for
the pre-trained and experience transferred Mixtutre of Score Machines (MSM) agent.

\begin{table}
\begin{center}
\begin{tabular}{ |l||c|c||c|c| }
	\hline
    \multicolumn{5}{|c|}{\textbf{Trading Agents Comparison Matrix: S\&P 500}} \\
    \hline
    \hline
    \textbf{Reward} & \multicolumn{2}{c||}{\textbf{Differential}} & \multicolumn{2}{c|}{\textbf{Log}} \\ 
	\textbf{Generating Function} & \multicolumn{2}{c||}{\textbf{Sharpe Ratio}} & \multicolumn{2}{c|}{\textbf{Returns}} \\
	\hline
    \hline
    & Cumulative & Sharpe & Cumulative & Sharpe \\
    & Returns (\%) & Ratio & Returns (\%) & Ratio \\
    \hline
    \hline
    % VAR & \textbf{VAR} & \textbf{RNN} & \textbf{DSRQN} & \textbf{REINFORCE} & \textbf{MSM} & \texttt{SPY} \\ 
    \textbf{SPY} & \cellcolor{gray!25}202.4 & \cellcolor{gray!25}1.95 & \cellcolor{gray!25}202.4 & \cellcolor{gray!25}1.95 \\
    \hline
    \textbf{VAR} & \cred 110.7 & \cred 1.30 & \cred 119.3 & \cred 0.78 \\
    \hline
    \textbf{RNN} & \cred 142.3 & \cred 1.49 & \cred 146.2 & \cred 0.91 \\
    \hline
    \textbf{DSRQN} & \cyellow 237.1 & \cred 1.96 & \cyellow 221.5 & \cred 1.12 \\
    \hline
    \textbf{REINFORCE} & \cgreen 307.5 & \cgreen 2.24 & \cyellow 241.3 & \cred 1.32 \\
    \hline
    \textbf{MSM} & \cgreen 310.8 & \cgreen 2.53 & \cyellow 251.9 & \cred 1.21 \\
    \hline
    \hline
    \textbf{REINFORCE \& PT} & \cgreen 353.6 & \cgreen 2.29 & \cgreen 272.7 & \cred 1.33 \\
    \hline
    \textbf{MSM \& PT} & \cgreen 363.6 & \cgreen 2.72 & \cgreen 277.1 & \cred 1.34 \\
    \hline
    \hline
    \textbf{REINFORCE \& PT \& TL} & \cgreen 364.2 & \cgreen 2.33 & \cgreen 280.9 & \cred 1.38 \\
    \hline
    \textbf{MSM \& PT \& TL} & \cgreen $\mathbf{381.7}$ & \cgreen $\mathbf{2.77}$ & \cgreen 291.0 & \cred 1.36 \\
    \hline
\end{tabular}
\end{center}
\caption{Comprehensive comparison of evaluation metrics of reinforcement learning trading algorithms
and their variants, namely pre-training (PL) and transfer learning (TL).}
\label{tab:s&p500}
\end{table}

\begin{remark}
The simulations confirm the superiority of the universal model-free
reinforcement learning agents, Mixture of Score Machines (MSM), in asset allocation, with the achieved performance
gain of as much as $9.2\%$ in cumulative returns and $13.4\%$
in Sharpe Ratio, compared to the most recent models in \citep{paper:drl-pm} in the same universe.
\end{remark}

\section{EURO STOXX 50} \label{sec:euro-stoxx-50}

Similar to S\&P 500, the EURO STOXX 50 (SX5E) is a benchmark for the
50 largest publicly traded companies by market value in countries
of Eurozone. In this section, the universality of 
the Mixture of Score Machines (MSM) agent is assessed against the Markowitz model (see Section \ref{sec:markowitz-model}).

\subsection{Sequential Markowitz Model} \label{sec:sequential-markowitz-model}

A universal baseline agent is developed, based on the Sharpe Ratio
with transaction costs (see optimization problem \ref{opt:sharpe-ratio-transaction-costs})
extension of the Markowitz model, from Section \ref{sec:transaction-cost-optimization}.
Therefore, a \textbf{Sequential Markowitz Model} (SMM) agent is derived by iteratively
applying the one-step optimization program solver for each time step $t$.
The Markowitz model is obviously a universal portfolio optimizer, since
it does not make assumptions about the universe (i.e., underlying assets) it is applied upon.

\subsection{Results}

\begin{figure}[h]
    \centering
    \includegraphics[width=\textwidth]{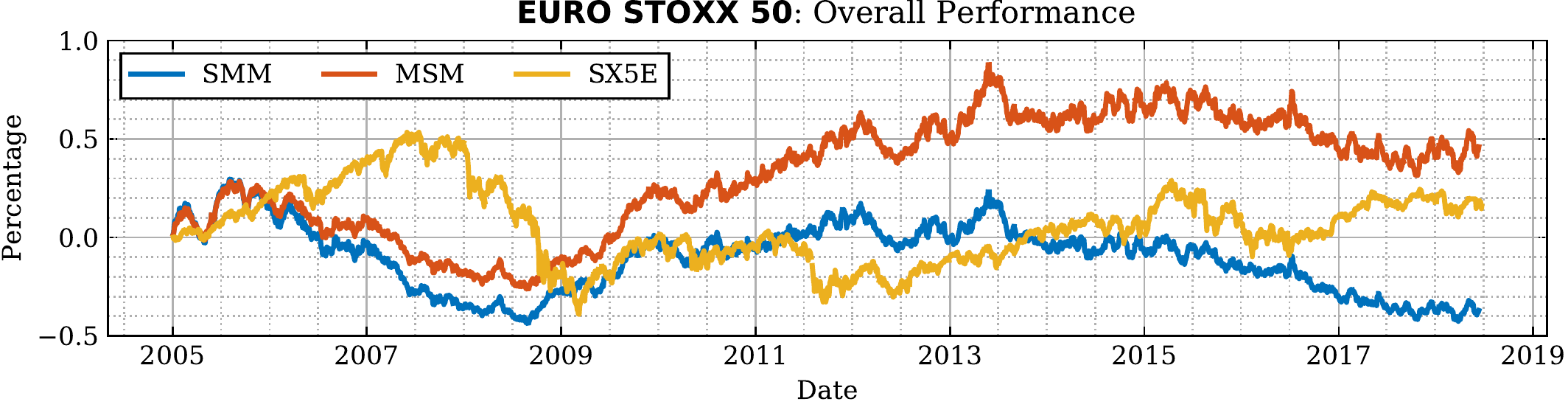}
    \caption{Cumulative Returns of Mixture of Score Machines (MSM) agent, trained on S\&P 500 market
    and transferred experience to EURO STOXX 50 market (SX5E), along with
    the traditional Sequential Markowitz Model (SMM).}
    \label{fig:euro-stoxx-50}
\end{figure}

Given the EURO STOXX 50 market, \textit{transfer learning} is performed
for the \textit{MSM agent trained on the S\&P 500} (i.e., only the
Mixture network is replaced and trained, while the parameters of the
Score Machine networks are frozen).
Figure \ref{fig:euro-stoxx-50} illustrates the cumulative returns of the market index (SX5E),
the Sequential Markowitz Model (SMM) agent and the Mixture of Score Machines (MSM) agent.

\begin{remark}
As desired, the MSM agent outperformed both the market index (SX5E) and the SMM agent,
reflecting the universality of the MSM learned strategies, which are
both successful in the S\&P 500 and EURO STOXX 50 markets.
\end{remark}

It is worth also noting that the cumulative returns of the MSM
and the SMM agents were correlated, however, the MSM performed
better, especially after 2009, when the SMM followed the declining
market and the MSM became profitable. This fact can be
attributed to the pre-training stage of the MSM agent, since
during this stage, the policy gradient network converges to the Markowitz
model, or effectively mimics the SMM strategies. Then, the reinforcement
episodic training allows the MSM to improve itself so that it
outperforms its initial strategy, the SMM.

\chapter{Conclusion} \label{ch:conclusion}

The main objective of this report was to investigate the
effectiveness of Reinforcement Learning agents on Sequential Portfolio Management. 
To achieve this, many concepts from the fields of Signal Processing, Control Theory,
Machine Intelligence and Finance have been explored, extended and combined.
In this chapter, the contributions and achievements of the project
are summarized, along with possible axis for future research.

% In this chapter, we take the time to revisit what has been
% achieved in this thesis and to sum up our original
% contributions to the reinforcement learning literature. Moreover, we draw some
% conclusions about the impact that these modern optimization techniques can have
% on the financial industry and the challenges that must be overcome
% to make it possible. Finally, we suggest some possible axis for future research.

% Many concepts from the fields of finance, computational geometry, and control engineering
% have been explored, extended, and tested in this project. The main objective of the work was
% to establish methods allowing an investor to optimally allocate capital to a range of assets from
% knowledge of their historical returns. In this chapter the developments made in the project will
% be summarised, along with all of the procedures, relaxations, and extensions that were used
% and presented. Additionally, a critical evaluation of the work conducted during the project will
% be included, noting the achievements and challenges that were faced during the execution of
% the work. Finally, due to the temporal constraints and scope of the project, there were a number
% of extensions and additional tests that were not conducted during the lifespan of this project -
% a brief outline of the proposed additions is presented in the Future Works section.

%
\section{Contributions} \label{sec:contributions}

To enable episodic reinforcement learning, a mathematical formulation of financial markets
as discrete-time stochastic dynamical systems is provided, giving rise to
a unified, versatile framework for training agents and investment strategies.

A comprehensive account of reinforcement agents has been developed, including traditional,
baseline agents from system identification (i.e., model-based methods) as well as context
agnostic agents (i.e., model-free methods). A universal model-free reinforcement learning family
of agents has been introduced, which was able to reduce
the model computational and memory complexity (i.e., linear scaling with universe
size) and to generalize strategies across assets and markets, regardless of
the training universe. It also outperformed all trading agents, found
in the open literature, in the S\&P 500 and the EURO STOXX 50 markets.

Lastly, model pre-training, data augmentation and simulations enabled robust training of
deep neural network architectures, even with a limited number of available real market data.

\section{Future Work} \label{sec:future-work}

Despite the performance gain of the developed strategies, the lack of
interpretability \citep{rico1994continuous} and the inability to exhaustively test the deep architectures used
(i.e., Deep Neural Networks) discourage practitioner from adopting these solutions.
As a consequence, it is worth investigating and interpreting the learned
strategies by opening the deep "black box" and being able to reason for its decisions.

In addition, exploiting the large number of degrees of freedom given by the
framework formulation of financial markets, further research on reward generating
functions and state representation could improve the convergence properties
and overall performance of the agents. A valid approach would
be to construct the indicators typically used in the technical
analysis of financial instruments \citep{king1992financial}. These measures would embed the expert knowledge
acquired by financial analysts over decades of activity and could help
in guiding the agent towards better decisions \citep{finance:quant-finance}.

Furthermore, the flexible architecture of Mixtures of Score
Machines (MSM) agent, the best-scoring universal trading agents,
allows experimentation with both the Score Machines (SM) networks and their
model-order, as well as the Mixture network, which, ideally, should
be universally used without transfer learning.

The trading agents in this report are based on point
estimates, provided by a deep neural network. However, due to the uncertainty
of the financial signals, it would be appropriate to also
model this uncertainty, incorporating it in the decision making process.
Bayesian Inference, or more tractable variants of it, including Variation Inference \citep{titsias2010bayesian},
can be used to train probabilistic models, capable of capturing the environment uncertainty \citep{vlassis2012bayesian}.

Last but not least, motivated by the recent\footnote{At the
time that report is submitted this paper has not been
presented, but it is accepted in International Conference on Machine
Learning (ICML) 2018.} publication by \citep{fourier2018icml} on exact calculation of the
policy gradient by operating on the Fourier domain, employing an
exact policy gradient method could eliminate the estimate variance and
accelerate training.

\appendix

\backmatter

\printbibliography

\end{document}